\documentclass[10pt,a4paper,twoside,openright]{memoir}
\usepackage[T1]{fontenc}
\usepackage[utf8]{inputenc}
\usepackage[english]{babel}
\chapterstyle{madsen}
\usepackage{geometry}
\usepackage{amsmath}
\usepackage{amsfonts}
\usepackage{amssymb}
\usepackage{graphicx}
\usepackage[table]{xcolor}
\usepackage{subcaption}
\usepackage{pdfpages}
\usepackage{booktabs}
\usepackage{multirow}
\usepackage{booktabs} 
\usepackage{color}
\usepackage{url}
\usepackage{multirow}
\usepackage{siunitx}
\usepackage{array}
\usepackage{subcaption}
\usepackage{float}
\usepackage{rotating}
\usepackage{adjustbox}
\usepackage{algorithmic}
\usepackage{algorithm}
\usepackage{todonotes}
\usepackage{etoolbox}
\usepackage{wrapfig}
\robustify\bfseries
\usepackage{colortbl}
\newfloat{algorithm}{t}{lop}
\usepackage{tabularx}
\usepackage{microtype}

\newcolumntype{C}[1]{>{\centering\arraybackslash}p{#1}}

\author{
	\Large{Niels Dalum Hansen} \\
	\texttt{zhv641@ku.dk}
	\vspace{5cm} \\
	\large{Principal Advisor: {\em Christina Lioma}}\\
	\large{Advisor: {\em Ingemar Johansson Cox}}\\
	\large{Advisor: {\em Kåre Mølbak}}\\
	\large{Company advisor: {\em Peter Bredsdorff Lange}}
}

\title{\Large{PhD Thesis\vspace{1cm}} \\ \LARGE{\textbf{Web data mining for public health purposes}}
}

\date{\vspace{0.5cm}13. December 2017\\\vspace{1cm}This thesis has been submitted to\\The PhD School of The Faculty of Science, University of Copenhagen}

\def \ColourPDF {images/science-logo.pdf}
\def \TitlePDF {images/science-english.pdf}

\begin{document}

\frontmatter

\AddToShipoutPicture*{\put(0,0){\includegraphics*[viewport=0 0 700 600]{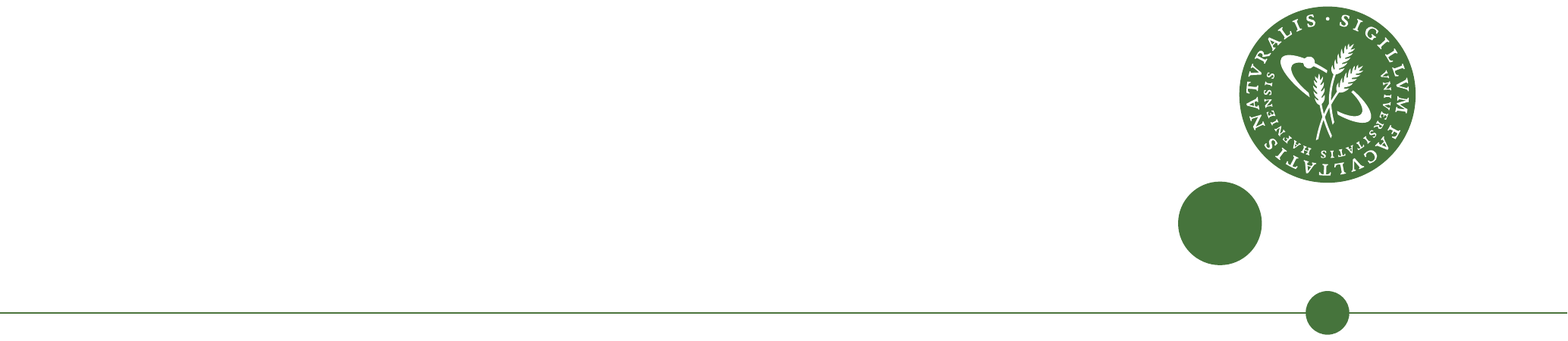}}}
\AddToShipoutPicture*{\put(0,602){\includegraphics*[viewport=0 600 700 1600]{\ColourPDF}}}
\AddToShipoutPicture*{\put(0,0){\includegraphics*{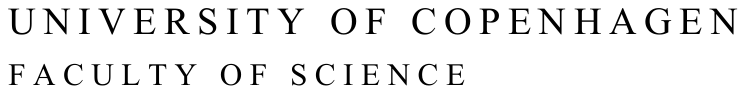}}}

\maketitle
\newpage

\tableofcontents
\newpage

\chapter{Abstract}

For a long time, public health events, such as disease incidence or vaccination activity, have been monitored to keep track of the health status of the population, allowing to evaluate the effect of public health initiatives and to decide where resources for improving public health are best spent. This thesis investigates the use of web data mining for public health monitoring, and makes contributions in the following two areas:

(I) New approaches for predicting public health events from web mined data. These include: (i) An online learning method that can automatically adapt to sudden temporal changes in the underlying signal. Health events often show temporal stability through many years and historical data is therefore often a good predictor, but in the case of sudden changes, this assumption no longer holds. Our online learning method aims at addressing this problem by automatically adjusting to temporal changes. (ii) Prediction models factoring event seasonality. We show how the expected seasonal variation can be used to optimize the usage of web mined data. (iii) Novel web data mining strategies that make it possible to target different population groups and reduce spurious correlations.

(II) Novel applications of web mined data. These cover: (i) Prediction using web mined data of health events, such as preventive measures and drug consumption. Prior research has primarily focused on prediction of contagious diseases, but public health institutions are also responsible for monitoring several other types of health events. Our extensions to preventive measures and drug consumption show that the potential of web mined data is far from fully utilized. (ii) Understanding the relationship between news media and vaccination uptake.  With the constant availability of news, both online and in print, understanding the effect of news media on public health events is important for designing accurate health monitoring systems. Increased understanding can, in addition, be useful in a variety of public health tasks, e.g.\ designing outreach campaigns. 

\chapter{Acknowledgments}

Many things have happened during the course of my PhD and I would not have been able to make it without help from family, friends, and colleagues. I will start by thanking my academic advisors Christina, Ingemar, and Kåre for teaching me how to be a researcher and for a fruitful collaboration. Especially, I would like to thank Christina, she was the one talking me into doing a PhD and she has made sure that I made it all the way through. From IBM I thank Peter for his great support and our interesting talks on our Monday meetings. 

Doing a PhD is not all work, and my colleagues at the image section at DIKU have been helpful in making sure that being at work, not necessarily always meant working. I have met so many nice and interesting people, and I think that I might end up with a complete list of employees if I start naming them all. The running club deserves special praise for making sure I did not get too fat during my studies. My office mates Jan, Oswin, Leise, and Kristoffer deserve special thanks, sharing an office with you has been great fun. The final two people from DIKU that I will mention are Casper and Brian, I have really enjoyed the time we have spent together, especially eating steaks and watching basketball. 

I have not only spent time at DIKU but also at the department of infectious disease epidemiology at SSI. Time spent at SSI has always been fun due to the wonderful people working there. A special thanks is for Camilla HS who I have not only enjoyed working with but also coffee drinking and interesting conversations.  

My parents should not go unmentioned, they have supported me throughout everything and have helped pick up my daughter from kindergarten and entertained her on weekends so that I had time to finish the last articles and the thesis. Finally, I would like to thank my daughter, for being herself and making everything worthwhile. 



\chapter{Preface}

During my time as a PhD student, I have sometimes wondered what direction I was actually heading. Many of the initial plans and ideas turned out to be dead ends and weeks of work often led to seemingly nothing but wasted time. Despite of this, when going through the work I have done a consistent pattern became clear. This thesis will hopefully make this clear to you too, as it has become to me. 

Before going into the actual substance of the thesis I will describe the setup in which I have written this thesis. This PhD is a three year Industrial PhD, meaning that it is a project funded in part by a private company, in this case, IBM Denmark, and the Innovation Fund Denmark.  From a scientific point of view, the project has been a collaboration between IBM Denmark, the Department of Computer Science at the University of Copenhagen (DIKU), and Statens Serum Institut (SSI), i.e.\ the Danish center for disease control and prevention. IBM Denmark has been contributing with methodological knowledge, while SSI has been supplying the epidemiological domain knowledge and problems of epidemiological relevance. The role of DIKU has been to assess the scientific relevance of the problems from a computer science perspective, to define the problem formulations and to contribute with the expert knowledge necessary for developing state-of-the-art solutions to the problems. SSI has been an important partner in this project, because of their access to and knowledge about the Danish health registries. This data has been essential for the studies included in this thesis. These registries make Denmark a well-suited country for epidemiological studies because the nationwide registries cover numerous health events ranging from visits to general practitioners and hospitals, drug sales, vaccinations and many others. 

The initial focus of the PhD was to use methods from computer science, and specifically web data mining, for disease surveillance. Even though disease surveillance was the focus, it became clear to me\footnote{For a public health professional this would probably have been clear from the start.} that for a public health institute to succeed in their task of disease control and prevention, they also have to keep track of many other types of events, such as monitoring compliance with drug use and preventive measures, such as vaccination uptake. This became very clear to me, because during my PhD we experienced a big, if not the biggest, crisis in the Danish vaccination program, namely the fear of adverse reactions to the HPV vaccine. The people at SSI who are responsible for the vaccination program spent a lot of energy on monitoring how the vaccination uptake developed from month to month, and analyzing how it changed with outreach campaigns and media coverage. Illustrating, that without timely and accurate monitoring data their work would have been performed in the dark. This showed me how accuracy and timeliness is not only important for surveillance of contagious diseases but also for other types of health events. This raises numerous questions: What types of health events can we actually monitor using web mined data? How can web mined data and traditional registry data be combined? How accurately can these predictions be performed? And to what extent do changes in web data reflect changes in the health events? These are some of the questions that this thesis addresses.

The thesis is written as a compilation of research articles. This means that the thesis starts with an introductory chapter followed by a chapter for each of the seven included papers.

In addition to publishing papers on web mined data for public health purposes, I have also published papers on topics ranging from semantic non-compositionally detection in text to epidemiological studies. All of these papers could not fit into one coherent thesis, and some of the papers have therefore not been included. 
Below is a list of all scientific publications I have either authored or co-authored in the last four years. Papers in bold are included in this thesis:

\begin{itemize}
	\item[\cite{hansen2014temporal}] Niels Dalum Hansen, Christina Lioma, Birger Larsen, Stephen Alstrup. ``Temporal Context for Authorship Attribution''. In \textit{Proceedings of the Information Retrieval Facility Conference}, 2014, p. 22-40.
	\item[\cite{Lioma2015}] Christina Lioma, Jakob Grue Simonsen, Birger Larsen, Niels Dalum Hansen. ``Non-compositional term dependence for information retrieval''. In \textit{Proceedings of the 38th International ACM SIGIR Conference on Research and Development in Information Retrieval}, 2015, p. 595-604.
	\item[\cite{molbak2016pre}] \textbf{Kåre Mølbak, Niels Dalum Hansen, Palle Valentiner-Branth. ``Pre-vaccination care-seeking in females reporting severe adverse reactions to HPV vaccine. A registry based case-control study''. In \textit{PLoS One}, 2016. }   
	\item[\cite{DalumHansen:2016:ELV:2983323.2983882}]\textbf{Niels Dalum Hansen, Christina Lioma, Kåre Mølbak. ``Ensemble learned vaccination uptake prediction using web search queries''. In \textit{Proceedings of the 25th ACM International Conference on Information and Knowledge Management}, 2016, p. 1953-1956.}
	\item[\cite{www17}] \textbf{Niels Dalum Hansen, Kåre Mølbak, Ingemar Johansson Cox, Christina Lioma. ``Time-Series Adaptive Estimation of Vaccination Uptake Using Web Search Queries''. In \textit{Proceedings of the 26th International Conference on World Wide Web}, 2017, p. 773-774.}
	\item[\cite{dalum2017seasonal}] \textbf{Niels Dalum Hansen, Kåre Mølbak, Ingemar Johansson Cox, Christina Lioma. ``Seasonal Web Search Query Selection for Influenza-Like Illness (ILI) Estimation''. In \textit{Proceedings of the 40th International ACM SIGIR Conference on Research and Development in Information Retrieval}, 2017, p. 1197-1200.}
	\item[\cite{Lioma2017}] Christina Lioma, Niels Dalum Hansen. ``A study of metrics of distance and correlation between ranked lists for compositionality detection''. In \textit{Cognitive Systems Research}, 2017, v. 44, p. 40-49.    
\end{itemize}

The following list of unpublished, but submitted papers are also included in the thesis.

\begin{itemize}
	\item[Ch. \ref{cha:antimicrobial}] \textbf{Niels Dalum Hansen, Kåre Mølbak, Ingemar Johansson Cox, Christina Lioma. ``Predicting antimicrobial drug consumption using web search data''.}
	\item[Ch. \ref{cha:MMR_Media}] \textbf{Niels Dalum Hansen, Kåre Mølbak, Ingemar Johansson Cox, Christina Lioma, ``An investigation of the relationship between media coverage and vaccination uptake in Denmark''.}
	\item[Ch. \ref{cha:HPV_media}] \textbf{Camilla Hiul Suppli, Niels Dalum Hansen, Mette Rasmussen, Palle Valentiner-Branth, Tyra Grove Krause, Kåre Mølbak. ``Decline in HPV-vaccination uptake in Denmark - The association between HPV-related media coverage and HPV-vaccination''.}
\end{itemize}

\mainmatter

\addtocontents{toc}{\protect\setcounter{tocdepth}{3}}
\setcounter{secnumdepth}{3}

\chapter{Introduction}

Web data mining for public health purposes is a broad topic. In this thesis, we focus on a few selected areas, such as how web mined data is used in relation to prediction of health events and the relationship between web mined data and health events. Our definition of public health events includes all health events where there is an interest in monitoring the event on a population scale.\footnote{Public health literature often uses the term \textit{event} for unstructured data and \textit{indicator} for structured data \cite{paquet2006epidemic}, in this thesis we use the term \textit{event} for both types.} This means that when we are talking about public health events we are referring to a wide range of event types ranging from diseases, preventive measures, to drug consumption. Similarly, our definition of web mined data is broad and includes both online user behavior, such as queries submitted to search engines, and information available online, such as online news media articles. 

We begin this thesis by stating  the objectives and contributions in Section \ref{sec:objectives_and_research_goals} which highlights the scientific contributions of the papers included in this thesis as Chapters 2--8. 
The majority of the included papers are on prediction or estimation of health events using web search query frequency data. To evaluate the usability of these new approaches it is useful to know the current state of public health monitoring systems. Section \ref{sec:monitoring_health_events} therefore contains an overview of three different types of health events and their corresponding surveillance systems. 
With an understanding of how health events are currently monitored, we are ready to discuss the topic of health event prediction using web search query frequencies in Section \ref{sec:prediction_with_query_frequency_data}.  Finally, we end this introductory chapter with a section on future work and perspectives.

\section{Objectives and contributions}\label{sec:objectives_and_research_goals}

Monitoring health events, such as disease incidence, preventive measures, or drug consumption,  is an essential part of a national public health program. Surveillance is necessary to assess the change in morbidity or mortality due interventions, or to detect changes in public health that call for governmental action. Timeliness, accuracy, and cost are important factors when designing a monitoring program. As reading news, seeking information and other everyday activities move online, we have an opportunity to rethink the problem of health event monitoring. Approaching the web as a data source that we can mine for information has been shown to be a powerful new tool. The web gives us access to information on user behavior through online search activity, to knowledge resources through online encyclopedias, to the news media that is shaping the public agenda, and much more.
The goals of this  thesis are: i) to present novel approaches to existing problems where web mined data is used for public health purposes, such as prediction of influenza-like illness (ILI); and ii) to address public health problems that have not previously been attempted solved using web mined data, such as prediction of vaccination uptake or antimicrobial drug consumption.

This thesis contains the following five research objectives and seven scientific contributions. 

\paragraph{What type of health events can be predicted using web mined data?} 
When using web mined data for predicting health events, the type of event has often been the incidence of a contagious disease. While this is of great interest, our \textbf{first objective} is to ask whether web mined data can also be used for prediction of other types of health events. Public health institutions are responsible for monitoring many types of events, such as drug consumption and health risk behavior. Some of these events make people seek information; What to do when faced with a specific disease symptom? What decision to make when asked about immunization of one's child? For some events the connection between information seeking and event is more indirect, for example, web searches indicating that a person has a specific disease could predict a further need for treatment of complications. If more health events can be predicted based on information seeking behavior,  it opens up for increased use of web mined data in surveillance programs. This has the possibility of leading to reduced cost and increased timeliness.
The \textbf{first contribution} of our work is to expand the types of health events that have been predicted using web mined data by successfully predicting both drug consumption in the form of antimicrobial drugs (Chapter \ref{cha:antimicrobial}), and compliance with preventive health measures in the form of vaccination activity (Chapters \ref{cha:ensemble_learned} and \ref{cha:time_series_adaptive}). Both represent a change from the traditional focus on contagious diseases to a more broad understanding of health events.

\paragraph{How do we mine data from the web?} The web consists of enormous amounts of data, which from a computational perspective is easily accessible, hence, making web data mining an attractive approach for information extraction. One aspect of web data mining is collection of user behavior for prediction, an example of this is prediction using web search query frequency data. Even if we just consider Google, there are billions of searches performed daily. Selecting which of those search queries to use in a prediction model is an essential part of predicting with web search query frequency data, because all available queries cannot be included. The \textbf{second objective} of this thesis is improving query selection, either by developing new methods or improving existing ones. One popular approach is to use the historical correlation between the health event and the query frequencies to select relevant queries. This approach has been shown to be vulnerable to spurious correlations due to overlaps in seasonality between different events. Our \textbf{second contribution} is to show that by modeling expected seasonal variations of the health events we can reduce the number of spurious correlations. This contribution is presented in Chapter \ref{cha:seasonal_web_search_query_selection}. Our \textbf{third contribution}, also within query selection, is to use knowledge resources to extract queries relevant to the health event. We show how this can be done in Chapters \ref{cha:antimicrobial} and \ref{cha:ensemble_learned}. An advantage of using knowledge resources is that it is possible to target a specific demographic by selecting a knowledge resource designed for that demographic. This approach also illustrates a second aspect of web data mining: mining information from online available knowledge resources. In this specific context, online resources are especially well suited for query extraction, because the user behavior we observe through web search queries, occurs because the users are searching for information on the web. In other words, we use the information we expect the user is looking for, to select queries the user would use to find the information.

\paragraph{How do we predict with web mined data?} When a set of queries has been selected as features for the prediction model, it is necessary to decide on the prediction approach. This leads us to the \textbf{third objective} of this thesis, improving predictions using web search query frequency data. In temperate countries, many health events exhibit seasonality, for example, yearly peaks in the winter for influenza or reduction in vaccination activity during holidays. The seasonality means that the health events often show a stable temporal pattern.  Occasionally, these stable patterns are interrupted by unexpected behavior, e.g.\ an off-season disease outbreak.  
Our \textbf{fourth contribution} aims at automatically adapting to such temporal changes in the health event. The method is presented in Chapter \ref{cha:time_series_adaptive} and is an online learning algorithm used for predicting vaccination uptake using web search query frequency data. The temporal nature of health events is also the motivation for the \textbf{fifth contribution}. Here we utilize the observation that it is not only the health event that shows seasonal variations but also the web search query frequencies. The contribution consists of modeling the user behavior, not only using the most recent query frequencies but also using lagged versions of the query frequency data. This approach is presented in Chapter \ref{cha:antimicrobial} and gives us the ability to take advantage of the seasonal variations in search patterns, which in our experiments have shown to improve prediction accuracy.  



\paragraph{How reliable are web mined data for health event prediction?} 
The reliability of web mined data for health event prediction is important when assessing whether the predictions should function as a supplement to existing surveillance methods or if they can replace existing methods. Our \textbf{fourth objective} is therefore to evaluate the reliability of predictions using web mined data. Reliability can be assessed in several ways, one is the accuracy of the predictions, i.e.\ how close to the target signal is the prediction. Another could be to evaluate how robust the predictions are, i.e.\ in the case of unexpected behavior how does the prediction model behave. Our \textbf{sixth contribution} is to evaluate the accuracy of predictions made using web mined data. When predicting antimicrobial drug consumption using web mined data, we observe that the predictions were close to the ones made using historical antimicrobial drug consumption data (Chapter \ref{cha:antimicrobial}). This indicates that web mined data might be a possible replacement for historical data. Similarly, for two vaccines in the Danish vaccination program, we observed a lower prediction error using web mined data, compared to historical vaccination activity data (Chapter \ref{cha:ensemble_learned}). While these results are encouraging, extensive periods of time are necessary for assessing real-life reliability, e.g.\ the robustness of the prediction models. An example of this is ILI prediction.  ILI is the most researched health event with respect to web search query frequency prediction and has twice been subject to noticeable mis-predictions from web data due to unexpected behavior of the ILI activity \cite{butler2013google}. We describe the problem of prediction reliability in more detail in Section \ref{subsec:problems_with_frequencies}.

\paragraph{Is there a relationship between news media coverage and health events?} It is not only our information seeking that is moving online; our news consumption is also moving from printed papers to online websites or digital newspapers. The \textbf{fifth objective} is to investigate if there is a relationship between news media coverage and health events. For some health events, such as, vaccination activity, it is not unlikely that people's actions are affected by the information that is disseminated in the news.  If this is the case, then news media coverage is a potential source of information for a health event monitoring system, and it could help guide the PR strategy of public health institutions. The \textbf{seventh contribution} of this thesis is two studies on the relationship between news media coverage and vaccination activity. We look both at the relationship between media coverage and the measles, mumps and rubella vaccine (Chapter \ref{cha:MMR_Media}) and for the human papillomavirus (HPV) vaccine (Chapter \ref{cha:HPV_media}). In both cases, we observe a correlation between vaccination activity and news media coverage under certain conditions.

\section{Public health surveillance systems}
\label{sec:monitoring_health_events}

Public health institutes rely on population-level health event data to assess the general public health and the quality of health initiatives, such as childhood immunization programs, or campaigns to reduce drug misuse. This data is collected using public health surveillance which is \textit{``the systematic, ongoing collection, management, analysis, and interpretation of data followed by the dissemination of these data to public health programs to stimulate public health action''} \cite{thacker2012public}. Current systems vary from manually maintained paper archives to computerized databases, resulting in vastly different possibilities with respect to data analysis and timeliness. This thesis includes work on three types of health events: vaccinations (a preventive measure), the incidence influenza-like illness (a contagious disease), and antimicrobial drug consumption. These three health events have been selected because they represent conceptually different health event types and because all are important in current public health programs. In the following, we go through the state-of-the-art in monitoring such health events. This overview is intended to illustrate why health event monitoring using web mined data is an interesting research area, but should also give a basis for evaluating the usefulness of the new approaches presented in this thesis.

\subsection{Vaccination uptake}
\label{subsec:vaccinationuptake}

Vaccines are designed to immunize an individual and thereby protect the vaccinated person from illness. For national immunization programs, an additional important effect is herd immunity. Herd immunity occurs when sufficiently many people in a population are immune to a disease, such that the disease can no longer spread in that population. This means that a vaccination program not only protects the vaccinated but also the people who can not be vaccinated, e.g.\ because of age. Depending on the contagiousness of the disease, different percentages of the population need to be vaccinated to obtain herd immunity. For very contagious diseases, such as measles, the recommended percentage of people who must be vaccinated to break the chain of transmission is 95\%. Monitoring vaccination coverage is therefore important to assess if enough people are vaccinated and to take the necessary actions if not. 

Monitoring vaccination uptake varies significantly between countries and sometimes even within countries \cite{Venice2007}. It is, therefore, a good example of how difficult it can be to roll out a health monitoring program in a country and thereby illustrates some of the benefits of national surveillance using web mined data. We begin with descriptions of four traditional surveillance systems and afterward some concrete examples of how vaccination uptake monitoring is performed in different countries.

\paragraph{Surveillance methods for vaccination uptake}

In their review of surveillance methods for vaccination coverage Haverkate et al.\ \cite{haverkate2011assessing} list four different approaches: 

\begin{itemize}
	
	\item \textbf{Administrative methods} \quad The number of administered vaccines is divided by the size of the target population. The number of administered vaccines can be based on the number of vaccines distributed, or number of subjects vaccinated, for example, based on school or daycare records, or similar data sources \cite{Venice2007}. 
	
	\item \textbf{Surveys} \quad Based on surveys it is possible to estimate the vaccination coverage in a population. Surveys are primarily used to provide estimates that can validate the vaccination coverage found using the administrative method. Surveys include, among other things: telephone interviews, mail surveys and focus groups. 
	
	\item \textbf{Seroprevalence surveys} \quad While the survey based methods presented above can be subject to bias, e.g.\ recall bias, serological tests give verifiable information about the antibodies present in a person based on blood samples. It is often not possible to distinguish between antibodies due to disease or vaccination, but serological tests will nevertheless give information about the immunity in the population. 
	
	\item \textbf{Immunization registries} \quad These registries are computerized population-based registries with registrations on an individual level. This is not only useful for assessing the vaccination coverage, but can for example also be used for generating vaccination notifications for individual people, and for research purposes.
	
\end{itemize}

Next, we give concrete examples of how these methods are used.

\paragraph{Surveillance of vaccination uptake in different countries}

To give the reader a feeling for the variations of surveillance systems we will now give examples of how vaccination coverage is measured in three countries. The three countries are selected because they represent three high income and highly developed countries, while they at the same time have very different healthcare systems.

\begin{itemize}
	\item[] \textbf{Denmark} \quad The Danish immunization information system contains data on all childhood vaccinations from 1996 and onwards, and for all administered vaccinations from 2015 \cite{grove2011danish}. It is mandatory for health professionals to register vaccinations, and the information they submit to the immunization information system is instantly available. 
	\item[] \textbf{Germany} \quad For Germany the vaccination coverage is assessed at school start (age $\approx$6), meaning that the information on vaccination coverage lags behind $\approx$5 years \cite{haverkate2011assessing}. Other initiatives for more timely vaccination monitoring have therefore been set in place. These initiatives include, for example, using physician's billing data to health insurance companies. Additionally, nationwide surveys are performed to assess the vaccination coverage.
	\item[] \textbf{USA}    \quad The National Immunization Survey is responsible for monitoring vaccination coverage in the USA \cite{haverkate2011assessing}. The surveys are performed as random-digit-dialing telephone surveys targeting families with children aged 19-35 months. In addition to the responder's answers, the immunization providers of the children are contacted by mail and requested to fill out a separate survey. The answers are compared to ensure accuracy. Estimates of the vaccination coverage are calculated based on the survey results. Similar surveys are performed for teenagers and adults. The USA has started to implement immunization information systems and are aiming for nationwide coverage \cite{haverkate2011assessing}.
\end{itemize}

As can be seen from the three examples above, there is a substantial variation in methods for vaccination monitoring between comparable countries, i.e. high income and highly developed countries. Establishing a real-time monitoring system as the one in Denmark can be difficult due to legal and technical challenges. Support/acceptance in the healthcare sector might also be limited, because mandatory registration increases the workload on the immunization provider, as reported when the Danish immunization information system was deployed \cite{grove2011danish}. In addition to seeing three conceptually different systems, we also see a considerable difference in the time between vaccination and registration. The German system has a five-year delay, while vaccinations in Denmark are registered shortly after the immunization. In addition to looking at the time between vaccination and registration, it is also interesting to look at how long time there is between assessments of national vaccination coverage. The VENICE report on vaccination coverage assessment in Europe \cite{Venice2007} from 2007, reports that for European countries these intervals vary from monthly to every 5 years. For years with missing data, interpolation between data points is sometimes used to fill out the missing pieces \cite{haverkate2011assessing}.

\subsection{Influenza-like illness incidence}
\label{subsec:ILI}

Influenza-like illness (ILI) has a prominent position when it comes to using web search frequency data for prediction and estimation. As will be described in Section \ref{sec:prediction_with_query_frequency_data} several papers have studied how and to what extent ILI can be predicted using web search frequency data. At the same time, governments around the world are using several different methods to predict ILI activity. We go through the most common methods being used and describe how the ILI surveillance system works in different countries.

The methods used for monitoring health events depend on the type of event, e.g.\ is it the incidence of an infectious disease, is it a treatment, or a preventive measure? For very contagious diseases with a high burden of morbidity and mortality, such as measles or meningitis, the danger and consequences of outbreaks mean that the healthcare sector needs to be informed about every case, and the tolerance for false negatives is very low. In these cases the diseases are typically notifiable, meaning that the healthcare professionals have to report all cases to the authorities. In the case of less dangerous and less contagious diseases, such as influenza, the authorities are still interested in monitoring them, but the commonness and typically mild course of disease make it difficult and expensive to monitor on a per case basis, i.e.\ registering each sick person individually. Finding the correct trade-off between cost and accuracy is hard, resulting in countries running several concurrent monitoring systems to get as high accuracy as possible within the allowed budget.

\paragraph{Surveillance methods for ILI incidence}

In the following, we will describe some of the methods used for ILI surveillance.

\begin{itemize}
	\item \textbf{Sentinel system} \quad The system consists of a sample of general practitioners, and sometimes other health professionals, who report the number of new cases of ILI or acute respiratory infection. In Europe, the sentinel doctors typically represent 1-5\% of the doctors working in each country \cite{ecdc_sentinell_surveillance}. The surveillance data is typically reported on a weekly basis.
	
	\item \textbf{Laboratory surveillance of influenza} \quad For some patients nose and/or throat swabs are taken to test for influenza virus \cite{ecdc_sentinell_surveillance}. The results from the tests are subsequently used in the surveillance system.
	
	\item \textbf{Self-reporting} \quad Some countries have online systems where voluntary citizens can report every week if they are suffering from an ILI \cite{doi:10.1093/infdis/jiw280}. This data improves coverage of people who are not seeking medical advice for their illness or who are otherwise absent in the sentinel system.
	
	\item \textbf{Other} \quad In addition to the above mentioned traditional systems for ILI surveillance alternative approaches also exist: Over-the-counter drug sales, emergency visits, absenteeism, health advice calls and telephone triage are all data sources that have been used as indicators for ILI surveillance \cite{dailey2007timeliness}. 
\end{itemize}

The list above illustrates the effort used for building ILI surveillance systems that can accurately and quickly detect ILI activity. The commonness of the disease makes accurate surveillance difficult since many persons will not seek any help from the healthcare system if they catch an ILI. A study of the 2010/2011 influenza season in the USA showed that only 45\% of adults with an ILI sought healthcare \cite{biggerstaff2014influenza}. 

\paragraph{Surveillance of ILI incidence in different countries}

As can be seen from the examples below countries use a variety of different methods to get the best quality information as possible.

\begin{itemize}
	\item[] \textbf{Denmark} \quad  The ILI surveillance system consists of several systems. (i) From October until the middle of May sentinel doctors report the weekly number of patients with an ILI. Additionally, for a sample of the patients, the doctors take nose and/or throat swabs which are tested for influenza \cite{ssi_influenza}. In general, all microbiological tests for influenza are included in the surveillance system.  (ii) During the whole year the Danish medical on-call service registers if patients report ILI symptoms \cite{harder2011electronic}. (iii) For the capital region, approximately 31\% of the population, the emergency on-call system reports consultations with patients having ILI symptoms \cite{doi:10.3109/23744235.2015.1122224}. (iv) A web-based self-reporting system where voluntary citizens can report ILI symptoms from November to the middle of May \cite{doi:10.3109/23744235.2015.1122224}. (v) Hospital admissions due to influenza, and (vi) deaths due to influenza \cite{ssi_alvorlig_influenza}.
	\item[] \textbf{Germany} \quad The system in Germany is based on three systems \cite{rki_ili_2015/2016}. (i) A sentinel system monitoring acute respiratory illnesses; (ii) laboratory test of patients with ILI; and (iii) voluntary self-reporting by citizens via a web-based system. 
	\item[] \textbf{USA}    \quad In the USA influenza surveillance is performed year round using five different monitoring systems \cite{cdc_ili_surveillance}. (i) Laboratory tests from public health or clinical laboratories, (ii) sentinel doctors, (iii) mortality data, i.e.\ deaths due to influenza, (iv) laboratory tests from hospitals, and (v) reports from state health departments about the estimated spread of influenza in different geographical regions.
\end{itemize}

None of the described methods give a perfect picture of the ILI activity and hence each country has to combine data from various sources. The latency of the systems is comparable, i.e.\ around one week, because timeliness is necessary for the ILI monitoring to be of practical use. 

\subsection{Antimicrobial drug consumption}
\label{subsec:antimicrobial}

Antimicrobial drug consumption is closely linked to antimicrobial resistance \cite{antimicrobial2017}. Monitoring of consumption is, therefore, an important component in addressing the problem of antimicrobial resistance. Compared to the previous two health events, ILI and vaccination activity, the monitoring of antimicrobial drug consumption is not as well established, and many countries are only in the process of building nationwide surveillance systems. 
The focus on antimicrobial drug consumption has been increasing in recent years. The Wellcome Trust and the UK Department of Health published in 2016 a report on the financial consequences of antimicrobial resistance stating that antimicrobial resistance is a major challenge, not only for the healthcare system but also for economic growth and welfare \cite{oneill_review}. Later in 2017, an EU report listed the use of antimicrobials as one of the main factors responsible for the development, selection, and spread of antimicrobial drug resistance \cite{antimicrobial2017}. Below we go through a couple of the methods used for monitoring antimicrobial drug consumption, and afterward, we describe how they are used in our three case countries. 

\paragraph{Surveillance methods for antimicrobial drugs}

The European Centre for Disease Prevention and Control (ECDC) monitors antimicrobial drug use for humans in two areas: primary care, e.g.\ general practitioners, and hospitals \cite{antimicrobial2012}. Depending on the country, either the number of packages of antimicrobial drugs and/or the number of defined daily doses\footnote{Defined daily doses (DDD) are the assumed average maintenance dose per day for adults for the condition the drug is registered for.} are reported. The surveillance data used in the European countries are collected in two ways:

\begin{itemize}
	\item \textbf{Sales} \quad Data from pharmacies, hospitals or other entities entitled to distribute antimicrobial drugs. 
	
	\item \textbf{Reimbursement} \quad Reimbursement data from health insurance companies or government agencies that are paying for or subsidizing the antimicrobial drugs. 
	
\end{itemize}

The ECDC encourages countries to report quarterly usage, but some countries only report yearly usage \cite{antimicrobial2012}.

\paragraph{Surveillance of antimicrobial drugs in different countries}

While Denmark and Germany both report consumption statistics to the ECDC and hence have established surveillance systems, the USA still does not have any national surveillance system and is therefore omitted in the list below.

\begin{itemize}
	\item[] \textbf{Denmark} \quad  All antimicrobials consumed in primary care are prescription only drugs and are therefore only sold in pharmacies. All pharmacies report their sales data to the Register of Medicinal Product Statistics \cite{danmap}. All hospitals in Denmark report the antimicrobial drug consumption, except for private hospitals and clinics, psychiatric hospitals, specialized non-acute care clinics, rehabilitation centers and hospices. These exceptions account for approximately 3\% of the antimicrobial drug consumption in hospitals \cite{danmap}. From all sources, the data is reported to the Danish authorities every month. 
	\item[] \textbf{Germany} \quad Germany reports to the ECDC  reimbursement data from health insurance companies, covering 85\% of the population. This data covers only primary care \cite{antimicrobial2012}. Projects for monitoring hospital usage are in development \cite{avs_germany}.
\end{itemize}

As illustrated above, the difference in, or lack of, surveillance systems account for large differences in the knowledge about antimicrobial drug consumption in the three case counties. Assessing the impact of new guidelines or information campaigns is difficult without any measurement of consumption and countries are therefore working on establishing accurate surveillance systems.

\subsection{Summary}

The examples presented above show that the methods used for monitoring health events vary based on the event type and resources available. Even between developed countries such as Denmark, Germany, and the USA, we observe significant differences in the surveillance systems. While the vaccination status for each individual Danish child is known as soon as the immunization provider completes the mandatory reporting, Germany only has the information on a population level with a five-year delay and without any identification of the vaccinated person. On the other hand, we see much more comparability in ILI surveillance; here all countries rely heavily on a sentinel system, coupled with auxiliary monitoring systems. Finally, for antimicrobial drug consumption, we see great variation. The centralized healthcare system is likely one of the contributing factors to why Denmark can have a detailed monitoring of antimicrobial drug consumption, while Germany and the USA are still in the process of designing their countrywide monitoring systems. 

In many countries, the systems for vaccination uptake surveillance have very large delays, up to several years. This delay makes it difficult to respond  to sudden changes in the vaccination uptake, as was witnessed in the UK with the autism scare in 1998 and in Denmark with the fear of adverse reactions to the HPV vaccine in 2013. In those cases, timely predictions based on web search query frequencies would be a valuable resource. Additionally, for countries with very sparse measurements, it is very likely that estimates based on query frequency data could replace the use of interpolation to estimate the coverage in years without data. 

For vaccination monitoring, the reduction in delay could be several years. A similar reduction in delay will not be obtainable for ILI activity because many countries have surveillance systems with weekly reporting. But since the risk of epidemic influenza is high compared to drops in vaccination uptake, a small reduction in the delay would still be useful. For ILI surveillance one problem is that only a fraction of the ill people have any contact with the healthcare sector, and ways to monitor this part of the population are therefore interesting. This is evident from the numerous projects where citizens are asked to report their ILI status online to the health authorities. 

Antimicrobial resistance might be one of the great threats to our healthcare sector and surveillance of antimicrobial drug consumption is one of the components in addressing this problem. A system based on web mined data could give countries an early start on this task, and work as a supplement when more traditional nationwide systems are in place. With the mobility between countries and continents, antimicrobial resistance is a global problem and methods for monitoring consumption in low resource countries will be a valuable tool. Prediction using online behavior might be one of those tools.

%

\section{Public health event prediction with web search query frequency data}
\chaptermark{Prediction with query frequency data}
\label{sec:prediction_with_query_frequency_data}

Web search engines were originally designed to help users navigate through the vast amounts of online web pages. As users utilize search engines in their pursuit of information, they leave a trail of submitted queries, called query logs. This trail of search queries has proven to be a great source of information about the users.  The ubiquity of web search means that people now search on the web instead of consulting a book or calling a friend, meaning that query logs now constitute a rich source of up-to-date information on people's information needs. The widespread use of web search engines throughout the population means that this information is not just on an individual level, but on a population level. Traditionally, analysis of query logs was performed within the area of information retrieval, where the focus was on improving the user's search experience. As has become apparent in the last 10 or more years, this is a rather restricted use of such a rich data source. From a public health perspective, this data is of huge interest, because it gives us a chance to gather nationwide information on the population in real time. More than 10 years ago people started using this data source for public health purposes, namely, monitoring of influenza-like illness (ILI)  \cite{eysenbach2006infodemiology}. 

Query frequency data is a popular choice of web mined data when performing predictions, as was illustrated by Bernardo et al.\ \cite{Bernardo} in a survey from 2013. They looked at the number of papers published using search queries or social media for disease surveillance. In their survey, they identified 32 research papers in the period 2006 to 2011 of which 21 used web search data, while 10 used Twitter data. This illustrates the dominance of web search frequency data for health event prediction. 

Prediction using query frequency data includes two general steps: (i) query selection, where web searches predictive of the health event are identified, and (ii) prediction using web searches, where time series of query frequencies are used as input in a prediction model. 
The following sections start with a description of query selection for prediction models (Section \ref{subsubsec:selecting_queries_for_prediction}), then moving on to methods used for making the predictions with query frequency data (Section \ref{sec:mathince_learning}), and finally a treatment of the problems that arise when using web search query frequencies for prediction (Section \ref{subsec:problems_with_frequencies}).

\subsection{Selecting queries for prediction}\label{subsubsec:selecting_queries_for_prediction}

Methods for selecting queries vary. A popular choice is the correlation-based approach, but other methods also exist. Some of the initial work on query selection relied on manual query selection, and we will, therefore, begin this section by describing this approach. Afterwards, we will discuss query selection by correlation and finally, we will cover how queries can be mined from knowledge resources. 

\paragraph{Manual query selection}

The number of searches submitted to a search engine is enormous (estimates from 2012 put Google at 1.2 trillion searches per day \cite{livestates_google}). Going through all unique queries manually is impossible, and traditional feature selection approaches are infeasible due to time constraints. To validate the potential of query search frequencies for prediction, some of the early work, therefore, used a few keywords and selected the queries where they occurred. Below are two examples of such approaches.

Eysenbach's 2006 paper \cite{eysenbach2006infodemiology} on the correlation between search activity and the 2004/2005 flu season in Canada is one of the pioneering papers on health event prediction using query frequency data. In 2006, access to query log data was limited to the people working at the search engine companies. To study the relationship between query frequency and the flu, Eysenbach, therefore, had to design an experiment that would give him similar information without having actual access to the data. Eysenbach's experiment consisted of buying an online ad on Google, which would be triggered by searches for one of two phrases: \textit{``flu''} or \textit{``flu symptoms''}. The ad consisted of the text: \textit{``Do you have the flu? Fever, Chest discomfort, Weakness, Aches, Headache, Cough.''} Clicks would redirect the user to a web page with general recommendations for people having the flu. Instead of counting actual searches he counted clicks on the ad. Eysenbach observed a high correlation between clicks and the influenza cases in the following week. This led to the conclusion that data on information demand has a potential for syndromic surveillance. 

A few years later Polgreen et al.\ published a paper \cite{polgreen2008using} where they used the actual query log data to predict the number of positive lab tests for influenza and the mortality due to pneumonia and influenza in the USA. Polgreen et al.\ manually selected a set of keywords (``\textit{influenza}'' and ``\textit{flu}'') which were used to retrieve all queries submitted to Yahoo!\ from March 2004 to May 2008 that contained one of the words. Based on this list of queries two new lists were created: one where all queries including ``\textit{bird}'', \textit{``avian''} and \textit{``pandemic''} were removed, and a second where \textit{``bird''}, \textit{``avian''}, \textit{``pandemic''}, \textit{``vaccine''}, \textit{``vaccination''} and \textit{``shot''} where removed. The two lists were then represented as two time series of weekly query counts, i.e.\ weekly query frequency. The decision to remove specific queries was an attempt to control for searches related to the avian flu or the seasonal influenza vaccine, both of which were judged to bias the results. Their results confirmed the results by Eysenbach.

Comparing the query selection approaches in the above two papers illustrates some interesting problems. For example, what criteria should be used for keyword selection? Eysenbach used \textit{``flu''} and \textit{``flu symptoms''} while Polgreen et al.\ used \textit{``influenza''} and \textit{``flu''}. 
Eysenbach's setup meant that it was very likely that only people who actually had an ILI would click on the sponsored link, thereby automatically filtering for queries such as \textit{``flu vaccine''} or other textually related queries that are non-informative about current disease levels. Polgreen et al.\ attempted to achieve the same effect by removing queries related to avian flu and the influenza vaccine. With the huge number of possible queries, it is very difficult to evaluate to what extent they have removed all biasing queries. Examples such as \textit{``stomach flu''} would still be included in their query frequency counts. The problem of manually deciding on what to include and exclude is addressed by the following two approaches.

\paragraph{Correlation-based query selection}

The introduction of the correlation-based query selection method addresses some of the problems of manual query selection. The high-level idea is to calculate the correlation between a historical time series for the event of interest and all query frequency time series. Queries are then sorted according to the correlation and queries with the highest correlation are selected as features for the prediction model. The method was presented by Ginsberg et al.\ in a 2009 paper \cite{ginsberg2009detecting}. More specifically, their method consists of selecting the 50 million most frequent queries submitted to Google between 2003-2008 and for each extract weekly query frequencies. For all queries, the log-odds between the weekly query frequency and the percentage of ILI visits were calculated (they focused on ILI prediction).
They then created a ranked list of queries, based on how well they predicted ILI in different regions of the USA. The number of queries to include in the final prediction model was found by forward feature selection, i.e.\ they continuously added queries from the top of the ranked list to the set of queries used in their model, until they observed a drop in predictive performance. To avoid overfitting, due to too many explanatory variables, they combined the query frequencies for all selected queries into one variable, which was used in a linear model with the ILI activity as the response variable. The peak of the performance was found when using the top 45 queries. The method used by Ginsberg et al.\ avoided the problem of manually selecting and subsequently filtering queries as was the case for the manual query selection methods. But, as we describe in Section \ref{subsec:problems_with_frequencies}, it introduced new problems, for example, spurious correlations due to queries following the same seasonal pattern as ILI. 

The correlation-based approach is simple and can easily be applied to all types of time series. The one major caveat is that it is necessary to have access to the query logs and have substantial computational power to apply the method. This problem was addressed in 2011 when Google launched \textit{Google Correlate} \cite{google_correlate} a service that made it possible for everybody to upload a time series and get a list of correlated queries back in return. Additionally, Google also launched the \textit{Google Trends} service which allows people to submit a query and get back a time series of query frequencies for the query. These two services have been an important part of many of the subsequent publications on health event prediction.

\paragraph{Mining queries from online knowledge resources}

While Google Correlate is helpful for research in prediction with query frequencies, it still only has limited coverage in less popular language domains, such as Danish. Other services such as \textit{Google Health Trends} allow for better coverage but rely on a list of input queries for retrieval of query frequency time series. This quickly leads back to manual query selection, as described above. One way of avoiding this problem is by mining online knowledge resources to generate queries.

For most public health events there are many online resources describing the events. These resources include information websites from health officials, or health information websites like \url{www.webmd.com}. Using these resources has several advantages: experts with domain knowledge are not necessary for query generation; query sets on many different topics can be generated automatically, and queries aimed at specific target audiences can be found. Below we will describe two ways of utilizing these resources for query selection.

Co-occurrence of words in a set of texts is one of the ways that written descriptions can be used for query selection. First textual descriptions of the health event are found and subsequently the text is processed to generate a set of keywords. The initial step is removal of stop words (these are common words like ``have'', ``is'', etc.). Having removed stop words, the queries are selected as words that co-occur in at least $n$ texts. With this approach, it is possible to select texts that have a specific target audience, and thereby get queries that are likely to be used by that audience. Text targeting laymen could be used to generate queries that the general public would use in their information search (Chapters \ref{cha:antimicrobial} and \ref{cha:ensemble_learned}). Though the method addresses some of the problems of manually selecting queries, e.g.\ reliance on domain experts, it has a tendency to generate queries that are too general. This is illustrated by some of the queries used in Chapter \ref{cha:ensemble_learned} for vaccination uptake estimation of the HPV vaccine. There the method generates queries such as ``the girls'' or ``the effect''. In context, these queries are meaningful, because the vaccine targets girls and the effect of the vaccine is an important item to describe. However, for vaccination uptake estimation, these queries are likely not very good estimators, because they are too general and imprecise. 

A second way to use knowledge resource for query generation is to take advantage of knowledge bases, e.g. online encyclopedias and other types of reference work. One example is to use an encyclopedia on diseases to extract the names of diseases that are treatable with a specific drug as done in Chapter \ref{cha:antimicrobial}. Compared to the co-occurrence based method, the method based on knowledge resources reduces the number of uninformative queries. 

\subsection{Prediction models for query frequencies}\label{sec:mathince_learning}

As outlined in Section \ref{subsubsec:selecting_queries_for_prediction} several papers on prediction using query frequency data have been published. While a few papers only focus on the correlation between query frequencies and the health event of interest, most papers apply some sort of statistical or machine learning approach to model and predict the health event. The following section goes through a few of the main methods. The first section describes how linear models have been used for prediction using query frequency data, and is followed by a section on non-linear models.

\paragraph{Linear models}

Linear models are a popular choice for prediction using query frequencies because they are easy to fit and interpret.

Some of the early models proposed were single variable models such as the one proposed by Ginsberg et al.\ \cite{ginsberg2009detecting} or Polgreen et al.\ \cite{polgreen2008using}. In both cases they grouped a set of query frequency counts into one variable and fitted a linear model:

\begin{equation}\label{eq:linear_simple}
y_t = \mu + \alpha Q_t + \beta t + \epsilon_t,
\end{equation}

\noindent here $y_t$ is the target value at time $t$, $\mu$ is the intercept, $Q_t$ is the query frequency at time $t$, $\alpha$ and $\beta$ are coefficients,  and $\epsilon_t$ is the prediction error at time $t$. The term $\beta t$ models a linear trend in the data, i.e.\ a consistent increase or decrease.\footnote{This term is only included in Polgreen et al.\  \cite{polgreen2008using}.}

The simple linear approach from Equation \ref{eq:linear_simple} does not take advantage of the fact that historical data on the health event is most often available. Lazer et al.\ \cite{lazer2014parable} achieved a big improvement on ILI prediction in the USA by combining the Google Flu Trends prediction with historical ILI activity reported by the Centers for Disease Control and Prevention. This was done by adding the Google Flu Trends estimate to an autoregressive model:

\begin{equation}\label{eq:linear_ar_gt}
y_t = \mu + \sum^p_{i=1} \theta_i y_{t-i} + \alpha GT_t + \epsilon_t,
\end{equation}

\noindent where $\sum^p_{i=1} \theta_i y_{t-i}$ denotes the sum of autoregressive terms, with $\theta$ being the coefficients and $p$ is the number of autoregressive terms. $GT_t$ is the Google Flu Trends prediction for time $t$, and the rest of the notation is the same as for Equation \ref{eq:linear_simple}. The use of autoregressive terms is a traditional approach for predicting time series with auto-correlation and serial dependencies. The ILI activity has known seasonality and, because Lazer et al.\ used weekly data, they set $p=52$ (i.e. the number of weeks in a year) to capture the yearly variation.  

The method presented above by Lazer et al.\ relied on Google Flu Trends for the prediction using web query data. With the introduction of the Google Correlate service, this is no longer necessary, because any time series can be uploaded, after which you can download the top 100 correlated queries together with their query frequencies. 
This led to the expansion of Equation \ref{eq:linear_ar_gt} with individual terms for each query. A popular approach is defined by Yang et al.\ \cite{yang2015accurate} as:

\begin{equation}\label{eq:linear_ar_qf}
y_t = \mu + \sum_{i=1}^p \theta_i y_{t-i} + \sum_{j=1}^k \alpha_j Q_{j,t} + \epsilon_t,
\end{equation}

\noindent where $k$ is the number of queries, $Q_{j,t}$ is the frequency for query $j$ at time $t$, and the rest of the notation is the same as for Equation \ref{eq:linear_ar_gt}. In Yang et al.'s setup $k=100$ and $p=52$. To capture changes in people's search behavior Yang et al.\ refit the model for every time step using only data from the two most recent years. Because they use weekly ILI data, they end up with 153 coefficients that need to be fitted using only 104 data points. Ginsberg et al.\ \cite{ginsberg2009detecting} did not model the individual query terms in order to avoid overfitting the model. Yang et al.\ overcame this problem by applying regularization. Their chosen form of regularization is LASSO, which consists of minimizing the following expression:

\begin{equation}
\underset{\theta, \alpha}{argmin}=\sum_t \left(y_t - \mu -\sum_{i=1}^p \theta_i y_{t-i} - \sum_{j=1}^k \alpha_j Q_{j,t} \right)^2 + \lambda \lVert \theta \rVert_1 + \lambda \lVert \alpha \rVert_1,
\end{equation}

\noindent where $\lambda$ is a hyperparameter controlling the amount of regularization and $\lVert \cdot \rVert_1$ denotes the L1 norm. LASSO regularization typically induces a sparse solution by only keeping one non-zero coefficient for strongly correlated features \cite{zou2005regularization}. Only maintaining one correlated feature is not necessarily an advantage, because one of the correlated features might be a false positive, i.e.\ it might not remain correlated in the future. A solution to this can be to use Elastic Net regularization instead, which groups correlated features and either includes or removes the whole group \cite{zou2005regularization}. Elastic Net regularization is defined as:

\begin{equation}
\begin{split}
\underset{\theta, \alpha}{argmin}=&\sum_t \left(y_t - \mu -\sum_{i=1}^p \theta_i y_{t-i} - \sum_{j=1}^k \alpha_j Q_{j,t} \right)^2 \\ &+ \lambda \lVert \theta \rVert_1 + \lambda \lVert \alpha \rVert_1 + \eta \lVert \theta \rVert_2^2 + \eta \lVert \alpha \rVert_2^2,
\end{split}
\end{equation}

\noindent where $\eta$ is a hyperparameter and $\lVert \cdot \rVert^2_2$ is the squared L2 norm. Elastic Net regularization is for example used by Lampos et al.\ \cite{lampos2015advances} for prediction of ILI and by Hansen et al.\ (Chapter \ref{cha:antimicrobial}) for prediction of antimicrobial drug consumption.

\paragraph{Non-linear models}

The simplicity of the linear models makes them appealing, but they lack the ability to capture non-linear relationships. Two non-linear approaches that have been used for prediction of health events using query frequency data are Gaussian processes and random forests. Both will be described below.

\subparagraph{Gaussian processes (GPs)}

One approach to capture non-linearities is to use GPs, as has been done with respect to ILI activity by Lampos et al.\ \cite{lampos2015advances} and antimicrobial drug consumption by Hansen et al.\ in Chapter \ref{cha:antimicrobial}. We refer to Chapter \ref{cha:antimicrobial} for more details on how GPs are used for predicting health events with query frequency data.

Gaussian processes are probability distributions over functions, where any finite set of function values have a joint Gaussian distribution \cite{Rasmussen2004}. Gaussian processes map from an input space, $X \in \mathbb{R}^n$, to an output space, $Y \in \mathbb{R}^m$. Two components are necessary to describe a GP: a mean function and a covariance function. The mean function is defined as:

\begin{equation}
E[f(x)] = \mu(x),    
\end{equation}

\noindent where $x$ is the input data and $\mu$ denotes the mean of the distribution of functions at point $x$. 

The covariance function is defined as

\begin{equation}
\text{Cov}[f(x), f(x')] = k(x, x'),
\end{equation}

\noindent where $x$ and $x'$ are two input vectors and $k$ is a kernel function \cite{Rasmussen2004}. 

\begin{figure}
	\centering
	
	\begin{subfigure}[b]{0.46\textwidth}
		\includegraphics[width=\textwidth]{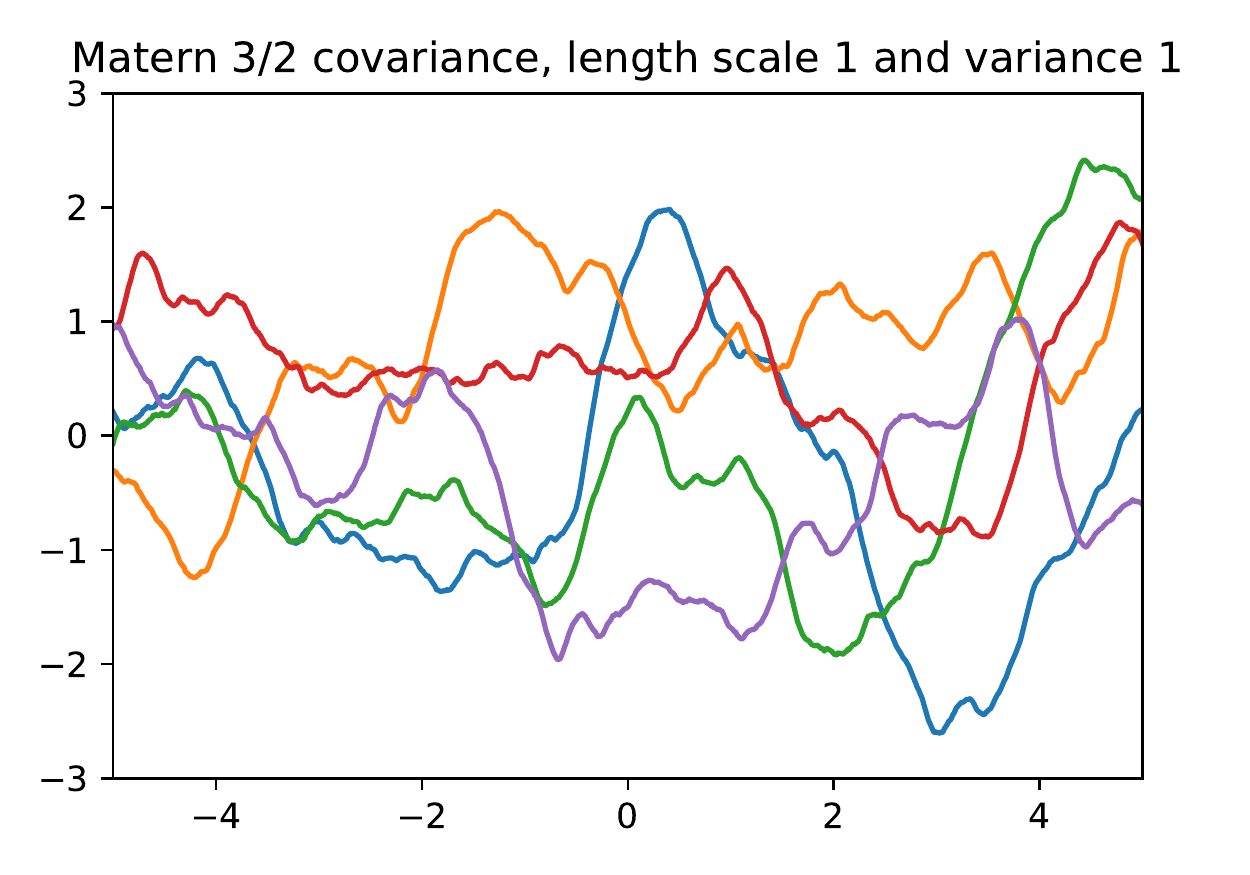}
	\end{subfigure}
	\begin{subfigure}[b]{0.46\textwidth}
		\includegraphics[width=\textwidth]{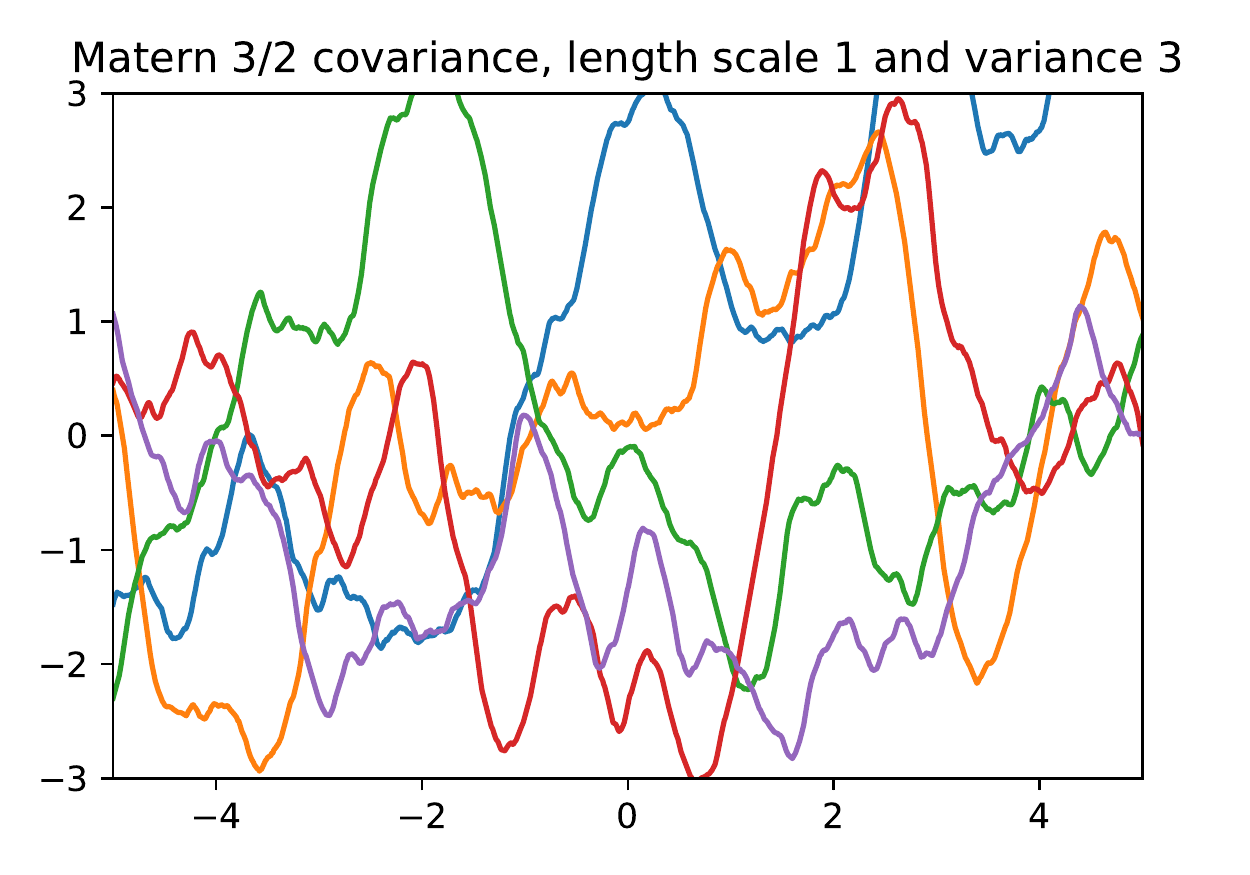}
	\end{subfigure}
	
	\begin{subfigure}[b]{0.46\textwidth}
		\includegraphics[width=\textwidth]{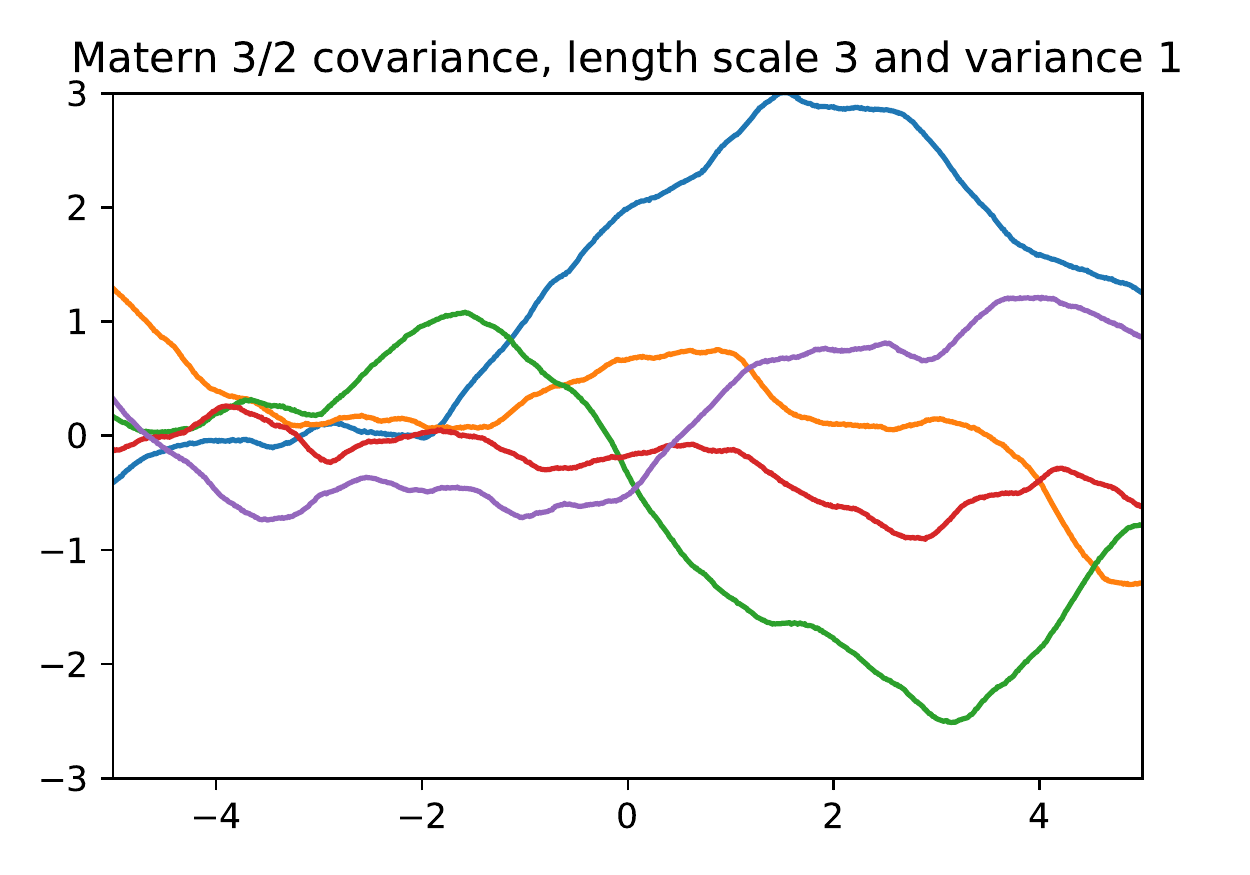}
	\end{subfigure}
	\begin{subfigure}[b]{0.46\textwidth}
		\includegraphics[width=\textwidth]{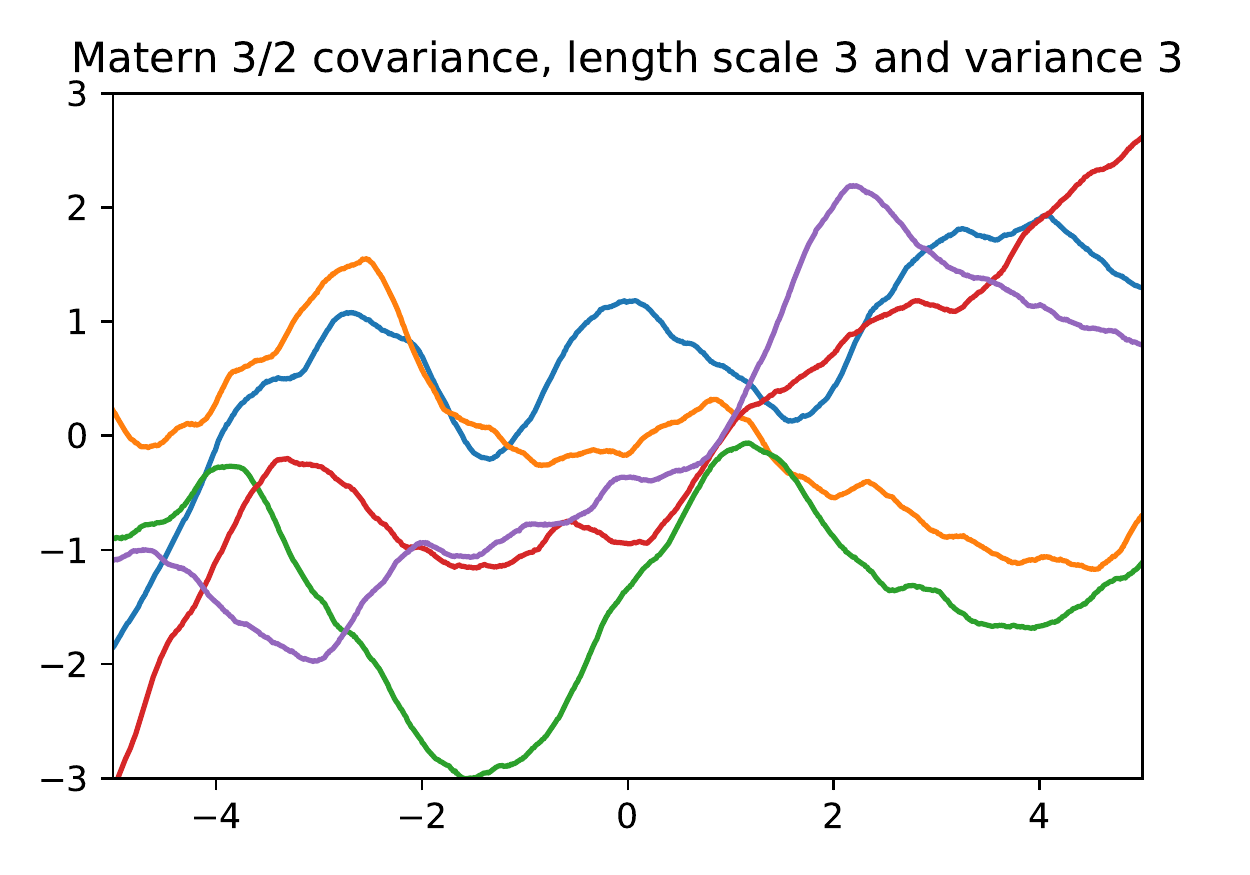}
	\end{subfigure}
	
	\caption{Difference in behavior of a GP with Matern($\nu=3/2$) covariance function when the length scale or variance parameter is changed.}
	\label{fig:gaussian_se_example}
\end{figure}

When working with GPs it is customary to assume that the mean value is zero, and focus only on the covariance function/kernel. The covariance function defines prior covariance between two input values and is typically controlled using two parameters: length scale and variance. An example of a covariance function is the Matern covariance function, in the case where $\nu=3/2$ it is defined as:



\begin{equation}
K_{(\nu = 3/2)}(r) = \sigma^2 \left(1+\frac{\sqrt{3}r}{l}\right)\exp\left(-\frac{\sqrt{3}r}{l}\right),
\end{equation}

\noindent where $r$ is $|x-x'|$, $l$ is the length scale, $\sigma^2$ is the variance and $\nu$ a hyperparameter \cite{Rasmussen2004}.
The Matern covariance function is popular because it can capture abrupt changes in the input variables, which is typically a more realistic assumption than the very smooth model imposed by a Gaussian covariance function. For the Matern covariance function, the length scale defines how quickly the underlying signal changes, the variance defines the magnitude of changes, and $\nu$ controls the smoothness. As $\nu \rightarrow \infty$, the covariance function approaches the Gaussian covariance function \cite{Rasmussen2004}. Figure \ref{fig:gaussian_se_example} shows an example of what happens when the length scale and variance are changed for the Matern covariance function with $\nu=3/2$. A feature of GPs is that covariance functions can be combined. This property can be used to create new covariance functions that can capture more aspects of the data. For example, combining covariance functions with different length scales could be used for modeling slow changes and quick changes. This feature is for example used by Hansen et al.\ (Chapter \ref{cha:antimicrobial}) for predicting antimicrobial drug consumption using 10 Matern covariance functions.

\begin{figure}
	\centering
	
	\begin{subfigure}[b]{0.46\textwidth}
		\includegraphics[width=\textwidth]{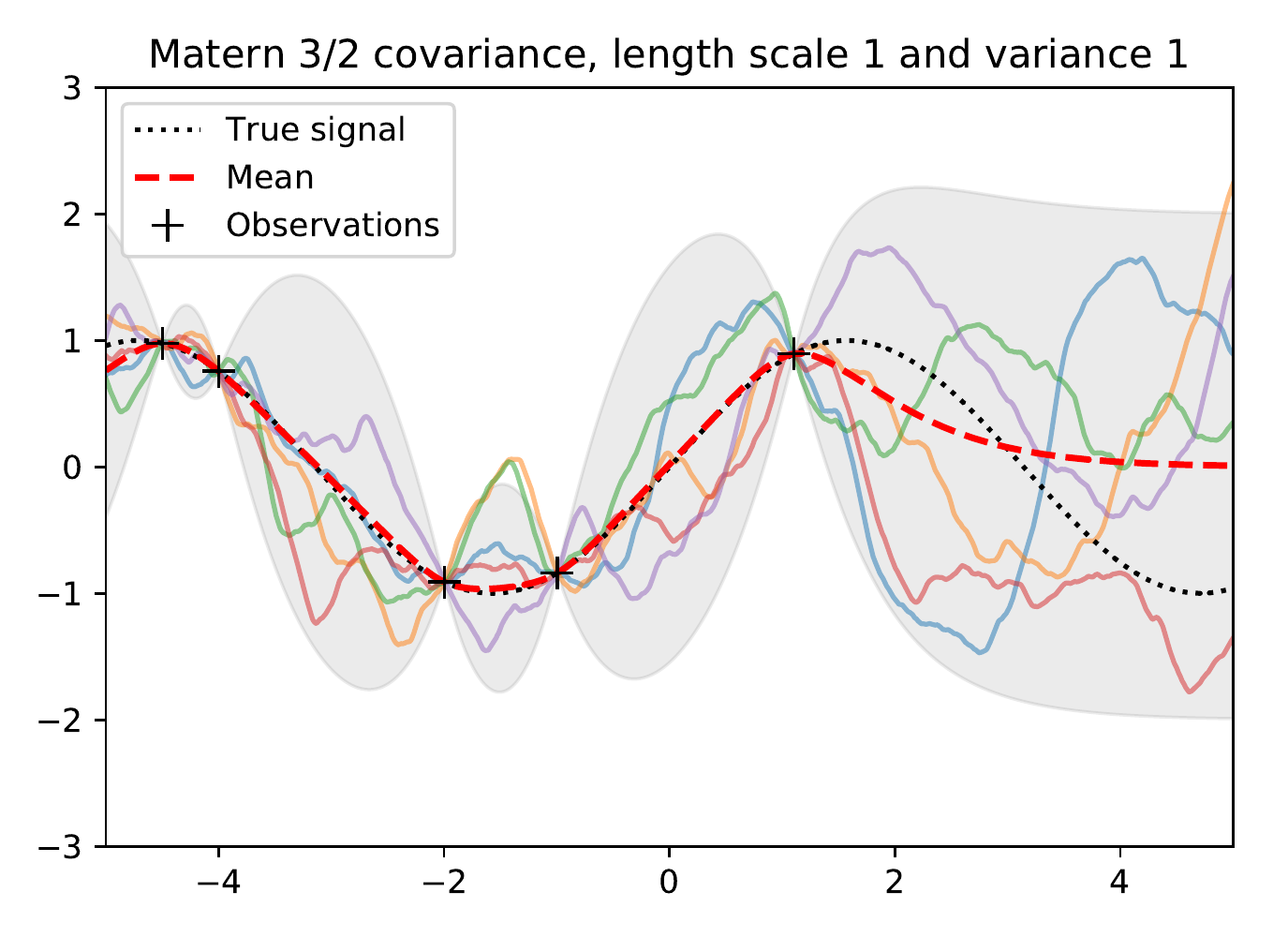}
	\end{subfigure}
	\begin{subfigure}[b]{0.46\textwidth}
		\includegraphics[width=\textwidth]{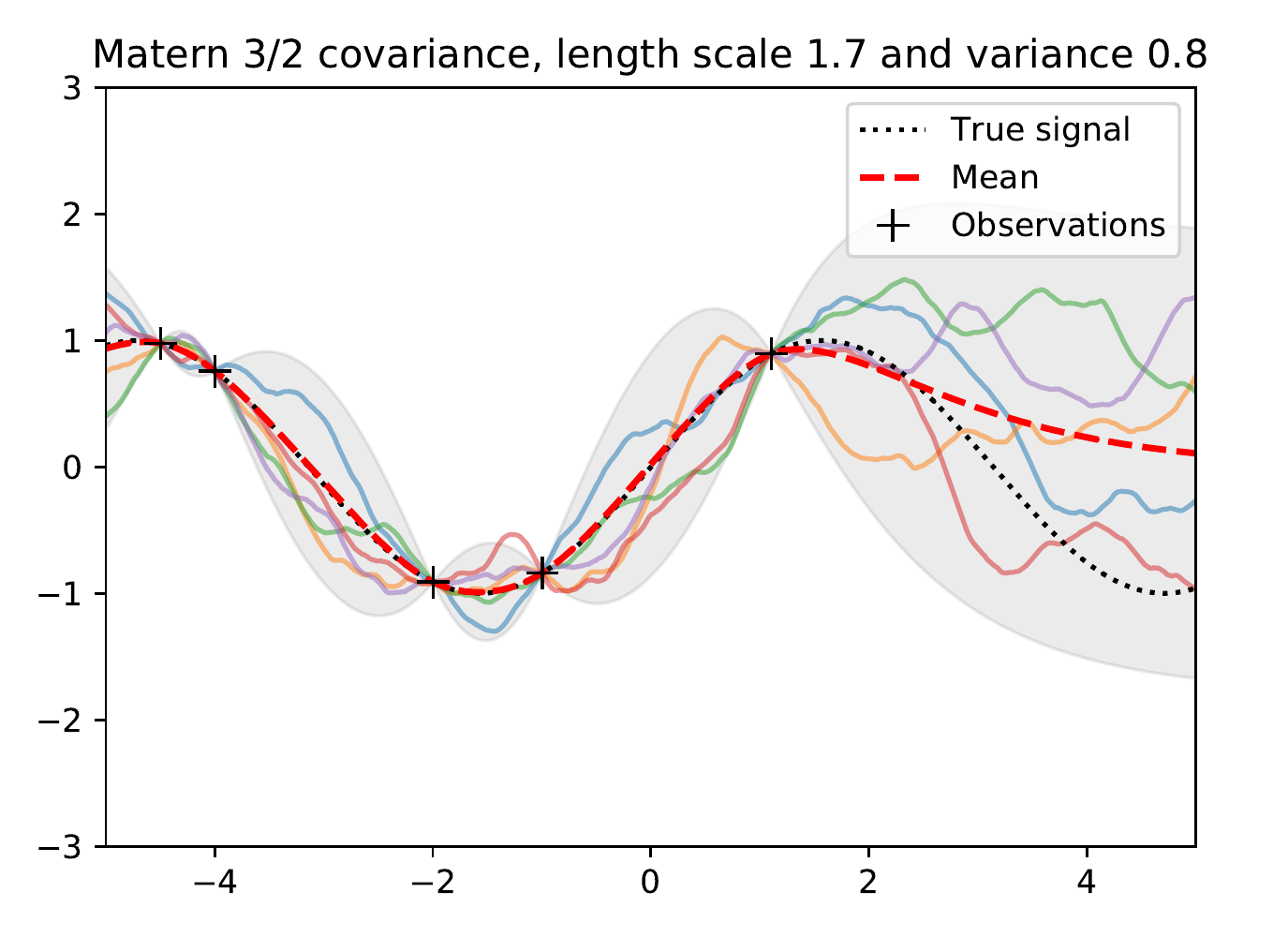}
	\end{subfigure}
	
	\caption{GP with Matern($\nu=3/2$) covariance function fitted to a sinusoidal function. The red dashed line denotes the GP mean at each point and the gray area denotes two standard deviations from the mean. The difference in the fitted model is due to the difference in length scale and variance. Parameters on the right have been found by maximizing the marginal likelihood.}
	\label{fig:gaussian_post_example}    
\end{figure}

The covariance function and hyperparameters represent the model which can be fitted to training data. Figure \ref{fig:gaussian_post_example} shows a GP with Matern($\nu=3/2$) covariance function fitted to a sinusoidal function. The optimal hyperparameters can be found by maximizing the marginal likelihood of the hyperparameters given the training data.

\subparagraph{Random forests}
\label{subsubsec:random_forest}

Another non-linear prediction model used for predicting health events is random forests. It is used by Hansen et al.\ \cite{www17} (Chapter \ref{cha:time_series_adaptive}) in a modified version for prediction of vaccination uptake, but it has also been used for predicting healthcare visits based on web data by Agarwal et al.\ \cite{agarwal2016impact}, and for ILI prediction by Santillana et al.\ \cite{santillana2015combining} in the form of AdaBoost with regression trees. Below is a general description of random forests regression. For a concrete example of how it is applied to query frequency data, we refer to our paper \cite{www17} in Chapter \ref{cha:time_series_adaptive}.


The primary component of a random forests model is a collection of classification or regression trees. The following description focuses on regression trees \cite{martinez2007computational}, but similar principles hold for classification trees.  
A regression tree is a binary tree where the path from the root to leaf depends on the explanatory variables, $X$, and the value at the leaf corresponds to the response variable, $y$. The non-leaf nodes, i.e.\ the internal nodes, represent binary decisions based on the input variables. The tree is built based on a data set $\{D = {(x_i, y_i)| i=1,\dots, n}\}$. Initially, all data samples are in the root node; then nodes are recursively split until some stopping criterion is met,  typically a minimum number of samples in a leaf node or maximum depth of the tree. The splitting is based on an impurity measure; for regression, this could be the mean squared error. Splits are made based on features of the input data, and the split that reduces the impurity measure the most is selected. 
The predicted value, $\hat{y}$, in a leaf node is equal to the average value of the samples in the node.  

%
%
%
%
%
%
%
%

Given a collection of regression trees, it is possible to construct a random forest. For the collection to be a random forest it is necessary for the tree generation to be randomized in some fashion.  In traditional implementations, the randomization could, among other things, consist of picking a random split among the top $k$ splits when splitting nodes or selecting only a random subset of the features to use for splitting. The predicting of a random forest is performed by averaging the predictions among all the trees in the collection.

%
%

\subsection{Problems with using web search frequencies for health event prediction}
\label{subsec:problems_with_frequencies}

The use of query frequencies for prediction has been subject to some criticism and some noticeable mis-predictions. Already in 2006 Eysenbach \cite{eysenbach2006infodemiology} stated that the usage of search frequencies could be subject to a so-called ``\textit{epidemic of fear}'', meaning that if people fear a specific event they might start searching for information about it without being subject to the event themselves. Another potential problem is that people's search behavior might change for unusual disease outbreaks, such as avian flu or swine flu. These concerns were tested for the first time in 2009 with the flu pandemic. 

The 2009 pandemic was off-season, as pandemics usually are in the first wave. It started in the summer, and it caused a change in search behavior, for example with searches for ``swine flu'', which was not part of the Google Flu Trends model \cite{cook2011assessing}. As a result, the predictions were significantly lower than the ILI activity reported in the USA by the Center for Disease Control and Prevention (CDC). Google subsequently updated their model, which reduced the prediction error.
A few years later, for the 2012/2013 flu season in the USA, the Google Flu Trends predictions were off again, this time with an estimate almost twice what the CDC reported \cite{butler2013google}. Together with the results from the 2009 pandemic, this raised the question of to what degree web-based monitoring could replace traditional monitoring systems, such as sentinel networks, and if web-based monitoring could only be used as a supplementary data source \cite{butler2013google}. 

It is not only for contagious diseases that sudden temporal changes have been observed to reduce prediction accuracy. For prediction of vaccination uptake similar observations have been made, where a sudden decrease in HPV vaccinations resulted in low prediction accuracy compared to other vaccines \cite{DalumHansen:2016:ELV:2983323.2983882}. We have proposed that this problem could be reduced with an online learning algorithm that automatically adapts to changes in the predicted signal \cite{www17}. In the case of vaccination uptake prediction, there was a clear improvement in prediction of HPV vaccination uptake with the new method. 

Other concerns about the prediction with search frequency data have revolved around the popular query selection method based on historical correlations. As noted by Lazer et al.\ \cite{lazer2014parable} this approach can lead to spurious correlations. Lazer et al.\ mention that sports events such as high school basketball are sometimes included in the query set because of the temporal correlation with the influenza season. This problem has been addressed by Lampos et al.\ \cite{www17_lampos} by replacing the timely correlations with semantic similarity measures that used neural word embeddings to calculate the similarity between queries and a set of positive and negative context words.  Another approach has been presented by us in \cite{dalum2017seasonal} (Chapter \ref{cha:seasonal_web_search_query_selection}), which works by discounting for known seasonality. We observe that while it was possible to automatically select queries of higher relevance, these queries did not result in the best predictions.

Another problem or concern with surveillance systems that use query frequencies, is that they need to be calibrated before they can be used for reliable predictions. For many health events there is no prior monitoring statistics, as we have seen with antimicrobial drug consumption in Section \ref{subsec:antimicrobial}, and hence model coefficients can not be fitted using historical training data. If web mined data is supposed to be used as a monitoring tool in low resource areas, this can be a major obstacle. In the section about future work (Section \ref{sec:looking_ahead}) some potential solutions to this problem are discussed.

\subsection{Summary}

Section \ref{sec:prediction_with_query_frequency_data} has focused on health event prediction using query frequency data. We have in the previous sections divided this problem into two parts: (i) selection of queries and (ii) prediction using those queries. 

Query selection can be approached from many different directions. We have covered three general approaches: Manually, automatic generation from knowledge resources or correlation-based. The two first approaches generate the query set based on domain knowledge, either in the form of a domain expert who manually defines the query set, or automatically based on written material relevant to the health event, i.e.\ knowledge resources. Finally, the correlation-based method which approaches the task as a feature selection problem. The feature selection approach is appealing since it generalizes well between problems, but unfortunately, research into this requires access to the query logs, which currently is difficult to get. On the other hand, the approaches where the queries are generated based on domain knowledge have more room for independent development, because publicly available services make it possible to get aggregated query frequency data for submitted queries. Hence, there are fewer restrictions on data access.


Having obtained a set of time series for the query frequencies, the next problem is to generate predictions. Much of the published work on prediction with query frequency data is using linear models, though the assumption of a linear relationship between query frequency and the predicted health event is not very likely. Work on using non-linear models has started, with different methods being tested. Gaussian processes have successfully been applied to this prediction problem, and the customizability of the method means that the peak performance of this approach has likely not yet been found. 


%
%


%
%
%
%
%
%

\section{Perspectives on future work}\label{sec:looking_ahead}

We now move on from the current state-of-the-art in health event prediction presented in the previous sections to the future of health event prediction using web mined data. In this section we allow ourselves to disregard constraints of what is done, and instead, speculate on what might be done. The following four perspectives arise from some of the unanswered questions raised in the previous sections.

\paragraph{Monitoring actions or action motivators}

One of our research goals poses the question of whether or not there is a relationship between news media coverage and health events. As we show in Chapter \ref{cha:MMR_Media} and \ref{cha:HPV_media} there is a correlation between media coverage and vaccination activity, which likely means that news media coverage can be a feature in a prediction model for vaccination uptake. Using media data compared to query search frequencies has some potentially interesting properties, given the inherent difference between the two data types. Roughly, we can view web search frequencies as a type of data that tells us something about people's current actions, while news media might act as a primer affecting people's thoughts and hence, tell us something about their future actions. This leads us to two different types of predictions: (i) predicting based on people's actions, and (ii) predicting based on events that motivate people's actions. An example could be that articles about vaccination side effects would make people more skeptical about vaccinating their children, resulting in reduced vaccination activity. Many health events are directly or indirectly affected by people's actions. Another example could be sexually transmitted diseases, where transmission is dependent on people's adherence to safe sex guidelines. Designing surveillance programs that monitor action motivators, e.g.\ news media coverage,  has the potential for improving prediction or to serve as early warning systems that can flag a potential change in public discourse about a certain public health event.

\paragraph{Using web mined data to identify determinants of citizen's behavior} 

When using web mined data the goal has often been to predict a health event. This means that the query selection methods aim at identifying queries that are good predictors. This has the side effect of also identifying which queries are linked to a specific behavior. Likewise, we might identify specific types of media coverage that trigger certain behavior. This information can be used in a prediction model, but it is also valuable information in understanding the determinants of people's actions. This knowledge can be incorporated into the public health programs. For example, if predictors can be identified as indicators of drug misuse, this information tells us that people's reasoning, e.g.\ which drug to use for treatment of a specific disease, is out of sync with the current recommendations. Hence, information targeted at correcting this specific misconception can be designed and distributed to the citizens. 
The information could be distributed through traditional means, such as pamphlets at the pharmacies, but could also be through information targeting the individual in what we might call an online intervention. Such individual information could be delivered from online ads in the search engines triggered by the specific search terms that originally identified the misconception. 
%
%
%

\paragraph{Evaluation methods for query frequency predictions}

Work on usage of web search frequencies for prediction of health events focuses on either: (i) reducing the prediction error on previous prediction problems, or (ii) using web search frequencies for new prediction problems. In both cases, the evaluation metric is typically the same, e.g.\ root mean squared error, but this might not be the metric of most interest for the public health professionals. A low prediction error in a stable period is of less interest compared to low prediction error in an unstable period. Often though, the signals are relatively stable for long periods of time and have short infrequent periods with unstable behavior.   This means that methods that make near perfect predictions in the stable periods, might get chosen in favor of a method that has a higher error in the stable periods, but lower in the unstable. An illustration of this problem is the ILI predictions. The influenza season is for example very seasonal and stable but has occasionally unstable behavior. This has resulted in Google Flu Trends selecting models that fitted well with the stable periods, but made large errors in unstable periods, i.e.\ the 2009 swine flu pandemic and the 2012/2013 flu season in the USA. Designing evaluation metrics especially for health event prediction could lead to bigger improvements in model performance, than improving the prediction models when evaluated with existing evaluation metrics.  Improving our predictions might therefore not be a question of only building better and more accurate models, but also of designing evaluation metrics that evaluate the models in a realistic setting, such that the models most fitted to real-world usage are the ones selected. Development of new evaluation metrics might, therefore, be the next step in health event prediction.

\paragraph{Improving query selection}

Using correlation for query selection is vulnerable to changes in search behavior, for example, introduction of new terms. For ILI surveillance new terms, such as ``swine flu'', was blamed for the mis-predictions of the 2009 flu pandemic. Relying on correlation for query selection has the inherent problem that emerging trends will not be discovered before there is sufficient historical data. Combining the introduction of new query terms with an online learning approach, like the one we propose in Chapter \ref{cha:seasonal_web_search_query_selection}, would mean that as the new terms grow more common, they would automatically receive a larger weight in the prediction model. Additionally, it is not unlikely that different query selection approaches select different types of queries. It might, therefore, be useful to consider the type of health event, before deciding on the query selection approach. Some health events are predicted by people's immediate prior information need, e.g.\ influenza-like illness, in which case a correlation-based approach might yield good results. Other events, e.g.\ antimicrobial drug consumption, might be predicted more indirectly by increases in diseases with complications needing antimicrobial treatment. In this case, more domain knowledge might be necessary to identify good predictors. Understanding the difference between the query selection approaches will be an important step in improving the performance of health event prediction using web mined data.

\paragraph{Using web mined data without health event data resources}
An appealing aspect of web mined data is that it is available even for developing countries, which gives hope of implementing cheap nationwide surveillance even without an established healthcare system. One major problem in such a setup is access to historical health event data, which is necessary to fit the model. Several approaches might be used for addressing this problem, such as \textit{transfer learning}. Transfer learning is a technique within machine learning where knowledge from the solution to one problem is used to solve another related problem. An example could be that a country has data on childhood vaccinations, which means that a prediction model for childhood vaccines can be fitted. The knowledge about the relationship between search activity and childhood vaccination activity could then be transferred when designing a model for prediction of the vaccination activity for vaccines targeting teenagers.  Another possibility could be to use data from a small area to fit the initial model, for example from a single hospital or school district. This model could then be used for prediction on a country level.

\vspace{1cm}

The remainder of the thesis consists of seven chapters, one for each of the included papers. The first paper is on prediction of antimicrobial drug consumption, this paper is followed by a paper on query selection for ILI prediction. Afterwards follows five papers on vaccinations. The first paper is a traditional registry study of severe adverse reactions to the HPV vaccine. While this paper does not use any web mined data, it is included because the problem that it addresses, i.e.\ a sudden drop in vaccination uptake, is the main motivator for the subsequent four papers on vaccination uptake. Then follows two papers on estimation of vaccination uptake and finally are two papers on the relationship between news media coverage and vaccination uptake.

\addtocontents{toc}{\protect\setcounter{tocdepth}{0}}



\setcounter{secnumdepth}{3}

\chapter{Predicting Antimicrobial Drug Consumption using Web Search Data}
\chaptermark{Predicting antimicrobial drug consumption}
\label{cha:antimicrobial}

\begin{center}
Venue: In review

\vspace{1cm}

\textit{
	Niels Dalum Hansen$^{ab}$, Kåre Mølbak$^c$,\\Ingemar Johansson Cox$^a$, Christina Lioma$^a$
}

\vspace{0.5cm}

$^a$University of Copenhagen, Denmark.
$^b$IBM Denmark.\\
$^c$Statens Serum Institut, Denmark. 

\end{center}

\leftskip=1cm
\rightskip=1cm

\noindent\textit{Consumption of antimicrobial drugs, such as antibiotics, is linked with antimicrobial resistance. Surveillance of antimicrobial drug consumption is therefore an important element in dealing with antimicrobial resistance. Many countries lack sufficient surveillance systems. Usage of web mined data therefore has the potential to improve current surveillance methods. To this end, we study how well antimicrobial drug consumption can be predicted based on web search queries, compared to historical purchase data of antimicrobial drugs. We present two prediction models (linear Elastic Net, and non-linear Gaussian Processes), which we train and evaluate on almost 6 years of weekly antimicrobial drug consumption data from Denmark and web search data from Google Health Trends. We present a novel method of selecting web search queries by considering diseases and drugs linked to antimicrobials, as well as professional and layman descriptions of antimicrobial drugs, all of which we mine from the open web. We find that predictions based on web search data are marginally more erroneous but overall on a par with predictions based on purchases of antimicrobial drugs. This marginal difference corresponds to less than 1\% point mean absolute error in weekly usage. Best predictions are reported when combining both web search and purchase data.}

\textit{This study contributes a novel alternative solution to the real-life problem of predicting (and hence monitoring) antimicrobial drug consumption, which is particularly valuable in countries/states lacking centralised and timely surveillance systems.}

\leftskip=0pt\rightskip=0pt

%
%



\section{Introduction}
Surveillance of antimicrobial drug consumption, such as antibiotics, is an important element in dealing with antimicrobial resistance. Antimicrobial resistance is recognized as a major challenge, not only for the
health care system, but also for economic growth and welfare \cite{oneill_review}. Use of
antimicrobials is one of the main factors responsible for the development, selection and spread of antimicrobial
resistance \cite{antimicrobial2017}. This has become a serious threat to public health, notably because of the emergence and spread of
highly resistant bacteria, and because there are very few novel antimicrobial agents in the research and
development pipeline. Some countries have the possibility of doing real time monitoring of
antimicrobial drug consumption, whereas other countries do not have this possibility and depend on data from
annual summaries or even sample surveys from where extrapolation is done.

To improve surveillance and stimulate prudent use of antimicrobial drugs, timeliness is important. We hypothesize that
antimicrobial drug consumption can be predicted from online behavior, such as the queries submitted to web
search engines. This can benefit public health by: (i) allowing countries without access to real-time data to forecast time trends and seasonal patterns of consumption; (ii) allowing all countries to analyze determinants of use, e.g., which types of web search are important
as predictors of certain classes of antimicrobial drugs. This information can be used in communication efforts to
stimulate antimicrobial stewardship; (iii) complementing syndromic surveillance, e.g.\ web searches that are predictive of drugs 
for respiratory infections can be used as an indicator of these diseases.

In this paper we study how well antimicrobial drug consumption can be predicted based on web search queries, and specifically the number of submitted queries to online search engines, e.g.\ how frequently people have searched for ``fever'' on Google in a specific time interval. To our
knowledge such web search data has not been previously used to predict antimicrobial drug consumption. However, this type of web search data has
been used previously to predict other health events, e.g.\ influenza like illness (ILI) \cite{eysenbach2006infodemiology} or vaccination uptake \cite{www17}.
We compare web search based prediction to the more traditional method of predicting based on historical purchase data of antimicrobial drugs. We present two prediction models (a linear one, namely Elastic Net, and a non-linear one, namely Gaussian Processes), which we train and evaluate on almost 6 years of weekly antimicrobial drug consumption data from Denmark and web search data mined from Google Health Trends for the location of Denmark. We further present a novel method of selecting web search queries by considering diseases and drugs linked to antimicrobials, as well as professional and layman descriptions of antimicrobial drugs, all of which we mine from the open web.
We find that the prediction error of swapping historical antimicrobial drug purchase data to web search queries is overall negligible, across different prediction offsets.

%
%
%
\section{Related work}

There is a large amount of work on using web search data to predict health events, though not antimicrobial drug consumption.
Considerable effort has been applied to estimating the prevalence of various diseases based on web search query frequencies. 
Especially influenza like illness (ILI) prediction has been the subject of numerous papers \cite{eysenbach2006infodemiology, polgreen2008using, ginsberg2009detecting, lazer2014parable, santillana2014can, www17_lampos, www17}, but prediction of other infectious diseases from web data has also been researched, e.g.\ dengue fever, gastrointestinal diseases, HIV/AIDS, scarlet fever, tuberculosis \cite{Bernardo}. The general idea of predicting health events from web data has been applied to events beyond diseases, for example prediction of vaccination
uptake \cite{DalumHansen:2016:ELV:2983323.2983882, www17} and hospital admissions \cite{agarwal2016impact}.

Two primary problems when predicting events using web data are: (i) query selection and (ii) the prediction model. Query selection can for example be performed using hand picked seed words \cite{eysenbach2006infodemiology, polgreen2008using}, which are used to filter relevant from irrelevant searches, or using written descriptions of the event that is being predicted \cite{DalumHansen:2016:ELV:2983323.2983882, www17}. Perhaps the most popular approach is to use  the historic correlation between the query search frequency time series and the time series to be predicted \cite{ginsberg2009detecting, lazer2014parable, santillana2014can}. The most prevalent prediction models are linear models \cite{ginsberg2009detecting, lazer2014parable, santillana2014can}, but other non-linear models such as random forests \cite{www17, agarwal2016impact} or Gaussian Processes \cite{lampos2015advances, www17_lampos} have also been used. 
 While all these studies use web search data, other types of online data have also been used, e.g.\ social media data, such as Twitter messages \cite{Lampos2010a, paul2014twitter, Signorini2011}. 
This work use web search frequency data and make predictions using both a linear model and Gaussian Processes. We select queries based on a collection of written resources on antimicrobial drug consumption.

While there is, to our knowledge, no prior work on the prediction of antimicrobial drug consumption using web search data, there has
been computational epidemiological work regarding antibiotics on Twitter. In 2010 Scanfeld et al.\ \cite{scanfeld2010dissemination} analyzed 1000 tweets mentioning antibiotics and categorized them into 11
categories. The top three categories were: ``general use'', ``advice/information'' and ``side effects/negative
reactions''. Scanfeld et al.\ concluded that social media was used for sharing information about antibiotics and
that the tweets could be used to identify potential misuse and misunderstandings regarding antibiotics. Later, in
2014, Dyar et al.\ \cite{Dyar2014} made a large scale analysis of world wide Twitter
activity mentioning antibiotics in the period September 2012 to 2013. They limited their analysis to four peaks in
the twitter activity and examined the reason for those peaks. They concluded that the peaks were caused by
institutional events, such as public announcements from the UK Chief Medical Officer regarding antibiotics. The
peaks did not result in any sustained twitter activity, and activity was generally back to baseline level after two
days. Kendra et al.\ \cite{Kendra2015} showed in 2015 that tweets regarding antibiotic usage could be
categorized automatically using a neural network. Like Dyar et al.\ \cite{Dyar2014},
they also observed that peaks in activity were correlated with public events such as a speech by the British prime
minister and an executive order from the President of the US regarding antibiotic resistance. None of the studies
address a potential relationship between Twitter activity related to antimicrobials and antimicrobial drug consumption. Since none of the studies collected data for more than one year, long term relationships between online activity and
antimicrobial drug consumption have not been analyzed. We use web search data spanning 5 years and 10 months. 

Next we describe, first the data collected for our analysis (Section \ref{s:data}), and then our prediction methods (Section \ref{s:prediction}).

\section{Data}
\label{s:data}

Three categories of data are used in our study: (1) Sales of antimicrobials in Denmark collected by Danish health
officials; (2) Web search query frequency data from Denmark; (3) Freely available online material related to
antimicrobial drugs such as disease descriptions or information about antimicrobials. 
Each of the three categories is described next in more detail.

\subsection{Antimicrobial usage in Denmark}
\label{ss:antibioticData}

\begin{figure}
	\centering
	\includegraphics[width=10cm]{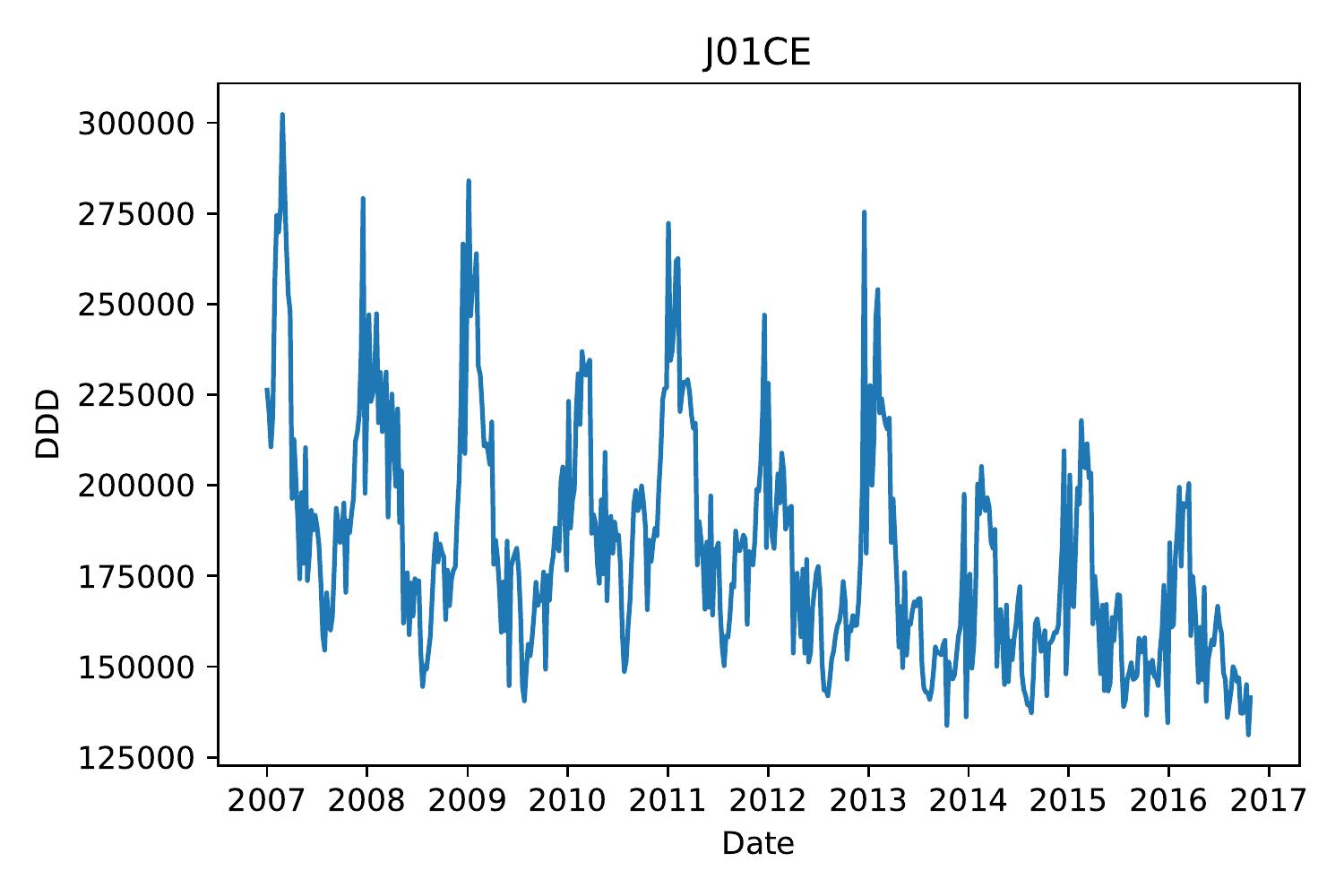}
	\caption{Usage of antimicrobial J01CE in Denmark from 2007 to 2016. DDD denotes \textit{defined daily doses}.}
	\label{fig:j0ce_usage}
\end{figure}

We use weekly data on purchases of antimicrobials for people in Denmark provided by the Danish Health Data Authority from the Register of Medicinal Product Statistics. This covers 96-97\% of all antimicrobial drug consumption in the primary care sector in Denmark.
The data spans the period 1 January 2007 -- 23 October 2016, inclusive. Due to limitations on the web frequency data we only use data from 2011 onwards in the evaluation. The data is collected from pharmacies in Denmark and consists of sales data for several antimicrobial subgroups. Different subgroups are used for different diseases. We focus on the largest subgroup, namely \textit{beta-lactamase sensitive penicillins}, with the Anatomical Therapeutic Chemical (ACT) classification system code J01CE. Usage is quantified as \textit{defined daily doses} (DDD), meaning assumed
average maintenance dose per day for adults for the condition the drug is registered for. Figure \ref{fig:j0ce_usage} shows the sales data of J01CE for the study period. We see that DDD tends to peak once or twice per year, and that from 2014 until 2016 the overall yearly range of DDD has decreased. There is a noticeable change in usage from 2014 onwards. This change is probably due to national campaigns aimed at reducing usage and change in usage to other antimicrobials. Seasonal variations are expected since many diseases treated with antimicrobials, e.g.\ pneumonia, show a seasonal pattern.

\subsection{Web search query frequency}
\label{ss:queryData}
Our second category of data consists of web search queries and their frequencies. We retrieve them from the Google Health Trends API. This is an API maintained
by Google that is similar to Google Trends. Google Health Trends makes it possible to submit a query and receive
aggregated weekly web search frequency data, i.e.\ a time series corresponding to how many times people have
searched for that specific query each week. The Google Health Trends API is based on a uniform sample
of 10\%-15\% of Google web searches. The results correspond to the probability of a short web search session
matching the submitted query. It is possible to restrict the search both with respect to a time period and
geographical region. We restrict our search to data from the period 2 January 2011 -- 23 October 2016, inclusive, and for the geographical region Denmark. We only use data from 2011 and onwards because
Google changed their geographical identification in 2011.


\subsection{Online antimicrobial material}
\label{sec:data_query_selection}
Our third and final category of data consists of freely available online information on antimicrobial drugs, and specifically on: (i) disease names, (ii) drug names, and (iii) descriptions of antimicrobials. For each of the three aspects, we extract data from the websites described below. The reason for collecting the online data on antimicrobials is to use them for selecting web search queries. We describe precisely how we do this in Section \ref{ss:querySelection}.

\subsubsection{Disease names:}\label{antib:diseaseNames} Descriptions of diseases are downloaded from two online resources, both maintained by sundhed.dk (ENG: health.dk) a governmental web site functioning as a digital gateway for citizens to health services, e.g.\ electronic patient journals, hospital treatment records, etc. The two websites, Patienthåndbogen\footnote{https://www.sundhed.dk/borger/patienthaandbogen/} (ENG: The Patient's Handbook) and Lægehåndbogen\footnote{https://www.sundhed.dk/sundhedsfaglig/laegehaandbogen/} (ENG: The Doctor's Handbook), are designed as encyclopedias of diseases. The target audience for The Patient's Handbook is laymen, and for The Doctor's Handbook health professionals.

\subsubsection{Drug names:}\label{antib:drugNames} The organization Dansk Lægemiddel Information A/S (ENG: Danish Drug Information) maintains two
websites, min.medicin\footnote{\url{http://min.medicin.dk/}} (ENG: My Medicine), and pro.medicin\footnote{\url{http://pro.medicin.dk/}} (ENG: Pro Medicine), with descriptions of drugs available on the Danish drug market. The organization is funded by the medical industry and the Danish government. My Medicine targets laymen, while Pro Medicine targets health professionals.

\subsubsection{Descriptions of antimicrobials:}\label{antib:descriptions} We choose descriptions from four websites describing to laymen what antimicrobials are and when to use them: \url{www.ssi.dk}, \url{www.netdoctor.dk}, \url{www.sundhed.dk}, and \url{www.antibiotikaellerej.dk}. The four websites are maintained by the following four groups: Danish Center for Disease Control, netdoctor.dk (a leading Danish health information website), sundhed.dk, and finally a collaboration of the Danish government, the pharmacist union, the doctors union, the society for general practitioners and the Danish Center for Disease Control. We consider all of the four groups to be authoritative and neutral.

\section{Prediction of antimicrobial drug\\ consumption}
\label{s:prediction}
The goal is to predict antimicrobial drug consumption using web search data. We do this in the following three steps: First, we select web search queries that are likely to indicate antimicrobial drug consumption (Section \ref{ss:querySelection}); then, for each query frequency time series we generate a number of lagged versions and decide which lags should be used for the prediction (Section \ref{sec:feature_selection}); and finally we use appropriate prediction models to infer antimicrobial drug consumption (Section \ref{ss:predictionModels}). 

\subsection{Query selection}
 \label{ss:querySelection}
 
For our analysis we use web search queries, and their frequencies, retrieved from Google Health Trends, as described in Section \ref{ss:queryData}. We select these queries based on the online antimicrobials material described in Section \ref{sec:data_query_selection} as follows. 

We start with an empty set of queries and a set of seed words. Our seed words are the ATC code ``J01CE'', and the individual
nouns of the antimicrobial name: ``penicillin'', ``penicilliner'' (plural form of penicillin) and ``beta-lactamase''. Using
these seed words, we populate the set of queries in the following way:

\paragraph{Disease names (described in Section \ref{antib:diseaseNames}):} A disease name is added to the set of queries if the description of the treatment for a disease mentions one of the seed words.

\paragraph{Drugs names (described in Section \ref{antib:drugNames}):} A drug name is added to the set of queries if the description of the active substances mentions one of the seed words, or if the sub-heading of the drug description page contains one of the seed words.

\paragraph{Description of antimicrobials (described in Section \ref{antib:descriptions}):} For each of the four descriptions (found in each of the four websites described in Section \ref{antib:descriptions}) we extract all unique words and remove stop words. 
We use as stop word set the 100 most frequent words in CorpusDK \cite{DSL2002}, a corpus of text representing written Danish around year 2000. Based on these four sets of words, we select two partially overlapping sets of queries: (i) Words that occur in at least two descriptions of antimicrobials or antimicrobial usage targeting laymen, and (ii) words that occur in at least three descriptions of antimicrobials or antimicrobial usage targeting laymen.

The above process results in the six sets of queries overviewed below. 
\begin{enumerate}
	\item \textbf{Disease names pro:} Disease names used by health professionals.
	\item \textbf{Disease names lay:} Disease names used by laymen.
	\item \textbf{Drug names pro:} Drug names used by health professionals.
	\item \textbf{Drug names lay:} Drug names used by laymen.
	\item \textbf{Descriptions lay:} Words co-occurring in two descriptions of antimicrobials for laymen.
	\item \textbf{Descriptions lay frequent:} Words co-occurring in three descriptions of antimicrobials for laymen.
\end{enumerate}

\begin{table}
	\centering
	\begin{tabular}{lr}
		\toprule
		Query sets & \#Queries \\
		\midrule
		Disease names pro & 47 \\
		Disease names lay & 11 \\
		Drug names pro    & 7 \\
		Drug names lay    & 8 \\
		Descriptions lay  & 72 \\
		Descriptions lay frequent & 18 \\
		\bottomrule
	\end{tabular}
	\caption{Number of queries in each query set.}
	\label{table:query_count}
\end{table}

Table \ref{table:query_count} displays the number of queries in each query set. We see that the highest number of queries is generated from \textit{Descriptions lay frequent}, followed by \textit{Disease names pro}. The queries generated from drug names (both \textit{Drug names pro} and \textit{Drug names lay}) are by far the fewest.

\subsection{Time lag selection}
\label{sec:feature_selection}
Each query in the query sets described above has an associated time series (time stamps and search frequency). It is not unlikely that there exist lagged effects, for example increased search activity in one week might correspond to increased antimicrobial drug consumption two weeks after.
To account for such effects we generate a number of lagged versions of each query frequency time series and include only a subset of them in our prediction models.
We select these by fitting a linear model with the antimicrobial drug consumption data as target variable and lagged versions of the query frequencies as predictors:

\begin{equation}
	y_t = \sum_{l=1}^L \sum_{i=1}^N \beta_{(l-1) N+i} Q_{t-l, i},
\end{equation}

\noindent where $y_t$ denotes the antimicrobial drug consumption at time $t$, $L$ is the number of lags, $N$ the number of queries, $Q_{t,i}$ is the query frequency at time $t$ for query $i$, and $\beta$ is the model coefficient. Each model coefficient is related to a variable, i.e.\ a query and a time offset, and the size of the coefficient defines the importance of the variable in the prediction model. We use the size of the coefficients for selecting queries and corresponding lags, which is described below.

To fit the model, we use Elastic Net which combines L1 and L2 regularization. Two hyperparameters, $\lambda_1$ and $\lambda_2$, control the L1 and L2 penalization for the Elastic Net regularization. Using matrix notation, the function being minimized can be written as: 

\begin{equation}\label{eq:elasticnet}
\| y- \beta X \|^2 + \lambda_1 \| \beta \|_1 + \lambda_2 \| \beta \|^2_2
\end{equation}

\noindent where $y$ is a vector with the target values, and $X$ is a matrix with lagged query frequency time series. Elastic Net is well suited for problems where the number of variables is much larger than the number of training samples \cite{zou2005regularization}, which can be the case in our setups if many lags are used. In addition, Elastic Net groups correlated features and, either keeps them in the model, i.e.\ non-zero coefficient, or leaves them out \cite{zou2005regularization}. This is a useful attribute for query selection, since we would like to select all correlated queries.
To select queries, we sort the queries according to the absolute value of the coefficients, and pick the 100 with highest absolute value. While Elastic Net can be used for query selection without any threshold, i.e.\ by removing features with zero valued coefficients, there is no upper bound on the number of features. We therefore enforce a hard threshold of 100 queries. 


\subsection{Prediction Models}
\label{ss:predictionModels}
Our goal is to predict antimicrobial drug consumption, i.e.\ the number of DDDs of antimicrobials being consumed per week. 
To this end, we use two types of prediction models: (i) Linear models with Elastic Net regularization presented in Section \ref{sec:linear}. This is a common approach often used when predicting with web search data \cite{ginsberg2009detecting, yang2015accurate}. (ii) Gaussian Processes, which are capable of capturing non-linearities in the data presented in Section \ref{sec:gp}. These models have successfully been applied to web search data to improve
predictive performance \cite{lampos2015advances, www17_lampos}. 

For both prediction models we use three setups for our predictions: 
(i) using only web search data,
(ii) using only using historic antimicrobial drug consumption data, and 
(iii) combining historic antimicrobial drug consumption data and web search data. 
We explain our prediction models next. 

\subsubsection{Linear models for antimicrobial drug consumption prediction}
\label{sec:linear}

We use linear models because they are easy to fit and to interpret, allowing us to draw direct inferences between their output and the real life prediction problem at hand. Using only web search data, the prediction model is defined as:

\begin{equation}
	y_{t+p} = \beta_0 + \sum_{i=1}^N \beta_i Q_{t,i}
\end{equation}

\noindent where $y_{t+p}$ is the antimicrobial drug consumption data at time $t$ with a prediction offset of $p$, $N$ is the number of queries, $Q_{t,i}$ is the query frequency at time $t$ for query $i$, and the $\beta$s are the model coefficients. 

When using only antimicrobial drug consumption data, we use a standard autoregressive model definition:

\begin{equation}
	y_{t+p} = \alpha_0 + \sum_{j=1}^M \alpha_j y_{t-j}
\end{equation}

\noindent where $M$ denotes the number of autoregressive terms, and the $\alpha$s are the model coefficients. This is a similar model to the one used in \cite{lazer2014parable}.

To make predictions using both web search data and historic antimicrobial drug consumption data, we combine the two above models into a single model closely resembling the approach described in \cite{yang2015accurate}. The model is as follows:

\begin{equation}
	 y_{t+p} = \theta_0 + \sum_{j=1}^M \theta_j y_{t-j} + \sum_{i=1}^N \theta_{i+M} Q_{t,i},
\end{equation}

\noindent where the $\theta$s are the model coefficients, and the remaining notation is as defined above. 

For all the linear models we use Elastic Net regularization for estimating the model coefficients, as described in Equation \ref{eq:elasticnet}. The two hyperparameters $\lambda_1$ and $\lambda_2$ are found using three fold cross-validation on the training data.

\subsubsection{Gaussian Processes for antimicrobial drug consumption prediction}
\label{sec:gp}

Gaussian Processes (GP) are probability distributions over functions, where any finite set of function values have a joint Gaussian distribution \cite{Rasmussen2004}. We focus on GP that learn functions that map from our input space of size $m$ to a single valued output, i.e.\ $f:R^m \rightarrow R$. The size of the input space is defined by either the number of queries, autoregressive terms, or a combination of the two. 

The functions drawn from a GP can be described by two functions: A mean function and a covariance function. The mean function is defined as:

\begin{equation}
	E[f(x)] = \mu(x)
\end{equation}

\noindent where $x$ is our input data, and $\mu$ denotes the mean of the function distribution at point $x$. The covariance function is defined as:

\begin{equation}
	Cov[f(x), f(x')] = k(x, x')
\end{equation}

\noindent where $x$ and $x'$ are two input vectors, and $k$ is a kernel function \cite{Rasmussen2004}.  When working with GP it is customary to assume that the mean value is zero, and focus only on the covariance/kernel. The covariance function defines prior covariance between two input values and is typically controlled using two parameters: length scale and variance. For the Squared Exponential covariance function that we use, the variance defines the average distance from the mean, and the length scale defines how quickly the underlying signal changes, i.e.\ the antimicrobial drug consumption. Given the input data and the covariance function, it is possible to automatically infer the optimal parameters of the model given the data. A feature of GP is that covariance functions can be combined. This property can be used to create new covariance functions that can capture several aspects of the data. For example, combining covariance functions with different length scales could be used to model slow changes and quick changes. We use this property below.

We use two different setups, one for experiments involving web search data, and one when only historical antimicrobials data is used. The covariance function we use for web data is the Matern covariance function which allows for adapting to non-smooth changes by varying the parameter $\nu$, and is defined as:

\begin{equation}
	k_m^\nu (x, x') = \sigma^2 \frac{2^{1-\nu}}{\Gamma(\nu)} \left( \frac{\sqrt{2\nu} r}{l} \right)^\nu K_\nu \left( \frac{\sqrt{2\nu} r}{l} \right),
\end{equation}

\noindent where $\nu$ is a parameter that in our case is set to $3/2$ (a common choice), $r$ is $|x-x'|$, $l$ is the length scale, $\sigma^2$ is the variance, and $K_\nu$ is a modified Bessel function \cite{Rasmussen2004}.  To model many different types of
behaviour we use the additive properties of the covariance
function and generate a new covariance function, $k_{web}$, for the web
search data consisting of 10 Matern functions:

\begin{equation}
	k_{web} (x, x') = \sum_{i=1}^{10} k_m^{\nu=3/2} (x, x'; \sigma_i, l_i)+N(\sigma_{11}),
\end{equation}

\noindent where $N(\sigma_{11})$ is Gaussian distributed noise.

When only working with historical antimicrobial drug consumption data, we use two other covariance functions: the linear and the squared exponential (SE). The linear covariance function can capture upwards or downwards trends in the data, and the SE function can capture short-term temporal variations in the data. The SE covariance function is defined as:

\begin{equation}
	k_{SE} (x, x') = \sigma^2 exp\left( \frac{-(x-x')^2}{2l^2} \right),
\end{equation}

\noindent where $l$ is the length scale, and $\sigma^2$ the variance. The linear kernel is defined as:

\begin{equation}
	k_{lin}(x, x') = \sigma^2 x^Tx',
\end{equation}

\noindent with $\sigma^2$ as variance.  Combing the two covariance functions we get a new
covariance function that we use for the antimicrobial drug consumption data. We denote the covariance function $k_{antimicrobial}$ and define it as follows:

\begin{equation}
	k_{antimicrobial}(x, x') = k_{SE} (x, x'; \sigma_1, l_1) + k_{lin}(x, x'; \sigma_2)+N(\sigma_3).
\end{equation}

Parameters for all models are found using gradient descent. Different co-variance functions have been tested, and we report the results of the setup with the lowest error.

\section{Experimental evaluation}

\subsection{Experimental setup}

To simulate a real world prediction situation we test the models in a leave-one-out fashion, where we re-train the prediction model after each time step such that all available data is used. This is a common setup in health event prediction \cite{yang2015accurate, www17, lampos2015advances}.

The number of autoregressive antimicrobial drug consumption terms varies between 4, 26 and 130 weeks. For the web search data we generate queries with a maximum lag of 4, 26 and 130 weeks. Four different prediction offsets are tested: 0, 4, 8, and 12 weeks. The prediction offset denotes how far into the future we are predicting. For example, an offset of 0 means that antimicrobial drug consumption in week $t$ is predicted using web data and historic antimicrobial drug consumption data from weeks prior to $t$. For an offset of 4 the antimicrobial drug consumption in week $t+4$ is predicted using web data and historic antimicrobial drug consumption data from weeks prior to $t$. 

The web search data covers 2 January 2011 -- 23 October 2016, in total 304 weeks of data. Queries are selected using the first 104 weeks of data. Each experiment uses as a minimum 104 weeks of training data for model fitting. 
With the 12 weeks of prediction offset and up 130 weeks of autoregressive terms, we end up with an evaluation period of the 58 weeks leading up to 23 October 2016. 


We evaluate predictions using the root mean squared error (RMSE) and mean absolute error (MAE). A feature of the RMSE is that large prediction errors receive a bigger penalty than small errors. This intuitively means that few large errors will result in a larger RMSE than many small errors. The MAE, on the other hand, assigns equal weight to all errors, and the final score is therefore easier to interpret. With respect to our data, a MAE of 10000 corresponds to the prediction on average being 10000 DDDs off on every weekly prediction. This corresponds to approximately 6\% of the weekly average of DDDs.

The RMSE is calculated as:

\begin{equation}
	RMSE = \sqrt{1/N \sum_{t=1}^N(y_t - \hat{y}_t)^2},
\end{equation}

\noindent where $y_t$ is the true value at time $t$, $\hat{y}_t$ is the predicted value at time $t$, and $N$ the number of predictions. The MAE is calculated as:

\begin{equation}
	MAE = 1/N \sum_{t=1}^N |y_t - \hat{y}_t|.
\end{equation}


\subsection{Experimental results}



\begin{table}
	\centering
	\begin{tabular}{p{1.1cm}lrr}
		\toprule
		Offset    & Data Source        &      GP & Elastic Net \\ \midrule
		\multirow{3}{*}{0 weeks}  & Web only           & 11011.0 &     11096.9 \\
		                    & Antimicrobial only & 11446.0 &      9980.9 \\
		                    & Web \& antimicrobial           & 10644.3 &      \textbf{9970.8}\vspace{0.1cm} \\
		\multirow{3}{*}{4 weeks}  & Web only           & 11270.7 &     11398.1 \\
		                    & Antimicrobial only & 11498.2 &      9990.3 \\
		                    & Web \& antimicrobial           & 10576.2 &      \textbf{9989.9}\vspace{0.1cm} \\
		\multirow{3}{*}{8 weeks}  & Web only           & 11026.1 &     11142.5 \\
		                    & Antimicrobial only & 11401.6 &     10301.0 \\
		                    & Web \& antimicrobial           & 10564.4 &      \textbf{9781.0}\vspace{0.1cm} \\
		\multirow{3}{*}{12 weeks} & Web only           & 10977.8 &     11189.1 \\
		                    & Antimicrobial only & 11231.0 &     10424.0 \\
		                    & Web \& antimicrobial           & 10249.1 &      \textbf{9644.3} \\ \bottomrule
	\end{tabular}
	\caption{RMSE for best performing prediction (among all query sets) with the non-linear (GP) and linear (Elastic Net) prediction model, using web, antimicrobial purchase data, and their combination. Lowest error per offset is in bold.}
	\label{tabel:rmse}
\end{table}

\begin{table}
	\centering
	\begin{tabular}{p{1.1cm}lrr}
		\toprule
		Offset      & Data Source        &     GP & Elastic Net \\ \midrule
		\multirow{3}{*}{0 weeks}  & Web only           & 8463.4 &      8507.7 \\
		                    & Antimicrobial only & 8571.0 &      7294.5 \\
		                    & Web \& antimicrobial            & 8105.3 &      \textbf{7282.7}\vspace{0.1cm} \\
		\multirow{3}{*}{4 weeks}  & Web only           & 8938.2 &      9040.8 \\
		                    & Antimicrobial only & 9265.8 &      \textbf{7989.9} \\
		                    & Web \& antimicrobial            & 8440.0 &      8073.2\vspace{0.1cm} \\
		\multirow{3}{*}{8 weeks}  & Web only           & 8596.4 &      8579.8 \\
		                    & Antimicrobial only & 9053.2 &      8200.7 \\
		                    & Web \& antimicrobial            & 8311.2 &      \textbf{7849.9}\vspace{0.1cm} \\
		\multirow{3}{*}{12 weeks} & Web only           & 8258.8 &      8463.5 \\
		                    & Antimicrobial only & 8997.7 &      8257.6 \\
		                    & Web \& antimicrobial            & 8056.7 &      \textbf{7750.0} \\ \bottomrule
	\end{tabular}
	\caption{MAE for best performing prediction (among all query sets) with the non-linear (GP) and linear (Elastic Net) prediction model, using web, antimicrobial purchase data, and their combination. Lowest error per offset is in bold.}	
	\label{tabel:mae}
\end{table}

\begin{table}
	\centering
	\begin{tabular}{p{1.1cm}lrr}
		\toprule
		Offset                    & Data Source          &    GP &                  Elastic Net \\ \midrule
		\multirow{3}{*}{0 weeks}  & Web only             & 5.4\% &                        5.4\% \\
		                          & Antimicrobial only   & 5.4\% &               \textbf{4.6}\% \\
		                          & Web \& antimicrobial & 5.1\% & \textbf{4.6}\%\vspace{0.1cm} \\
		\multirow{3}{*}{4 weeks}  & Web only             & 5.7\% &                        5.7\% \\
		                          & Antimicrobial only   & 5.9\% &               \textbf{5.1}\% \\
		                          & Web \& antimicrobial & 5.4\% & \textbf{5.1}\%\vspace{0.1cm} \\
		\multirow{3}{*}{8 weeks}  & Web only             & 5.5\% &                        5.5\% \\
		                          & Antimicrobial only   & 5.8\% &                        5.2\% \\
		                          & Web \& antimicrobial & 5.3\% & \textbf{5.0}\%\vspace{0.1cm} \\
		\multirow{3}{*}{12 weeks} & Web only             & 5.2\% &                        5.4\% \\
		                          & Antimicrobial only   & 5.7\% &                        5.2\% \\
		                          & Web \& antimicrobial & 5.1\% &               \textbf{4.9}\% \\ \bottomrule
	\end{tabular}
	\caption{MAE as a percentage of the average weekly antimicrobial usage for the 58 week evaluation period.}
	\label{tabel:mae_percentage}
\end{table}

\begin{figure}
	\centering
	\includegraphics[width=10cm]{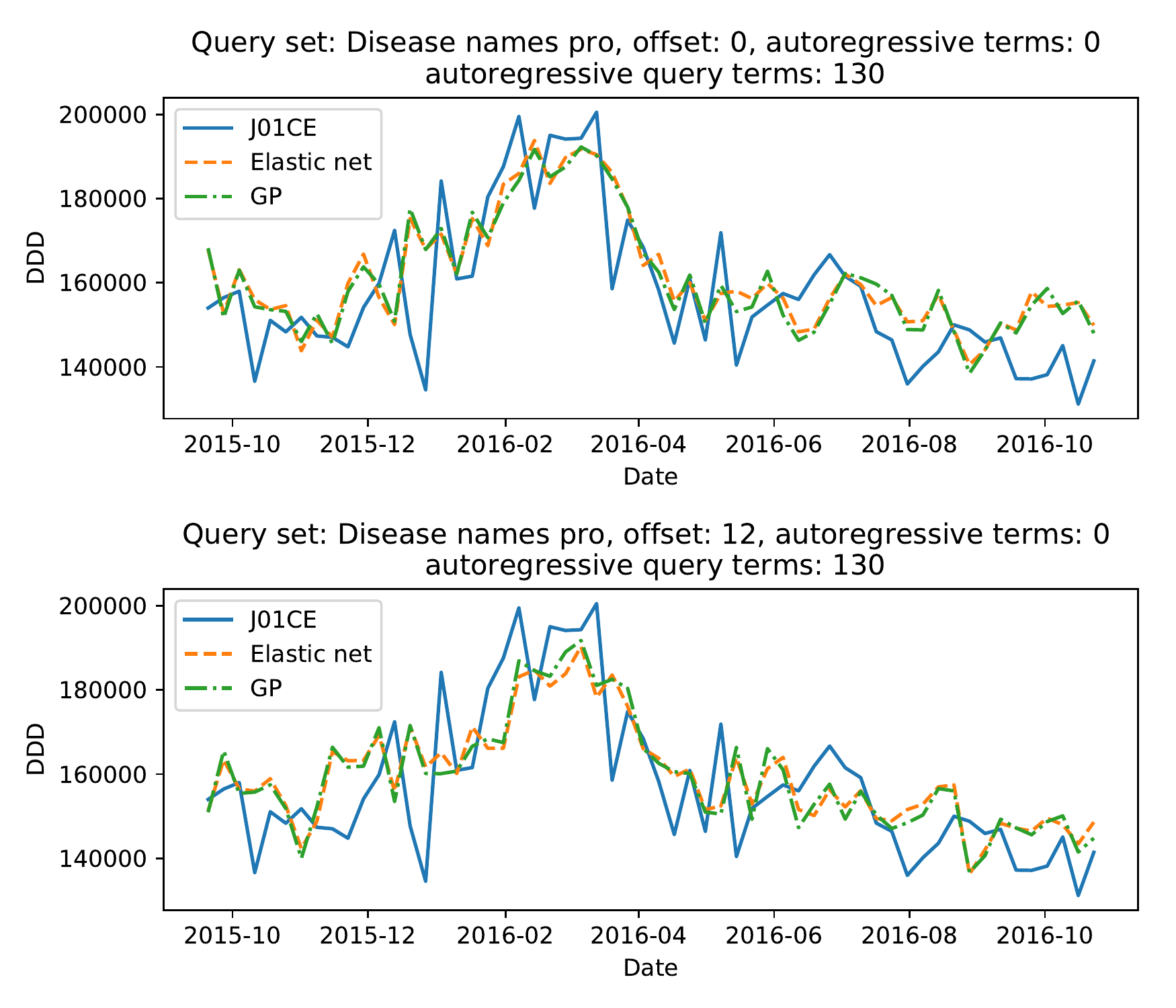}
	\caption{Prediction using only web search data.}
	\label{fig:prediction_web}
\end{figure}

\begin{figure}
	\centering
	\includegraphics[width=10cm]{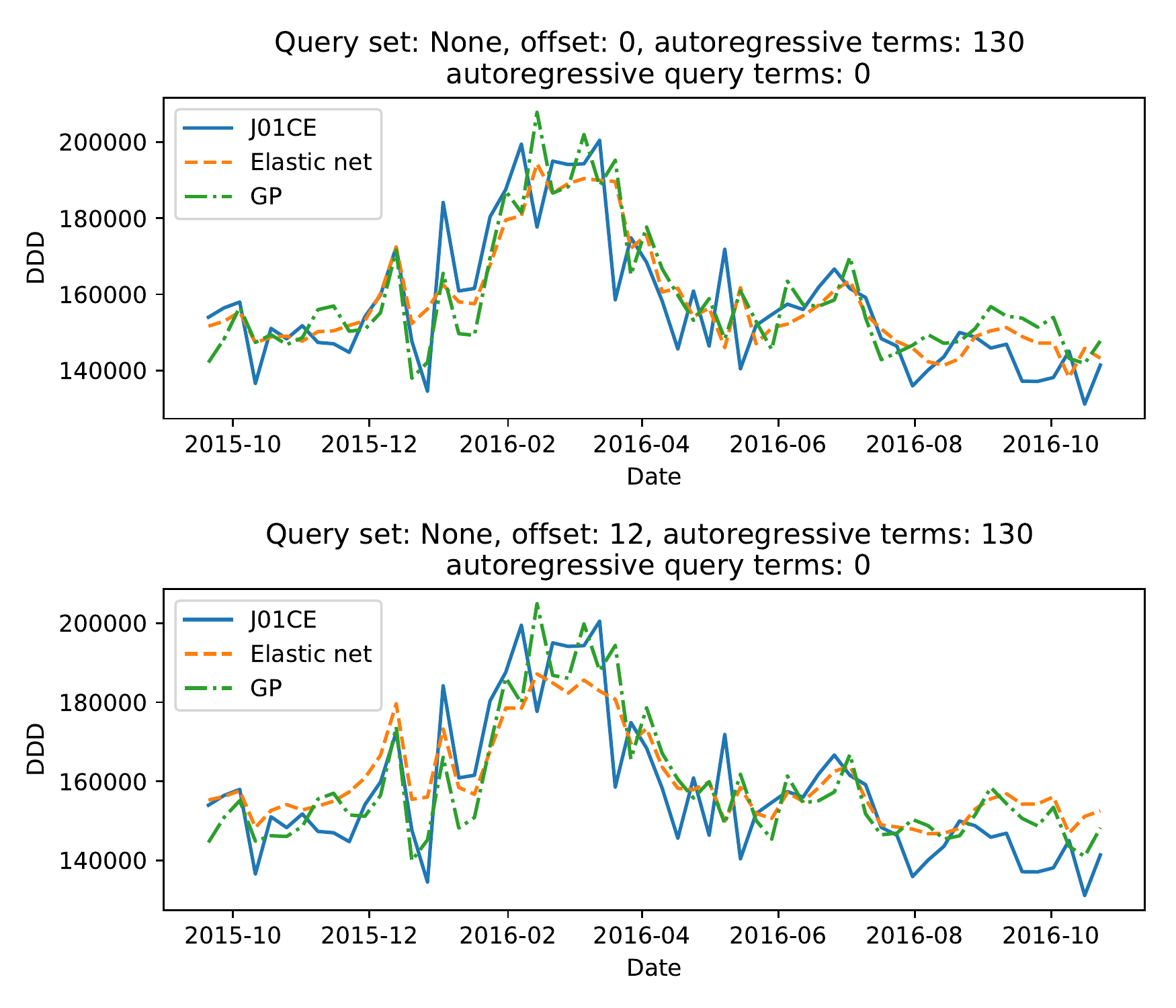}
	\caption{Prediction using only historic antimicrobial purchase data.}
	\label{fig:prediction_antimicrobial}
\end{figure}

\begin{figure}
	\centering
	\includegraphics[width=10cm]{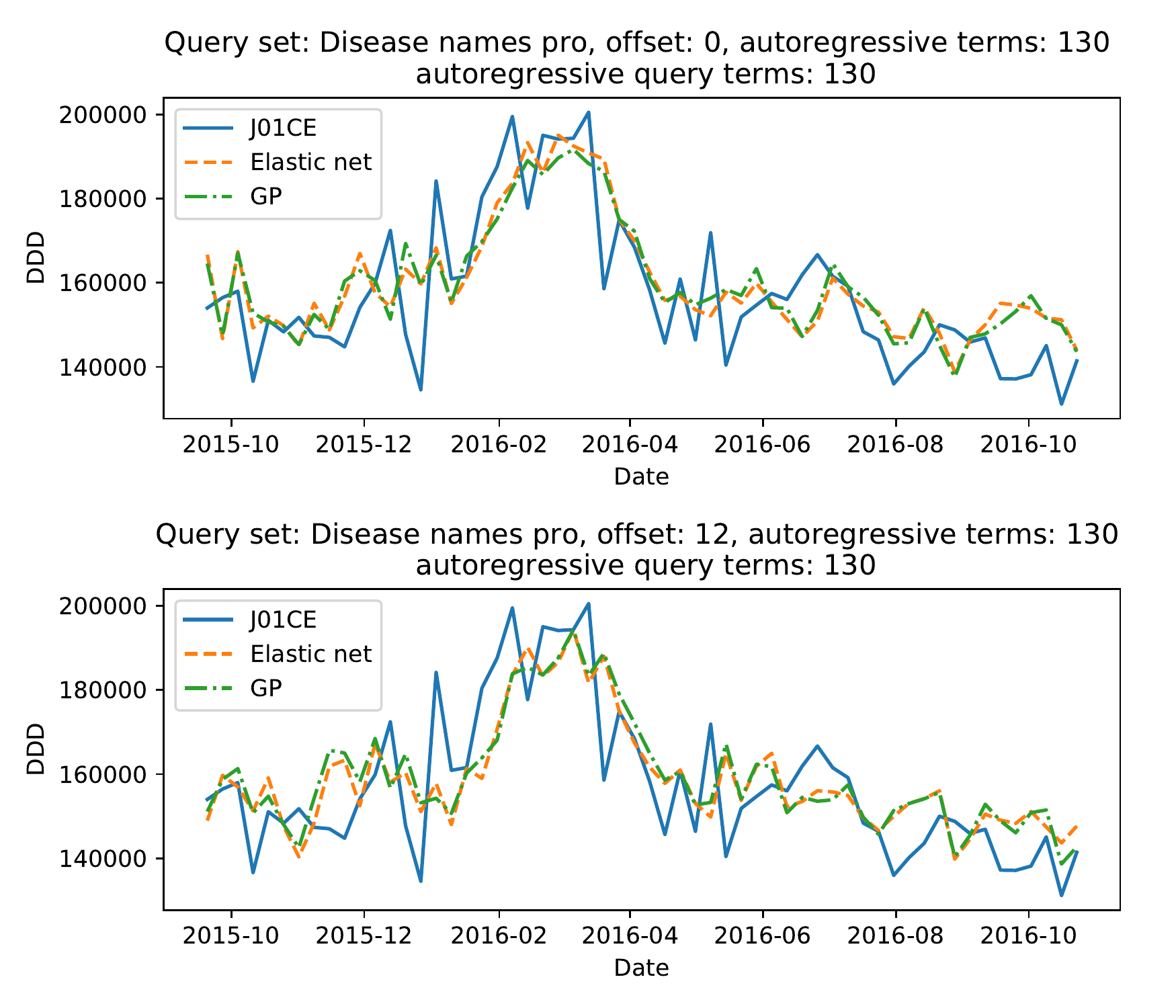}
	\caption{Prediction using both web and antimicrobial purchase data.}
	\label{fig:prediction_combined}
\end{figure}

Tables \ref{tabel:rmse} \& \ref{tabel:mae} show the RMSE and MAE when predicting antimicrobial drug consumption using (i) only web search data, (ii) only historic antimicrobial purchase data, and (iii) both web search and antimicrobial purchase data. Only the best performance (lowest error) is reported per data source, prediction model, and offset. 

We see that predictions based on a combination of web and antimicrobial purchase data give almost always the lowest error. With Gaussian Processes as the prediction model, predictions based only on web data outperform those based on antimicrobial purchase data; with Elastic Net, the situation is reversed: predictions based on antimicrobial purchase data outperform those based on web data.
Comparing the two prediction models (Gaussian Processes and Elastic Net), we only observe minor differences, generally favouring the linear models. These results fit well with the general high prevalence of linear models for prediction using web search data. Though it is curious why Gaussian Processes in other papers have outperformed Elastic Net on similar tasks \cite{www17_lampos, lampos2015advances}. One explanation could be that the feature selection described in Section \ref{sec:feature_selection} uses Elastic Net, which means that we are selecting features that have a linear relationship with the target variable, i.e.\ antimicrobial drug consumption. In other words, we have a priori selected features that are well suited for the Elastic Net model.

Overall, the fluctuations in error (both RMSE and MAE) across the different prediction models, data sources, and offsets are generally small. This is also illustrated in Table \ref{tabel:mae_percentage}, which displays the MAE scores of Table \ref{tabel:mae} as the percentage of average weekly consumption in the evaluation period.  We see that our predictions are, in the best case, off by 4.6\% of the average weekly consumption, and in the worst case by 5.9\%.

We also observe in Tables  \ref{tabel:rmse} -- \ref{tabel:mae_percentage} that error remains generally stable independent of offset. Looking back at Figure \ref{fig:j0ce_usage} we observe two things which might explain this: (i) The antimicrobial  drug consumption is strongly seasonal, therefore models that capture the latent seasonality will perform well even with a large offset. (ii) As will be described next, many of the queries with highest model coefficients have approximately 1 year lag. Combined with the fact that the consumption patterns in the last three years of the time series are very similar, we should expect to be able to predict the antimicrobial drug consumption relatively accurately one year into the future.


In Figures \ref{fig:prediction_web} -- \ref{fig:prediction_antimicrobial} we further plot the predictions by the two models (GP and Elastic Net) using only web or only antimicrobial purchase data against actual antimicrobial purchase data (J01CE). The precise settings of these four runs are stated in the figure titles. 
We see that, when using web data only, seasonal variations are generally captured; however, a drop in antimicrobial drug consumption in January 2016 is not captured. Visually, the difference between a 0 week offset and a 12 week offset is hard to spot. When using only antimicrobial purchase data, on the other hand, the GP model captures the drop in consumption in January 2016, both with a 0 week offset and a 12 week offset.  Again differences between 0 week offset and 12 week offset are negligible. Finally, Figure \ref{fig:prediction_combined} shows the combination of the two data sources. Here neither model captures the drop in January 2016. Differences between 0 week offset and 12 week offset are, as before, minor.

Overall, we find that the use of web data only gives predictions that are slightly more erroneous, but generally not that far off, from those made when using only historical antimicrobial purchase data. This is valuable for countries lacking timely access to centralised antimicrobial purchase data, because it means that we can approximate predictions that are roughly less than 1\% point erroneous compared to those using antimicrobial purchase data (for the same offset -- cf. Table \ref{tabel:mae_percentage}). This performance appears generally stable across different prediction offsets and linear (Elastic Net) vs non-linear (GP) prediction models.

Next we analyse the impact of web search query selection to prediction performance.

\subsubsection{Web search query analysis}
\label{sec:query_set_performance}

\begin{figure*}
	\begin{subfigure}[b]{0.32\textwidth}
		\includegraphics[width=\linewidth]{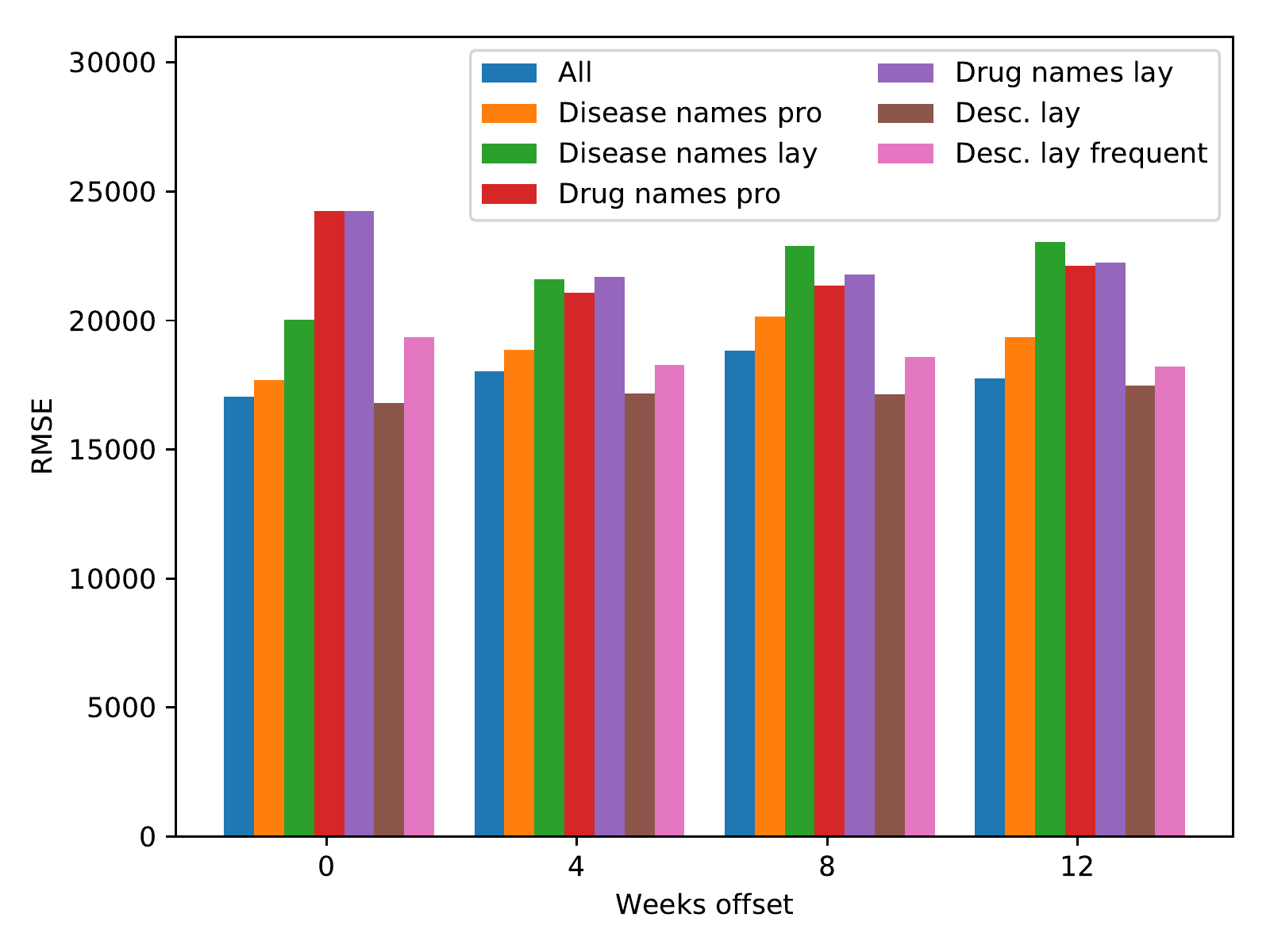}
		\caption{Up to 4 weeks lag}
	\end{subfigure}
	\begin{subfigure}[b]{0.32\textwidth}
		\includegraphics[width=\linewidth]{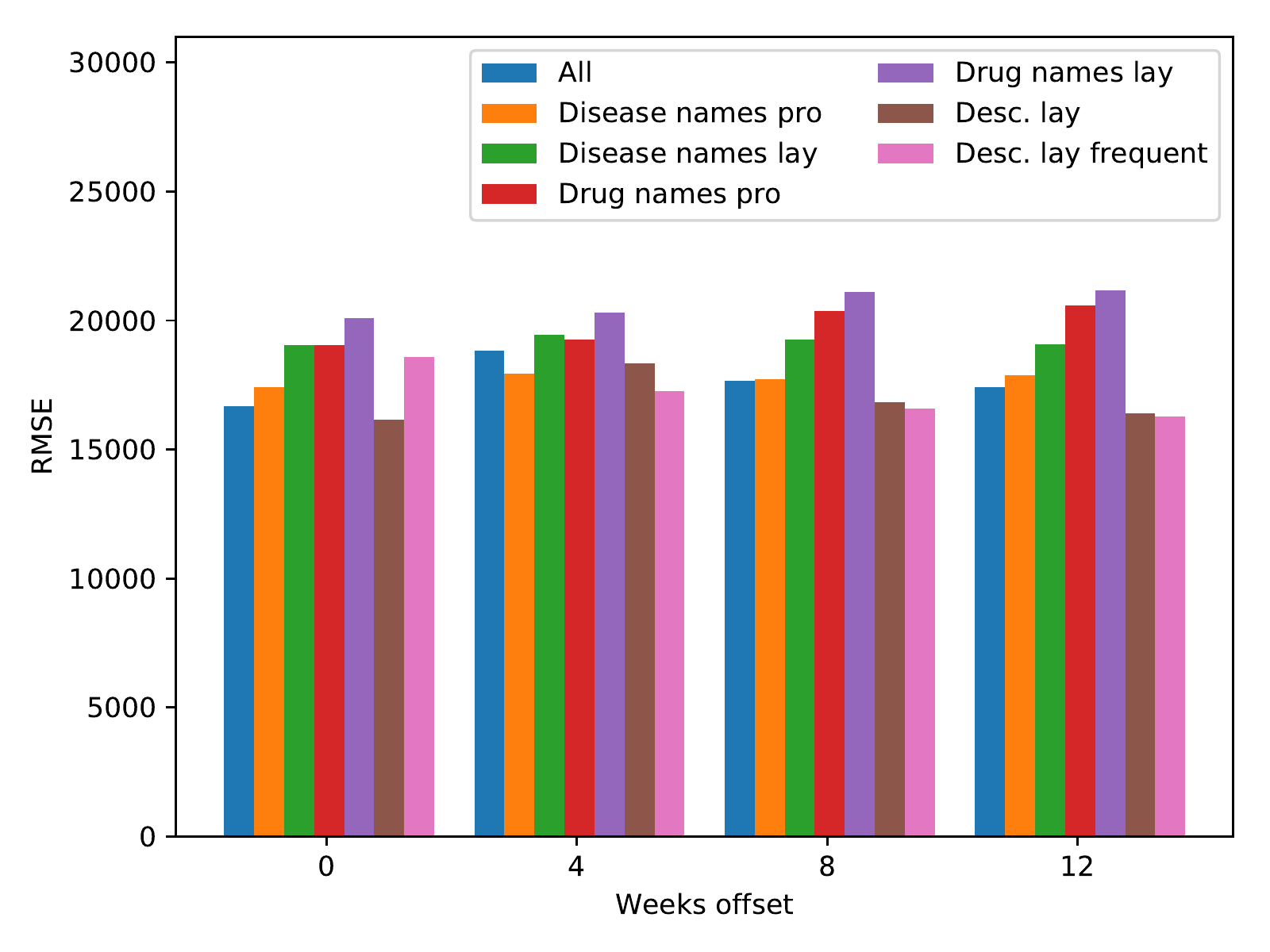}
		\caption{Up to 26 weeks lag}
	\end{subfigure}
	\begin{subfigure}[b]{0.32\textwidth}
		\includegraphics[width=\linewidth]{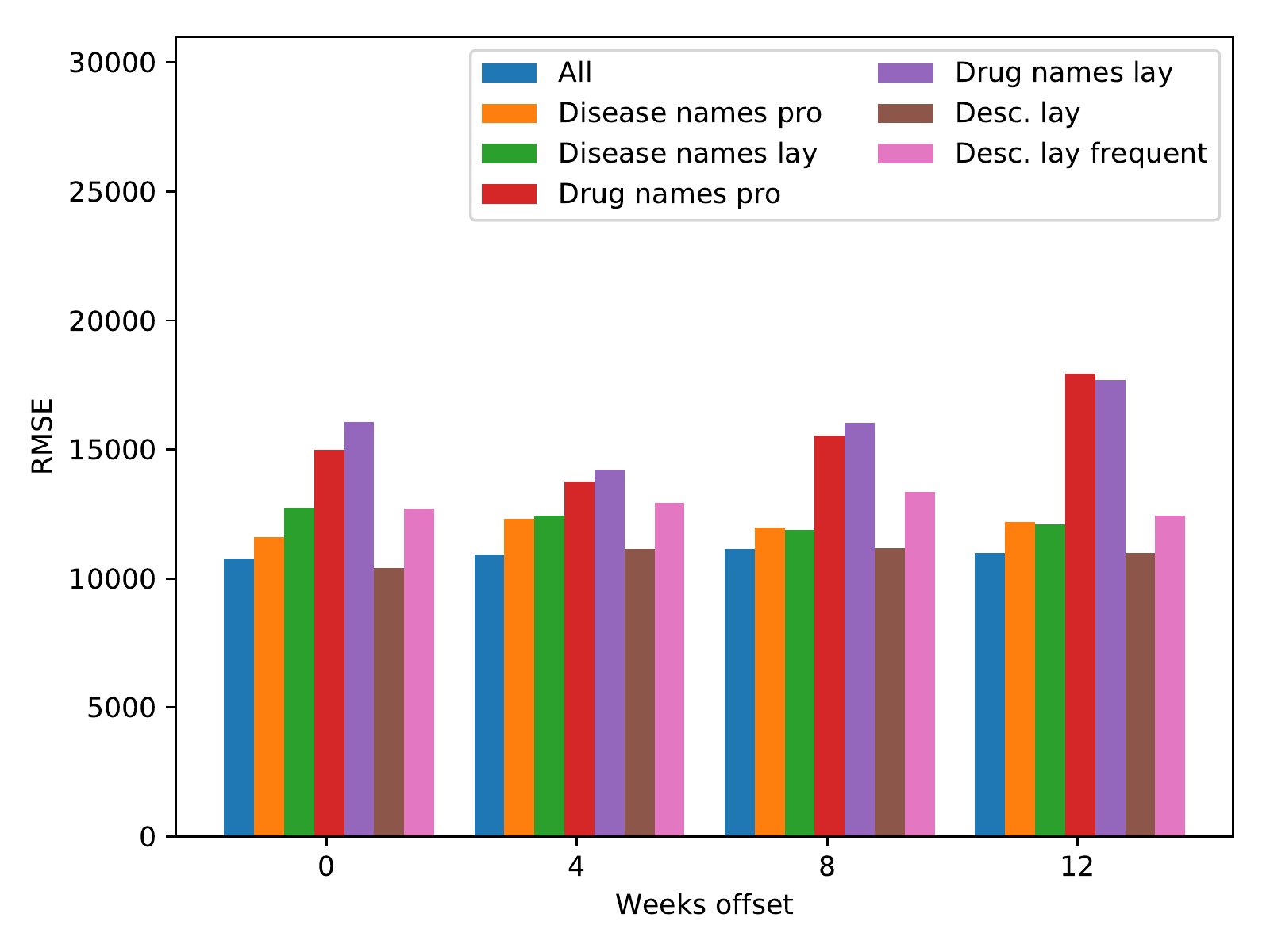}
		\caption{Up to 130 weeks lag}
	\end{subfigure}
	\caption{Prediction based on web search data with different query sets and lags, using Gaussian Processes.}
	\label{fig:query_set_performance_web-gp}

\end{figure*}

\begin{figure*}
	
	\begin{subfigure}[b]{0.32\textwidth}
		\includegraphics[width=\linewidth]{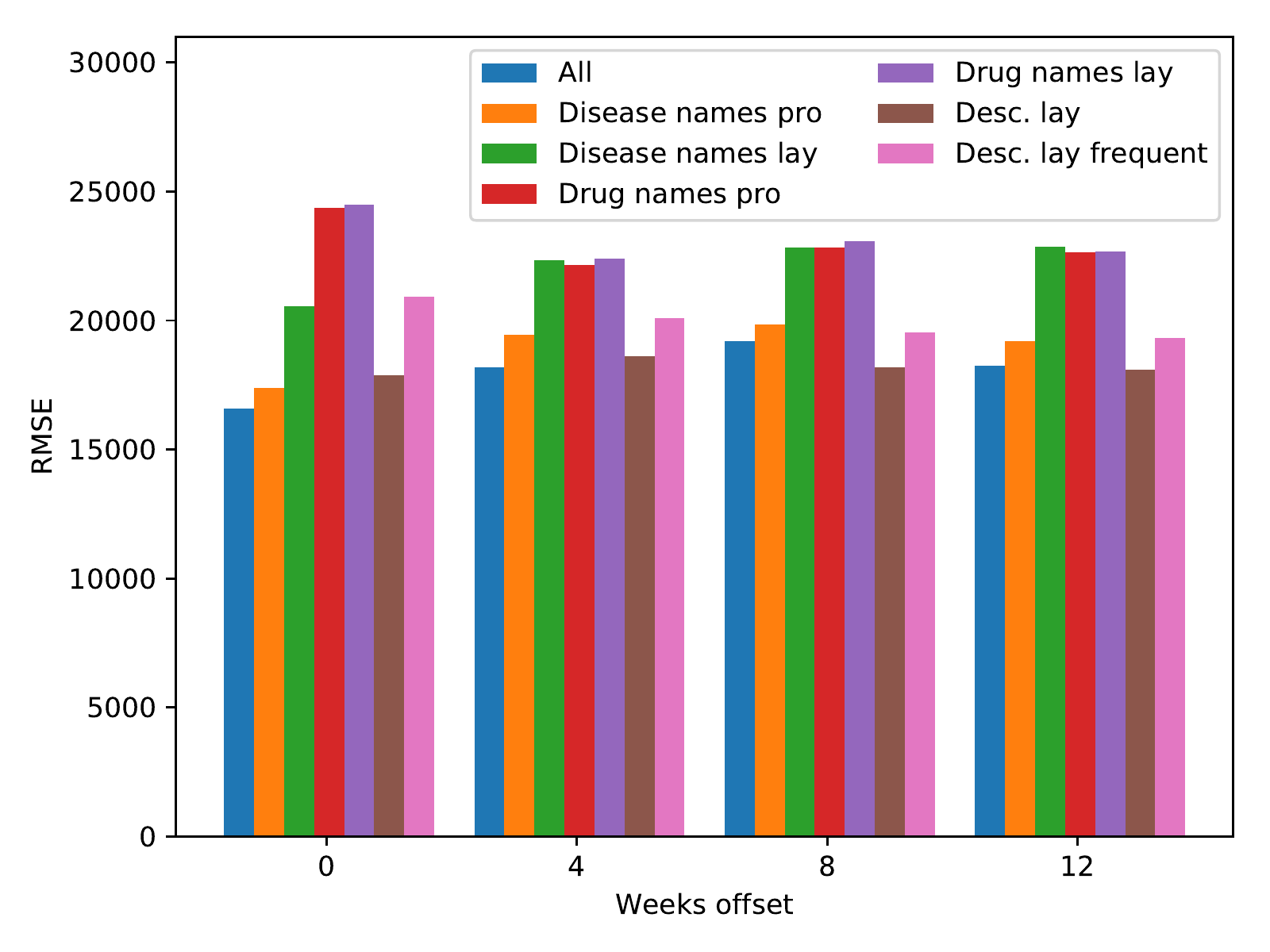}
		\caption{Up to 4 weeks lag}
	\end{subfigure}
	\begin{subfigure}[b]{0.32\textwidth}
		\includegraphics[width=\linewidth]{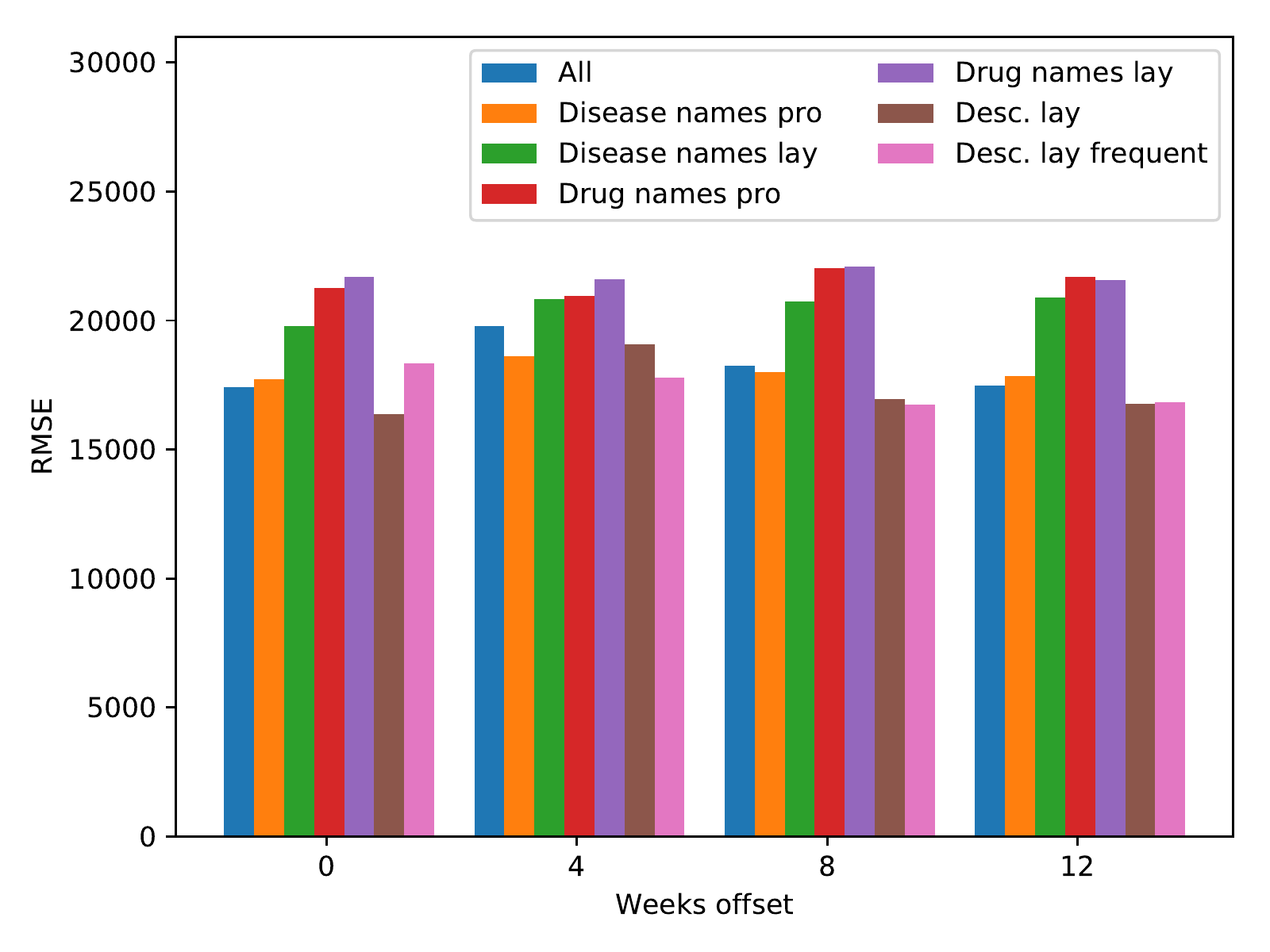}
		\caption{Up to 26 weeks lag}
	\end{subfigure}
	\begin{subfigure}[b]{0.32\textwidth}
		\includegraphics[width=\linewidth]{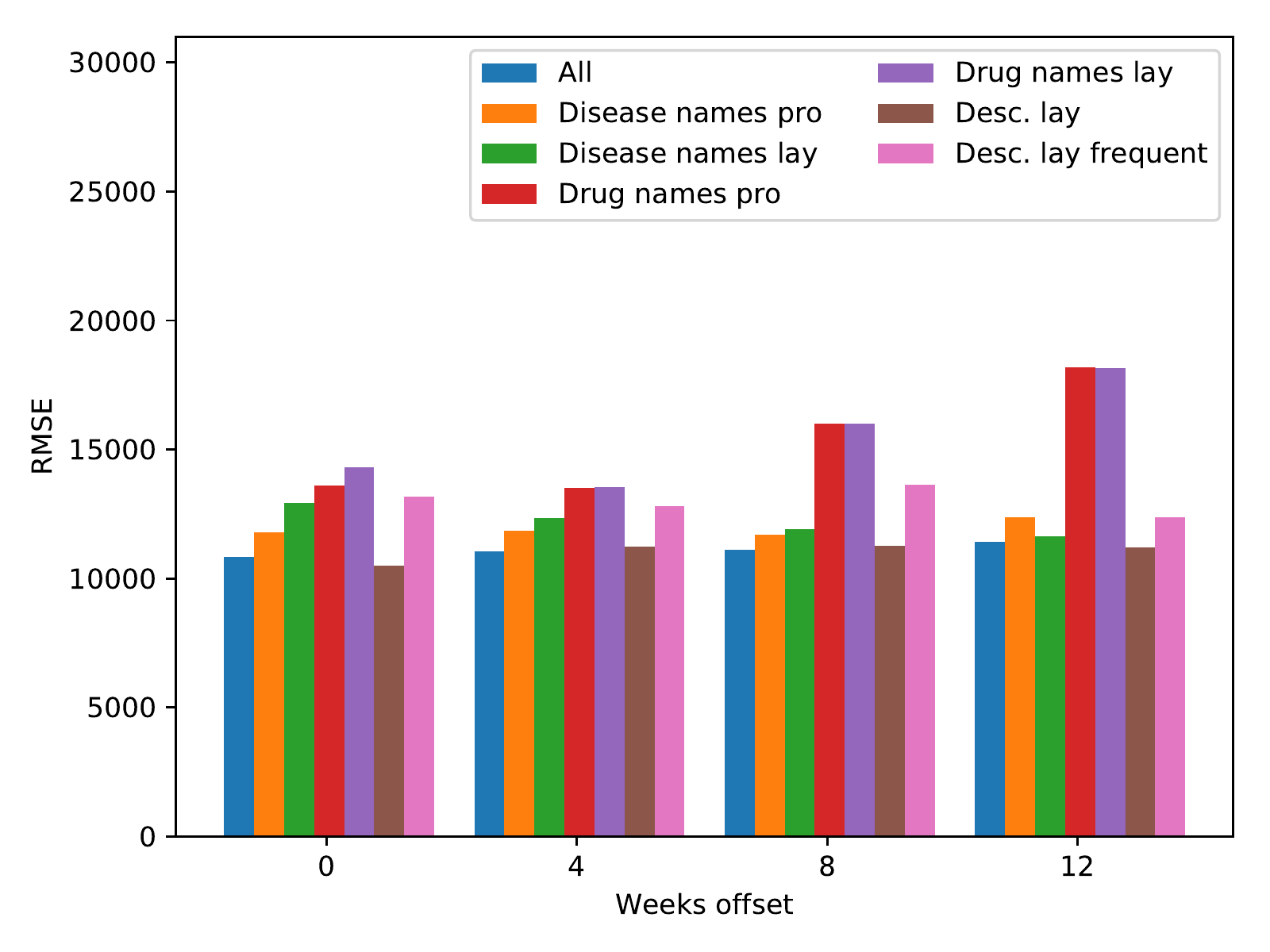}
		\caption{Up to 130 weeks lag}
	\end{subfigure}
	\caption{Prediction based on web search data with different query sets and lags, using Elastic Net.}
	\label{fig:query_set_performance_web-en}

\end{figure*}

\begin{figure*}
	
	\begin{subfigure}[b]{0.32\textwidth}
		\includegraphics[width=\linewidth]{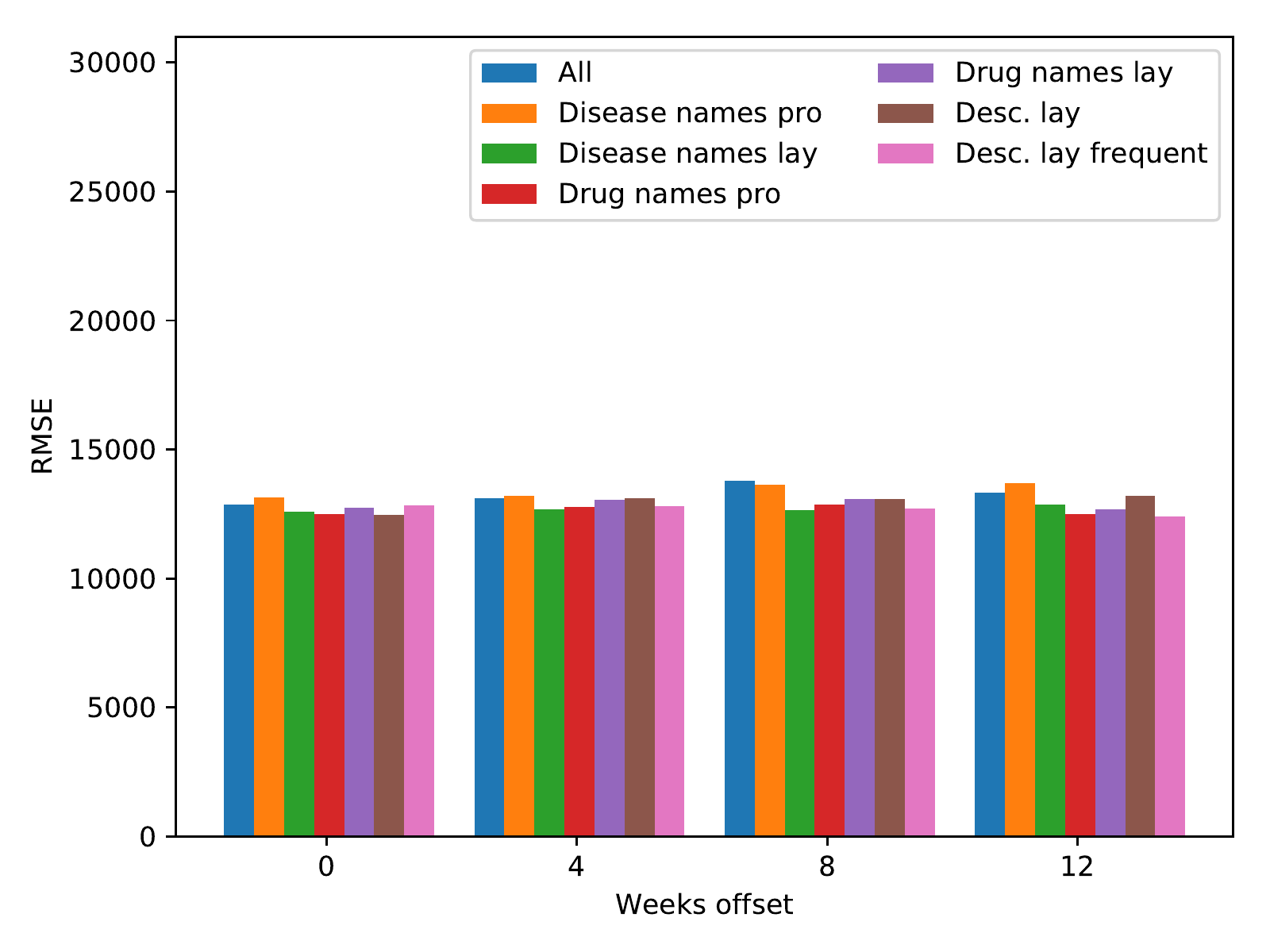}
		\caption{Up to 4 weeks lag}
	\end{subfigure}
	\begin{subfigure}[b]{0.32\textwidth}
		\includegraphics[width=\linewidth]{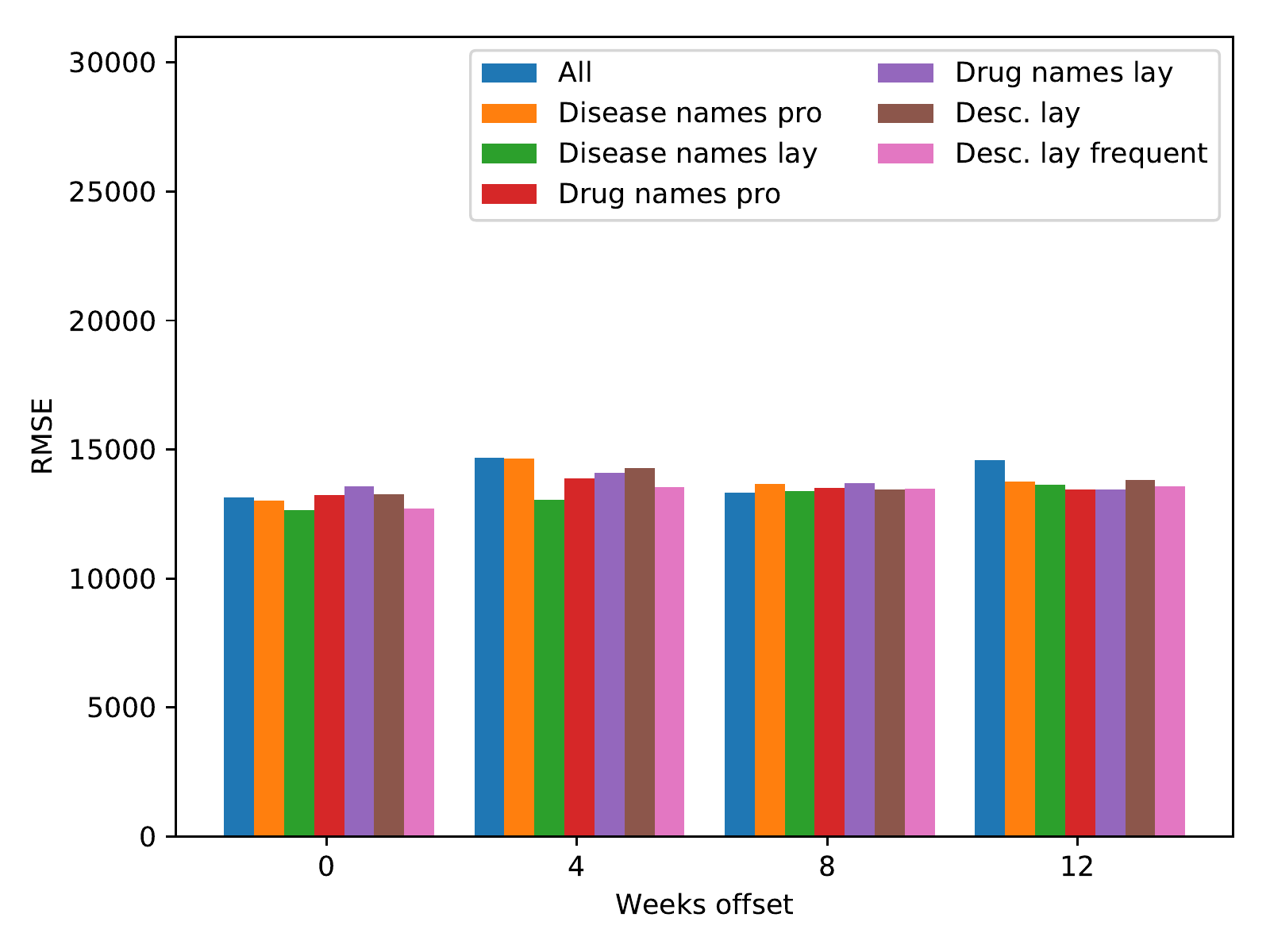}
		\caption{Up to 26 weeks lag}
	\end{subfigure}
	\begin{subfigure}[b]{0.32\textwidth}
		\includegraphics[width=\linewidth]{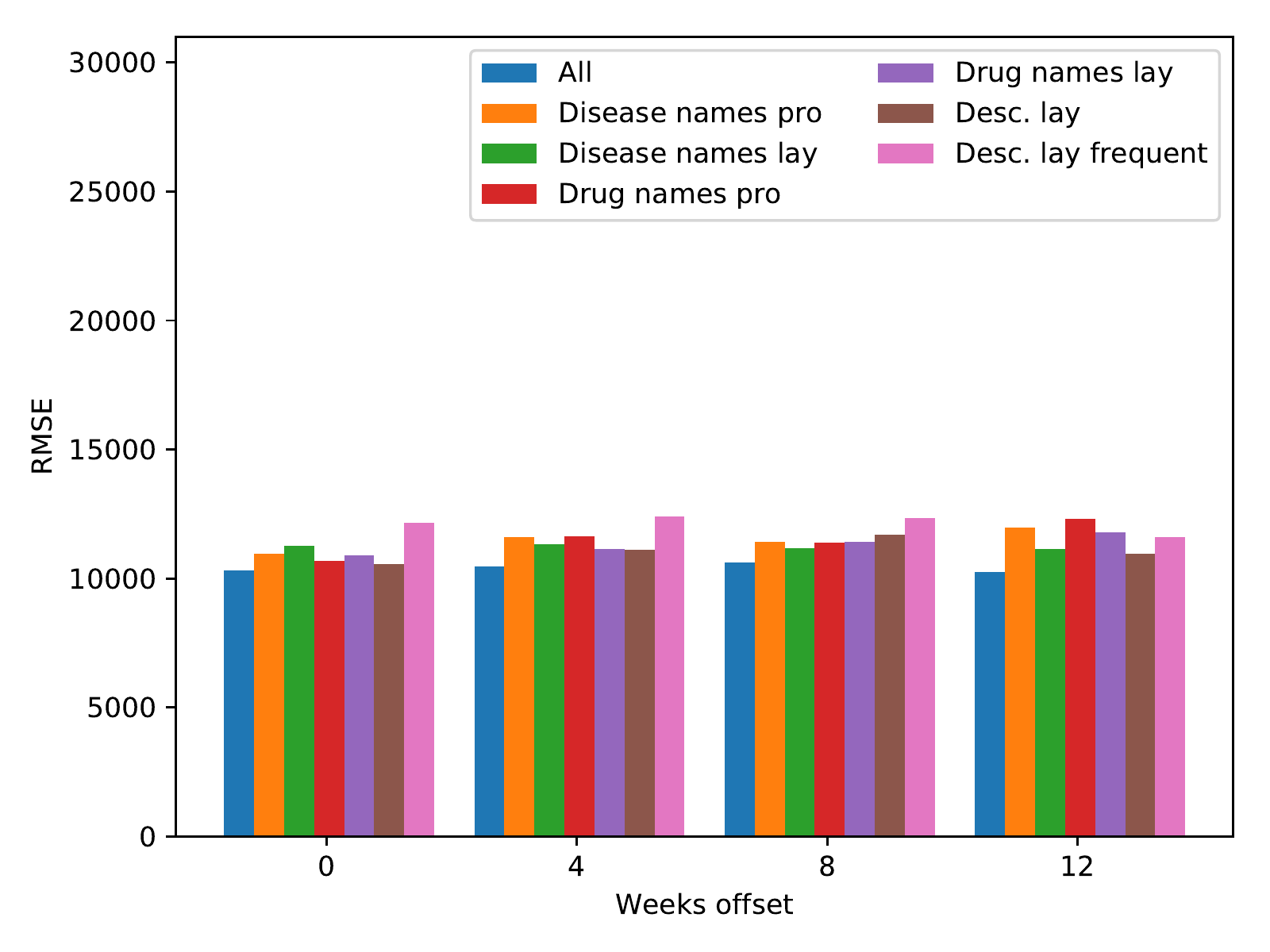}
		\caption{Up to 130 weeks lag}
	\end{subfigure}
	\caption{Prediction based on both web data and antimicrobial purchase data (130 autoregressive terms), with different query sets and lags, using Gaussian Processes.}
	\label{fig:query_set_performance_antibiotis-gp}
	
\end{figure*}

\begin{figure*}
	
	\begin{subfigure}[b]{0.32\textwidth}
		\includegraphics[width=\linewidth]{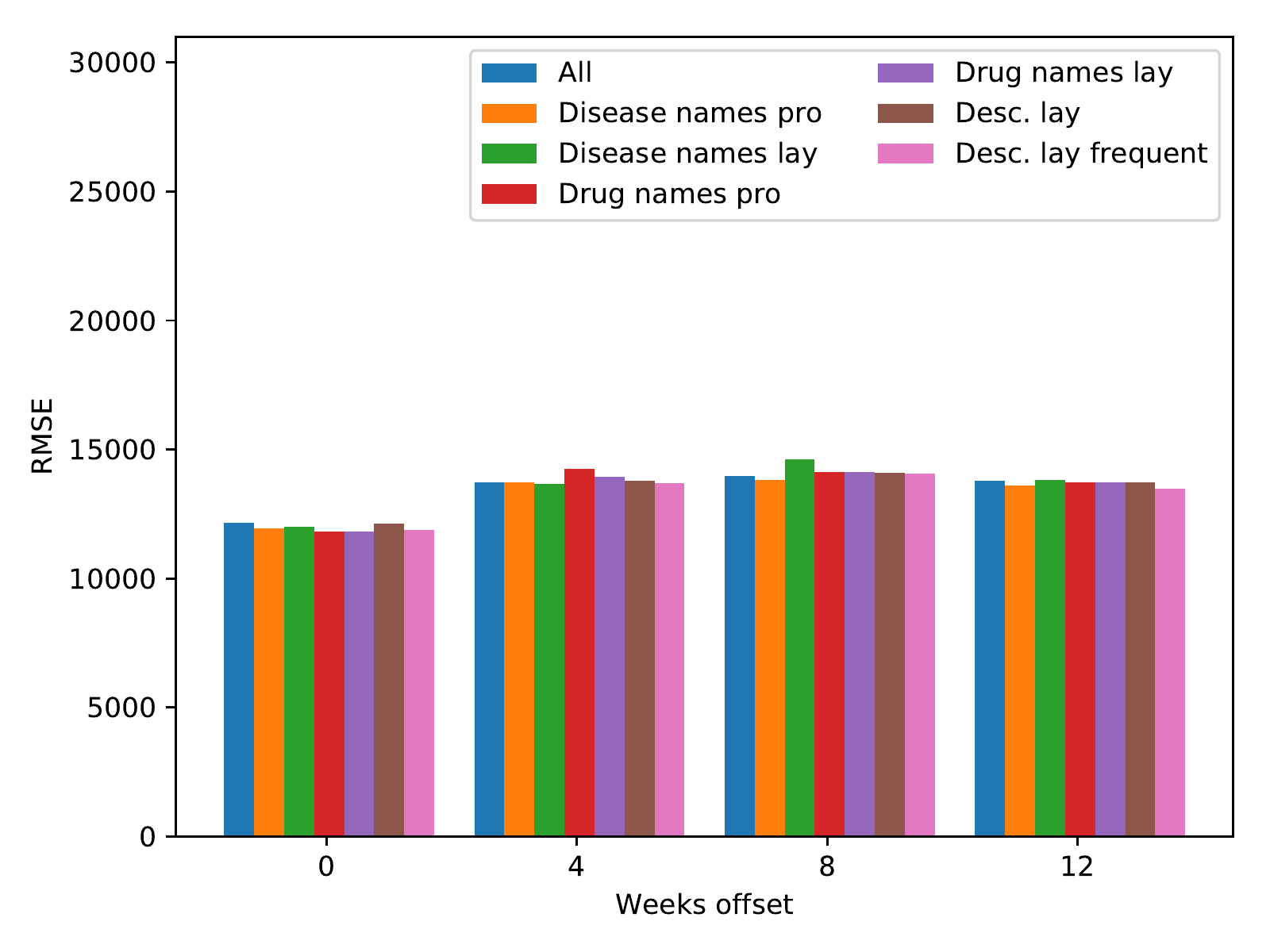}
		\caption{Up to 4 weeks lag}
	\end{subfigure}
	\begin{subfigure}[b]{0.32\textwidth}
		\includegraphics[width=\linewidth]{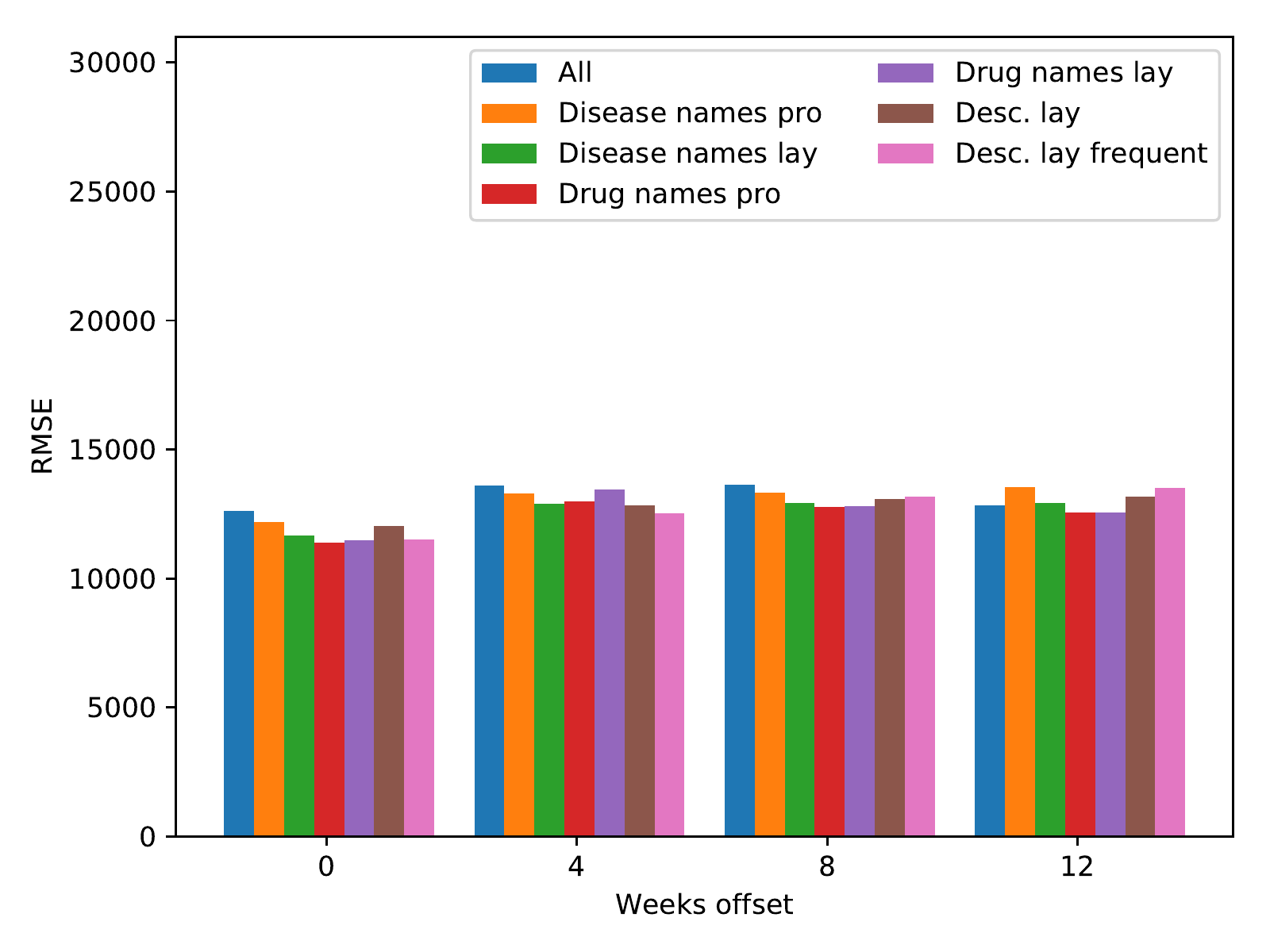}
		\caption{Up to 26 weeks lag}
	\end{subfigure}
	\begin{subfigure}[b]{0.32\textwidth}
		\includegraphics[width=\linewidth]{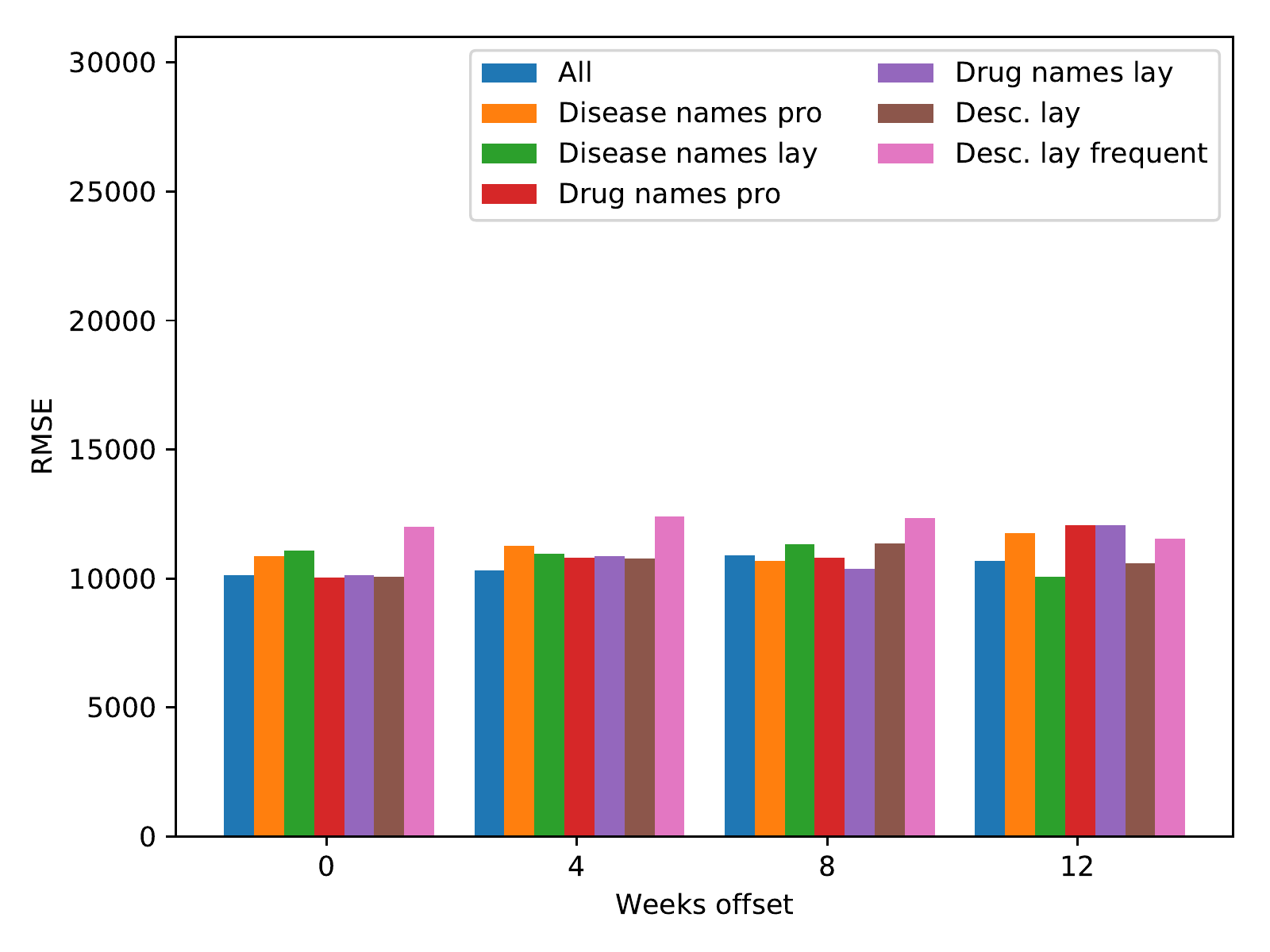}
		\caption{Up to 130 weeks lag}
	\end{subfigure}
	\caption{Prediction based on both web data and antimicrobial purchase data (130 autoregressive terms), with different query sets and lags, using Elastic Net.}
	\label{fig:query_set_performance_antibiotis-en}
	
\end{figure*}

Depending on the amount of maximum number of lags used, different queries are selected from the six query sets displayed in Table \ref{table:query_count}. We vary the maximum number of lags between 4, 26 and 130 weeks. Table \ref{tabel:top10_4weeks} shows the top 10 queries from each query set when using 4 weeks of historic antimicrobial data. The queries are selected using the method described in Section \ref{sec:feature_selection}. We see that most of the queries listed in Table \ref{tabel:top10_4weeks} are diseases curable with antimicrobials, such as Scarlet Fever and pneumonia. We also see diseases such as psoriasis, which itself is not treatable with antimicrobials, but which increases the risk of skin infections. It is interesting to note that even rare diseases, such as anthrax (typically a non-lethal skin infection) and syphilis, are in the top 10; this occurs because both of these diseases have antimicrobials as primary treatment. For the queries derived from the laymen descriptions of antimicrobials, there is a number of spurious correlations, e.g.\ words such as ``one'', ``effect'', etc. While these queries are apparently well correlated to antimicrobial drug consumption, they are semantically unrelated to antimicrobial usage, meaning that their generalisable discriminative strength is limited. 

\begin{table}

\adjustbox{center}{

	\begin{tabular}{lp{9cm}}
		\toprule
		Query set & Top 10 queries \\
		\midrule
		~Disease names pro & psoriasis$_{t-2}$, skarlagensfeber$_{t-3}$  (\textit{Scarlet Fever}), skarlagensfeber$_{t-2}$, lungebetændelse$_{t-4}$ (\textit{pneumonia}), prostatitis$_{t-4}$, diabetes - type 2$_{t-3}$, psoriasis$_{t-3}$, blindtarmsbetændelse$_{t-3}$ (\textit{appendicitis}), lungebetændelse$_{t-3}$, blyforgiftning$_{t-4}$ (\textit{lead poisoning})\\\rule{0pt}{5ex}
		Disease names lay &skarlagensfeber$_{t-2}$, skarlagensfeber$_{t-3}$, skarlagensfeber$_{t-1}$, endokardit$_{t-2}$ (\textit{endocarditis}), endokardit$_{t-1}$, endokardit$_{t-3}$, endokardit$_{t-4}$, brystbetændelse$_{t-1}$ (\textit{mastitis}), syfilis$_{t-3}$ (\textit{syphilis}), miltbrand$_{t-2}$ (\textit{anthrax})\\\rule{0pt}{5ex}
		Descriptions lay & lungebetændelse$_{t-3}$, én$_{t-2}$ (\textit{one}), lungebetændelse$_{t-2}$, én$_{t-4}$, virkning$_{t-3}$ (\textit{effect}), én$_{t-1}$, lungebetændelse$_{t-4}$, halsbetændelse$_{t-1}$ (\textit{sore throat}), stoffer$_{t-1}$ (\textit{drugs}), én$_{t-3}$ \\
		\bottomrule
	\end{tabular}}
	\caption{Top 10 queries for each query set generated with a maximum lag of 4 weeks and a 0 week prediction offset. The subscript denotes the offset from the current prediction point. \textit{Drug names lay}, \textit{drug names pro} and \textit{descriptions lay frequent} have zero valued coefficients and are therefore omitted from the Table. English translations in brackets.}
	\label{tabel:top10_4weeks}
\end{table}

\begin{table}
	
\adjustbox{center}{

	\begin{tabular}{lp{8.5cm}}
		\toprule
		Query set & Top 10 queries \\
		\midrule
		~Disease names pro & blindtarmsbetændelse$_{t-6}$ (\textit{appendicitis}), caries$_{t-1}$ , skarlagensfeber$_{t-52}$ (\textit{Scarlet Fever}), diabetes - type 2$_{t-121}$, kol$_{t-83}$ (\textit{COPD}), gasgangræn$_{t-38}$ (\textit{gas gangrene}), kol$_{t-3}$, caries$_{t-125}$, diabetisk neuropati$_{t-112}$, kol$_{t-82}$ \\\rule{0pt}{5ex}
		Disease names lay & skarlagensfeber$_{t-52}$, skarlagensfeber$_{t-53}$, skarlagensfeber$_{t-51}$, skarlagensfeber$_{t-50}$, skarlagensfeber$_{t-55}$, skarlagensfeber$_{t-54}$, gasgangræn$_{t-100}$, skarlagensfeber$_{t-3}$, gasgangræn$_{t-38}$, skarlagensfeber$_{t-103}$ \\\rule{0pt}{5ex}
		Drug names pro & novu$_{t-59}$, novu$_{t-111}$, novu$_{t-116}$, novu$_{t-30}$, novu$_{t-9}$, novu$_{t-62}$, novu$_{t-32}$, novu$_{t-7}$, novu$_{t-127}$, novu$_{t-82}$\\\rule{0pt}{5ex}
		Drug names lay & novu$_{t-59}$, novu$_{t-111}$, novu$_{t-62}$, novu$_{t-9}$, novu$_{t-30}$, novu$_{t-66}$, novu$_{t-7}$, novu$_{t-116}$, novu$_{t-33}$, novu$_{t-32}$ \\\rule{0pt}{5ex}
		Descriptions lay & skyldes$_{t-107}$ (\textit{due}), vækst$_{t-94}$ (\textit{growth}), udvikle$_{t-123}$ (\textit{develop}), immunforsvar$_{t-83}$ (\textit{immune system}), resistens$_{t-4}$ (\textit{resistance}), allergi$_{t-92}$ (\textit{allergy}), bivirkninger$_{t-9}$ (\textit{side-effect}), dræbe$_{t-99}$ (\textit{kill}), behandlingen$_{t-85}$ (\textit{treatment}), skyldes$_{t-104}$\\\rule{0pt}{5ex}
		Descriptions lay frequent & infektion$_{t-51}$ (\textit{infection}), vækst$_{t-55}$, vækst$_{t-94}$, infektion$_{t-78}$, medicin$_{t-4}$ (\textit{medicine}), virus$_{t-102}$, bakterier$_{t-83}$ (\textit{bacteria}), vækst$_{t-42}$, behandling$_{t-62}$, bakterierne$_{t-117}$ \\
		\bottomrule
	\end{tabular}}
	\caption{Top 10 queries for each query set with a maximum lag of 130 and a 0 week prediction offset. The subscript denotes the offset from the current prediction point. English translations in brackets.}
	\label{tabel:top10_130weeks}

\end{table}

Table \ref{tabel:top10_130weeks} further shows the top 10 queries for a lag of up to 130 weeks. In this case we clearly begin to see the effect the seasonality (seen in Figure \ref{fig:j0ce_usage}), because several of the diseases have lags of approximately one year, i.e.\ 52 weeks. In the top 10 for \textit{Disease names pro} are two chronic diseases: type-2 diabetes and COPD. For both of these the lag does not correspond to a yearly seasonality. This likely indicates that, it is not these precise diseases that are treated by antimicrobials; rather, patients of these diseases are likely to develop weaker immune systems, indicating that after one or two years they are more prone to complications needing antimicrobial treatment. 

Similarly to Table \ref{tabel:top10_4weeks}, we also observe in Table \ref{tabel:top10_130weeks}, that \textit{Descriptions lay} and \textit{Description lay frequent} yield words unrelated to antimicrobials as queries. Inspecting the lag of the unrelated words, e.g.\ 94 weeks for growth, we observe that they are spurious correlations, but not seasonal, as has been previously observed for the prediction of influenza like illnesses \cite{lazer2014parable, dalum2017seasonal}. 
Such spurious correlations cannot be expected to reliably model rapid changes in antimicrobial drug consumption, as discussed above. 

Figures \ref{fig:query_set_performance_web-gp} \& \ref{fig:query_set_performance_web-en} show the prediction error when only using web search data from the different query sets and for different lags, for Gaussian Processes and Elastic Net, respectively. While we previously saw in Tables \ref{tabel:top10_4weeks} \& \ref{tabel:top10_130weeks}, that \textit{Drug names pro} and \textit{Drug names lay} seemed as the data sources with the most semantically relevant queries, we now see that \textit{Descriptions lay} generally is the best performing query set. 
We previously observed that \textit{Descriptions lay} contained spurious
correlations, so it seems strange that this query set performs best.
Similar observations have been made with respect to ILI prediction,
where the semantically relevant query set was outperformed by
a less relevant one \cite{dalum2017seasonal}. This likely happens
for two reasons: (i) spurious correlations can model the expected
seasonality well, (ii) lack of evaluation data can make correlations
due to chance more likely. 

We also see in Figures \ref{fig:query_set_performance_web-gp} \& \ref{fig:query_set_performance_web-en} that there is a noticeable reduction in prediction error, for all query sets, when moving from a maximum lag of 26 weeks to 130 weeks. This is likely due to the modeling of seasonal variations that we noticed in Table \ref{tabel:top10_130weeks}. Even when predicting 12 weeks into the future, the prediction error is still relatively stable. As we saw in Table \ref{tabel:top10_130weeks}, there are long term effects of antimicrobial drug usage, either seasonal changes or long term predictors such as type-2 diabetes, and these are likely some of the reasons why prediction into the future works well.

The impact of query selection upon prediction performance
significantly diminishes when prediction is based on a combination
of web data and historical antimicrobial purchase data, i.e.\ when high quality time series data is available. We see that
in Figures \ref{fig:query_set_performance_antibiotis-gp} \& \ref{fig:query_set_performance_antibiotis-en}. 

Finally, as we noted previously, the consistent prediction performance across a prediction offset of 0 weeks and 12 weeks is remarkable. Figure 1 shows that the three last years of our antimicrobial consumption time series are very similar. This is likely one of the reasons for the consistent performance independent of the prediction offset. It is not unlikely that the prediction models will perform significantly worse in case of a sudden change, as was observed with Google Flu during the 2009 swine flu \cite{cook2011assessing}. In such a scenario we would expect the semantically relevant queries to remain correlated with the consumption, while the search pattern for the irrelevant queries should remain unchanged given changes in antimicrobial consumption. Given such a change in consumption, it is likely that the difference between the \textit{Disease names pro} query set and \textit{Descriptions lay} query set would become apparent.

%
%
%
%

\section{Conclusion}
We studied the extent to which consumption of antimicrobial drugs, such as antibiotics, can be predicted from web search data. We compared this to predictions based on more traditional historical purchase data of antimicrobial drugs. We experimented with different prediction models (Elastic Net and Gaussian Processes), and a novel method of selecting web search queries indicative of antimicrobial drug consumption by mining antimicrobial related information from publicly available descriptions of diseases and drugs linked to antimicrobials. Experiments with more than 9 years of weekly antimicrobial drug consumption data from Denmark showed that prediction using web search data are overall comparable and marginally more erroneous than predictions using antimicrobial drug sales data. The difference in error between the two is equivalent to 1\% point mean absolute error in weekly consumption. This performance was found to be relatively stable across variations in prediction offsets, prediction models, and query selection methods. 





\chapter{Seasonal Web Search Query Selection for Influenza-Like Illness (ILI) Estimation}
\chaptermark{Seasonal query selection}
\label{cha:seasonal_web_search_query_selection}

\begin{center}
	Venue: Conference on Research and Development in Information Retrieval (SIGIR),\\Tokyo 2017
	\vspace{1cm}
	
\textit{
	Niels Dalum Hansen$^{ab}$, Kåre Mølbak$^c$,\\Ingemar Johansson Cox$^a$, Christina Lioma$^a$
}

\vspace{0.5cm}

$^a$University of Copenhagen, Denmark.
$^b$IBM Denmark.\\
$^c$Statens Serum Institut, Denmark. 
	
\end{center}

\leftskip=1cm
\rightskip=1cm

\noindent Influenza-like illness\textit{ (ILI) estimation from web search data is an important web analytics task. The basic idea is to use the frequencies of queries in web search logs that are correlated with past ILI activity as features when estimating current ILI activity. It has been noted that since influenza is seasonal, this approach can lead to spurious correlations with features/queries that also exhibit seasonality, but have no relationship with ILI. Spurious correlations can, in turn, degrade performance. To address this issue, we propose modeling the seasonal variation in ILI activity and selecting queries that are correlated with the residual of the seasonal model and the observed ILI signal. Experimental results show that re-ranking queries obtained by Google Correlate based on their correlation with the residual strongly favours ILI-related queries.}

\leftskip=0pt\rightskip=0pt


%
%

\section{Introduction and background}

The frequency of queries in web search logs has been found useful in estimating the incidence of \textit{influenza-like illnesses} (ILIs) \cite{ginsberg2009detecting,lampos2015advances,lazer2014parable,santillana2015combining,yang2015accurate}. 
Current methods use two core discriminative features for ILI estimation: (i) past ILI activity, and (ii) the frequency of queries in web search logs that correlate strongly with past ILI activity.  There are two problems with this approach.

The \textit{first problem} is that not all queries whose frequency is strongly correlated to ILI activity are necessarily informative with respect to ILIs, and hence discriminative as a feature for ILI estimation.
At the most general level, this is an example of the fundamental issue of correlation not being causation. In the case of estimating ILI, this is exacerbated by the seasonal nature of influenza. In fact, it has been previously observed that previous methods can identify queries that have a very similar seasonality but are clearly not related to ILI. For example, the query \textit{``high school basketball''} has been found to have a high correlation with ILI activity \cite{lazer2014parable} even though it is obviously unrelated to ILI. The seasonality of high school basketball accounts for this correlation. 
Queries unrelated to the ILI activity will not be useful in the case of irregular ILI activity, e.g.\ an off season influenza outbreak. Additionally, changes in for example the high school basketball schedule would result in changes in the ILI estimates. 

The \textit{second problem} is that by using two types of features that are strongly correlated to each other (past ILI activity, and queries whose frequencies are strongly correlated to past ILI activity), we may compromise diversity in the representations one would expect from the features. Better estimations may be produced by using features that \textit{complement} each other, regardless of their between--feature correlation. 

Motivated by the above issues, we 
propose an alternative approach to selecting queries.
Our approach consists of two steps. (1) We model the seasonal variation of ILI activity, and (2) we select queries whose search frequency fits aspects of this seasonality. Specifically, we present two variations of our algorithm: select queries that 
correlate with our seasonal model of ILI,
and select queries that correlate with the residual between the seasonal model and observed ILI rates.

Our results are two fold. (i) Experimental evaluation of our seasonal query selection models for ILI estimation against strong recent baselines (of no seasonality) show that we can achieve performance that is overall more effective (reduced estimation error), and requires fewer queries as estimation features. With respect to error reduction we see that selecting queries fitting regular seasonal ILI variation is a better strategy than selecting queries fitting ILI outbreaks. (ii) Selecting queries that fit seasonal irregularities result in much more semantically relevant queries. These queries are surprisingly not the ones that result in the best predictions. 

Our main results are: (i) We demonstrate that Google Correlate retrieves many non-relevant queries that are highly correlated with a time series of historic ILI incidence, and that the ILI-related queries are not highly ranked; (ii) re-ranking these queries based on their correlation with a residual signal, i.e.\ the difference between a seasonal model and historic data, strongly favours ILI-related queries; (iii) the performance of a linear estimator is improved based on the re-ranked queries. 
To our knowledge, the seasonal examination of ILI activity for query selection in automatic ILI estimation is a novel contribution. Seasonal variation has, however, been studied for other medical informatics tasks, such as vaccination uptake estimation \cite{www17, DalumHansen:2016:ELV:2983323.2983882}.


\section{Problem Statement}
The goal is to estimate ILI activity at time $t$, denoted $y_t$, using observed historical ILI activity (reported e.g.\ by the Centers for Disease Control and Prevention (CDC) in US) and query frequencies in web search logs. This is most commonly done by (i) submitting to Google Correlate\footnote{\url{https://www.google.com/trends/correlate}} a file of historical ILI activity, and receiving as output web search queries (and their frequencies) that are most strongly correlated to the input ILI data. Then, $y_t$ can be estimated with a linear model that uses only web search frequencies \cite{ginsberg2009detecting} as follows:

\begin{equation}\label{eq:queries}
y_t = \alpha_0 + \sum^n_{i=1} \alpha_i Q_{t,i} + \epsilon,
\end{equation}

\noindent where $n$ is the number of queries, $Q_{t,i}$ is the frequency for query $i$ in the web search log at time $t$, the $\alpha$s are coefficients, and $\epsilon$ is the estimation error. 

Including historical ILI activity data can improve the estimations of Eq.\ \ref{eq:queries}, for instance with an autoregressive model \cite{yang2015accurate}, as follows:

\begin{equation}\label{eq:clinical+queries}
y_t = \beta_0 + \beta_1 t + \sum^m_{j=1} \beta_{j+1} \cdot y_{t-j} + \sum^n_{i=1} \beta_{i+m+1} \cdot Q_{t,i}+ \epsilon, 
\end{equation}

\noindent where $m$ is the number of autoregressive terms, and the $\beta$s are coefficients to be learned.
With $m=52$ and $n=100$, Eq.\ \ref{eq:clinical+queries} corresponds to the model presented by Yang et al.\ \cite{yang2015accurate}. 

Most ILI estimation methods (exceptions include \cite{www17_lampos}) that use web search frequencies use all queries found to be correlated to ILI activity, i.e.\ in Eq.\ \ref{eq:queries} and Eq.\ \ref{eq:clinical+queries} $n$ corresponds to \textit{all} strongly correlated queries, and query selection is typically left for the model regularisation, for example using LASSO regularisation. In the next section we present a novel way of selecting which among these correlated queries to include in the estimation of $y_t$ according to how well they fit the seasonal variation of ILI activity. 

\section{Seasonal Query Selection}\label{sec:seasonal_query_selection}

We reason that among the queries whose frequency is correlated with past ILI activity, some queries may fit the ILI seasonal variation better than others. This is supported by the literature \cite{lazer2014parable}. We further reason that this fit of queries to seasonal ILI variation may not be sufficiently captured by simply measuring the correlation between the frequency of those queries and ILI activity. Based on this, we (i) present two  models to represent seasonal variation of ILI activity, and (ii) select queries 
based on these seasonal models.

\subsection{Step 1: Model seasonal ILI variation}

\begin{figure}
\centering{
\includegraphics[width=10cm]{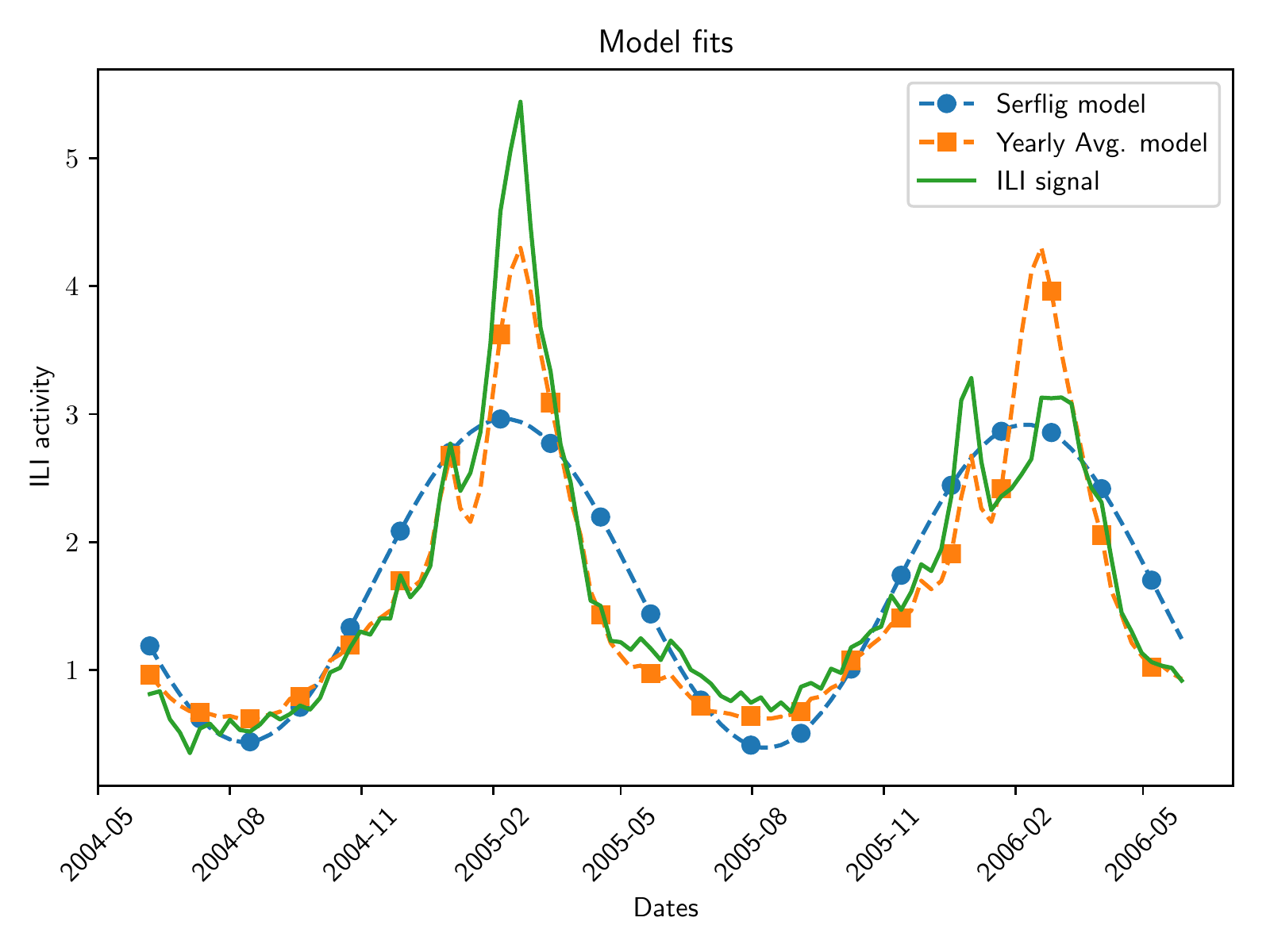}
}
\caption{Fit of the Serfling model (Eq.\ \ref{eq:serf}) and the Yearly Average model (Eq.\ \ref{eq:avg}) to historical ILI data (described in Section \ref{s:eval}).}
\label{fig:serflingfit}
\end{figure}


We model seasonal variation in two ways.
The first model is the Serfling model \cite{serfling1963methods}, chosen because of its simplicity and expressiveness. The Serfling model (Eq.\ \ref{eq:serf}) uses pairs of sine and cosine terms to model seasonality, and a term linear in time to model general upward or downward trends. We use this model with weekly data and one yearly cycle (details on data are given in Section \ref{s:eval}), resulting in the following ILI estimation model:

\begin{equation}\label{eq:serf}
y_t = \beta_0 + \beta_1 t + \beta_2 \sin\left(\frac{2 \pi t}{52}\right) + \beta_3 \cos\left(\frac{2 \pi t}{52}\right) + \epsilon,
\end{equation}

\noindent where the $\beta$s denote model coefficients and $\epsilon$ the error, i.e.\ residual. 

For the second model we use a yearly average (YA). Here the expected value of $y_t$ is calculated as the average value of $N$ seasons of ILI activity data, 

\begin{equation}\label{eq:avg}
\hat{y}_t = \frac{1}{N} \sum_{i=0}^{N-1} y_{(t \text{ mod } S) + i \cdot S},
\end{equation}

\noindent where $S$ is the season length in weeks, in our case 52.

Fig.\ \ref{fig:serflingfit} shows the fit of the two models, i.e.\ the Serfling model (Eq.\ \ref{eq:serf}) and the YA model (Eq.\ \ref{eq:avg}) to historical ILI activity data. We see that the Serfling model fits the general seasonality, but not more complex patterns representing higher transmission. It does not, for example, model differences between the start and end of the ILI season. This is better captured by the YA model. 


\subsection{Step 2: Query selection}\label{ss:qs}

Having modelled seasonal ILI variation, the second step is to approximate how well queries fit the seasonal variation of ILI activities modelled by Eq.\ \ref{eq:serf}-\ref{eq:avg}. We do this in two ways:

\textbf{Seasonal correlation.} We compute the Pearson correlation between the query frequencies and the ILI seasonal model, i.e.\ Eq.\ \ref{eq:serf} or \ref{eq:avg}. We then select queries that are most strongly correlated to the ILI activity model.

\textbf{Residual correlation.} We compute the Pearson correlation between the query frequencies and the residual between the ILI seasonal model and the historical ILI activity. We then select queries that are most strongly correlated to the residual, i.e.\ \textit{unexpected variations in ILI activity} (possible outbreaks). 

The four query selection methods are denoted (i) Seasonal (Serfling), (ii) Seasonal (YA), (iii) Residual (Serfling), and (iv) Residual (YA).

\section{Evaluation}
\label{s:eval}

\paragraph{\textbf{Experimental Setup}}
We experimentally evaluate our seasonality-based query selection methods using two types of data: weekly ILI activity data and Google search frequency data. The ILI activity data is from the US CDC for the period 2004-6-6 to 2015-7-11\footnote{\url{https://gis.cdc.gov/grasp/fluview/fluportaldashboard.html}} (inclusive). The CDC reports this in the form of weighted ILI activity across different US regions. ILI activity for a region corresponds to the number of ILI related visits to health care providers compared to non-influenza weeks, e.g.\ an ILI activity of 2 corresponds to twice as many visits as in non-epidemic weeks.

We retrieve query search frequencies from Google Correlate with the location set to the US. Specifically, we use the 100\footnote{This is a maximum number set by Google Correlate.} queries that have the highest correlation with the ILI activity from 2004-6-6 to 2009-3-29 according to Google Correlate. Google normalizes the search frequencies for each query to unit variance and zero mean, i.e.\ we do not know the actual search frequencies. We use the interval 2004-6-6 to 2009-3-29 because it represents a non-epidemic period (it excludes the 2009 pandemic of H1N1 influenza virus that caused highly irregular ILI activity). The 100 queries are shown in Tab.~\ref{tabel:ili_queries}. Only 21 of the 100 queries are related to ILI (in bold).

We compare our query selection methods (Section \ref{sec:seasonal_query_selection}) to the following three baselines: (Tab.\ \ref{tab:resultsQuery} baseline i) uses the top-$c$ queries to estimate ILI activity, where the top-$c$ are chosen to minimise the RMSE error, i.e.\ if we use $c+1$ queries the RMSE increases; (Tab.\ \ref{tab:resultsQuery} baseline ii) using \textbf{\textit{all}} 100 queries to estimate ILI activity; (Tab.\ \ref{tab:resultsQuery} baseline iii) using no queries, only past ILI activity, i.e.\ an autoregressive model.  For (i) and (ii), the query ranking is determined by Google Correlate. For (iii), two autoregressive models are fitted: one using 3 autoregressive terms \cite{lazer2014parable,yang2015accurate} and one with 52 autoregressive terms \cite{yang2015accurate}. This setup is similar to the setup of Yang et al.\ \cite{yang2015accurate}. We implement baseline (iii) using Eq.\ \ref{eq:clinical+queries} where $m$ is set to 3 and 52 terms, respectively, and $n=0$. Similarly to \cite{lazer2014parable,yang2015accurate}, we evaluate estimation performance by reporting the root mean squared error (RMSE) and Pearson correlation between the estimations and the observed historical ILI activity.

For all runs, we use data from 2004-6-1 to 2009-3-29 for training, and data from 2009-4-1 to 2015-7-11 for testing. The training data is used to fit Eq.\ \ref{eq:serf}-\ref{eq:avg}, and to calculate the correlation scores as described in Section \ref{ss:qs}. Estimations are made in a leave-one-out fashion where data prior to the data point being estimated is used to fit the estimation model. Each model is retrained for every time step using the 104 most recent data points (exactly as in Yang et al.\ \cite{yang2015accurate}). We determine the number of queries $n$ in Eq.\ \ref{eq:queries}-\ref{eq:serf} by iteratively adding the next highest ranked query, where query rank is given by either Google Correlate (for the baselines), or by the four variants of our algorithm, specifically (i) correlate seasonal (Serfling), (ii) correlate seasonal (YA), (iii) correlate residual (Serfling), and (iv) correlate residual (YA). The models are fitted using LASSO regression, where the hyperparameter is found using three fold cross-validation on the training set.

\paragraph{\textbf{Results}}

\begin{table}
	\centering
	\begin{tabular}{p{12cm}}
		\toprule
		florida spring, \textbf{influenza symptoms}, \textbf{symptoms of influenza}, new york yankees spring training, yankees spring training, \textbf{flu incubation}, \textbf{flu incubation period}, \textbf{flu duration}, florida spring training, spring training locations, \textbf{influenza incubation}, florida spring break, spring break dates, \textbf{flu fever}, sox spring training, new york mets spring training, \textbf{bronchitis}, red sox spring, spring training in florida, snow goose, spring break calendar, spring training in arizona, red sox spring training, mlb spring training, \textbf{flu report}, baseball spring training, mariners spring training, wrestling tournaments, spring training, golf dome, \textbf{flu recovery}, city spring, wrestling singlets, spring training sites, boys basketball, \textbf{type a influenza}, yankee spring training, spring training tickets, las vegas march, indoor track, harlem globe, spring break panama city, girls basketball, panama city spring break, cardinals spring training, ny mets spring training, ny yankees spring training, \textbf{flu symptoms}, minnesota twins spring training, concerts in march, spring training map, \textbf{tessalon}, boston red sox spring training, \textbf{flu contagious}, \textbf{symptoms of the flu}, events in march, seattle mariners spring training, singlets, \textbf{influenza contagious}, \textbf{influenza incubation period}, spring break schedule, spring vacation, \textbf{treating the flu}, college spring break, basketball boys, college spring break dates, boys basketball team, \textbf{respiratory flu}, atlanta braves spring training, \textbf{acute bronchitis}, march madness dates, spring break florida, braves spring training, college basketball standings, in march, braves spring training schedule, high school boys basketball, spring break ideas, spring break miami, banff film, addy awards, grapefruit league, spring clothing, spring collection, banff film festival, st. louis cardinals spring training, april weather, spring break family, red sox spring training schedule, miami spring break, nj wrestling, spring break getaways, spring break date, high school boys, march concerts, high school basketball, indoor track results, \textbf{tussionex}, globetrotters, orioles spring training\\
		\bottomrule
	\end{tabular}
	\caption{The 100 queries retrieved from Google Correlate. We treat queries in bold as ILI related.}
	\label{tabel:ili_queries}
\end{table}

\begin{figure}
	\centering
	\includegraphics[width=10cm]{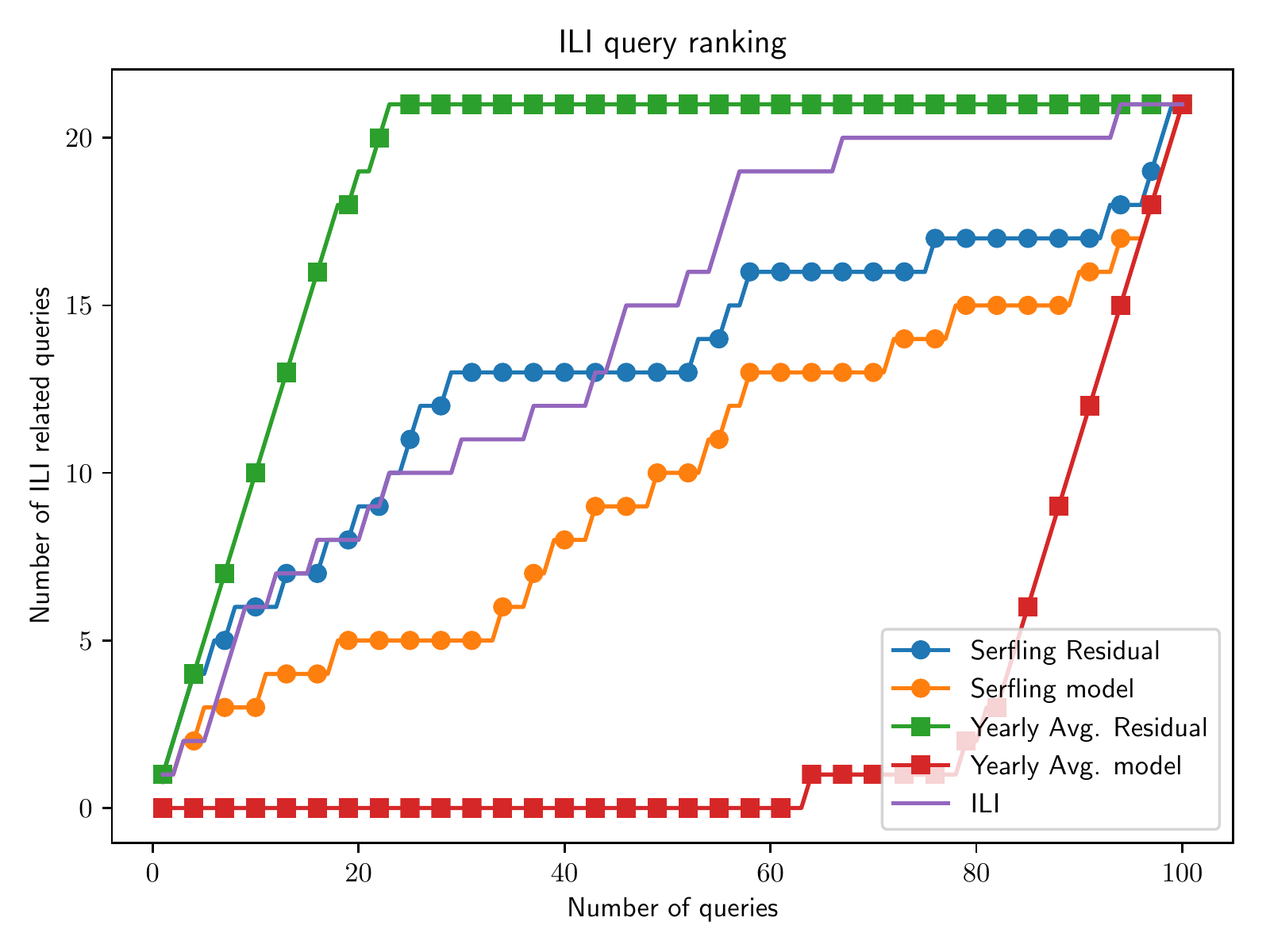}
	\caption{Portion of ILI related queries in the top $n$ queries for each of the five ranking methods.}
	\label{fig:ILIrelevant}
\end{figure}

As noted earlier, Google Correlate identifies the top-100 queries, but only 21 of these are ILI-related. Our four algorithms re-rank the 100 queries. Fig.~\ref{fig:ILIrelevant} plots the number of ILI-related queries as a function of the number of rank-ordered queries. The solid curve is based on the original ranking provided by Google Correlate. We observe that both the (i) Seasonal (Serfling) and (ii) Seasonal (YA), re-rank the queries such that, in general, the ILI-related queries are ranked worse. The Residual (Serfling) generally performs similarly to or worse than Google Correlate in favouring ILI-related queries. In contrast, Residual (YA) re-ranks the queries such that almost all ILI-related queries are favoured. Of the top-21 queries, 19 are ILI-related. All 21 ILI-related queries are within the top-23. The only two non-related queries in the top-23 are ranked at 19 and 21.  Clearly re-ranking queries based on Residual (YA) strongly favours ILI-related queries much more than Google Correlate or our other three variants.

For each ranking of the queries, we select the top-$n$ queries that either minimise the RMSE or maximise the Pearson correlation. This is done for the Linear model of Eq.~\ref{eq:queries} and for the autoregressive model of Eq.~\ref{eq:clinical+queries},
Tab.\ \ref{tab:resultsQuery} shows the results.
For the Linear model (column 1), we observe that Residual (YA) performs best w.r.t.\ RMSE and Pearson correlation, though the latter is not significant. Note that in both cases, (i) the number of queries needed by Residual (YA) is significantly less than for the other three variants and (ii) the two baselines performed worse.

For the autoregressive models, we observe that the Seasonal (Serfling) model performs best w.r.t.\ RMSE and Pearson correlation. This is achieved with relatively few queries (5, 9, or 11). However we note that of the top-5, -9 or -11 queries only 3, 3 or 4, resp.\ are ILI-related. In general, autoregressive models perform well when the signal has a strong autocorrelation. However, should the signal strongly deviate from seasonal patterns, then it is unlikely that the ILI estimates would be accurate.   


\begin{table}
\begin{adjustbox}{center}

	\begin{tabular}{l l rr rr rr}
	
	&  & \multicolumn{6}{c}{\textbf{RMSE} \textit{(the lower, the better)}} \\
\toprule
	& & Linear  & \#q. & Autoreg. 52 & \#q. & Autoreg 3  & \#q. \\
 \cmidrule(lr){3-4} \cmidrule(lr){5-6} \cmidrule(lr){7-8}
Seasonal (Serfling)      &  & 0.398 & 97 & \textbf{0.280}& 5 & \textbf{0.280} & 11 \\
Residual (Serfling)  &  & 0.407 & 98 & 0.309 & 98 & 0.312 & 98 \\
Seasonal (YA)     &  & 0.394 & 96 & 0.311 & 100 & 0.298 & 68 \\
Residual (YA)  &  & 0.390 & 79 & 0.309 & 49 & 0.310 & 47 \\
Not Seasonal        & \textit{baseline i} & 0.413 & 33 & 0.309 & 68 & 0.312 & 46\\
Not Seasonal all q. & \textit{baseline ii} & 0.416 &    & 0.310 &    & 0.314 &\\
ILI History         & \textit{baseline iii} & n/a   &    & 0.348 &    & 0.333 &\\
\\
	&  & \multicolumn{6}{c}{\textbf{Correlation} \textit{(the higher, the better)}} \\
\toprule
& & Linear & \#q. & Autoreg. 52  & \#q. & Autoreg. 3  & \#q. \\
\cmidrule(lr){3-4} \cmidrule(lr){5-6} \cmidrule(lr){7-8}
Seasonal (Serfling)      &  & 0.948 & 96 & 0.973  & 5 & \textbf{0.974} & 9 \\
Residual (Serfling)   &  & 0.946  & 98 &0.968 & 100 &  0.967  & 98 \\
Seasonal (YA)    &  & 0.948 & 99 & 0.969 & 99 & 0.971 & 68 \\
Residual (YA)  &  & 0.949 & 77 &  0.969 & 59 & 0.966 & 47 \\
Not Seasonal        & \textit{baseline i} & 0.942 & 86 & 0.968 & 68 & 0.967 & 46 \\
Not Seasonal all q. & \textit{baseline ii} & 0.941 &   & 0.967 &    & 0.967 & \\
ILI History         & \textit{baseline iii} & n/a   &  & 0.959  & & 0.962  &\\ 
\bottomrule
	\end{tabular}
\end{adjustbox}
	
\caption{Root mean squared error (RMSE) and Pearson Correlation of our seasonal ILI estimation methods and the three baselines. Bold marks the best score. \#q denotes the number of queries used in the estimation. The linear models are using Equation \ref{eq:queries} and the autoregressive are using Equation \ref{eq:clinical+queries}}
	\label{tab:resultsQuery}
\end{table}

\section{Conclusion}

The incidence of influenza-like illness (ILI) exhibits strong seasonal variations. These seasonal variations cause Google Correlate to identify a large number of non-relevant queries (80\%). Many of the relevant queries are not highly ranked. Estimating the incidence of ILI with non-relevant queries is likely to become problematic when ILI deviates significantly from its seasonal variation.

We proposed a new approach to ILI estimation using web search queries. The novelty of our approach consists of re-ranking queries derived from Google Correlate. We first developed two models of the seasonal variation in ILI. The first is an analytical Serfling model. The second is an empirical model based on yearly averages. Four methods of re-ranking queries were then examined. The first two re-rank the queries based on their correlation with the two seasonal models. The second two re-rank queries based on their correlation with the residual between the seasonal models and historical ILI activity.

Experimental results showed that re-ranking queries based on Residual (YA) strongly favoured ILI-related queries, but 
re-ranking queries based on the two seasonal models, Seasonal (Serfling) and Seasonal (YA) led to rankings that were worse than those of Google Correlate.

When ILI estimates were based on both queries and autoregression the best performance was obtained when queries were re-ranked based on Seasonal (Serfling). Future work is needed to determine why, but we reason that 
(i) autoregessive models perform better when the signal has strong autocorrelation, i.e.\ is strongly seasonal, and (ii) this strong seasonality was present in our dataset, i.e.\ there was little deviation from the seasonal models. If, however, strong deviations did arise, we expect that models based on autoregression and queries re-ranked based on correlation with seasonal models will perform much worse.

This work complements the use of information retrieval and machine learning methods in the wider area of medical and health informatics \cite{DragusinPLLJW11,DragusinPLLJCHIW13}.

\paragraph{Acknowledgments.} Partially supported by Denmark's Innovation Fund (grant no.\ 110976).



\chapter{Pre-Vaccination Care-Seeking in Females Reporting Severe Adverse Reactions to HPV Vaccine. A Registry Based Case-Control Study}
\chaptermark{Pre-vaccination care-seeking in females}
\label{cha:hpv_side_effects}

\begin{center}
	
	Venue: PLoS One, 2016
	
	\vspace{1cm}
	
	\textit{
		Kåre Mølbak$^a$, Niels Dalum Hansen$^{bc}$, Palle Valentiner-Branth$^{a}$
	}

\vspace{0.5cm}
	
	$^a$Statens Serum Institut, Denmark.\\
	$^b$University of Copenhagen, Denmark.
	$^c$IBM Denmark.
	
\end{center}

\leftskip=1cm
\rightskip=1cm

		\indent\textit{\textbf{Background}: Since 2013 the number of suspected adverse reactions to the quadrivalent human papillomavirus (HPV) vaccine reported to the Danish Medicines Agency (DMA) has increased. Due to the resulting public concerns about vaccine safety, the coverage of HPV vaccinations in the childhood vaccination programme has declined. The aim of the present study was to determine health care-seeking prior to the first HPV vaccination among females who suspected adverse reactions to HPV vaccine.}
		
		\indent\textit{\textbf{Methods:} In this registry-based case-control study, we included as cases vaccinated females with reports to the DMA of suspected severe adverse reactions. We selected controls without reports of adverse reactions from the Danish vaccination registry and matched by year of vaccination, age of vaccination, and municipality, and obtained from the Danish National Patient Registry and The National Health Insurance Service Register the history of health care usage two years prior to the first vaccine. We analysed the data by logistic regression while adjusting for the matching variables.}
		
		\indent\textit{\textbf{Results:} The study included 316 cases who received first HPV vaccine between 2006 and 2014. Age range of cases was 11 to 52 years, with a peak at 12 years, corresponding to the recommended age at vaccination, and another peak at 19 to 28 years, corresponding to a catch-up programme targeting young women. Compared with 163,910 controls, cases had increased care-seeking in the two years before receiving the first HPV vaccine. A multivariable model showed higher use of telephone/email consultations (OR 1.9; 95\% CI 1.2-3.2), physiotherapy (OR 2.1; 95\% CI 1.6-2.8) and psychologist/psychiatrist (OR 1.9; 95\% CI 1.3-2.7). Cases were more likely to have a diagnosis in the ICD-10 chapters of diseases of the digestive system (OR 1.6; 95\% CI 1.0-2.4), of the musculoskeletal system (OR 1.6; 95\% CI 1.1-2.2), symptoms or signs not classified elsewhere (OR 1.8; 95\% CI 1.3-2.5) as well as injuries (OR 1.5; 95\% CI 1.2-1.9).}
		
		\indent\textit{\textbf{Conclusion:} Before receiving the first HPV vaccination, females who suspected adverse reactions has symptoms and a health care-seeking pattern that is different from the matched population. Pre-vaccination morbidity should be taken into account in the evaluation of vaccine safety signals.}

		\leftskip=0pt\rightskip=0pt

		\section{Introduction}
		
		Vaccination against human papilloma virus (HPV) was introduced in Denmark in 2006 and a quadrivalent HPV vaccine was included in Danish childhood vaccination programme from 2009. The primary target group is girls at the age of 12 years, but various catch-up programmes has also been offered including a free and recommended programme for young women 19-28 years from August 2012 throughout 2013 \cite{annual_reports_childhood_vaccination_programme}. In addition, risk based self-paid HPV vaccination outside the target groups has been advocated by the Danish Society of Gynecology of Obstetrics, among others, and it is estimated that by the end of 2015 more than 0.5 million corresponding to one fifth of all Danish females has been vaccinated against HPV. Notably, 151,487 (89\%) of 170,630 girls born 1996-2000 has received at least one dose of HPV vaccine.
		
		From 2013, this programme was challenged by an increasing number of reported suspected adverse reactions \cite{danishHealth_hpv}. Some of the reported reactions were classified as postural orthostatic tachycardia syndrome (POTS) whereas others resembled chronic fatique syndrome including fatigue, long-lasting dizziness, headache, and syncope \cite{brinth2015suspected, brinth2015chronic}. The Danish safety signal was raised to the European Medicines Agency (EMA) Pharmacovigilance Risk Assessment Committee (PRAC) in September 2013. Furthermore, in July 2015, the European Commission requested PRAC to assess whether there was evidence of a causal association between HPV vaccines and POTS and/or complex regional pain syndrome (CRPS) as an association between HPV vaccine and CRPS was raised from Japan \cite{kinoshita2014peripheral}. PRAC concluded that the current evidence does not support that HPV vaccines causes CRPS or POTS \cite{prac_ema}. In December 2015 the WHO Global Advisory Committee on Vaccine Safety reported that it has not found any safety issue that would alter its recommendation of the use of HPV vaccines, and argued that despite the difficulties in diagnosing or fully characterizing CRPS and POTS, reviews of pre- and post-licensure data provide no evidence that these syndromes are associated with HPV vaccination \cite{gbcvs}.
		
		In spite of these reports, the Danish debate about safety of HPV vaccines has not been settled, and vaccination rates are declining \cite{ssi_hpv_nyhed} as also seen in France and Japan \cite{fonteneau2015use, hanley2015hpv}. It can be argued that many of the symptoms that have been reported by vaccinated females are non-specific and will not be captured in epidemiological studies where outcomes are based on hospital discharge diagnosis \cite{brinth2015chronic}. For example, headache, orthostatic intolerance, fatigue, nausea, and cognitive dysfunction were the five most commonly reported symptoms among 53 girls/women who suspected adverse events \cite{brinth2015suspected}. These symptoms will not be systematically recorded in registries and databases that are used for epidemiological studies. Hence, there is a need for additional studies to investigate the safety of HPV vaccines. In order to inform such studies, we aimed to elucidate the Danish signal and to obtain a better understanding of the epidemiology of the reported adverse events. The specific aim of the present study was to determine health care-seeking prior to the first HPV vaccination among females who suspected and reported adverse reactions to HPV vaccine.
		
		\section{Methods}
		
		We conducted a registry-based case-control study where cases were vaccinated females with reports of suspected severe adverse reactions, and controls were vaccinated females without known reports of suspected adverse reactions. From the DMA we obtained a line-list of individuals with suspected adverse reactions. These events were reported by the patients, the parents, or their doctors, and were classified as severe adverse reactions by established criteria (mainly because the symptoms affected everyday life, e.g., not being able to attend school or work as usual, and lead to ``persistent or significant disability/incapacity'') \cite{ichHarmonised}. The list was obtained in August 2015, and included data on 322 individuals with a valid Danish Civil Registry Number (CPR-number) which we needed for further linkage. These individuals represented all cases classified as severe and reported with a valid CPR-number at the time of the data retrieval. After excluding one female from Greenland and five males, the dataset consisted of 316 cases.

		Controls were selected from the Danish vaccination registry \cite{grove2011danish} and matched with cases in groups by year of vaccination (+/- 1 year), age at vaccination (+/- 1 year) and municipality. We included all available controls rather than aiming at a fixed number of controls per case. We obtained information on contacts to primary health care and hospital services in a two-year period prior to the first HPV vaccine. Primary health care data were obtained from The National Health Insurance Service Register. Based on reimbursement codes, contacts were categorized as (1) consultations including visits at general practitioners and specialists as well as home visits, (2) consultations by phone or e-mail, (3) laboratory analyses at a primary health care provider or by referral, (4) any type of physiotherapy or referral to chiropractor, (5) private specialist in psychiatry (including youth psychiatry) and psychologist, and (6) dental treatment.

		Contacts to hospitals were obtained from the National Patients Registry. We included only main diagnoses. We parametrized diagnoses according to chapters in ICD-10 except for Z00-Z99 which was a commonly used category that was separated according to frequency of contacts into Z01.6 (radiological examination, not elsewhere classified), Z03-9 (observation for suspected disease or condition, unspecified), Z30-Z39 (persons encountering health services related to reproduction), and other contacts in this chapter. 
		
		Both for primary health care contacts and hospital diagnosis we counted one activity per group rather than calculating the number of contacts or diagnoses, i.e., if a study subject had at least one visit at the general practitioner's office it would count as an exposure in that category and any follow-up visits were not included.

		We analysed the data by logistic regression while adjusting for the matching variables; this procedure was applied as cases and controls were matched on a group basis. Due to low numbers, municipality was aggregated to provinces (nomenclature of territorial units for statistics (NUTS) code 2), year of vaccination was categorized as 2006-2009, 2010, 2011, 2012, 2013, 2014-15, and age was fitted as years except 10-12, 30-39, and 40+ years that were treated as groups due to low numbers. All these parameters were parametrized by dummy variables. For the final multivariable model, we considered variables that were associated with being a case (p-value {\textless} 10 \%). These variables were subsequently eliminated from the logistic regression model if they were not significant at the 5 \% level. Excluded variables were one-by-one re-entered in the final model and once more eliminated based on significance testing. Finally, the significant exposures from the multivariable model were examined for statistical interaction with age categorised as  10-12 years (the youngest target age for the HPV vaccination programme), 13-18 years (old girls), 19-28 (young women in the catch-up programme), and 29+ (older women). The study was notified to the Danish data protection agency under the record number 2008-54-0474.

		\section{Results}
		
\begin{figure}
			\centering
			\includegraphics[width=0.9\textwidth]{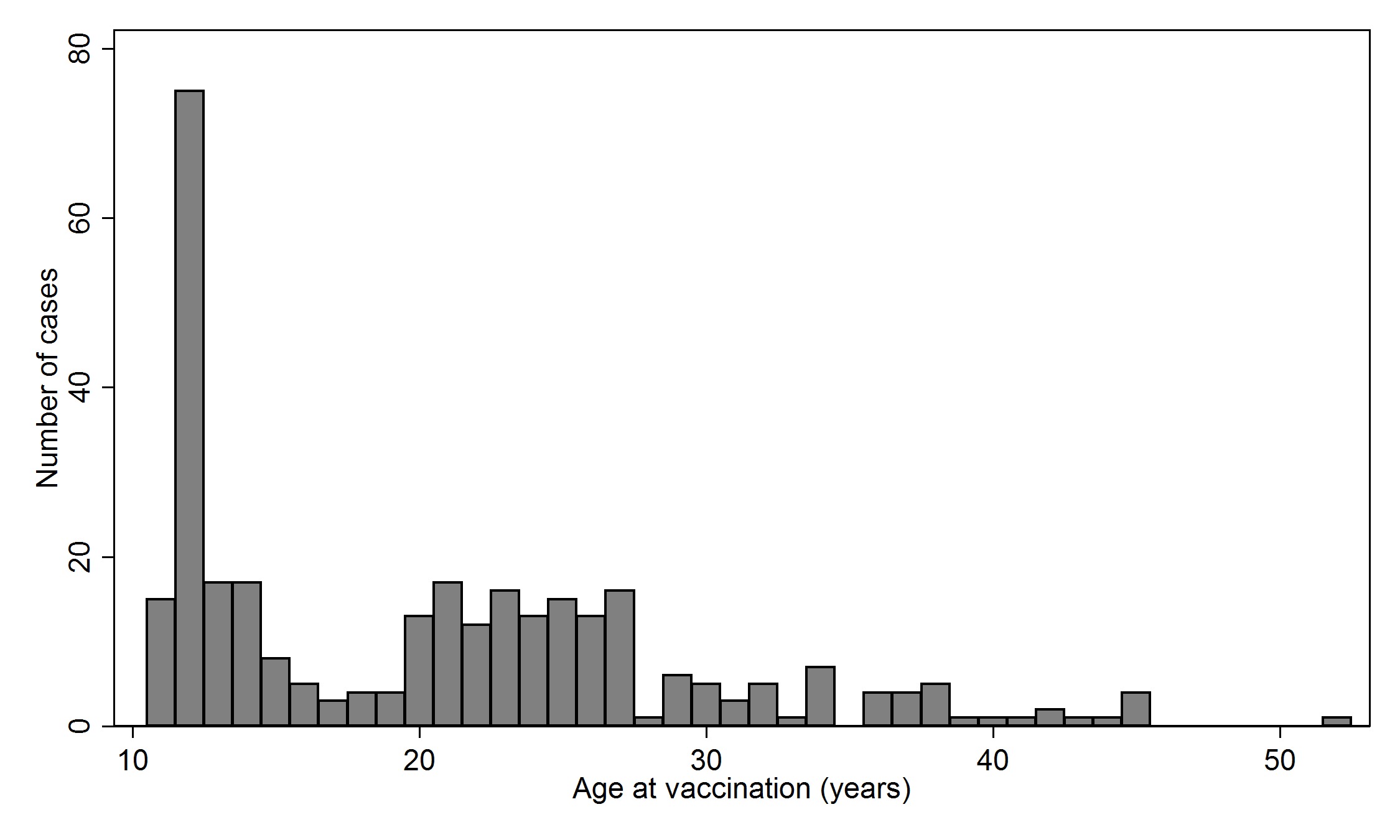}
			\caption{\textbf{Age at first HPV vaccination for 316 females who had suspected adverse events reported to the Danish Medicines Agency and served as cases for the registry based case-control study on care-seeking  prior to the first HPV vaccination.}}
			\label{fig:HPV_sideeffects_fig1}
		\end{figure}

		The 316 cases had a median age of 20 years (range 11-52 years) and the 163,910 controls a median age 19 years (range 10-52 years). Figure \ref{fig:HPV_sideeffects_fig1} shows a bimodal age distribution of the cases with one peak around 12 years of age when childhood vaccinations are offered and another 19-28 years of age corresponding to the catch-up programme. The cases received the first vaccine between 2006 and 2014; in total 168 (53.2\%) received it in 2012-2013 and 140 of these were {\textgreater} 18 years and therefore likely to be part of the catch-up programme offered in those two years.
		
Table \ref{table:hpvsideeffectstable1} shows contacts at the primary health care level in the two years prior to vaccination. Cases had increased odds of consultations by phone or e-mail and had more laboratory analyses requested than controls. Furthermore, cases were more often referred to physiotherapy/chiropractor and psychiatry/psychologist.

		\begin{table}
			\small
			\centering
			\begin{adjustbox}{center}
			\begin{tabular}{lrrrrp{1cm}c}
				\toprule
				\multirow{2}{*}{Type of contact before first vaccination} & \multicolumn{2}{c}{Cases} & \multicolumn{2}{c}{Controls} & \multirow{2}{1cm}{Odds ratio} & \multirow{2}{*}{95\% CL} \\
				& Number & Percent & Number & Percent & & \\
				\cmidrule(lr){1-1} \cmidrule(lr){2-3} \cmidrule(lr){4-5} \cmidrule(lr){6-6} \cmidrule(lr){7-7}
				Consultation at the office & 310 & 98.1 & 157853 & 96.3 & 1.74 & 0.77–3.92 \\
				Consultation by phone or e-mail & 299 & 94.6 & 142862 & 87.2 & 2.38 & 1.44–3.92 \\
				Laboratory analysis request & 249 & 78.8 & 110092 & 67.2 & 1.73 & 1.29–2.32 \\
				Physiotherapy or related & 96 & 30.4 & 20459 & 12.5 & 2.56 & 1.99–3.31 \\
				Psychologist, psychiatrist et cetera & 39 & 12.3 & 9751 & 5.9 & 2.18 & 1.54–3.11 \\
				Dentist & 134 & 42.4 & 62590 & 38.2 & 0.94 & 0.65–1.35 \\
				\bottomrule
			\end{tabular}
			\end{adjustbox}
			\caption{\textbf{Primary health care contacts (assessed by reimbursement codes) two year before the first HPV vaccination in 316 females who reported
					suspected adverse events to the vaccine and 163,910 matched controls.} Denmark, 2006 to 2015. Odds ratio is adjusted for age at vaccination, year of
				vaccination, and province (NUTS 2 area code).}
			\label{table:hpvsideeffectstable1}
		\end{table}
		
Table \ref{table:hpvsideeffectstable2} shows that in the two years prior to vaccination there was an increased odds of diagnosis in several chapters of the ICD-10 system including the respiratory system, the digestive system, the musculoskeletal system, injuries as well as factors influencing health status and contact with the health services and symptoms, signs and abnormal findings, not elsewhere classified.
		
		\begin{table}
			\centering
			\begin{adjustbox}{center}
			{\small
			\begin{tabular}{rp{5cm}rrrrC{1cm}c}
				\toprule
				\multicolumn{2}{l}{\multirow{2}{*}{Type of contact before first vaccination}} & \multicolumn{2}{c}{Cases} & \multicolumn{2}{c}{Controls} & \multirow{2}{1cm}{Odds ratio} & \multirow{2}{*}{95\% CI} \\
				& & Number & Percent & Number & Percent & & \\
				\cmidrule(lr){1-2} \cmidrule(lr){3-4} \cmidrule(lr){5-6} \cmidrule(lr){7-7} \cmidrule(lr){8-8}
				Z01.6 & Radiological examination, not elsewhere classified &68 &21.5 &18359 &11.2 &1.91 &1.45-2.53\\\rule{0pt}{2.5ex}
				Z03.9 & Observation for suspected disease or condition, unsp.  &26 &8.2 &5660 &3.5 &2.18 &1.45-3-27\\\rule{0pt}{2.5ex}
				Z30-Z39 & Persons encountering health services related to reproduction &37 &11.7 &9413 &5.7 &1.35 &0.92-1.98\\\rule{0pt}{2.5ex}
				Z00-Z99 & Factors influencing health status and contact with health services, \textit{not included above} &80 &25.3 &21835 &13.3 &1.82 &1.40-2.36\\\rule{0pt}{2.5ex}
				A00-B99 & Certain infections and parasitic diseases  &5 &1.6 &2400 &1.5 &1.07 &0.44-2.62\\\rule{0pt}{2.5ex}
				C00-D89 & Neoplasms, disease of the blood and blood forming organs and certain disorders involving immune mechanisms  &6 &1.9 &2319 &1.4 &0.93 &0.41-2.12\\\rule{0pt}{2.5ex}
				E00-E90 & Endocrine, nutritional and metabolic diseases &9 &2.9 &3368 &2.1 &1.18 &0.60-2.30\\\rule{0pt}{2.5ex}
				F00-F99 & Mental and behavioural disorders &15 &4.8 &6450 &3.9 &1.20 &0.71-2.03\\\rule{0pt}{2.5ex}
				G00-G99 & The nervous system &9 &2.9 &2561 &1.6 &1.55 &0.79-3.03\\\rule{0pt}{2.5ex}
				H00-H59 & The eye and adnexa &7 &2.2 &1806 &1.1 &1.79 &0.84-3.81\\\rule{0pt}{2.5ex}
				H60-H95 & The ear and mastoid process &3 &1.0 &1483 &0.9 &1.01 &0.32-3.19\\\rule{0pt}{2.5ex}
				I00-I99 & The circulatory system &3 &1.0 &1094 &0.7 &0.92 &0.29-2.92\\\rule{0pt}{2.5ex}
				J00-J99 & The respiratory system &14 &4.4 &3958 &2.4 &1.81 &1.05-3.10\\\rule{0pt}{2.5ex}
				K00-K93 & The digestive system &23 &7.3 &5420 &3.3 &2.08 &1.35-3.20\\\rule{0pt}{2.5ex}
				L00-L99 & The skin and subcutaneous tissue &6 &1.9 &2555 &1.6 &1.14 &0.50-2.57\\\rule{0pt}{2.5ex}
				M00-M99 & The musculoskeletal system and connective tissue &38 &12.0 &8516 &5.2 &2.28 &1.62-3.23\\\rule{0pt}{2.5ex}
				N00-N99 & The genitourinary system &15 &4.8 &5875 &3.6 &0.93 &0.55-1.59\\\rule{0pt}{2.5ex}
				O00-O00 & Pregnancy, childbirth and the puerperium &41 &13.0 &11037 &6.7 &1.37 &0.95-1.98\\\rule{0pt}{2.5ex}
				P00-P96 & Conditions originating in the perinatal period &0 &{}- & 53 & {\textless} 0.1 & {}- &  \\\rule{0pt}{2.5ex}
				Q00-Q99 & Congenital malformations, deformations and chromosomal abnormalities & 5 & 1.6 & 2310 & 1.4 & 1.10 & 0.45-2.66\\\rule{0pt}{2.5ex}
				R00-R99 & Symptoms, signs and abnormal clinical and laboratory findings, not elsewhere classified & 44 & 13.9 & 9709 & 5.9 & 2.33 & 1.68-3.22\\\rule{0pt}{2.5ex}
				S00-T14 & Injuries & 102 & 32.3 & 35693 & 21.8 & 1.81 & 1.42-2.30\\\rule{0pt}{2.5ex}
				T14-T98 & Other injuries, poisoning and certain other consequences of external causes not included in S00-T14 & 11 & 3.5 & 3952 & 2.4 & 1.36 & 0.74-2.49\\\rule{0pt}{2.5ex}
				X00-X90 & Various external causes of accidental injury & 1 & 0.3 & 93 & 0.1 & 6.77 & 0.93-49.45\\
				\bottomrule
			\end{tabular}}
			\end{adjustbox}
			\caption{\textbf{Hospital contacts two year before the first HPV vaccination in 316 females who reported suspected adverse events to the vaccine and
					163,910 matched controls.} Denmark, 2006 to 2015. Odds ratio is adjusted for age at vaccination, year of vaccination and province (NUTS 2 area code).}
			\label{table:hpvsideeffectstable2}
		\end{table}

The final multivariable model, Table \ref{table:hpvsideeffectstable3}, found independent effects of use of telephone/email consultations (Odds Ratio (OR) 1.9), physiotherapy/chiropractor (OR 2.1), and psychiatrist/psychologist (OR 1.9). Further, cases were more likely to have a diagnosis in the chapters of diseases of the digestive system (OR 1.6), of the musculoskeletal system (OR 1.6), symptoms or signs not classified elsewhere (OR 1.8), as well as injuries (OR 1.5). In total 203 (64.2 \%) of cases had one or more occurrences of the hospital diagnoses in Table \ref{table:hpvsideeffectstable3} or had received reimbursement for physiotherapy, chiropractor, psychiatry or psychologist, compared with 66,222 (40.4 \%) of controls.

\begin{table}
\centering
\begin{adjustbox}{center}
\begin{tabular}{p{7cm}C{4cm}c}
\toprule
\textbf{Type of contact before first HPV vaccination} & Multivariable odds ratio for care-seeking &  95\% CI\\
\midrule
Consultation at primary health care provider by phone or e-mail & 1.91 & 1.15-3.16\\
\rule{0pt}{4ex}Reimbursement of physiotherapy, chiropractor or related treatment & 2.13 & 1.64-2.76\\
\rule{0pt}{4ex}Reimbursement of psychologist, psychiatrist or related treatment & 1.87 & 1.31-2.66\\
\rule{0pt}{4ex}Hospital contact, ICD-10 code K00-K93, the digestive system & 1.57 & 1.01-2.45\\
\rule{0pt}{4ex}Hospital contact, ICD-10 code M00-M99, the musculoskeletal system and connective tissue & 1.56 & 1.09-2.23\\
\rule{0pt}{4ex}Hospital contact, ICD-10 code R00-R99, symptoms, signs and abnormal clinical and laboratory findings, not elsewhere classified & 1.77 & 1.27-2.48\\
\rule{0pt}{4ex}Hospital contact, ICD-10 code S00-T14, injuries & 1.51 & 1.18-1.93\\
\bottomrule
\end{tabular}
\end{adjustbox}
\caption{\textbf{Final multivariable model showing health care-seeking in the two years prior to vaccination in 316 Danish females who reported suspected adverse events to HPV vaccination compared with 163,910 matched controls.} Care-seeking was defined as contact to primary health care (from registry of reimbursements) or hospital (from the Danish National Patient Registry) in a two-year period before the first HPV vaccine. All estimates were adjusted for age at vaccination, year of vaccination and province (NUTS 2 area code). }
\label{table:hpvsideeffectstable3}
\end{table}
		
Table \ref{table:hpv_side_effects_table4} shows the age-stratified results. Overall, there was no effect modification by age as evaluated by a formal test for interaction (p-values ranging from 0.10 to 0.78, Table \ref{table:hpv_side_effects_table4}).  Thus, with the current sample size, findings were consistent across age groups. However, there was a tendency of a stronger association between exposure and outcome in the oldest age-category. In most exposure-categories, care-seeking tended to increase by increasing age with the notable exception of injuries which were more common in girls than in young women. Forty two percent of the cases 11-12 years of age had a hospital contact due to injuries compared with 27 \% cases, Table \ref{table:hpv_side_effects_table4}. 
		
\begin{table}
	
	\centering
	\begin{adjustbox}{center}
	\scalebox{0.8}{
	\begin{tabular}{p{5cm}p{2cm}rrrrC{1cm}cC{2cm}}
		\toprule
		& & \multicolumn{2}{c}{Cases} & \multicolumn{2}{c}{Controls} &  & & \\
		\cmidrule(lr){1-2} \cmidrule(lr){3-4} \cmidrule(lr){5-6} \cmidrule(lr){7-9}
		Type of contact before first vaccination & Age at first
		HPV vaccine & Number & Percent & Number & Percent & Odds ratio$^*$ & 96\% CI & p-value for interaction with age$^{**}$ \\
		\cmidrule(lr){1-2} \cmidrule(lr){3-4} \cmidrule(lr){5-6} \cmidrule(lr){7-9}
		\multirow{4}{5cm}{Consultation at primary health care provider by phone or e-mail} & 10-12 years & 78 & 86.7 & 39739 & 77.0 & 2.11 & 1.25-3.59 & 0.567 \\
		& 13-18 years & 51 & 94.4 & 23176 & 79.4 & 1.23 & 0.81-1.87 & ~ \\
		& 19-28 years & 118 & 98.3 & 77003 & 96.1 & 1.92 & 1.28-2.90 & ~ \\
		& 29 years + & 52 & 100.0 & 2944 & 99.2 & 23.54 & 15.21-36.46 & ~ \\
		
		\rule{0pt}{5ex}\multirow{4}{5cm}{Reimbursement of physiotherapy, chiropractor or related treatment} & 10-12 years & 12 & 13.3 & 3032 & 5.9 & 1.91 & 1.05-3.44 & 0.104 \\
		& 13-18 years & 14 & 25.9 & 2700 & 9.2 & 1.62 & 0.73-3.61 & ~ \\
		& 19-28 years & 49 & 40.8 & 13809 & 17.2 & 1.58 & 0.82-3.07 & ~ \\
		& 29 years + & 21 & 40.4 & 918 & 30.9 & 9.95 & 4.73-20.93 & ~ \\
		
		\rule{0pt}{5ex}\multirow{4}{5cm}{Reimbursement of psychologist, psychiatrist or related treatment} & 10-12 years & 3 & 3.3 & 520 & 1.0 & 2.54 & 0.81-7.97 & 0.486\\
		& 13-18 years & 2 & 3.7 & 494 & 1.7 & 0.88 & 0.15-5.36 & ~ \\
		& 19-28 years & 27 & 22.5 & 8409 & 10.5 & 0.87 & 0.26-2.93 & ~ \\
		& 29 years + & 7 & 13.5 & 328 & 11.1 & 5.80 & 1.46-22.99 & ~ \\
	
		\rule{0pt}{5ex}\multirow{4}{5cm}{ICD 10 code K00-K93, the digestive system} & 10-12 years & 2 & 2.2 & 909 & 1.8 & 0.93 & 0.23-3.76 & 0.782\\
		& 13-18 years & 2 & 3.7 & 652 & 2.2 & 1.71 & 0.24-12.27 & ~ \\
		& 19-28 years & 14 & 11.7 & 3726 & 4.7 & 2.48 & 0.56-11.06 & ~ \\
		& 29 years + & 5 & 9.6 & 133 & 4.5 & 23.32 & 4.41-123.31 & ~ \\

		\rule{0pt}{5ex}\multirow{4}{5cm}{ICD 10 code M00-M99, the musculoskeletal system and connective tissue} & 10-12 years & 4 & 4.4 & 1575 & 3.1 & 1.13 & 0.42-3.05 & 0.629\\
		& 13-18 years & 4 & 7.4 & 1306 & 4.5 & 1.51 & 0.37-6.12 & ~ \\
		& 19-28 years & 23 & 19.2 & 5409 & 6.8 & 2.48 & 0.85-7.30 & ~ \\
		& 29 years + & 7 & 13.5 & 226 & 7.6 & 17.97 & 5.12-63.06 & ~ \\

		\rule{0pt}{5ex}\multirow{4}{5cm}{ICD 10 code R00-R99, symptoms, signs and abnormal clinical and laboratory findings, not elsewhere classified} & 10-12 years & 9 & 10.0 & 2052 & 4.0 & 1.95 & 1.00-3.81 & 0.465 \\
		& 13-18 years & 7 & 13.0 & 1220 & 4.2 & 1.62 & 0.59-4.42 & ~ \\
		& 19-28 years & 21 & 17.5 & 6210 & 7.8 & 1.07 & 0.48-2.39 & ~ \\
		& 29 years + & 7 & 13.5 & 227 & 7.7 & 9.77 & 3.54-26.99 & ~ \\

		\rule{0pt}{5ex}\multirow{4}{5cm}{ICD 10 code S00-T14, injuries} & 10-12 years & 38 & 42.2 & 13945 & 27.0 & 1.25 & 0.87-1.79 & 0.681 \\
		& 13-18 years & 24 & 44.4 & 7352 & 25.2 & 1.53 & 0.89-2.64 & ~ \\
		& 19-28 years & 30 & 25.0 & 14028 & 17.5 & 1.14 & 0.68-1.91 & ~ \\
		& 29 years + & 10 & 19.2 & 368 & 15.6 & 3.95 & 7.42-32.59 & ~ \\

		\bottomrule
	\end{tabular}}
\end{adjustbox}

\caption{\textbf{Age specific health care-seeking in the two years prior to vaccination in 316 Danish females who reported suspected adverse events to HPV vaccination compared with 163,910 matched controls.} Care-seeking was defined as contact to primary health care (from registry of reimbursements) or hospital (from the Danish National Patient Registry) in a two-year period before the first HPV vaccine. The analyses were restricted to significant variables from the final multivariable model, see Table \ref{table:hpvsideeffectstable3}.\newline
\textit{Note:} Cases/controls included: 10-12 years: 90/51,583; 13-18 years: 54/29,205; 19-28 years: 120/80,154; 29 years +: 52/2,968. \newline
* Odds ratio from a logistic regression model adjusting for year of vaccination and province.  \newline
** Log-likelihood ratio test comparing main effects model with a model including main effects and an interaction term between age group and care-seeking}

\label{table:hpv_side_effects_table4}
\end{table}

In the national patient registry, cases had in total 1030 discharge diagnoses (mean 3.3 per case) compared with 267,912 diagnoses among the controls (mean 1.6 per case). There were 34 diagnoses given to the 23 cases with a diagnosis of the digestive system, including haemorrhage of anus and rectum (4 occurrences) as well as  functional dyspepsia and maxillary hypoplasia (3 occurrences each), gastro-oesophageal reflux disease, irritable bowel syndrome, unspecified functional intestinal disorder, unspecified anal fissure, and calculus of gallbladder without cholecystitis (all 2 occurrences). There were 73 diagnoses given to the 38 cases with a diagnosis of the musculoskeletal system included epicondylitis lateralis (7 occurrences), chondromalacia patellae, pains in limb, other chronic osteomyelitis (4 occurrences each), unspecified shoulder lesion, muscle strain, dislocation and subluxation of joint, juvenile rheumatoid arthritis, and other psoriatic arthropathies (all 3 occurrences each). There were 99 diagnoses given to the 44 cases with a code from the R-chapter, including 47 diagnoses of abdominal and pelvic pain, complex chronic non-malignant pain (7 occurrences), syncope and collapse (6 occurrences) and other and unspecified symptoms and signs involving the nervous and musculoskeletal systems (7 occurrences). There were 200 diagnoses given to the 102 cases with a diagnosis  from the S00 to T14 (injuries). The most commonly reported diagnosis was sprain and strain of ankle (18 occurrences), various contusions (44 occurrences), soar and strains of fingers, wrist or knee (21 occurrences), open wound of finger (7 occurrences) and fracture of finger (5 occurrences). 

\section{Discussion}

We found that girls or women with reports of suspected adverse reactions from HPV vaccination already before their first vaccination had a health care-seeking pattern different from a matched population who did not report adverse events. For example, there was increased odds of having had injuries and being in contact with physiotherapy and chiropractor. This is compatible with the hypothesis that many  cases had high levels of physical activity before onset of symptoms, as suggested by Brinth et al.\ \cite{brinth2015suspected}. In a case-series (without a reference group) of 53 patients with possible adverse events Brinth et al.\ found that 67 \% had a high and 33 \% had a moderate physical activity level before symptom onset. Five patients were competing on a national or international level in their sport \cite{brinth2015suspected}. Our observation of increased number of injuries, musculoskeletal complaints and referral to physiotherapy can be interpreted as an agreement with their findings.

It is, however, unlikely that high physical activity can explain all our observations. Frequent health care attendance and the increased odds of having symptoms in the R-chapter of the ICD-10 system, including abdominal pains, syncope and other symptoms from the autonomic nervous system may point towards another hypothesis namely the presence of medically unexplained physical symptoms coinciding with the time of vaccination. Medically unexplained physical symptoms and somatoform disorders are among the most common causes of contacts with health care and is particularly common among frequent health care attenders \cite{peveler1997medically, reid2001medically}. Because of the high prevalence, medically unexplained physical symptoms are likely to be present in a considerable proportion of vaccinated females. The temporal association may, in the mindset of the parents or the vaccinated, be linked to vaccination.

Furthermore, cases had an increased odds ratio of contact to psychologist or psychiatrist but not an increased odds of hospital diagnoses in the F-chapter of ICD-10, mental and behavioural disorders. The most common psychiatric diagnosis was emotionally unstable personality disorder including borderline disorder but this diagnosis was not more frequent in cases than in controls. 

Further studies based on larger datasets are needed to examine whether increased care-seeking is related to specific disorders in order to understand whether any specific factors may be associated with the experience of suspected adverse events.

Notably, of the first 107 claims that were reported to The Danish Patient Compensation Association, only three were recognized as adverse reactions and received compensation whereas all others were dismissed. In several of the dismissed claims, the symptoms were present before the vaccination \cite{danishPatient}.

One important strength of the study was that the assessment of care-seeking was obtained by an ongoing and unbiased collection of data to national registries. The figures are therefore reliable. Also, a number of outcomes occurred at equal odds including for example physical primary health consultations, contacts related to pregnancy and childbirths, psychiatric diagnoses given at a hospital as well as appointment at the dentists. This suggests that the case- and the control-groups were comparable. Furthermore, all reported estimates were adjusted for age at vaccination, year at vaccination and place of residence.

The study is subject to some limitations. Since August 2015 when we received the data, the number of reported possible adverse events have increased furthermore, and by the end of 2015 a total of 2019 possible adverse events have been reported to the DMA of which 823 have been classified as severe. At the present, we do not know if the 316 cases included in our study are representative of all reported cases. The primary health care data contained limited clinical information that represents another limitation. The assessment of the care-seeking behaviour to primary health care was crude and data on diagnoses and medications issued in primary health care would have added further value to the study. Furthermore, the data from primary health care was limited to activities that were reimbursed as part of the National Health Insurance Service. We did not obtain information on activities such as self-paid contacts to psychologist or school psychologists as well as physiotherapy offered in organisations such as sports clubs. Hence, the proportion of females with this type of care are minimum estimates.

The study was based on data prior to the first HPV vaccine and does not exclude that some of the females who reported suspected adverse events indeed experienced reactions caused by the vaccine. However, the study shows that the case-population as a whole was a selected population; this may in particular be the situation for those above 28 years of age. In most cases, some symptoms were present before the start of HPV vaccination. As mentioned, 140 cases (44 \%) had received the vaccination in the catch-up programme 2012-13 and in total 52 (16 \%) were older than 28 years at first vaccination and thus outside of the official programme. This may indicate that more perceived adverse events can be expected to be reported when the vaccines are offered outside of the core target group, i.e., girls before onset of sexual activity. We propose that the high vaccination-uptake in particular in the catch-up programme and beyond by coincidence led to considerable number of temporal associations between vaccination and medically unexplained physical symptoms. Teenagers and young women have, as seen in the control population of the present study, a higher proportion of ICD-10 diagnosis in the K, M and R chapters of the ICD-10 system (Table \ref{table:hpv_side_effects_table4}), and therefore a catch-up programme may have a higher risk of being challenged by the perception of adverse events than a steady routine vaccination at 12-years old.

The observed excess morbidity and excess care seeking does not rule out that the vaccine in certain situations may have triggered a course that resulted in deterioration of symptoms in some individuals in a perhaps vulnerable subpopulation, and further research should address this possible explanation. The aim was not to disentangle these and other possible trajectories; our aim was to explore the care-seeking behaviour before the first vaccination and emphasise that any conclusions as regards the safety of the vaccine should be taken based on an understanding of the characteristics of the group from which adverse events were reported.

A number of post-licensure epidemiological studies have addressed the safety of HPV vaccines, including the expected number of immune mediated disorders in girls 12-15 years prior to introduction of HPV vaccination \cite{callreus2009human}. After the introduction of the HPV vaccine the risk of a number of defined outcomes was not found to be associated with HPV vaccination in several studies \cite{arnheim2013autoimmune, scheller2015quadrivalent, cameron2016adverse}. A large cohort study from France did not find an increased incidence of the autoimmune disorders studied, with the exception of  Guillain{}-Barré syndrome for which there was a small increased risk in the HPV vaccinated girls ($\sim$1 per 100,000 vaccinated girls) \cite{gbcvs}. Only few studies have looked at non-specific outcomes \cite{klein2012safety, donegan2013bivalent}, and we hope that our study will provide inspiration for additional epidemiological studies to address the safety of HPV vaccines, with a strong suggestion that pre-vaccination morbidity should be taken into account in the evaluation of safety signals.

\section{Acknowledgements }

We thank Line Michan at the Danish Medicines Agency, Ina Tapager, The Danish Health Data Authority, Tyra Grove Krause and Michael Galle, Statens Serum Institut, for collaboration and assistance.

\chapter{Ensemble Learned Vaccination Uptake Prediction using Web Search Queries}
\chaptermark{Ensemble Learned Vaccination Uptake}
\label{cha:ensemble_learned}

\begin{center}
	
	Venue: Conference on Information and Knowledge Management (CIKM), Indianapolis 2016
	
	\vspace{1cm}
	
\textit{
	Niels Dalum Hansen$^{ab}$, Kåre Mølbak$^c$,  Christina Lioma$^a$
}

\vspace{0.5cm}

$^a$University of Copenhagen, Denmark.
$^b$IBM Denmark.\\
$^c$Statens Serum Institut, Denmark.

\end{center}

\leftskip=1cm
\rightskip=1cm

\noindent\textit{We present a method that uses ensemble learning to combine clinical and web-mined time series data in order to predict future vaccination uptake. 
The clinical data is official vaccination registries, and the web data is query frequencies collected from Google Trends. 
Experiments with official vaccine records show that our method predicts vaccination uptake effectively (4.7 Root Mean Squared Error). Whereas performance is best when combining clinical and web data, using solely web data yields comparative performance.} 
\textit{To our knowledge, this is the first study to predict vaccination uptake using web data (with and without clinical data).}

\leftskip=0pt\rightskip=0pt

\section{Introduction and Related Work}
Predicting public health events, e.g.~how many people may get vaccinated in the near future, 
can reduce the reaction time of public health professionals, resulting in more efficient services and 
improved public health. 
Traditionally, public health event prediction relied on \textit{clinical data} (e.g.~microbiological results or patient registries) 
that was collected from designated bodies. In the last decade however, non-clinical \textit{web data} (e.g.~search engine queries or microblog messages), 
has been shown useful to the task of predicting public health events. Clinical and web data are complementary sources of evidence: 
Whereas clinical data contributes expert and curated information to the prediction, web data contributes near real-time information on a large scale about e.g.~symptoms or health concerns that may go undetected or unreported by the official clinical channels. 

We present a method for predicting vaccination uptake by combining clinical and web data using ensemble learning. 
Combining such clinical and web search data for vaccination uptake prediction is novel. 
So far, research on vaccination uptake has focused on the effect of physician recommendations on vaccination 
uptake \cite{Gargano2013}; how combined sources of information (e.g. physician, television, friends) 
influence people's decisions about vaccination \cite{gargano2015influence}; and the effects of media coverage on vaccination uptake with respect to influenza vaccination \cite{ma2006influenza}, HPV vaccination \cite{kelly2009hpv}, 
and MMR vaccination \cite{
	smith2008media}. To our knowledge, our study is the first to predict vaccination uptake using web data (with and without clinical data). 


Web and/or clinical data have been used before for other types of health event predictions, 
e.g.~influenza activity 
\cite{ginsberg2009detecting, mciver2014wikipedia, paul2014twitter, polgreen2008using, santillana2014using, yuan2013monitoring}, dengue fever \cite{chan2011using} and cholera \cite{chunara2012social}.  How the different types of data should be handled has evolved from using a unified model for  both web and clinical data \cite{lazer2014parable, yang2015accurate}, to using ensemble methods that model separately clinical and web data and then combine the outputs \cite{santillana2015combining}.
When web search query frequencies are used for prediction \cite{santillana2015combining, santillana2014using, yang2015accurate}, 
a single linear model is used to combine the query frequencies into a prediction. 
Methods using query frequencies select queries either by (i) timely correlation between query search frequency and 
the health event \cite{ginsberg2009detecting, santillana2015combining, santillana2014using, yang2015accurate}, or by (ii) expert selection of 
queries \cite{chan2011using, polgreen2008using, yuan2013monitoring}. 
Both approaches have disadvantages. Approach (i) relies on calculating the correlation between the health event time series and all queries, which is computationally expensive. It also assumes that historic correlation equals predictive power in the future, which may not always be the case. Approach (ii) relies on human experts, which is costly and does not scale well. In this work we propose a third approach: We select queries based on web descriptions of the health event, in our case of the vaccine in question, and we use an ensemble learning approach, specifically stacking, to predict vaccination uptake. 

\section{Ensemble Learning Prediction}\label{s:predicting_health}



Vaccination uptake prediction with time series data can be formulated as: 
$\hat{E}(t) \approx E(t-1)$, 
where $\hat{E}(t)$ is the predicted vaccination uptake at time $t$, and $E(t-1)$ is the observed vaccination uptake at time $t-1$. 
We compute $\hat{E}(t)$ using ensemble learning by combining separate predictions on vaccination uptake based on clinical 
and web data into one prediction. 
Ensemble learning combines predictions from an ensemble of level-0 models into one prediction using a level-1 meta model. 
We use an ensemble method called \textit{stacking}. 
First, all level-0 models are trained. 
Then, a level-1 model is trained to make a final prediction using all the predictions of the level-0 models as input. 
We experiment with three different types of level-1 models: a linear model, support vector regression (SVR) with a linear kernel, 
and SVR  with a Gaussian kernel. Both our clinical and web data are time series, i.e.~each data point has a temporal reference.

\subsection{Level-1 models}
\noindent\textbf{Stacking with linear model.} We define a linear model with two explanatory variables:
$\hat{E}(t) = \mu +  \beta_1 \hat{E}_c(t) + \beta_2 \hat{E}_w(t)$,
where $\hat{E}_c(t)$ is the prediction based on clinical data at time $t$,   
$\hat{E}_w(t)$ is the prediction based on web data at time $t$, and $\mu$, $\beta_1$ and $\beta_2$ denote the coefficients that need to be optimized. We use ordinary least squares to find the coefficients that minimize:
$\underset{\mu, \beta_1, \beta_2}{min} \sum_t \left(E(t) - \mu - \beta_1 \hat{E}_c(t) - \beta_2 \hat{E}_w(t)\right)^2$.

\noindent\textbf{Stacking with SVR.} SVR solves the same problem as the linear model presented above, but with the possibility of using kernels to transform the input into another feature space.
In addition $\mu$, $\beta_1$ and $\beta_2$ are selected to minimize the following:
$\underset{\mu, \beta_1, \beta_2}{min} \sum_t V(E(t) - \mu - \beta_1 \hat{E}_c(t) - \beta_2 \hat{E}_w(t)) + \frac{\lambda}{2} (\mu^2 + \beta_1^2 + \beta_2^2)$,
where $\lambda$ is a hyperparameter controlling the penalty for large coefficients, and $V(r)$ is defined as $0$ if $|r| < \epsilon$ and otherwise $|r|-\epsilon$. The parameter $\epsilon$ controls how precise the prediction has to be before it is treated as correct.

We experiment with an SVR with linear kernel and with a Gaussian kernel defined as:
$K(x,x') = \exp(-\gamma ||x-x'||^2)$,
where $\gamma$ is a hyperparameter.

\subsection{Level-0 models}
\noindent \textbf{Prediction with clinical data.} As level-0 models we use 
three well-known time series methods: autoregressive (AR) models \cite{yang2015accurate,lazer2014parable}, ARIMA and Holt Winters (HW). 

AR models estimate $\hat{E}(t)$ as: 
$\hat{E}(t) = \mu + \sum^m_{i=1} \beta_i E(t-i)$ 
where $m$ is the number of autoregressive terms, $\mu$ is the intercept, and the $\beta$s control the weight that each past observation has on the prediction. AR models assume that future values of $E$ can be predicted by a linear combination of the $m$ most recently observed values of $E$. 
With enough autoregressive terms AR models can handle seasonal changes, but not general upwards or downwards trends.


An extension of the AR models are the ARIMA (AutoRegressive Integrated Moving Average) models. In addition to the autoregressive terms, these models also include a moving average, which is a weighted sum of the $q$ most recent forecasting errors. 
Let $m$ denote the number of autoregressive terms and $q$ the number of moving averages; then:
$\hat{E}(t) = \mu + \sum^m_{i=1} \beta_i E(t-i) + \sum^q_{j=1} \phi_j \epsilon_{t-j} + \epsilon_t$,
where $\epsilon_t = E(t) - \hat{E}(t)$. To handle trend, the original signal $E$ can be differentiated one or more times \cite{chatfield2016analysis}.


HW forecasting is defined by three recursive equations controlling: level, trend and seasonality. HW can forecast time series with both trend and seasonal changes. 
Each equation is defined as a weighted sum in which the weight of historic observations decreases exponentially with time. 
HW forecasting with level, trend and seasonality is recursively defined as:

\begin{equation}\label{eq:hw}
\begin{aligned}
\text{level} & \quad &a_t &= \alpha (E(t) - s_{t-l}) + (1-\alpha) (a_{t-1} + b_{t-1})\\
\text{trend} & \quad &b_t &= \beta (a_t - a_{t-1}) + (1-\beta) b_{t-1}\\
\text{seasonality} & \quad& s_t &= \gamma (E(t) - a_t) + (1-\gamma) s_{t-l}\\
\end{aligned}
\end{equation}

\noindent where $l$ is the length of the season and $\alpha$, $\beta$ and $\gamma$ are the smoothing parameters which control the influence of the historic level, trend and seasonality. 
Predictions are made by combining level, trend and seasonality:
$\hat{E}(t) = a_{t-1} + b_{t-1} + s_{t - l + 1}$. 



\noindent \textbf{Prediction with web data.} As level-0 models we use a linear model, bagging and weighted majority. 
Our web data consists of time-stamped query frequencies (described in Section \ref{s_ensemble_eval}).

\noindent\textit{Linear model.} Given a collection of $n$ query frequency time series, denoted $Q$, we define a simple linear model as:
$\hat{E}(t) = \mu + \sum_{i=1}^n \alpha_i Q_i(t)$,
where $\mu$ and $\alpha$ are coefficients to be estimated. Such a model can be fitted using any of 
several methods, 
the most common being ordinary least squares. 
Another approach is to use LASSO regularization which is commonly used for making predictions using query frequencies \cite{santillana2015combining, santillana2014using, yang2015accurate}. This approach adds an additional constraint to the optimization, namely that the sum of the coefficients should also be minimized. 
The weight of this sum is controlled by the hyperparameter $\lambda$. This approach can be used to avoid overfitting and to reduce the coefficients of non-informative 
features to zero and thereby induce a sparse model. 
This is a useful property in this context because the collection of queries might 
contain non-informative terms.

\noindent\textit{Bagging.} With bagging, we consider the average of the predictions made on subsets of the training data. This helps to reduce variance and overfitting. We generate subsets of the training data by uniformly sampling with replacement $n$ datasets of size $m$. For each dataset a linear model, as defined above, is fitted using LASSO regularization, 
where the parameter $\lambda$ is found using 3-fold cross-validation. The prediction of the ensemble is the average of the $n$ predictions.

\noindent\textit{Weighted Majority.} We extend the bagging approach to a boosting approach using a weighted majority (WM) algorithm \cite{littlestone1994weighted}. 
The WM algorithm works by combining predictions from a collection of models using a weighted average. Each model is associated with its own weight related to its 
previous predictive performance. If the overall prediction is wrong by a constant $\epsilon$, the weights are updated. The updating works as follows: 
if the individual prediction of a model has an error $> \epsilon$, a new weight is calculated as $w_i = w_i \exp(-\eta)$, where $w_i$ is the weight for 
model $i$ and $\eta$ is a hyperparameter controlling the penalty for making wrong predictions. Our collection of models is identical to the models used for the 
bagging approach described above.

%
%

\begin{table}
	\begin{tabular}{lp{10cm}}
		  \textbf{Vaccine} & \textbf{Terms in Danish (English)}\\ 
		  \hline
		   MMR &levende (alive), mæslinger (measles), vaccine, vaccinen (the vaccine), udbrud (outbreak), alvorlige (serious),  fåresyge (mumps), måneders (months), undersøgelser (examinations)\\
		   &beskyttelse (protection), voksne (adults), gravid (pregnant), kombineret (combined), dosis, hunde (dogs), alderen (the age), hjernebetændelse (inflammation of the brain)\\
		   &lungebetændelse (pneumonia), gives (is given), mfr (mmr), røde (red)\\
		   DiTeKiPol & mæslinger (measles), vaccinen (the vaccine), alvorlige (serious), beskyttelse (protection), kombineret (combined), vaccination, indeholder (contains), type, beskytter (protects)\\ 		&sygdomme (illness), meningitis, forårsaget (caused), dræbte (killed), b, kighoste (whooping cough), vare (lasts), polio, difteri (diphtheria), mindst (least), stivkrampe (tetanus)                  \\
		   PCV & vaccinen (the vaccine), alvorlige (serious), alderen (the age), lungebetændelse (pneumonia), vaccination,  infektioner (infections), sygdomme (illness), forebygger (prevents)\\ 				&meningitis, forårsaget (caused), antal (number), blodforgiftning (blood poisoning)  \\
		   HPV & beskyttelse (protection), gives (is given), vaccination, tilbuddet (the offer), kondylomer (condyloma), doser (doses), kønsvorter (genital warts), tilbydes (is offered), piger (girls)\\ 	&livmoderhalskræft (cervical cancer), forventes (is expected), indeholder (contains), januar (january), langvarig (long term), indført (introduced), tilbud (offer), type, human\\ 
		   &beskytter (protects), effekten (the effect), skyldes (caused by), hpv, pigerne (the girls)\\
	\end{tabular} 
	
	\caption{Our 58 queries.}
\label{tab:queries}

\end{table}

\section{Experimental Evaluation} \label{s_ensemble_eval}


\noindent\textbf{Data\footnote{\scriptsize{All our data is freely available at: \url{https://sid.erda.dk/share_redirect/c7j6MdrscL}}}.} We evaluate the effectiveness of our approach in predicting vaccination uptake in Denmark for 
all official children vaccines: DiTeKiPol-1, DiTeKiPol-2, DiTeKiPol-3, DiTeKiPol-4, PCV-1, 
PCV-2, PCV-3, MMR-1, MMR-2(4), MMR-2(12), HPV-1, HPV-2 and HPV-3. We use as clinical data the actual vaccination uptake recorded by the country's official body, the State Serum Institut. Specifically, the vaccination uptake is the total number of vaccines given in a month divided by the number of people expected to be vaccinated that month 
(based on the size of the monthly birth cohorts published by Statistics Denmark).

We use as web data web search queries that are related to each vaccine. We generate these queries from descriptions of each vaccine in: \textit{www.ssi.dk}, \textit{www.patienthåndbogen.dk}, and \textit{www.min.medicin.dk} (authoritative medical health portals). 
We remove stopwords and collect terms that occur in at least two different descriptions of each vaccine. 
 We treat each term as a query (i.e.\ we use only single term queries) and we submit it to Google Trends using Denmark as the geographical region and with the time period set to January 2011 - September 2015   
(only limited coverage of Denmark is available prior to 2011). 
Only 58 out of 85 queries had enough coverage in Google Trends to return a result. We use these 58 queries for our predictions (shown in Table \ref{tab:queries}).

\noindent\textbf{Training.} We use as training data all data which is available prior to the data point being predicted. 
Hence if we are predicting the vaccination uptake in February 2014 we train on data from January 2011 -- January 2014. 
All models are refitted for each time step. We use monthly time steps. 
To allow for inference of seasonality, the level-0 models are initialized with 24 months of available data (January 2011 -- December 2012) as training data. 
For the level-1 models we start by using 12 months of data (January 2013 -- December 2013). 
We evaluate our predictions using the 
root mean squared error (RMSE), 
which penalizes large errors more than small. 


Our prediction methods are fitted using R packages with default settings at all times, except for the starting point for HW, where we manually select a starting point 
of the optimization if it cannot be completed with the default value. The AR model is trained using 12 autoregressive terms to capture seasonal variations.
For bagging and weighted majority we use as many subsets as there are queries, each subset contains 10 randomly sampled queries. 
For the weighted majority we use $\eta=5$ and $\epsilon=2$ for all experiments.

\noindent \textbf{Results.} Table \ref{tab:individual} shows the results when predicting vaccination uptake using either clinical or web data only (with the methods presented in Section 2).
 ``Naive'' refers to our naive baseline $\hat{E}(t) = E(t-1)$. 
Our methods outperform the naive baseline except for the HPV vaccines. This might be due to an intense debate in Denmark regarding the safety of this particular vaccine. Such a debate is likely to boost query frequencies but not necessarily vaccination uptake (the fact that many more people talk about HPV does not mean that many more HPV vaccines are given). We see that methods using clinical data outperform the methods using web data for the majority of the vaccines. But interestingly this difference is not very big and for the vaccines DiTeKiPol-3 and DiTeKiPol-4 the methods based on web data perform best. DiTeKiPol-4 is especially interesting since a shortage in 2013 resulted in unusual vaccination behaviour for a few months. When making predictions from web data our two new approaches (bagging and WM) perform best for 9 of the 13 vaccines.


Table \ref{tab:ensemble} shows the results for the ensemble predictions using clinical and web data. Except for the three HPV vaccines, the ensemble approaches outperform all other methods using only one data source. We see that when using an SVR with a Gaussian kernel as level-1 model we obtain the best results, i.e.~7/13 lowest RMSE. When comparing within the methods using an SVR with a Gaussian kernel, the HW+WM is the best performing method. The most improvements are obtained when combining predictions based on web data with either predictions from HW or AR12. 
  
\begin{table}[!h]
\centering
\adjustbox{center}{
{\small
\begin{tabular}{lrrrrrrrr}
\toprule
 & & \multicolumn{3}{c}{Clinical data} & \multicolumn{4}{c}{Web data} \\
             &   Naive &     HW &   AR12 &   ARIMA &     WM &      B & L &     O \\
\cmidrule(lr){2-2} \cmidrule(lr){3-5} \cmidrule(lr){6-9}

 MMR-1       &  20.704                   & \textbf{18.149} & \textbf{18.606} & \cellcolor{blue!24} \textbf{15.574} & \textbf{16.609} & \textbf{16.597} & \textbf{16.605} & 30.387 \\
 MMR-2 (4)   &  20.582                   & \cellcolor{blue!24}\textbf{13.110} & \textbf{16.566} &  \textbf{16.284} & \textbf{15.841} & \textbf{15.635} & \textbf{15.500} & 29.288 \\
 MMR-2 (12)  &  20.637                   & \textbf{19.592} & \textbf{20.600}   &  \cellcolor{blue!24}\textbf{18.726} & 21.631 & 20.815 & \textbf{21.112} & 31.897 \\
 HPV-1       & \cellcolor{blue!24} 8.080 & 11.291          & 11.192 &   9.871 & 13.474 & 14.320  & 12.701 & 11.547 \\
 HPV-2       & \cellcolor{blue!24} 8.704 & 12.522          & 12.806 &  11.276 & 18.154 & 18.025 & 18.423 & 15.404 \\
 HPV-3       & \cellcolor{blue!24} 6.579 &  9.161          & 13.958 &   9.418 & 24.239 & 23.494 & 23.074  & 17.317 \\
 DiTeKiPol-1 &  14.091                   &  \textbf{6.700} &   \textbf{5.185} &   \cellcolor{blue!24}\textbf{5.097} &  \textbf{8.067} &  \textbf{8.058} & \textbf{8.069 }& 15.913 \\
 DiTeKiPol-2 &  17.693                   &  \cellcolor{blue!24}\textbf{7.520} &    \textbf{8.030} &   \textbf{8.064} & \textbf{10.003} &  \textbf{9.941} & \textbf{9.951} & 20.082 \\
 DiTeKiPol-3 &  17.884                   & \textbf{17.596} &  20.936 &  19.459 & \textbf{17.160}  & \textbf{17.160}  &\cellcolor{blue!24} \textbf{17.158} & 30.424 \\
 DiTeKiPol-4 &  21.676                   & 26.103          &  21.676 &  23.385 &\cellcolor{blue!24} \textbf{15.414} & \textbf{15.535} & \textbf{15.934} & 33.888 \\
 PCV-1       &  13.323                   &  \textbf{6.897}  &  \cellcolor{blue!24}\textbf{6.394} &   \textbf{6.623} &  \textbf{7.745} &  \textbf{7.797} & \textbf{7.845} & 14.014 \\
 PCV-2       &  17.533                   & \cellcolor{blue!24} \textbf{7.266}  &  \textbf{8.845} &   \textbf{8.353} &  \textbf{9.679} &  \textbf{9.796} & \textbf{9.770} & \textbf{16.027} \\
 PCV-3       &  18.405                   &  \textbf{7.877}  &  \textbf{7.781} &  \cellcolor{blue!24} \textbf{7.634} & \textbf{10.410}  & \textbf{10.364} & \textbf{10.368} & \textbf{15.582} \\
 \bottomrule
\end{tabular}}}
\caption{RMSE of predictions with only clinical or web data. WM: weighted majority, B: Bagging, L: linear model w. LASSO and O: linear model w. OLS. Blue: lowest RMSE per vaccine. Bold: better than naive.}
\label{tab:individual}
\end{table}

\begin{table}[!h]
\centering
\adjustbox{center}{
\scalebox{0.6}{
\begin{tabular}{lrrrrrrrrrrrr}
& \multicolumn{12}{c}{\large{OLS}} \\
\toprule
             &   HW+WM &   HW+B &   HW+L &   HW+O &  AR12+WM &   AR12+B &   AR12+L &   AR12+O &   ARIMA+WM &   ARIMA+B &   ARIMA+L &   ARIMA+O \\
\cmidrule(lr){2-5} \cmidrule(lr){6-9} \cmidrule(lr){10-13}
 MMR-1       &  \textbf{15.190} & \textbf{15.476} & \cellcolor{blue!24}\textbf{15.187} & \textbf{16.842} &   \textbf{17.697} & 17.572          &    17.457          &   \textbf{18.151} &   16.296          &   16.492          &  15.968          &     17.877 \\
 MMR-2 (4)   &  \cellcolor{blue!24}\textbf{12.875} & 13.497          & 13.349          &  13.121         &  16.305          & 16.108          &    16.195          &   \textbf{16.032} &   18.872          &   16.220          &  21.085          &     \textbf{15.871} \\
 MMR-2 (12)  &  \textbf{18.082} & \textbf{17.650} & \textbf{17.541} & \cellcolor{blue!24}\textbf{17.221} &  \textbf{18.711} & \textbf{19.523} &    \textbf{18.762} &   \textbf{18.909} &   \textbf{18.469} &   19.409          &  18.961          &     \textbf{18.693} \\
 HPV-1       &  \textbf{10.552} & \textbf{10.435} & \textbf{10.960} & 11.516          &   \cellcolor{blue!24}\textbf{9.377} &  \textbf{9.690} &    \textbf{10.130} &   \textbf{10.281} &   10.348          &   10.080          &   9.992          &     10.080 \\
 HPV-2       &  12.743          & \textbf{12.923} & 14.384          & \textbf{12.191} &  \textbf{11.883} & \textbf{12.201} &    13.240          &   \textbf{10.503} &   \textbf{10.708} &   \textbf{10.655} &  \textbf{10.279} &     \cellcolor{blue!24} \textbf{9.220} \\
 HPV-3       &   \cellcolor{blue!24}\textbf{8.743} &  \textbf{8.771} & 10.231          &  9.987          &  \textbf{11.321} & \textbf{11.063} &    \textbf{12.110} &   \textbf{12.151} &    9.893          &    \textbf{9.237} &   9.818          &      9.918 \\
 DiTeKiPol-1 &   \textbf{6.416} &  \textbf{6.875} &  \textbf{6.477} &  \textbf{5.498} &   \textbf{4.835} &  \textbf{4.831} &   \cellcolor{blue!24}  \textbf{4.829} &    \textbf{5.082} &    6.072          &    5.690          &   5.625          &      5.584 \\
 DiTeKiPol-2 &   9.094          &  8.216          &  8.967          &  7.956          &   \textbf{7.686} &  \cellcolor{blue!24}\textbf{7.343} &     8.116          &    \textbf{8.019} &    9.461          &    8.989          &  15.485          &      9.057 \\
 DiTeKiPol-3 &  18.478          & 17.891          & 18.410          & \cellcolor{blue!24}\textbf{16.662} &  17.529          & 18.225          &    17.550          &   \textbf{18.439} &   17.168          &   17.227          &  \textbf{17.137} &     \textbf{19.076} \\
 DiTeKiPol-4 &  \cellcolor{blue!24}15.812          & 17.977          & 17.495          & \textbf{19.860} &  17.290          & 19.891          &    16.537          &   \textbf{19.849} &   24.391          &   45.079          &  36.403          &     24.220 \\
 PCV-1       &   7.042          &  \textbf{5.783} &  \textbf{5.716} & \cellcolor{blue!24} \textbf{5.391} &   \textbf{6.201} &  \textbf{5.950} &     \textbf{5.830} &    \textbf{5.973} &   10.785          &    \textbf{6.569} &   9.174          &      \textbf{6.169} \\
 PCV-2       &   \textbf{8.317} &  8.236          &  9.283          &  \cellcolor{blue!24}7.553          &  10.135          &  \textbf{8.670} &    10.401          &    \textbf{8.284} &    8.395          &    8.399          &   9.681          &      \textbf{8.330} \\
 PCV-3       &   \textbf{7.345} &  \textbf{7.436} &  8.199          &  \textbf{7.759} &   \textbf{6.825} &  \textbf{7.014} &     \textbf{7.108} &  \cellcolor{blue!24}  \textbf{6.736} &    7.931          &    8.364          &   8.240          &      7.670 \\
\bottomrule
\\
& \multicolumn{12}{c}{\large{SVR linear}} \\
\toprule
             &   HW+WM &   HW+B &   HW+L &   HW+O  &   AR12+WM &   AR12+B &   AR12+L &   AR12+O &   ARIMA+WM &   ARIMA+B &   ARIMA+L &   ARIMA+O \\
\cmidrule(lr){2-5} \cmidrule(lr){6-9} \cmidrule(lr){10-13}
 MMR-1       &  \textbf{16.541} & \textbf{15.969} & \cellcolor{blue!24}\textbf{15.478}                     & \textbf{16.905} &  17.298          & 17.252          &    17.399                             &   \textbf{18.496}                    &   19.406          &   16.793          &  16.857          &     \textbf{18.204} \\
 MMR-2 (4)   &  \textbf{12.648} & \textbf{12.981} & \cellcolor{blue!24} \textbf{12.388} & \textbf{12.952} &  16.148          & \textbf{15.373} &    16.663                             &   \textbf{15.095}                    &   \textbf{14.969} &   16.101          &  \textbf{15.419} &     \textbf{15.620} \\
 MMR-2 (12)  &  \textbf{17.906} & \textbf{18.054} & \textbf{17.872}                     & \cellcolor{blue!24}\textbf{17.374} &  \textbf{19.302} & \textbf{19.020} &    \textbf{18.248}                    &   \textbf{19.075}                    &   \textbf{19.123} &   \textbf{18.193} &  19.195          &     \textbf{17.731} \\
 HPV-1       &  \textbf{10.530} & \textbf{10.649} & \textbf{10.922}                     & \textbf{11.146} &  \textbf{10.486} & \textbf{10.494} &    \textbf{10.688}                    &   \textbf{10.657}                    &   \cellcolor{blue!24}10.331          &   10.789          &  10.810          &     10.448 \\
 HPV-2       &  13.489          & \textbf{12.344} & \textbf{12.130}                     & \textbf{12.425} &  \textbf{10.147} & \textbf{10.467} &    \textbf{11.984}                    &   \textbf{11.062}                    &    \textbf{9.738} &    \textbf{9.906} &  \textbf{10.482} &     \cellcolor{blue!24} \textbf{8.727} \\
 HPV-3       &   \textbf{8.903} & \cellcolor{blue!24} \textbf{8.260} &  \textbf{8.501}                     & 10.674          &  \textbf{11.732} & \textbf{11.718} &    \textbf{13.049}                    &   \textbf{12.617}                    &    9.806          &   10.001          &   9.484          &     10.411 \\
 DiTeKiPol-1 &   6.926          &  \textbf{5.993} &  \textbf{5.739}                     &  \textbf{6.500} &   \textbf{4.740} &  \textbf{4.758} &    \cellcolor{blue!24} \textbf{4.626} &    \textbf{5.077}                    &    \textbf{5.090} &    5.354          &   5.287          &      5.507 \\
 DiTeKiPol-2 &   9.808          &  9.476          &  9.006                              &  9.100          &   8.476          &  8.563          &    10.503                             &    \cellcolor{blue!24}\textbf{7.734}                    &    9.938          &    8.480          &   9.872          &      9.591 \\
 DiTeKiPol-3 &  22.546          & 22.599          & 22.546                              & \cellcolor{blue!24}\textbf{17.433} &  21.402          & 21.925          &    21.154                             &   \textbf{19.003}                    &   22.244          &   21.319          &  22.104          &     \textbf{19.568} \\
 DiTeKiPol-4 &  16.694          & 37.909          & 17.357                              & \cellcolor{blue!24} \textbf{14.461} &  15.922          & 16.405          &    16.124                             &   \textbf{16.164}                    &   22.072          &   26.591          &  18.984          &     \textbf{17.609} \\
 PCV-1       &   7.351          &  7.186          &  \textbf{6.175}                     &  \textbf{6.198} &   \cellcolor{blue!24}\textbf{5.282} &  \textbf{6.143} &     \textbf{6.335}                    &    \textbf{5.510}                    &    6.765          &    7.413          &   6.840          &      6.830 \\
 PCV-2       &   \cellcolor{blue!24}\textbf{7.689} &  7.946          & 15.613                              &  7.955          &   9.029          &  8.879          &     9.104                             &    \textbf{8.823}                    &    8.794          &   11.662          &  14.559          &      9.024 \\
 PCV-3       &   \textbf{7.648} &  8.261          &  8.388                              &  \textbf{7.784} &   \textbf{6.904} &  \textbf{6.758} &     \textbf{6.994}                    &   \cellcolor{blue!24} \textbf{6.633} &    9.649          &    9.384          &   9.058          &      8.491 \\
\bottomrule
\\
& \multicolumn{12}{c}{\large{SVR Gaussian}} \\
\toprule
             &   HW+WM &   HW+B &   HW+L &   HW+O  &   AR12+WM &   AR12+B &   AR12+L &   AR12+O &   ARIMA+WM &   ARIMA+B &   ARIMA+L &   ARIMA+O \\
\cmidrule(lr){2-5} \cmidrule(lr){6-9} \cmidrule(lr){10-13}
 MMR-1       &  \cellcolor{blue!24} \textbf{14.928}& 16.694          & \textbf{16.355}                     & \textbf{16.198}                     & 17.770                             & 18.207          &    18.117          &   \textbf{16.927} &   16.703          &   17.490          &  17.569          &     17.560 \\
 MMR-2 (4)   &  14.377                             & \cellcolor{blue!24}\textbf{12.870} & 14.094                              & 13.122                              & \textbf{15.709}                    & \textbf{14.780} &    \textbf{15.024} &   \textbf{15.468} &   \textbf{14.973} &   \textbf{15.109} &  16.625          &     \textbf{15.838} \\
 MMR-2 (12)  &  \textbf{18.007}                    & \textbf{17.446} & \textbf{18.972}                     & \cellcolor{blue!24} \textbf{16.530} & \textbf{17.945}                    & \textbf{19.115} &    \textbf{18.406} &   \textbf{18.176} &   \textbf{18.385} &   \textbf{18.553} &  19.625          &     19.041 \\
 HPV-1       &  \textbf{10.748}                    & 11.606          & 11.918                              & \textbf{11.289}                     & \textbf{11.002}                    & \textbf{10.902} &    11.403          &    \textbf{9.130} &   11.664          &   10.684          &  11.505          &    \cellcolor{blue!24}  9.987 \\
 HPV-2       &  13.513                             & \textbf{11.958} & \textbf{12.304}                     & \textbf{12.097}                     & \textbf{10.789}                    & \textbf{12.376} &    \cellcolor{blue!24} 9.964          &   \textbf{12.483} &   \textbf{10.537} &   \textbf{11.249} &  11.715          &     \textbf{10.961} \\
 HPV-3       &  12.889                             & 13.784          & 14.775                              & 13.352                              & \textbf{13.016}                    & 14.521          &    15.222          &   14.374          &   12.818          &   12.172          &  \cellcolor{blue!24}12.147          &     12.682 \\
 DiTeKiPol-1 &   \textbf{5.204}                    & \cellcolor{blue!24} \textbf{5.008} &  \textbf{5.098}                     &  \textbf{5.331}                     &  5.486                             &  5.467          &     5.486          &    6.227          &    5.725          &    6.393          &   5.577          &      6.615 \\
 DiTeKiPol-2 &   7.927                             &  8.017          &  8.172                              &  9.661                              & \cellcolor{blue!24} \textbf{7.149} &  \textbf{7.443} &     \textbf{7.818} &    8.078          &    9.009          &    9.870          &   9.848          &      8.721 \\
 DiTeKiPol-3 &  \textbf{16.639}                    & \textbf{16.448} & \cellcolor{blue!24} \textbf{16.433} & \textbf{17.275}                     & 18.650                             & 18.442          &    19.009          &   \textbf{18.355} &   18.380          &   17.545          &  18.298          &     \textbf{18.962} \\
 DiTeKiPol-4 &  15.616                             & \cellcolor{blue!24}\textbf{14.877} & \textbf{15.246}                     & \textbf{16.543}                     & 16.865                             & 16.038          &    16.687          &   \textbf{15.653} &   15.938          &   15.647          &  \textbf{15.923} &     \textbf{15.932} \\
 PCV-1       &   \cellcolor{blue!24}\textbf{5.256} &  \textbf{5.808} &  \textbf{5.769}                     &  \textbf{5.664}                     &  \textbf{5.450}                    &  \textbf{6.358} &     \textbf{6.363} &    \textbf{6.614} &    \textbf{6.724} &    7.160          &   \textbf{6.611} &      7.091 \\
 PCV-2       &   \cellcolor{blue!24}\textbf{6.463} &  7.665          &  7.366                              &  7.470                              &  \textbf{8.811}                    &  \textbf{7.450} &     \textbf{6.952} &    9.026          &    8.672          &    9.062          &   8.991          &      9.148 \\
 PCV-3       &   \cellcolor{blue!24}\textbf{7.121}                    &  \textbf{7.665} &  8.396                              &  \textbf{7.798}                     &  9.022                             &  9.556          &    \textbf{10.008} &    7.871          &    8.616          &    8.148          &   9.527          &      8.176 \\
\bottomrule
\end{tabular}}}
\caption{RMSE of ensemble predictions (clinical and web data). Blue: lowest RMSE per vaccine. Bold: lower RMSE than for the individual ensemble components in Table 1.}
\label{tab:ensemble}
\end{table}
	
\section{Conclusions}

We presented a method that uses ensemble learning to combine clinical and web-mined time series data to make predictions about future vaccination uptake. As clinical data we used official registries of vaccines in Denmark. As web data we used query frequencies collected from Google Trends. We created those queries by extracting terms from publicly available descriptions of the vaccines on the web. Experiments using all officially recommended children vaccines in Denmark for the period January 2011 -- September 2015 showed that for 10/13 vaccines our ensemble learning methods that combined clinical with web data for prediction outperformed predictions using either clinical or web data alone.
Though this combination yields the lowest overall error, using only web data gives predictions with an error only slightly worse than for the predictions made using only clinical data. This indicates the potential usefulness of web data, such as query frequencies, to predict vaccination uptake in countries where there is no national vaccination registry. This work complements wider efforts in tackling medical and health problems computationally with machine learning or retrieval 
\cite{DragusinPLLJW11, DragusinPLLJCHIW13}.

%


\chapter{Time-Series Adaptive Estimation of Vaccination Uptake Using Web Search Queries}
\chaptermark{Time-Series Adaptive Estimation}
\label{cha:time_series_adaptive}

\begin{center}

Venue: The World Wide Web Conference (WWW), Perth 2017

\vspace{1cm}
	
\textit{
	Niels Dalum Hansen$^{ab}$, Kåre Mølbak$^c$,\\Ingemar Johansson Cox$^a$, Christina Lioma$^a$
}

\vspace{0.5cm}

$^a$University of Copenhagen, Denmark.
$^b$IBM Denmark.\\
$^c$Statens Serum Institut, Denmark. 
	
\end{center}

\leftskip=1cm
\rightskip=1cm

\noindent\textit{Estimating vaccination uptake is an integral part of ensuring public health. 
It was recently shown that vaccination uptake can be estimated automatically from web data, instead of slowly collected clinical records or population surveys \cite{DalumHansen:2016:ELV:2983323.2983882}. All prior work in this area assumes that features of vaccination uptake  collected from the web are temporally regular. We present the first ever method to remove this assumption from vaccination uptake estimation: our method dynamically adapts to temporal fluctuations in time series web data used to estimate vaccination uptake. We show our method to outperform the state of the art compared to competitive baselines that use not only web data but also curated clinical data. This performance improvement is more pronounced for vaccines whose uptake has been irregular due to negative media attention (HPV-1 and HPV-2), problems in vaccine supply (DiTeKiPol), and targeted at children of 12 years old (whose vaccination is more irregular compared to younger children).}

\leftskip=0pt\rightskip=0pt

%
%

%
%

%
%



\section{Introduction and related work}

Vaccination programs are an efficient and cost effective method to improve public health. 
With sufficiently many people vaccinated the population gains herd immunity, meaning the disease cannot spread. Timely actions to avoid drops in vaccination coverage are therefore of great importance. Many countries have no registries of timely vaccination uptake information, but rely for example on yearly surveys. In such countries estimations of near real-time vaccination uptake based solely on web data are valuable. We extend prior work in this area \cite{DalumHansen:2016:ELV:2983323.2983882}, which showed that vaccination uptake can be estimated sufficiently accurately from web search data. 
Our extension consists of a new estimation method that adapts dynamically to temporal fluctuations in the signal (web search queries in our case) instead of assuming temporal stationarity as in \cite{DalumHansen:2016:ELV:2983323.2983882}. This contribution is novel within vaccination uptake estimation.


Linear models have been used previously to estimate health events, for instance by  combining data from multiple sources with an ensemble of decision trees \cite{santillana2015combining}, or, closer to our work, by using query frequencies for influenza like illness \cite{yang2015accurate} or vaccination uptake estimation \cite{DalumHansen:2016:ELV:2983323.2983882}. These approaches are designed for stationary time series analysis, i.e.\  they assume data is generated by a stationary stochastic process. Our motivation is that vaccination uptake often does not follow stationary seasonal patterns. External events such as disease outbreaks, suspicion of adverse effects, or temporary vaccine shortages can alter uptake patterns for shorter or longer periods of time. Hence, while historical data is a good estimator in stable periods, as shown in \cite{DalumHansen:2016:ELV:2983323.2983882}, we reason that adapting the estimation to any unstability can reduce estimation error. We experimentally confirm this on all official children vaccines data used in Denmark between 2011 - 2016.  
%




\section{Aggregation with regression\\ trees for time series adaptation}
To account for seasonal non-stationarity, we use an online learning method, Aggregation Algorithm (AA) \cite{vovk2001competitive}, designed to automatically reduce estimation error in a changing environment. AA was recently used in time series prediction combined with an ensemble of ARIMA models \cite{jamil2016aggregation}. However, we reason that ARIMA models, or other traditional time series models, are not likely to be sufficient for vaccination uptake estimation in cases where: (i) there is more than one data source, e.g.\  vaccine uptake data and search frequency data, and (ii) when the time series data to be estimated are assumed to be unavailable (near real-time). 
To address these challenges, we combine AA with regression trees, motivated by recent research showing that random forests outperform ARIMA models on avian influenza prediction \cite{kane2014comparison}. A random forest, i.e.\  an ensemble of decision trees, is well suited for our problem since it is easy to extend to multiple data sources. 

Our method works as follows: We initially generate a set of regression trees. For each time step the ensemble of regression trees is retrained based on the initial set of trees and a weighted sum is used to make the estimation. AA is used to continuously update the weights of each tree. 
Each regression tree is trained based on a set of features and training samples. For each tree a feature set is drawn with replacement from the complete feature set.
Training samples are selected based on time-relative indices, where index 0 corresponds to the current time step. The indices are uniformly drawn with replacement from the interval $[0:s]$, where $s$ is a window size. We use trees with different window sizes to account for stationarity and non-stationarity of the signal. 

Our adaptive vaccination estimation algorithm is shown in Algorithm \ref{alg1}, where $\eta$ is the learning rate, RT a set of $N$ regression trees, $i$ the amount of initial training data and $y$ the vaccination uptake.
{ \footnotesize
\begin{algorithm}
	\caption{Adaptive time series estimation}
	\label{alg1}
	\begin{algorithmic}[1]
	\REQUIRE RT, $\eta$, $i$
	\STATE $W \leftarrow$ list with weights, initialize to be uniform
	\STATE $X \leftarrow$ list with the first $i$ training samples
	\STATE $Y \leftarrow$ list with the first $i$ observations of $y$
	\STATE $\hat{Y} \leftarrow$ empty list of estimations
	\STATE $t \leftarrow$ current time step, starting at $i+1$
	\WHILE {True}
	\STATE $x_t \leftarrow$ receive new observation from data stream
	\FOR {$n=0$ \TO $N$}
	\STATE Train RT$[n]$ using $X$ and $Y$
	\STATE $\hat{Y}_{temp}[n] \leftarrow$ estimation of RT$[n]$ given $x_t$
	\ENDFOR 
	\STATE $\hat{Y}[t] \leftarrow$ $\sum_{n=0}^N W[n]\cdot\hat{Y}_{temp}[n]$
	\STATE $Y[t] \leftarrow $ observed $y$ at time $t$
	\FOR{$n=0$ \TO $N$}
	\STATE $W[n] \leftarrow W[n]\cdot\exp(-\eta\cdot(\hat{Y}_{temp}[n]-Y[t])^2))$ 
	\ENDFOR 
	\STATE $W \leftarrow$ normalize $W$
	\STATE $X[t] \leftarrow x_t$
	\STATE $t \leftarrow t + 1$
	\ENDWHILE
	\end{algorithmic}
\end{algorithm}
}

\section{Evaluation}
To facilitate direct comparison, we evaluate our method on the same data as \cite{DalumHansen:2016:ELV:2983323.2983882}: monthly vaccination uptake of all official children vaccines in Denmark from January 2011 - June 2016. Vaccination uptake is defined as the total number of people vaccinated in a month divided by the birth cohort for that month. To estimate vaccination uptake, we use frequencies of web search queries extracted from Google Trends. We use the exact same frequencies of single term queries provided by \cite{DalumHansen:2016:ELV:2983323.2983882}. 


We compare to two baselines: (1) Linear regression with lasso regularization where the hyper-parameter is found using three fold cross-validation on the training data; (2) Linear regression with Elastic Net regularization where the two hyper-parameters are also selected using three fold cross-validation. 
We also include for reference two upper bounds corresponding to the best score reported in \cite{DalumHansen:2016:ELV:2983323.2983882} when using (i) only web data, and (ii) web data combined with clinical data. These scores are not theoretical upper bounds, but just the best scores across all methods evaluated in \cite{DalumHansen:2016:ELV:2983323.2983882}. We treat them as performance upper bounds because they do not correspond to any individual method, but to the best score per vaccine across all methods reported in \cite{DalumHansen:2016:ELV:2983323.2983882}. Neither baselines or upper bounds account for time series adaptation, i.e.\  they all assume data stationarity.


The initial number of training samples, $i$, is set to 24. All algorithms are evaluated in a leave-one-out fashion, where all data prior to the data point being estimated is used for training. For our algorithm (ATSE) a parameter search is performed by randomly sampling from the following intervals: Window size interval 1-46, number of features derived from vaccination data 0-45, number of features derived from web data 0-30, number of regression trees 500-10000, $\eta$ between 0.001-0.25.

\begin{table}
    \centering
    \scalebox{1}{
    \begin{tabular}{lrrrrr}
    	\toprule
    	Vaccine     & {LASS}      & {EN}    & {ATSE}     & {UBW \cite{DalumHansen:2016:ELV:2983323.2983882}} & {UBWC \cite{DalumHansen:2016:ELV:2983323.2983882}} \\ \midrule
    	HPV-1       & 14.6        & 13.8    & \textbf{10.0{*}} & 11.5                                              & 9.3                                                \\
    	HPV-2       & 15.9        & 16.1    & \textbf{10.1}    & 15.4                                              & 8.7                                                \\
    	MMR-1       & 12.9        & 12.9    & \textbf{12.6{*}} & 16.5                                              & 14.9                                               \\
    	MMR-2(4)    & 15.5        & 14.7    & \textbf{14.2}    & 12.4                                              & 12.3                                               \\
    	MMR-2(12)   & 21.7        & 21.4    & \textbf{16.0{*}} & 20.8                                              & 16.5                                               \\
    	DiTeKiPol-1 & 16.2        & 16.2    & \textbf{10.8}    & 8.0                                               & 4.6                                                \\
    	DiTeKiPol-2 & 14.1        & 14.2    & \textbf{12.4}    & 9.9                                               & 7.1                                                \\
    	DiTeKiPol-3 & 10.8        & 11.1    & \textbf{10.0{*}} & 17.1                                              & 16.4                                               \\
    	DiTeKiPol-4 & \textbf{13.7{*}} & 14.3{*} & 14.4{*}    & 15.4                                              & 14.4                                               \\
    	PCV-1       & \textbf{7.5}     & 7.8     & 10.0       & 7.7                                               & 5.2                                                \\
    	PCV-2       & 9.6*        & \textbf{9.5}  & 10.0       & 9.6                                               & 6.4                                                \\
    	PCV-3       & \textbf{9.4{*}}  & 9.5{*}  & 10.1{*}    & 10.3                                              & 6.6                                                \\ \bottomrule
    \end{tabular}
    }
    \caption{Estimation error when estimating vaccination uptake from web search queries with our method (ATSE), Lasso (LASS), Elastic Net (EN), and the two performance upper bounds of \cite{DalumHansen:2016:ELV:2983323.2983882} with web search (UBW) and web search and clinical data (UBWC). Bold marks best (excluding upper bounds). Asterisk marks better or equal to any upper bound.}
	\label{tab:results}
\end{table}

 Table~\ref{tab:results} displays the root mean squared error (RMSE) between the estimated vaccination uptake and the real vaccination uptake for all methods. 
Our method yields the overall best performance compared to the baselines (it outperforms all baselines for 8 out of 12 vaccines). Our method also outperforms the upper bounds of \cite{DalumHansen:2016:ELV:2983323.2983882} (any of the two) for 6 vaccines. This supports our reasoning that adapting the estimation to temporal fluctuations is a better strategy than assuming data stationarity. Our method yields the strongest performance improvements for HPV-1, HPV-2, MMR-2(12) and DiTeKiPol. All of these vaccines have temporally irregular uptake patterns, as explained next.
HPV-1 and HPV-2 have been subject to a heavy media debate in Denmark, with a subsequent drop in vaccinations. MMR-2(12) denotes the second MMR vaccine targeted 12 year-olds. As children grow, parents are less likely to follow the recommended vaccination schedule and fluctuations correlated with measles outbreaks are observed, thus making the time series less stationary. Lastly, in recent years there have been problems obtaining a sufficient supply of certain DiTeKiPol vaccines in Denmark, which might have forced people to postpone the initial vaccination, hence introducing irregularities in the signal. For PCV vaccines there have been no noted irregularities in their uptake patterns, which explains the slight drop in performance by our method compared to the baselines.


\section{Conclusion}

We presented an automatic method for near real time estimation of health events using web search query data. Our method combines an Aggregation Algorithm (AA) to automatically reduce estimation error in changing environments with regression trees. We applied our method to estimate vaccination uptake in all official Danish children vaccines, following  \cite{DalumHansen:2016:ELV:2983323.2983882}, and showed that our approach overall outperformed strong baselines that assumed data to be temporally regular. Our method was particularly strong estimating uptake for vaccines with known irregularities in their usage, such as HPV-1, HPV-2, MMR-2(12) and DiTeKiPol.

This work confirms recent findings that vaccination uptake can be automatically estimated only from web data, and further extends this area by accounting for irregular uptake patterns.

\chapter{Investigating the Relationship Between Media Coverage and Vaccination Uptake in Denmark}
\chaptermark{Media coverage and vaccination uptake}
\label{cha:MMR_Media}

\begin{center}
Venue: In review

\vspace{1cm}

\textit{
	Niels Dalum Hansen$^{ab}$, Kåre Mølbak$^c$,\\Ingemar Johansson Cox$^a$, Christina Lioma$^a$
}

\vspace{0.5cm}

$^a$University of Copenhagen, Denmark.
$^b$IBM Denmark. \\
$^c$Statens Serum Institut, Denmark.

\end{center}

\leftskip=1cm
\rightskip=1cm

\textit{\textbf{Background:} Understanding the influence of media coverage upon vaccination activity is a potential resource for
timely vaccination surveillance and similarly, might be an important factor when designing outreach campaigns. The
necessity and safety of the measles, mumps and rubella (MMR) vaccine has been debated for many years, making it a
suitable candidate for studying the interplay between media and vaccinations.}

\textit{\textbf{Objective:} Study the relationship between media coverage, incidence of measles, and vaccination activity of the
MMR vaccine in Denmark.}

\textit{\textbf{Methods:} The cross-correlations between media coverage (1,622 articles), vaccination activity (2 million
individual registrations), and incidence of measles are analyzed for the period 1997-2014. All 1,622 news media
articles are annotated as being pro-vaccination, anti-vaccination, or of neutral stance.
}

\textit{\textbf{Results:} The majority of anti-vaccination media coverage (65\% of total anti-vaccination coverage) is observed
in the period 1997-2004, immediately prior to and following the 1998 publication of the falsely claimed link between
autism and the MMR vaccine. For the period 1998-2004 we observe a statistically significant positive correlation
between the first MMR vaccine (targeting children aged 15 months), and pro-vaccination coverage (r=.49,
\textit{P}=.004), and between the first MMR vaccine and neutral media coverage (r=.45, \textit{P}=.003). For the first
MMR vaccine during the full period 1997-2014, we observe a statistically significant positive correlation with the
measles incidence (r=.31, \textit{P}=.005) with a lag of one month, indicating an increase in vaccinations following
measles outbreaks. Looking at the whole period, 1997-2014, we observe no significant correlations between vaccination
activity and media coverage.}

\textit{\textbf{Conclusions:} While there is no correlation between vaccination uptake and media coverage for the full period
1997-2014, there is a statistically significant positive correlation between pro-vaccination and neutral media coverage
and vaccination activity for the period following the falsely claimed link between autism and the MMR vaccine, in
1998-2004. The fact that a correlation was only observed during a period of controversy suggests that people are more
susceptible to media influence when presented with diverging opinions. Additionally, this correlation was only observed
for the first MMR vaccine, indicating that the influence of media is stronger on parents when they are deciding on the
first vaccine of their children, than on the subsequent vaccines.
}

\leftskip=0pt\rightskip=0pt

\section{Introduction}
The increased digitalization of traditional news media and the emergence of online-only media outlets creates
opportunities for real time computerized access and thereby more active utilization of media coverage in designing
public health strategies. Understanding the interplay between media coverage and public health events is the first step
in designing such strategies. We focus on vaccination programs because they are an important area within public health,
and also because they have historically proven vulnerable to media attention, e.g.\  the autism scare for the measles,
mumps and rubella (MMR) vaccine, and more recently the scare of adverse reactions to the human papillomavirus vaccine
\cite{molbak2016pre}. The public debate about vaccination safety and the necessity of vaccination programs has been going on since the
1800s \cite{larson2011addressing}. With the introduction of the internet this debate is no longer restricted to a small group of printed
newspapers, radio stations and TV-stations. This has resulted in a huge increase in the amount of stories being
published on this topic. In addition, social media platforms, web search engines, blogs and online discussion forums
contribute to the pool of online available information. Understanding the influence of media on vaccination uptake is
therefore important for designing public health strategies. This understanding can also be used to design automated
monitoring systems. These systems could for example alert health professionals when they need to actively participate
in the public debate or predict vaccination uptake faster than current census-based systems.

We focus on the MMR vaccine because reaching all children with two doses of a measles containing vaccine is an important
aim of all national immunization programs. In spite of that, many countries have difficulties achieving the declared
aim of measles elimination, i.e.\ 95\% coverage with both doses of the measles vaccine. According to WHO statistics for
2016, only 41/160 countries have a coverage of 95\% for the second MMR vaccine \cite{who_immunization}. In this respect, dissemination of
relevant news in public media may have both positive and negative impacts on the vaccination coverage, e.g.\  stories
about outbreaks may serve as reminders for parents to have their children vaccinated, whereas concerns about vaccine
safety may discourage parents from making appointments for vaccination. In this paper we focus on the potential
interplay between the uptake of the MMR vaccine, measles incidents and written media coverage for a 18 year period
(1997-1-1 to 2014-12-31) in Denmark. We use data from digital versions of traditional newspapers, online newspapers and
web only media sources. All data has been annotated as pro-vaccination, anti-vaccination or of neutral stance.

\subsection{Related work}
Understanding public health behavior is an important tool for planning and evaluating intervention programs. The effect
of news media coverage on people's behavior has been studied for many years both in general and specifically with
respect to vaccinations. In the 1970s during the presidential elections in the USA, McCombs and Shaw \cite{mccombs1972agenda} observed a
correlation between people's news consumption and their political opinions. With respect to the causal relationship,
McCombs and Shaw's results suggested an agenda-setting effect, meaning that the news affected people's opinion. Since
then, further studies have been conducted that research and confirm the agenda-setting effect of news media in a number
of settings, ranging from presidential elections to people's perception of crime rates and many more \cite{mccombs2009news}. The
agenda-setting effect depends on the issue at hand. If the issue affects people directly, e.g.\  raising gas prices, the
effect will be minimal; however, for more abstract issues, e.g.\  trade deficits or balancing the national budget, the
effect will be strong \cite{mccombs2009news}.

The effect of media coverage on vaccination uptake has been studied with respect to the influenza vaccine \cite{Yoo2010}, HPV vaccine \cite{kelly2009hpv}, and MMR vaccine \cite{smith2008media, mason2000impact}. Smith et al.\ \cite{smith2008media} focused on selective MMR non-recipients, meaning children who
received all recommended vaccinations except the MMR vaccine, and concentrated on media related to Wakefield et al.'s
1998 paper \cite{wakefield1998retracted} and its now discredited link between the MMR vaccine and autism. They concluded that there was a
limited influence of mainstream media on MMR vaccinations in the USA Mason and Donnely \cite{mason2000impact} observed that in the
period 1997-1998 the vaccination uptake in Wales was lower in areas where a series of anti-MMR vaccine articles had
been published, than areas where they had not been published. Ma et al.\ \cite{ma2006influenza} concluded that media coverage together
with recommendations from physicians were associated with increased influenza vaccination coverage in young children.
Finally Kelly et al.\ \cite{kelly2009hpv} looked at the relationship between media exposure and knowledge about the HPV vaccine. They
found that people exposed to health related media had more knowledge about HPV than people with less exposure. These
results indicate that, to some extent, there is an agenda-setting effect from media on people's vaccination behavior.

\section{Methods}

\begin{figure}
	\centering
	\includegraphics[width=\linewidth]{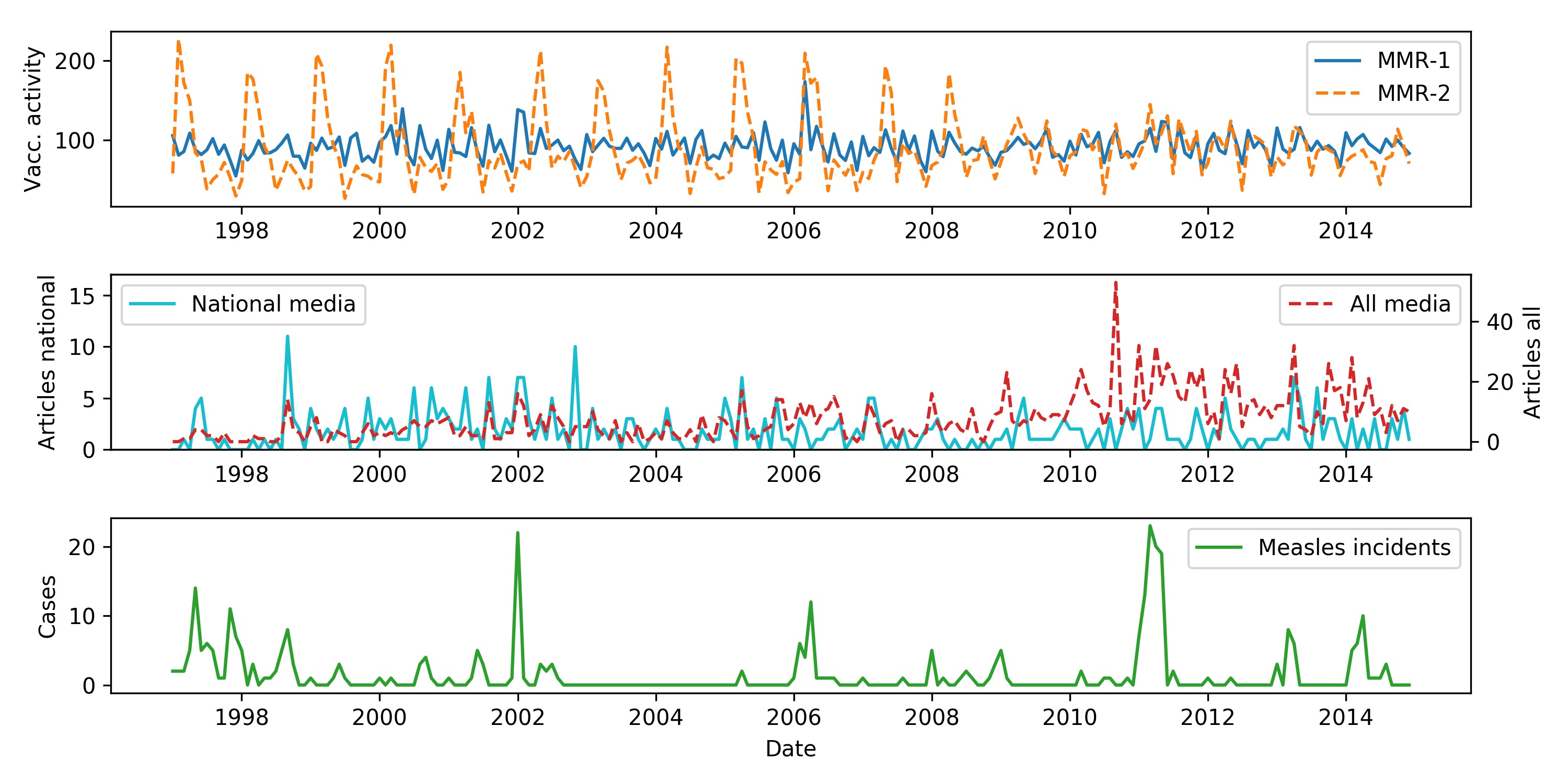}
	\caption{Plot of monthly vaccination activity, media coverage and measles incidents.}
	\label{fig:mmr+media_fig1}
\end{figure}

\subsection{Vaccination and measles incidence data}
The MMR vaccination program was introduced in Denmark on 1 January 1987 \cite{epinyt2015}. The vaccination program consists of two
vaccinations: one targeted at 15 month old children (MMR-1), and one targeted at 12 year old children (MMR-2). As of 1
April 2008 the MMR-2 vaccination schedule has changed to target 4 year old children \cite{epinyt2008}. Every time a general
practitioner vaccinates a child, the date and civil registry number (CPR) of the child are recorded in order for the
doctor to receive a reimbursement \cite{grove2011danish}. These reports are saved in the childhood vaccination database, an immunization
information system containing reports from 1997 onwards. Using the CPR number we looked up the birthday of the vaccinee
and calculated the child's age when receiving the vaccine. We separated the registered MMR vaccines into groups based
on the recommended vaccination schedule of 15 months, 4 years or 12 years. Each registered vaccine was assigned the
group where the age of the child at vaccination was closest to the target age of the group. We excluded data on the 4
year old children, because they were not represented in the full study period (this corresponds to 374,867
vaccinations). Using this approach, we had 1,098,389 registered vaccinations for MMR-1 and 1,108,205 registered
vaccinations for MMR-2. In the following we use MMR-1 to denote vaccinations of the 15 month olds and MMR-2 for 12 year
olds. Figure \ref{fig:mmr+media_fig1} (top plot) shows the vaccination activity for the two vaccines.

The above may data contain some reporting errors, mainly when doctors report the date of the reimbursement claim instead
of the vaccination date. We therefore aggregated data on a monthly basis.

To evaluate to what extent MMR vaccination numbers and media coverage about MMR were correlated with the number of
measles incidents, we also retrieved information about the number of measles incidents during the study period. Measles
is a notifiable disease and each reported case is registered in a central database. We aggregated data on a monthly
basis, which is shown in Figure \ref{fig:mmr+media_fig1} (bottom plot).

\subsection{Media coverage of MMR mined from the web}

To determine media coverage of MMR, we used infomedia \cite{infomedia}, an online Danish news archive. The archive covers 9 major
Danish newspapers, as well as a variety of national, regional and online publications. The number of sources indexed is
continuously expanding as regional newspapers, niche magazines and web pages are added to the database.

To measure media coverage related to the MMR vaccine, we constructed a query to retrieve relevant articles from the
infomedia database (this is standard practice when mining health information from the web \cite{DragusinPLLJW11,DragusinPLLJCHIW13}). The query, which we
will refer to as the MMR-query, is the following:

\begin{quotation}
\textit{((mæslinger OR mæslinge OR fåresyge OR ``røde hunde'' OR mfr) AND vaccine) OR 	mæslingevaccine OR fåresygevaccine OR ``røde hunde-vaccine'' OR ``mfr-vaccine''}
\end{quotation}

\noindent where \textit{mæslinger} is the Danish word for measles, \textit{fåresyge} means mumps, \textit{røde hunde} means
rubella and \textit{mfr} is the Danish abbreviation for MMR. This query will match all articles mentioning
\textit{mæslinger} or \textit{mæslinge} (plural or singular) or \textit{fåresyge} or \textit{røde hunde} together with
\textit{vaccine} or articles where either one of the compound phrases (as shown in the second line of the MMR-query) is
present. We then count the number of articles returned for this query, for each month of our study period. This type of
analysis, which is based on article frequency counts, is inspired by computational epidemiology approaches that use web
search frequencies to predict health events, e.g.\  influenza like illness \cite{ginsberg2009detecting}, vaccination coverage \cite{www17}.

The infomedia archive has expanded throughout the 18 year study period. In 1997 the infomedia \ archive indexed articles
from 20 sources, while in 2014 this number was 1,389. As the number of news sources grows, the number of new articles
added to the archive each month also grows. To accommodate for this change in archive size we use two sets of article
frequency counts; (i) 8 major nationwide newspapers that have been present in the full duration of the study. (ii) All
news sources in the database. We refer to approach (i) as \textit{national media} and (ii) as \textit{all media}. The
middle plot in Figure \ref{fig:mmr+media_fig1} shows the monthly number of articles retrieved using the MMR-query for each approach. For the
\textit{national media} we retrieved in total 390 articles and for \textit{all media} a total of 1,622 articles.

\subsection{Vaccination attitude annotation}
The first author of this work annotated each article mined from infomedia with respect to its attitude towards
vaccination, using these three vaccination attitude categories:

\textbf{Pro-vaccination:} Articles expressing positive views about the vaccine and/or encouraging people to get
vaccinated.

\textbf{Anti-vaccination:} Articles expressing negative views about the vaccine and/or discouraging people to get
vaccinated.

\textbf{Neutral:} Neutral information about the vaccine, e.g.\  reports on the number of people vaccinated per year or
diseases covered by the vaccine.

An article could be categorized into zero or more categories. For example, an article with an interview of an
anti-vaccination group accompanied by comments from a doctor explaining the medical reasons and benefits of getting
vaccinated would be categorized as both pro and anti-vaccination. Articles whose main focus is not the MMR vaccine
(e.g.\  vaccines for pets, annual accounts of vaccination producers, charities for developing countries, etc.) would
receive zero categories.

\subsection{Data analysis}

The data described above is time series, i.e.\ it consists of MMR/measles signals that have timestamps. We analyze this
data as follows. First we remove any seasonality to avoid general seasonal trends biasing the results, and then we
quantify the relationship between the MMR and media (or measles) signals.

\paragraph{Adjusting for seasonal correlations}
Any seasonality or serial dependencies in the signals is removed by fitting an autoregressive model to the signal and
subsequently using the residual of the fitted model. An autoregressive model is defined as:

\begin{equation}\label{eq:mmr+media_eq1}
	x_t = \alpha_0 + \sum_{i=1}^p \alpha_i X_{t-i} + \epsilon_t
\end{equation}

Where $X$ is a time series, $t$ is a time point, $p$ is the number of autoregressive terms, the
$\alpha$'s are the model coefficients, and $\epsilon_t $ is the
residual at time $t$.

To quantify seasonality and serial dependencies we calculate the auto-correlation and partial auto-correlation for all
signals. Auto-correlation refers to calculating the Pearson's correlation (Pearson's r) between the signal and a lagged
version of itself. Pearson's correlation for two time series, $X$ and $Y$, with mean $\mu $ and
length $n$ is defined as:

\begin{equation}\label{eq:mmr+media_eq2}
	r(X, Y) = \frac{\sum^{n}_{i=1}(X_i - \mu_X)(Y_i - \mu_Y)}{\sqrt{\sum_{i=1}^{n}(X_i-\mu_X)}\sqrt{\sum_{i=1}^n (Y_i - \mu_Y)}}
\end{equation}

The partial auto-correlation consists of calculating the correlation between the signal $X$, and a version of
itself with a lag of $k$, $X_k$, while at the same time controlling for the
auto-correlation of the $k-1$ previous lags \cite{chatfield2016analysis}. The partial auto-correlation at lag $k$ can be
calculated by fitting an auto-regressive model, as defined in Equation \ref{eq:mmr+media_eq1}, with $k$ auto-regressive terms. The
value of the $k$'th coefficient, i.e.\ $\alpha_k$, corresponds to the partial
auto-correlation at lag $k$. The partial auto-correlation can be used to determine the value of $p$ in
Equation \ref{eq:mmr+media_eq1}, because as the partial auto-correlation approaches zero, the value of additional autoregressive terms is
reduced.

\paragraph{Quantifying the relationship between signals}

To quantify the relationship between two signals we use the cross-correlation. The cross-correlation consists of
calculating the Pearson's correlation (Equation \ref{eq:mmr+media_eq2}) between two signals using different lags. A cross-correlation of 1
means perfect positive correlation, while a correlation of -1 corresponds to perfect negative correlation.

To measure the significance of the correlations we treat a series of $n$ cross-correlations as random variables
from a student's t-distribution with degrees of freedom $n-1$. We report the \textit{P-}value for correlations
that exceed a 99\% confidence interval.

\paragraph{Software}
The Python packages \textit{statsmodels} version 0.8.0 and \textit{scipy} version 0.19.0 were used for calculating
auto-correlation, partial auto-correlation and cross-correlation.

\section{Results}

\begin{figure}
	\centering
	\includegraphics[width=\linewidth]{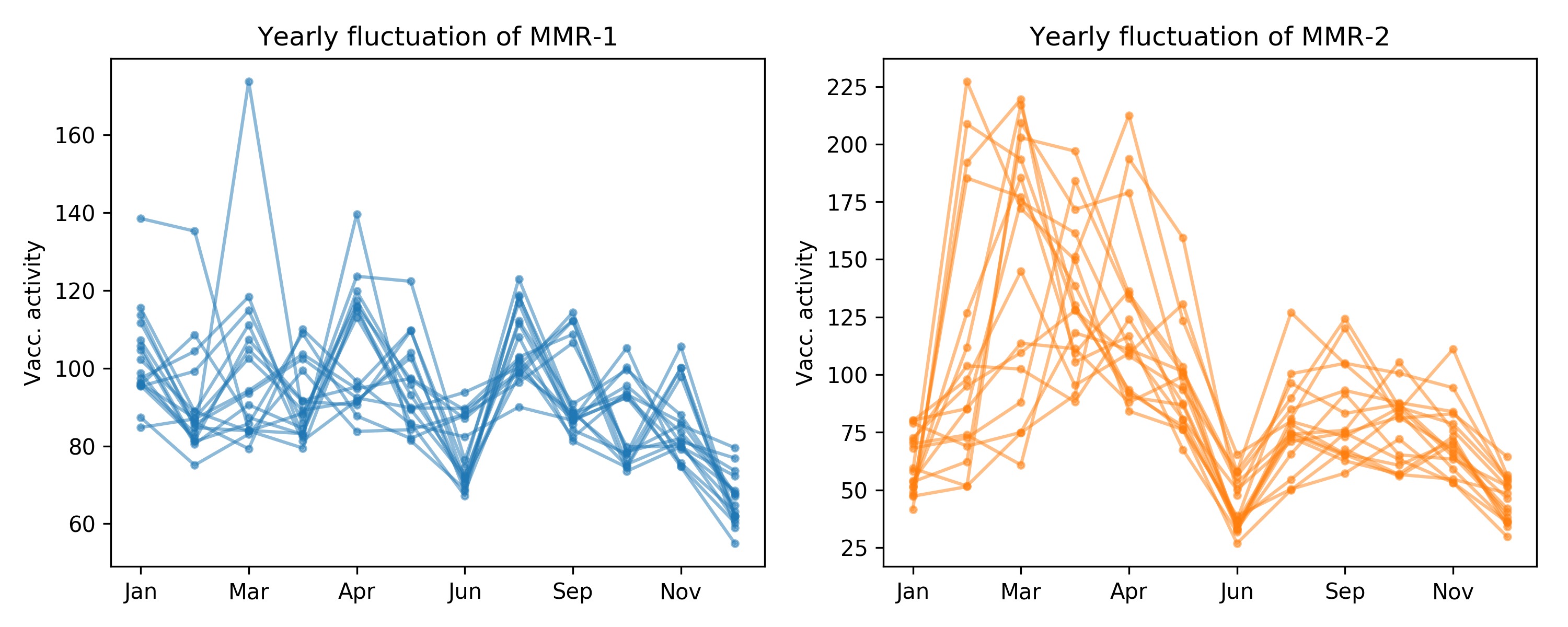}
	\caption{Vaccination activity plotted by month, lines denote data points from the same year.}
	\label{fig:mmr+media_figure2}
\end{figure}

\subsection{Vaccination activity}
To control for the change in birth cohort both within a year and between years, we report the vaccination numbers as
percentage of eligible children who have been vaccinated. This number is henceforth referred to as the
\textit{vaccination activity}. This is the number plotted in Figure \ref{fig:mmr+media_fig1}. We define as eligible the children who should be
vaccinated in a given month according to the recommended vaccination schedule. The plot in Figure \ref{fig:mmr+media_fig1} shows the monthly
vaccination activity for both MMR-1 and MMR-2 from 1997-2014 as recorded in the childhood vaccination registry \cite{grove2011danish}. Values greater than 100\% occur because people do not necessarily follow the recommended vaccination schedule.
Figure \ref{fig:mmr+media_figure2} shows seasonal variations both for MMR-1 and MMR-2 appearing as a reduction in vaccination activity around the
summer holidays in June and around Christmas and New Year. However, most striking is the periodicity of the number of
MMR-2 vaccinations and the visible change in vaccination pattern around 2009, see Figure \ref{fig:mmr+media_fig1}. From 1997-2008, inclusive,
a reminder letter was sent at the beginning of the year to all children turning 12 that year. The letter was sent at
the beginning of each year and we assume that it was responsible for the annual peak around March.

\subsection{Adjusting for seasonal correlations}


\begin{figure}
	\centering
	\begin{minipage}{.45\textwidth}
		\centering
		\includegraphics[width=\textwidth]{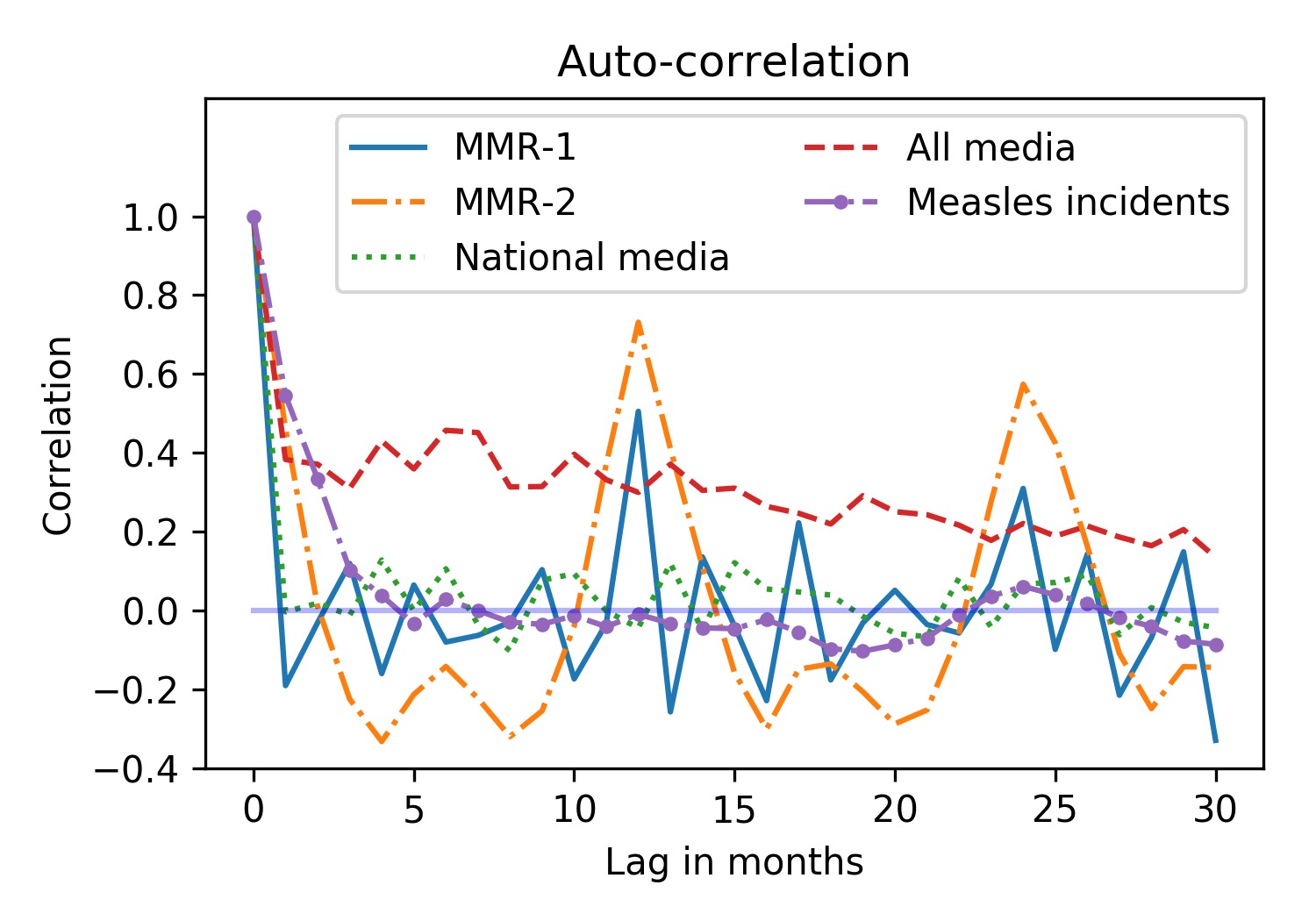}
		\caption{Auto-correlation for MMR-1, MMR-2, national media coverage, all media coverage and the number of measles incidents.}
		\label{fig:mmr+media_figure3}
	\end{minipage}%
	\quad
	\begin{minipage}{.45\textwidth}
		\centering
		\includegraphics[width=\textwidth]{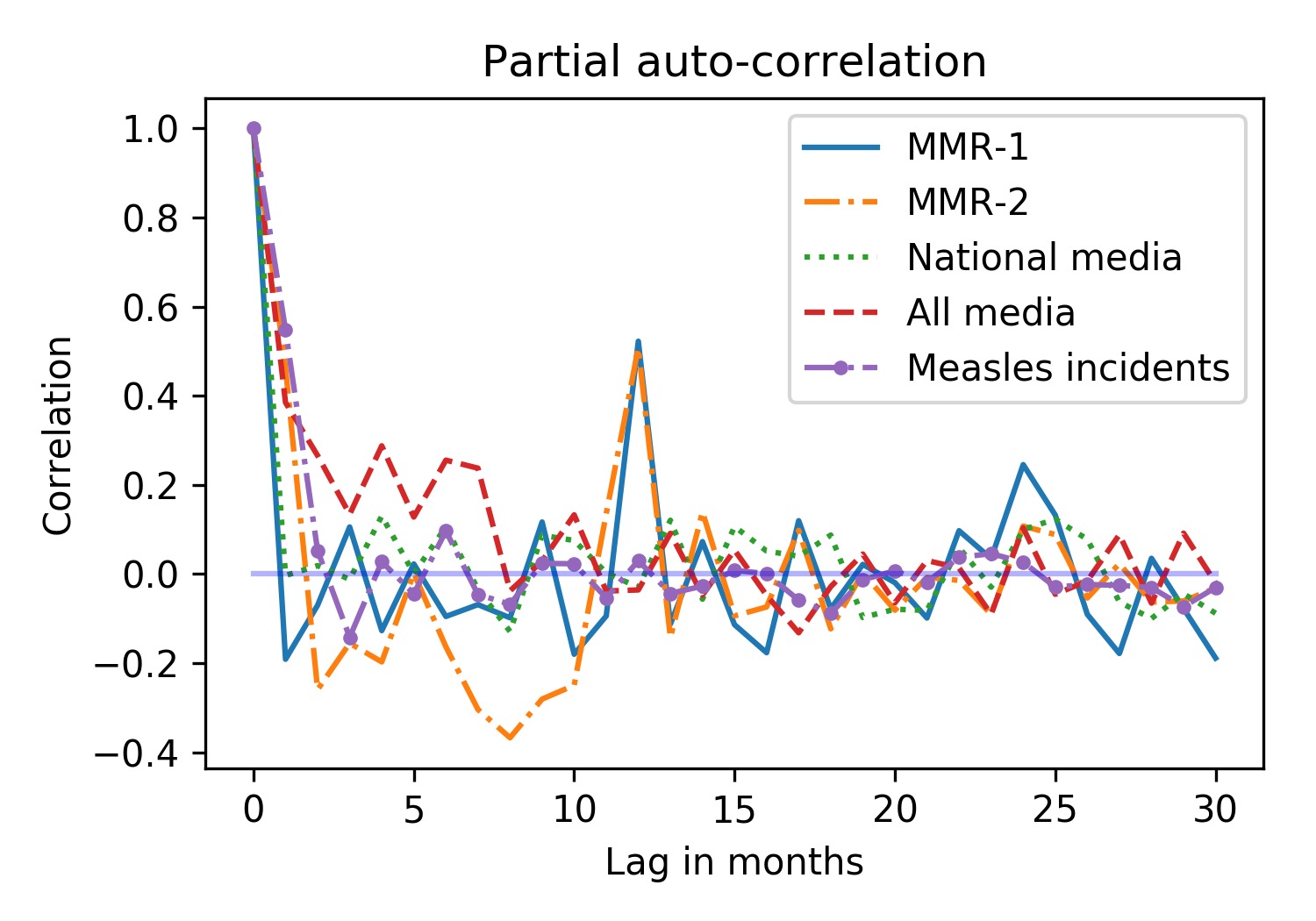}
		\caption{Partial auto-correlation for MMR-1, MMR-2, national media coverage, all media coverage and the number of measles incidents.}
		\label{fig:mmr+media_figure4}
	\end{minipage}
\end{figure}

Figure \ref{fig:mmr+media_figure3} shows the result of calculating the auto-correlation of vaccination activity, media coverage and measles
incidents. From the auto-correlation we see that the vaccination activity has a peak at 12 months, corresponding with
the observed seasonality of the vaccination activity illustrated in Figure \ref{fig:mmr+media_figure2}. For MMR-2 this annual correlation is more
pronounced than for MMR-1. For \textit{all media} we observe consistent high auto-correlation due to a steady increase
in the number of retrieved articles each month throughout the study period. Since this increase is not observed for the
\textit{national media}, it is likely explained by the increasing number of media sources in the infomedia archive.

Figure \ref{fig:mmr+media_figure4} shows the partial auto-correlation for the vaccination activity, media coverage and measles incidents. The
partial auto-correlation can be used to identify the number of autoregressive terms (i.e.\ \textit{p} in Equation \ref{eq:mmr+media_eq1}) in
the autoregressive models used to control for seasonality and serial dependencies. The partial auto-correlation for
MMR-1 and MMR-2 quickly drops after the first lag and subsequently peaks again at a 12 months lag. For \textit{all
media,} partial auto-correlation remains close to 0.2 until a 7 month lag after which it fluctuates around zero.

To control for the seasonal and serial dependencies we use an autoregressive model with 12 terms, corresponding to the
peak in partial auto-correlation for the vaccination activity. We fit the model both to the vaccination activity and
media coverage time series, since \textit{all media} was also auto-correlated. The residual, i.e.\ \textit{$\varepsilon
$}\textit{\textsubscript{t}} from Equation \ref{eq:mmr+media_eq1}, will be used in the remaining analysis, since this part of the signal is
not accounted for by seasonality or serial dependencies.

\subsection{Quantifying the relationship between signals}

\begin{figure}
	\centering
	\includegraphics[width=0.7\textwidth]{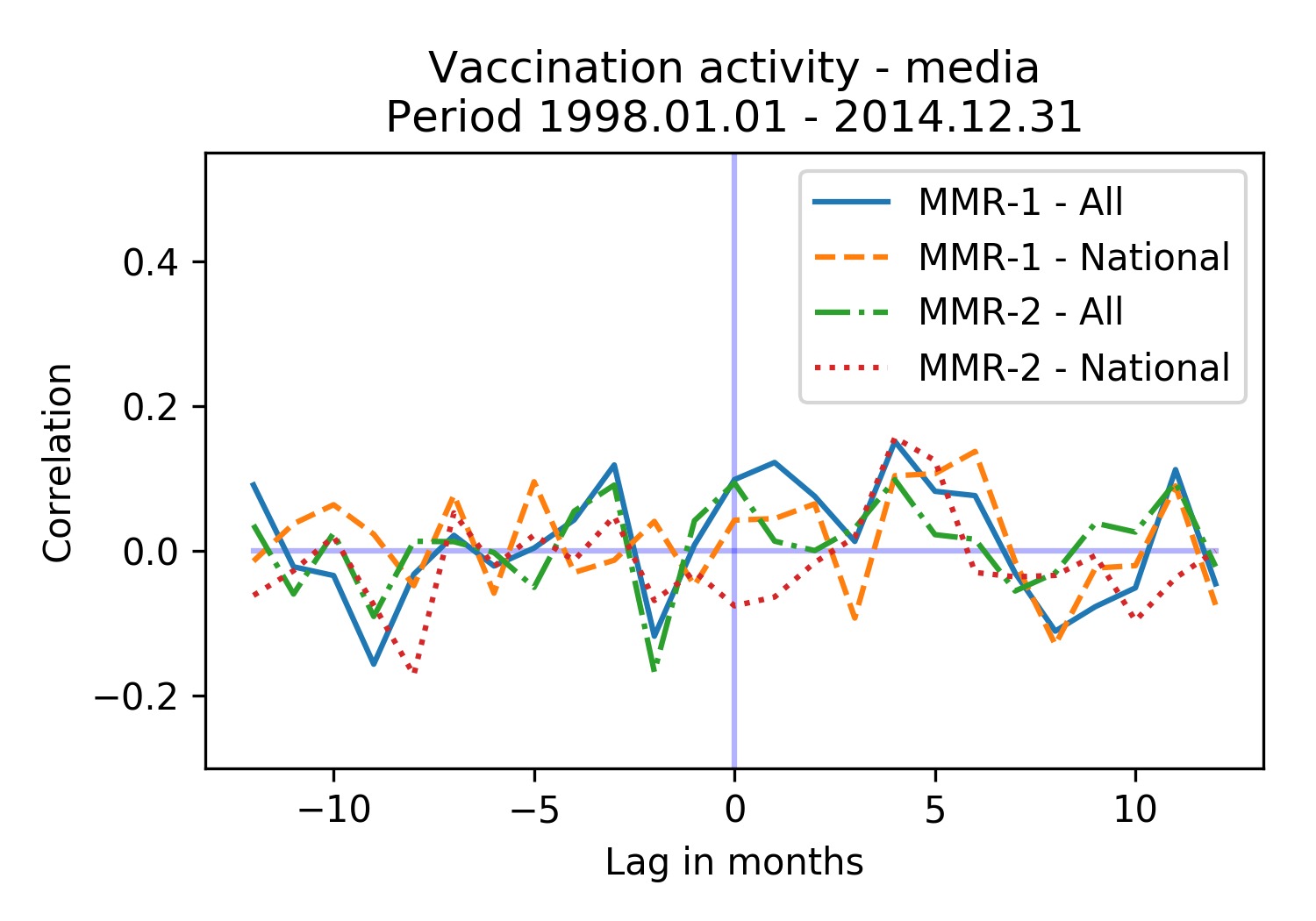}
	\caption{Cross-correlation between media and vaccination activity.}
	\label{fig:mmr+media_figure5}
\end{figure}

\begin{figure}
	\centering
	\includegraphics[width=0.7\textwidth]{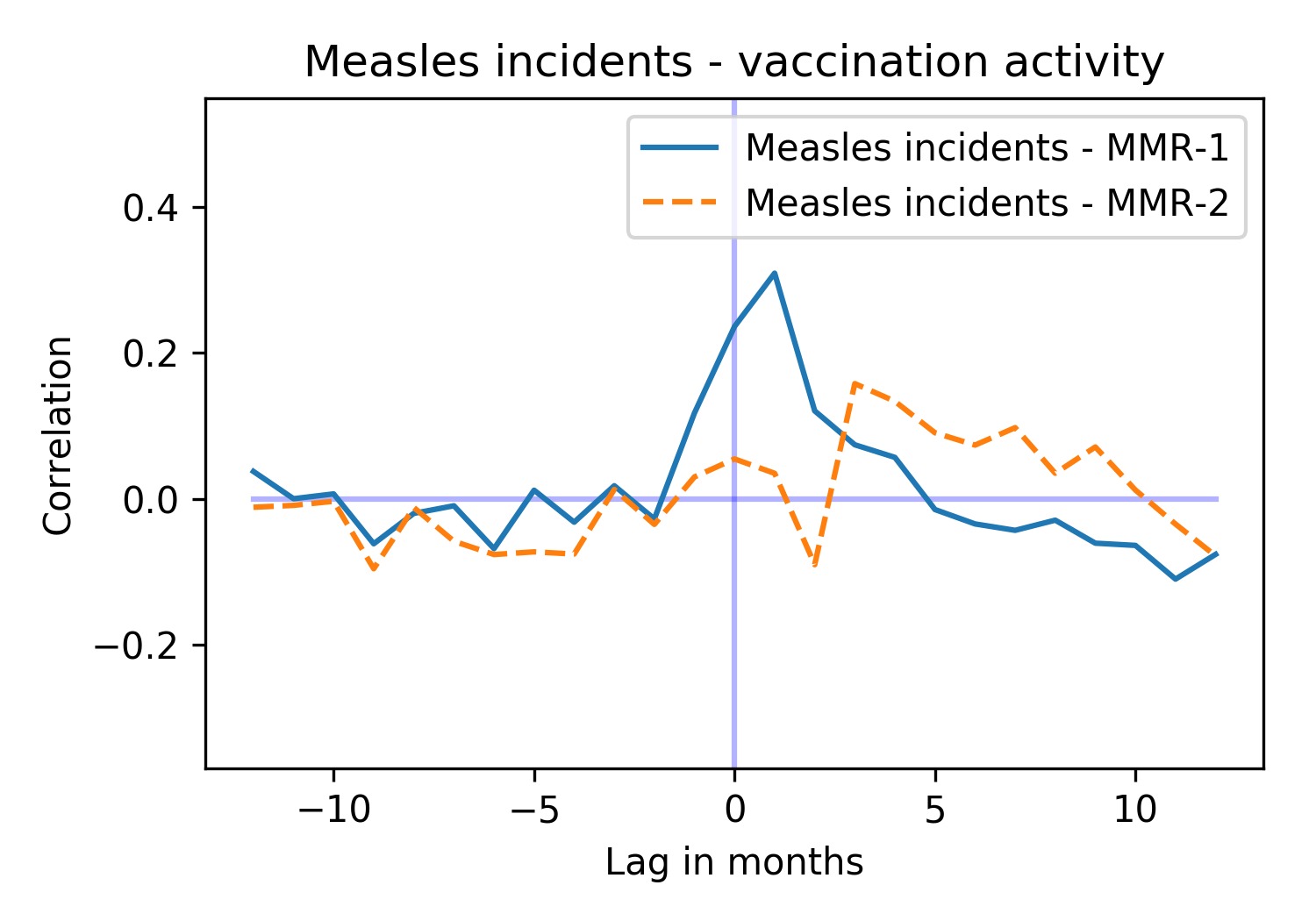}
	\caption{Cross-correlation between measles incidents and vaccination activity.}
	\label{fig:mmr+media_figure6}
\end{figure}

Figure \ref{fig:mmr+media_figure5} shows the cross-correlation between MMR-1 and media coverage and MMR-2 and media coverage. In all cases
correlation is insignificant. Figure \ref{fig:mmr+media_figure6} shows the correlation between MMR-1 and measles incidents and MMR-2 and measles
incidents. In this case the correlation between measles incidents and MMR-1 (r=.31, \textit{P}=.005) is statistically
significant at shift 1, meaning that an increase in measles incidents is followed by an increase in MMR-1
vaccinations.

\subsubsection{Vaccination attitude}

\begin{figure}
	\centering
	\includegraphics[width=0.7\linewidth]{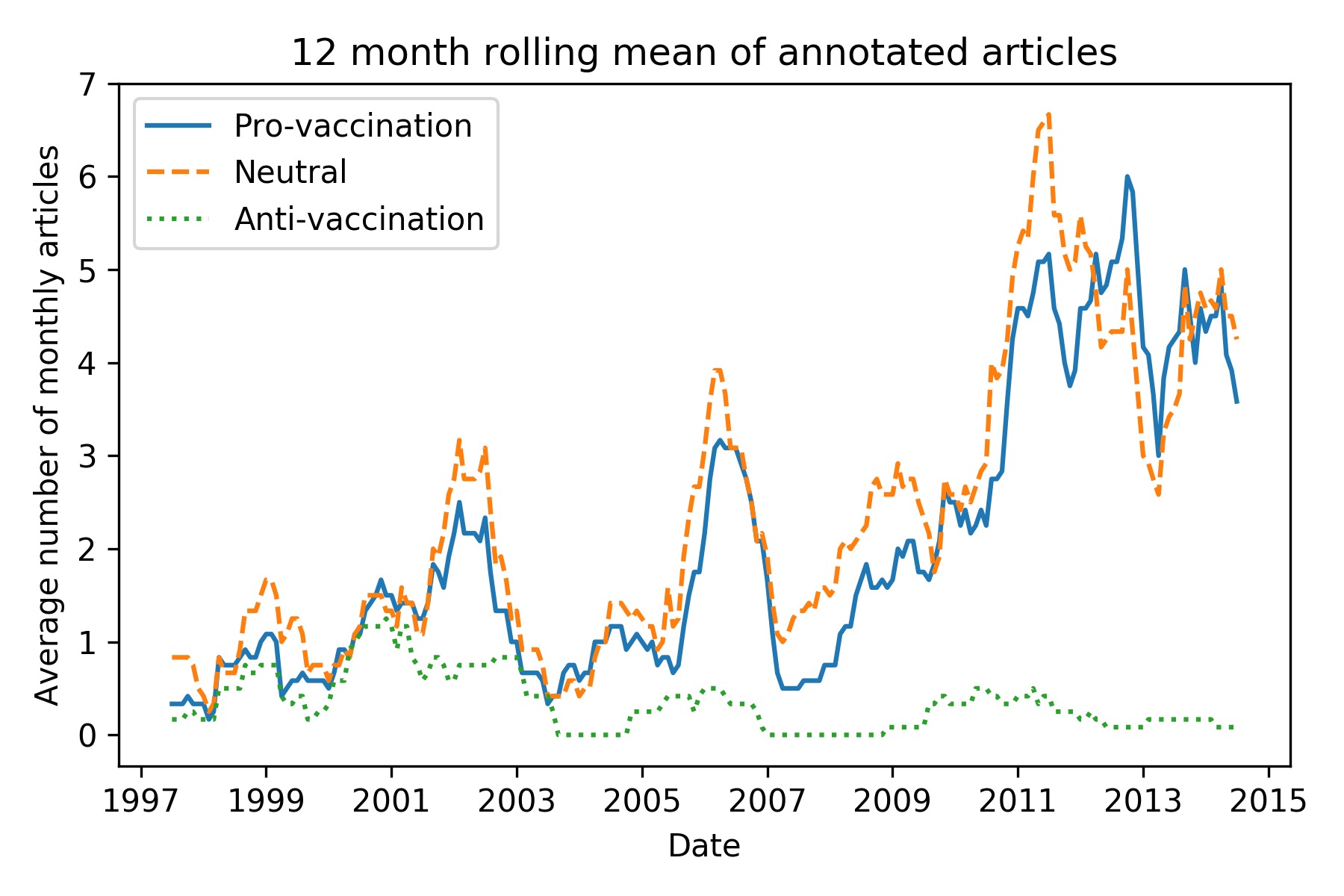}
	\caption{Vaccination attitude in media. For readability we plot a 12 months rolling mean. The rolling mean is calculated based on the number of articles published in a window of the 6 months before and after a given data point.}
	\label{fig:mmr+media_figure7}
\end{figure}

\begin{figure}
	\centering
	\includegraphics[width=\linewidth]{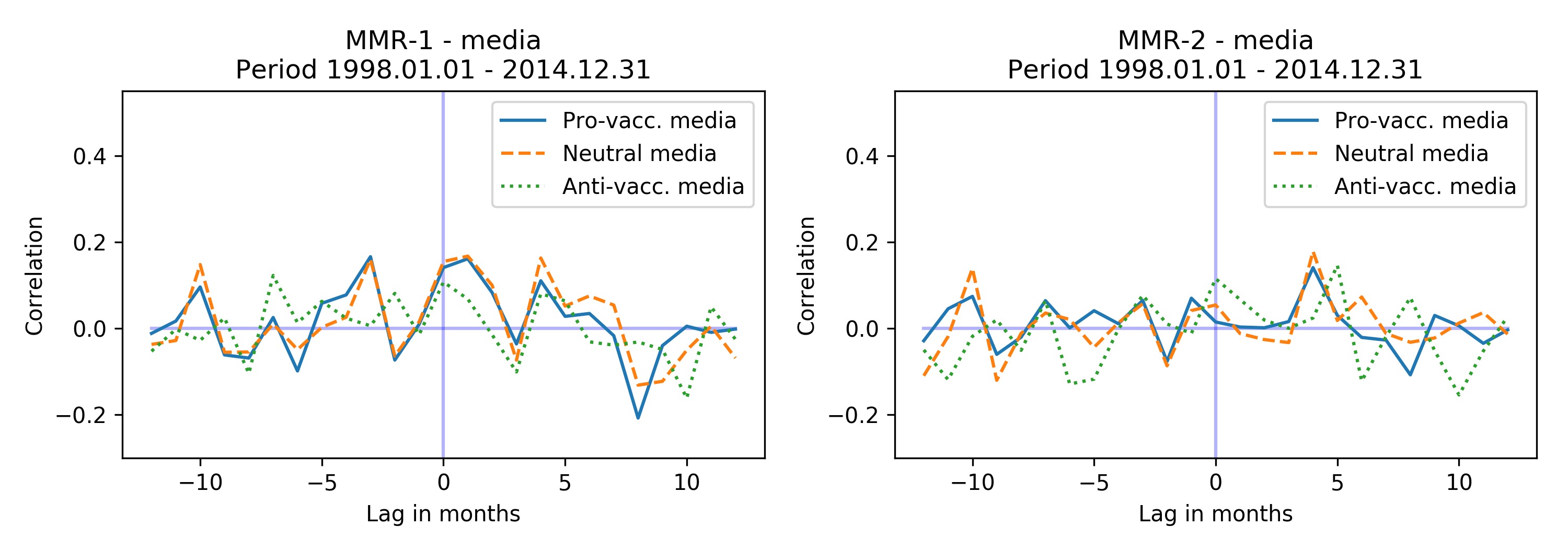}
	\caption{Cross-correlation between annotated media and vaccination activity.}
	\label{fig:mmr+media_figure8}
\end{figure}

\begin{table}
	\centering
	\begin{tabular}{lr}
		\toprule
		 & Article count \\
		 \midrule
		 Pro-vaccination & 430 \\
		 Neutral         & 500 \\
		 Anti-vaccination & 72 \\
		 &\\
		 Total number of articles & 1622 \\
		 MMR relevant articles & 681 \\
		 \bottomrule
	\end{tabular}
	\caption{Overview of articles.}
	\label{tab:mmr_media_tab1}
\end{table}

Table \ref{tab:mmr_media_tab1} shows statistics on the number of annotated articles. Out of 1622 articles, 941 were irrelevant to MMR.
Irrelevant articles include reports on yearly accounts from vaccine producing companies or articles about charities
collecting money for vaccination programs in developing countries, etc. Of the 681 articles relevant to MMR, the
majority of articles contained pro-vaccination and/or neutral content. Figure \ref{fig:mmr+media_figure7} shows the distribution of each category
during the study period. The peaks in pro-vaccination and neutral information in 2002, 2006 and 2011 correspond to
measles outbreaks. The majority of anti-vaccination articles occurred in the period 1997-2004. This coincides with the
retracted study by Wakefield et al.\ linking autism to the MMR vaccine published in 1998. The anti-vaccination articles
were primarily about the now falsified link between autism and MMR, but also about Danish court cases on allegations of
adverse reactions to the MMR vaccine.

\begin{figure}
	\centering
	\includegraphics[width=0.7\textwidth]{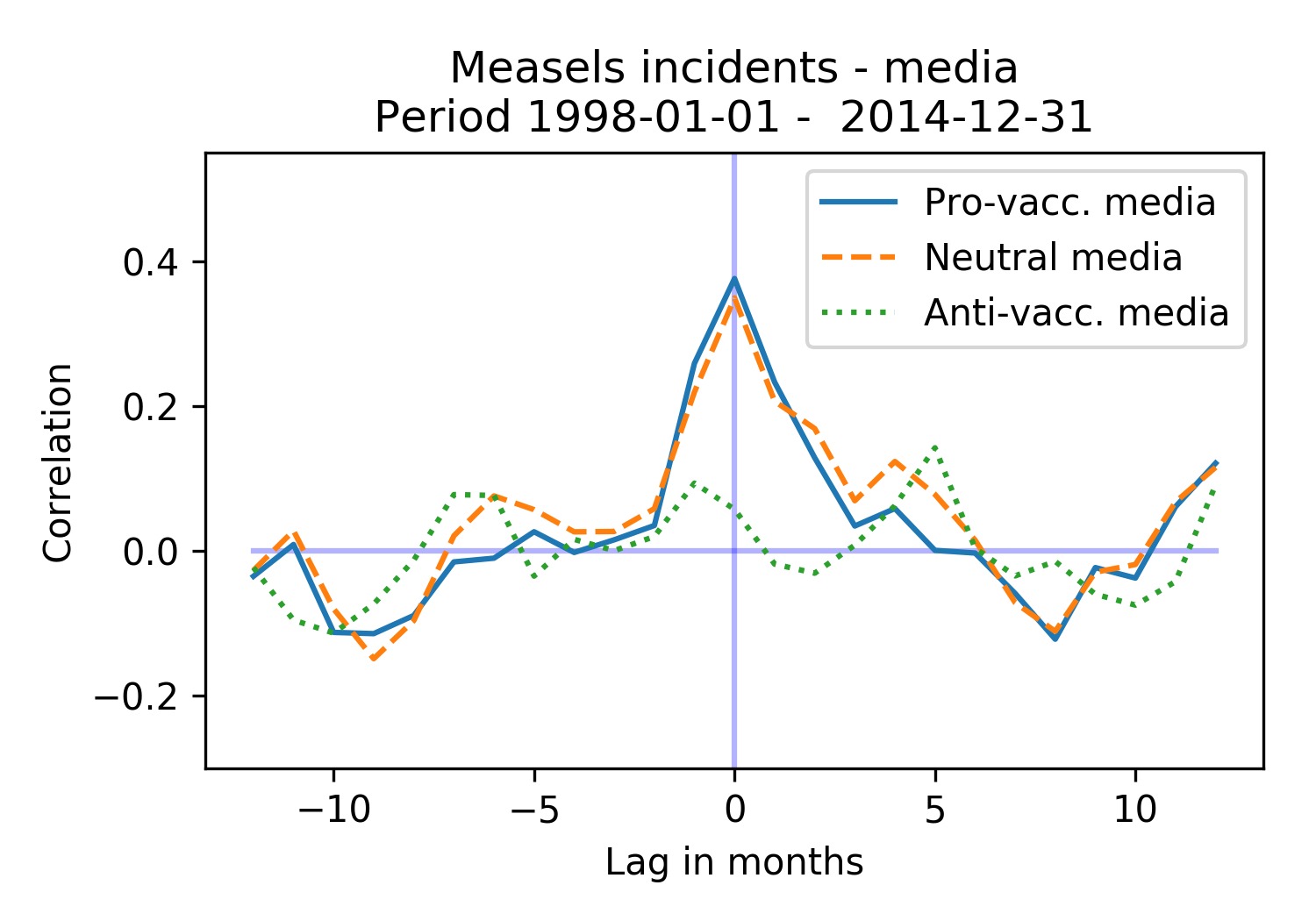}
	\caption{Cross-correlation between measles incidents and annotated media.}
	\label{fig:mmr+media_figure9}
\end{figure}

Figure \ref{fig:mmr+media_figure8} shows the cross-correlation between MMR-1 and the annotated articles, and MMR-2 and the
annotated articles. For both vaccines there are no significant correlations. The correlation between measles incidents
and the annotated articles is shown in Figure \ref{fig:mmr+media_figure9}. In this case there is a statistically significant correlation at lag 0
between pro-vaccination media and the number of measles incidents (r=.38,
\textit{P}=.007). Though not statistically significant, the correlation between
neutral media and measles incidents is also relatively high.

\subsubsection{During and after the autism controversy}
\begin{figure}
	\centering
	\includegraphics[width=\linewidth]{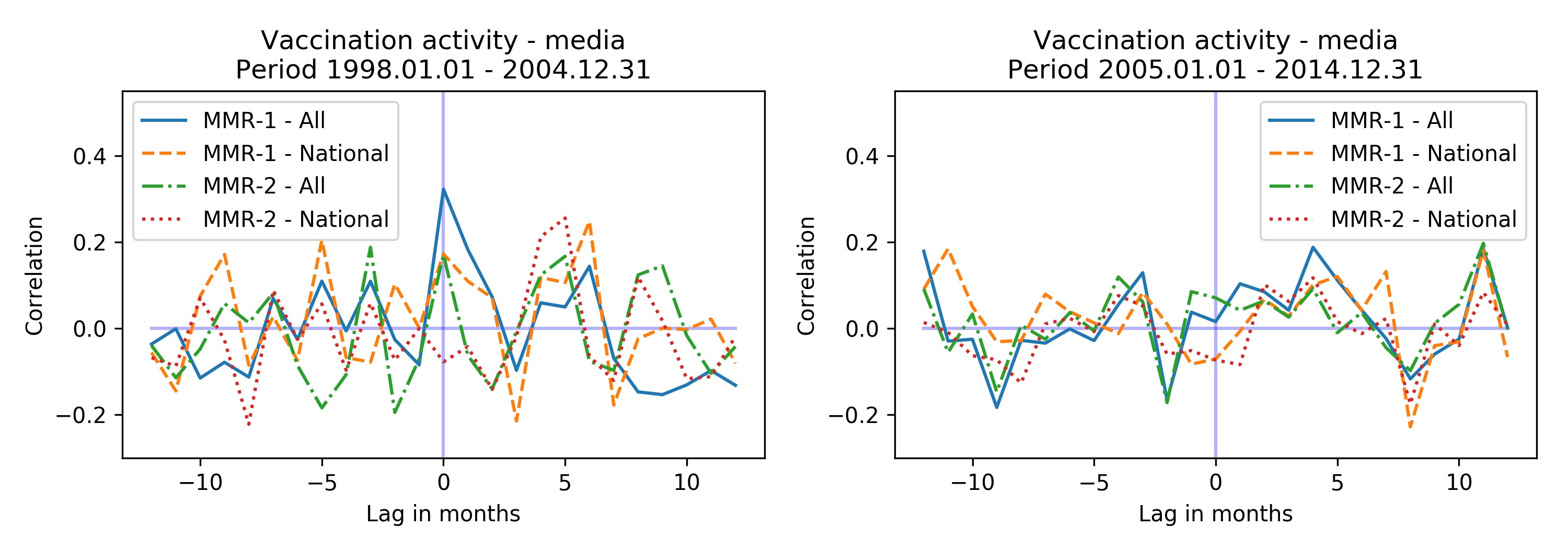}
	\caption{Cross-correlation between media coverage and vaccination activity for the two periods.}
	\label{fig:mmr+media_figure10}
\end{figure}

It is evident from Figure \ref{fig:mmr+media_figure7} that the majority of negative media coverage occurred in the period 1997-2004 due to the
fear of adverse reactions to the MMR vaccine. To evaluate if people in this period were more susceptible to media
influence than in the following period we split the data set into two: 1998-2004 and 2005-2014 (1997 is omitted because
we use 12 months autoregressive models to control for the seasonal changes and serial dependencies). Figure \ref{fig:mmr+media_figure10} shows
the cross-correlation between MMR-1 and media coverage and MMR-2 and media coverage. For the period 1998-2004 there is
a statistically significant correlation at lag 0 between MMR-1 and \textit{all media} (r=.32, \textit{P}=.009). For the
second period, 2005-2014, there is a statistically significant negative correlation between MMR-1 and \textit{national
media} coverage at lag 8 (r=-.23, \textit{P}=.009).

\begin{figure}
	\centering
	\includegraphics[width=\linewidth]{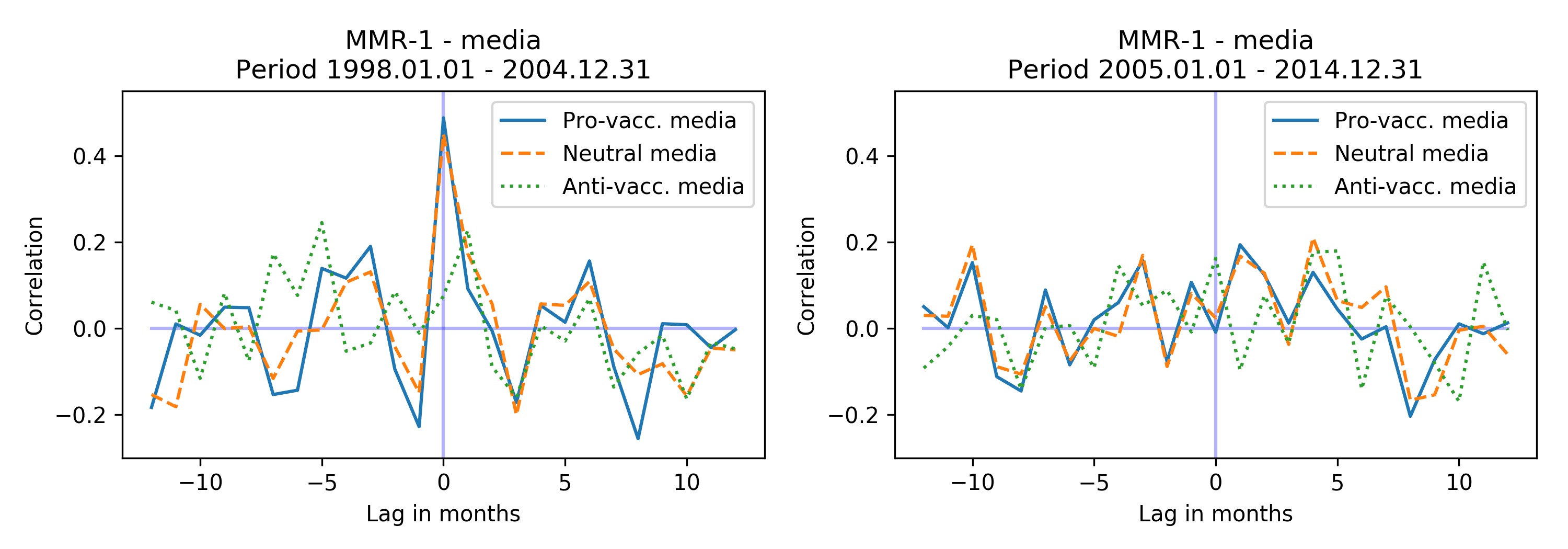}
	\caption{Cross-correlation for vaccination activity of MMR-1 and annotated media data for the two periods.}
	\label{fig:mmr+media_figure11}
\end{figure}

\begin{figure}
	\centering
	\includegraphics[width=\linewidth]{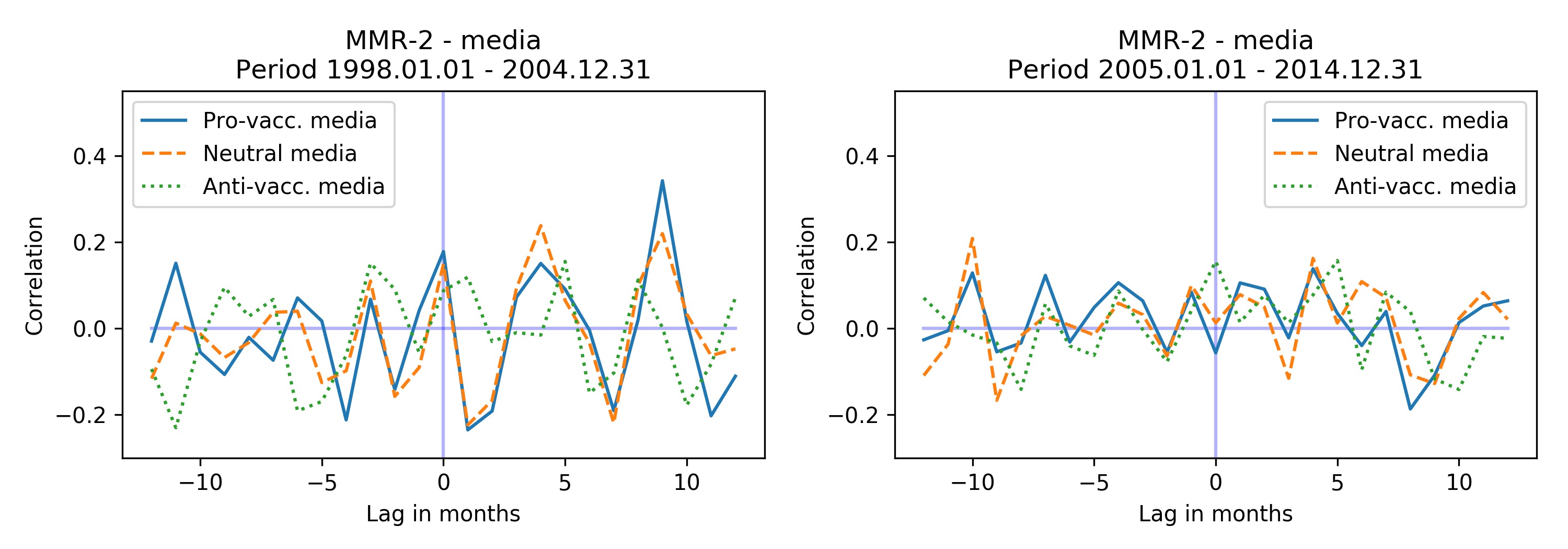}
	\caption{Cross-correlation for vaccination activity of MMR-1 and annotated media data for the two periods.}
	\label{fig:mmr+media_figure12}
\end{figure}

\begin{figure}
	\centering
	\includegraphics[width=\linewidth]{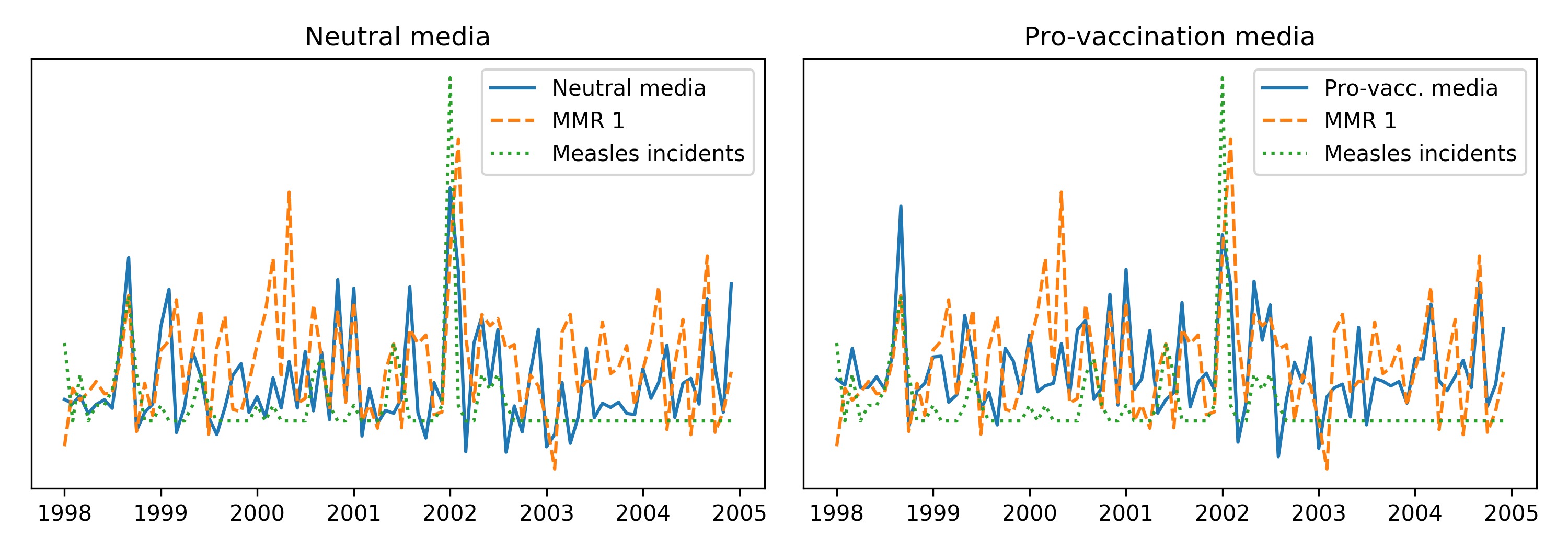}
	\caption{Left: the residual of MMR-1, residual of neutral media and measles incidents. Right: residual of MMR-1, residual of pro-vaccination media and measles incidents. Values have been normalized for comparison.}
	\label{fig:mmr+media_figure13}
\end{figure}

Figure \ref{fig:mmr+media_figure11} shows the cross-correlation for MMR-1 and the annotated media for the two periods. For the period 1998-2004
there is a statistically significant correlation between pro-vaccination media and MMR-1 vaccination activity (r=.49,
\textit{P}=.004). Similar is the correlation between neutral media and MMR-1 vaccination activity (r=.45,
\textit{P}=.003).

The cross-correlation between MMR-2 and the annotated media for the two periods is shown in Figure \ref{fig:mmr+media_figure12}. In this case
there are no significant correlations. The difference between MMR-1 and MMR-2 in Figures \ref{fig:mmr+media_figure11} and \ref{fig:mmr+media_figure12} is interesting and
might be related to the fact that one vaccine is the first that children receive, while the other is the second one.
This will be discussed in more detail in the next section.

To visualize the relationship between pro-vaccination, neutral media, and measles incidents, with the MMR-1 vaccination
activity the residual of each signal has been plotted. Figure \ref{fig:mmr+media_figure13} shows on the left a plot of the residual of the MMR-1
vaccination activity, the residual of the neutral media coverage and the measles incidents. On the right plot is the
residual of the MMR-1 vaccination activity, the residual of pro-vaccination media coverage and the measles incidents.
We observe that peaks in media coverage are most often correlated with a peak in vaccination activity, while the
reverse is not the case. Similarly we observe that measles incidents correlate with peaks in media and vaccination
activity, but not the reverse. This indicates that media is not the only factor affecting vaccination activity.

\section{Discussion}
Our study investigates the relationship between written media coverage and vaccination activity for the MMR vaccine in
the period 1997-2014 in Denmark. The media coverage is quantified both as total volume of written articles per month
and as articles with pro-vaccination, anti-vaccination and neutral content. While treating the whole period as one
reveals no relationship between media and vaccination activity, we observe a relationship between vaccination activity
and media coverage when looking at the period immediately following the discredited article by Wakefield et al.\ \cite{wakefield1998retracted},
i.e.\ 1998-2004. During this period there is a statistically significant positive correlation between both
pro-vaccination media and vaccination activity for MMR-1, and between neutral media coverage and vaccination activity
for MMR-1. In the period afterwards (2005-2014) there is no observed correlation.

Though the correlations in the period 1998-2004 between pro-vaccination media and MMR-1 vaccination activity (r=.49,
\textit{P}=.004) and between neutral media and MMR-1 vaccination activity (r=.45, \textit{P}=.003) are statistically
significant, they are small, indicating only a limited influence between media coverage and vaccination activity. We
can visually confirm this when plotting the media coverage on top of the vaccination activity, Figure \ref{fig:mmr+media_figure13}, where we
observe that peaks in media coverage correspond to peaks in vaccination activity, but peaks in vaccination activity
might occur without any peak in media. This aligns with the 2008 study by Smith et al.\ \cite{smith2008media} which observed only a
limited effect of the autism controversy on vaccination uptake in the USA. Additionally, we only observe a
significant correlation for the first MMR vaccine, indicating that for the first vaccination, parents are more
susceptible to media influence, than for the second MMR vaccine.

When analyzing the relationship between measles incidents and vaccination activity we observe for the whole study period
a statistically significant, but small, positive correlation (r=.31, \textit{P}=.005) between measles incidents and
MMR-1 vaccination activity with a lag of one month. This means that one month after an increase in measles incidents
there is an increase in the number of MMR-1 vaccinations. As expected, the pro-vaccination and neutral media is also
positively correlated with the measles incidents, the pro-vaccination media having a significant positive correlation
of r=.38 (\textit{P}=.007).

The period 1997-2004 was a time when the public was presented with opposing views on the safety and necessity of the MMR
vaccine. After that period the lack of anti-vaccination media coverage suggests a common consensus on the topic. Our
results indicate that the agenda-setting effect of media is only present in the period when people are actually
debating whether or not to vaccinate. When the consensus was reached the effect disappeared. Similar observations have
been made with respect to political debates \cite{mccombs2009news}, where a correlation between media coverage and people's opinions was
observed for countries where the politicians did not agree, but no correlation was observed if the politicians agreed.
From a public health point of view these results imply that monitoring controversies in the media could be used as an
indicator as to when additional resources should be spent on public outreach. Our results show that a weak signal can
be found simply by counting all articles matching a query about MMR, i.e.\ the correlation between MMR-1 and \textit{all
media} for 1998-2004 (r=.32, \textit{P}=.009). But for a stronger signal it is necessary to analyze the opinions
expressed in the articles. Traditional sentiment detection will likely not suffice, since articles do not necessarily
express negative views about the vaccine, but could for example emphasise benefits of ``natural'' immunization, i.e.
getting infected by measles. A related approach, namely stance detection \cite{hasan2013stance}, aims at automatically determining the
stance expressed in ideological debates. Such approaches could potentially be used for detecting changes in attitudes
expressed in the continuous stream of published media.

\subsection{Strengths and limitations}
The long study period of 18 years strengthens the study since the dynamics between media coverage and vaccination uptake
could be studied both in a period with debate and in one without. The Danish vaccination register ensures very reliable
vaccination data on a per person level, which allows us to investigate timely changes in the vaccination activity. This
is not possible with vaccination uptake data accumulated for each birth cohort.

There are some limitations to the study design. First of all only written media data were included, but social media
and/or broadcast media are likely to also influence people's choices. We know from other studies on the relationship
between social media and news media during a measles outbreak in the Netherlands, that correlation between social media
and news media is very high \cite{mollema2015disease}. We stipulate that the correlation between news coverage in the written media and
broadcast media is also high. Another limitation is that the articles have only been annotated by one person, the first
author. It has therefore been impossible to calculate annotator agreement and the consistency of the annotations can
therefore not be assessed. \ Finally, though we observe a correlation between media coverage and vaccination activity,
the general vaccination activity throughout the period is relatively stable, indicating a priori that external events
only have a limited effect on the vaccination activity.

\section{Conclusion}
To conclude, this is to our knowledge the first national assessment of the overall effect of media coverage on the rate
of the measles, mumps and rubella vaccination in Denmark during the period 1997-2014. The study shows that while for
the whole period 1997-2014 there is no correlation between vaccination uptake and media coverage, there is for the
period following the falsely claimed link between autism and the MMR vaccine a positive correlation between
pro-vaccination and neutral media coverage and vaccination activity. This correlation was only observed for the first
MMR vaccine targeted at 15 month old children. The results indicate two things: (i) That the influence of media is
stronger on parents when they are deciding on the first vaccine, and (ii) that the effect of media coverage depends on
the amount of consensus on the topic.

\section{Acknowledgments}
The study was in part supported by a grant from Innovation Fund Denmark and IBM Denmark.

\section{Conflicts of Interest}
There are no conflicts of interest to report.

\chapter{Decline in HPV-vaccination Uptake in Denmark  --  The Association Between HPV-Related Media Coverage and HPV-Vaccination}
\chaptermark{Decline in HPV-vaccination uptake}
\label{cha:HPV_media}

\begin{center}

Venue: In review

\vspace{1cm}
	
	\textit{
		Camilla Hiul Suppli$^a$, Niels Dalum Hansen$^{b,c}$, Mette Rasmussend$^d$,\\Palle Valentiner-Branth$^a$, Tyra Grove Krause$^a$, Kåre Mølbak$^a$	
	}

	\vspace{0.5cm}
	
	$^a$Statens Serum Institut, Denmark.
	$^b$University of Copenhagen, Denmark.\\
	$^c$IBM Denmark.
	$^d$National Institute of Public Health, University of Southern Denmark
	
\end{center}

\leftskip=1cm
\rightskip=1cm

\textit{\textbf{Background:} With nearly 400 women annually diagnosed with cervical cancer in Denmark, the incidence remains high among Western European countries. Accordingly, Denmark was one of the first countries to implement public funded vaccination against human papillomavirus (HPV) in 2009. Initially the vaccine received positive public attention, but media turned increasingly critical as the number of reported suspected adverse events rose. \ From an uptake above 90\% for girls born in 2000 there was a decline to 54\% for girls born in 2003. The collapse coincided with increasing suspected adverse event reporting to the Danish Medicines Agency. The aim of this study is to describe the HPV-vaccination uptake, to quantify relevant HPV-related written media coverage, and analyse the relation between media coverage and HPV-vaccination acceptance in Denmark in year 2009-2016.}

\textit{\textbf{Methods}: We obtained data on vaccination uptake in 243,415 girls from the national immunization registry while written media coverage of HPV-vaccination (8,524 articles) was retrieved from the Infomedia database comprising the majority of published written media in Denmark. \ The tipping point between vaccination uptake and media was identified by iteratively calculating the Pearson's correlation between media attention and vaccination activity. We analysed the period before and after using a linear regression model.}

\textit{\textbf{Results: }We found no significant relationship between media coverage and vaccination uptake in the first part of the time series (2010 to June 2013), whereas there was a significant negative regression coefficient with a Pearson's r of -0.54 from July 2013 and onwards.
}

\textit{\textbf{Conclusions}: Following a successful launch of the HPV-vaccination programme, concerns about vaccine safety dominated public opinion and the press. The noticeable shift in correlation between vaccination uptake and media coverage before and after July 2013 suggest that increased media coverage influenced the decline in vaccination uptake. Media monitoring may represent an important tool in future monitoring and assessment of confidence in vaccination programmes.
}

\leftskip=0pt\rightskip=0pt

\section{Background}
With nearly 400 women diagnosed with cervical cancer and 100 annual deaths in Denmark (data including 2014), the incidence rate remains high among Western European countries \cite{engholm2014nordcan}. Accordingly, Denmark was one of the first countries to implement publicly funded vaccination against human papillomavirus (HPV). The programme was launched in 2008 primarily targeting 12-year-old girls with a quadrivalent HPV-vaccine. Effectiveness and impact of the quadrivalent HPV-vaccine is well established \cite{garland2016impact}.

In the first years after the launch, the HPV-vaccine was received positively resulting in a high uptake ({$>$90\%). From 2013, the programme was challenged by an increasing number of reported suspected adverse events \cite{danishHealth_hpv} and subsequently the media attention increased and shifted in content. \ The Danish media started to recount case-stories of suspected adverse events from Danish girls and women. This accelerated after a television documentary, March 2015, describing a group of girls with a wide range of disabling symptoms apparently following HPV-vaccination. The documentary was widely shared and reported on social- and written media. Subsequently an additional number of similar case-stories and anecdotes were disseminated in various media sources, including social media. Until now, no epidemiological study have been able to substantiate an increased risk of the alleged adverse events with a history of HPV-vaccination. Nevertheless, this shift in media attention was followed by a marked decline in HPV-vaccination uptake.

When analysing large amounts of textual data a qualitative analysis is infeasible. In other domains such as analysis of web searches, this problem is generally addressed by applying a frequency based approach. A popular example is Google flu trends \cite{ginsberg2009detecting} where the frequency of web searches is used for predicting influenza. More closely related to our problem, frequencies of web searches has successfully been used for predicting vaccination uptake in Denmark \cite{DalumHansen:2016:ELV:2983323.2983882}, and the frequency of articles matching a query has been used to quantify media attention with respect to the measles, mumps and rubella vaccine in the US \cite{smith2008media}. For our qualitative analysis we apply a similar idea where we use frequency of a HPV related query in a news media database to quantify media attention.

The aim of this study is to describe this collapse in the uptake of HPV-vaccination, the timing of selected events in Denmark that affected the media and the correlation between media coverage and vaccination uptake. Furthermore, we aim to lay the basis for future in-depth qualitative analysis of the relation between media coverage and vaccination uptake.

\section{Methods}
\paragraph{Vaccination data}

HPV-vaccination is offered free of charge in the childhood vaccination programme in Denmark. Vaccines are administered by general practitioners who report the vaccinations to the Danish Vaccination Register (DDV) \cite{grove2011danish}. Vaccination uptake by birth cohort as well as monthly vaccination activity (see below) was calculated from these data including data on 243,415 girls (retrieved October 2016).

To quantify timely effects of media on vaccination uptake we used the notion ``vaccination activity''. Vaccination activity is defined as the monthly number of girls who received the first dose of HPV-vaccine divided by the number of girls at the target age, i.e.\ girls with their 12-years birthday in that month. This measure includes the monthly variation of the birth cohorts, and will be greater than 100 in periods with high vaccination activity due to, e.g.\ catch-up activities.

\paragraph{Written media coverage data}

As a proxy for public attention, we extracted all articles from the media database, Infomedia.dk \cite{infomedia}, matching a HPV-relevant query, see below, English translations in parentheses:

\begin{quote}
\textit{``HPV AND livmoderhalskræft (cervical cancer)'' OR ``HPV AND cervix cancer'' OR ``HPV AND cancer'' OR ``HPV AND kræft (cancer)'' OR ``HPV AND POTS'' OR ``HPV AND CRPS'' OR ``HPV AND kønsvorter (genital warts)'' OR ``HPV AND kondylomer (condyloma)'' OR ``HPV AND condyloma'' OR ``HPV AND sygdom (disease)'' OR ``HPV AND kønssygdom (sexually-transmitted disease)''}
\end{quote}

The query was designed to return as many HPV-related articles as possible. We excluded product names of HPV-vaccines (Gardasil and Cervarix) since we were uninterested in product specific effects. POTS (Postural Orthostatic Hypotensive Syndrome) and CRPS (Chronic Regional Pain Syndrome) are abbreviations commonly used in Danish media to describe suspected adverse events.

Infomedia.dk retrieved a total of 8,524 unique articles, ranging from a minimum of 1 article per month to a maximum of 507 articles in one month. An increase in articles matching the query was expected, as additional data sources are included in the database over time. We normalized by the total number of articles containing the Danish word for ``and'' (og), on the assumption that all articles contain this word, this is referred to as the \textit{article percentage}.

To confirm the initial hypothesis, i.e.\ that media coverage of the HPV-vaccine turned increasingly critical, a small sample of articles were annotated as being neutral, focusing on benefits or focusing on adverse reactions. For each year, the 20 articles that best matched the HPV query (according to Infomedias retrieval function) were selected. Three annotators independently performed the hypothesis-generating annotation of a sample of articles retrieved by the query.

\paragraph{Statistical analysis}

We applied a two-step analysis: (i) identifying the tipping point, where the relationship between media coverage and vaccination activity changed. (ii) Analysing the association between media coverage and vaccination activity using a linear regression model for the period before and after the tipping point, respectively. In order to identify the tipping point, we iteratively separated the data, using one month intervals, into two subsets and calculated Pearson's correlation for both time periods, ranging from January 2010 to January 2015. We defined the tipping point as the point in time with the highest change in correlation. To estimate the association between media coverage and vaccination activity, we used linear regression with article percentage as independent variable and vaccination activity as dependent. While there is likely no linear relation between the two variables, the results will nevertheless indicate the existence of a possible relationship. Two regression models were fitted: Before and after the tipping point. The goodness of fit was evaluated using R-squared.

\section{Results}

\subsection{HPV-vaccination uptake in Denmark}

Figure \ref{fig:HPV_media_1} shows that the vaccination uptake for the first dose of HPV-vaccine increased from 80\% to 92\% for the birth cohorts of 1993 to 2000. For girls born after 2000, the initiation of HPV-vaccination decreased for successive birth cohorts, reaching the lowest uptake of 42\% for the birth cohort of 2004 (data per June 2017).

\begin{figure}
\includegraphics[width=\textwidth]{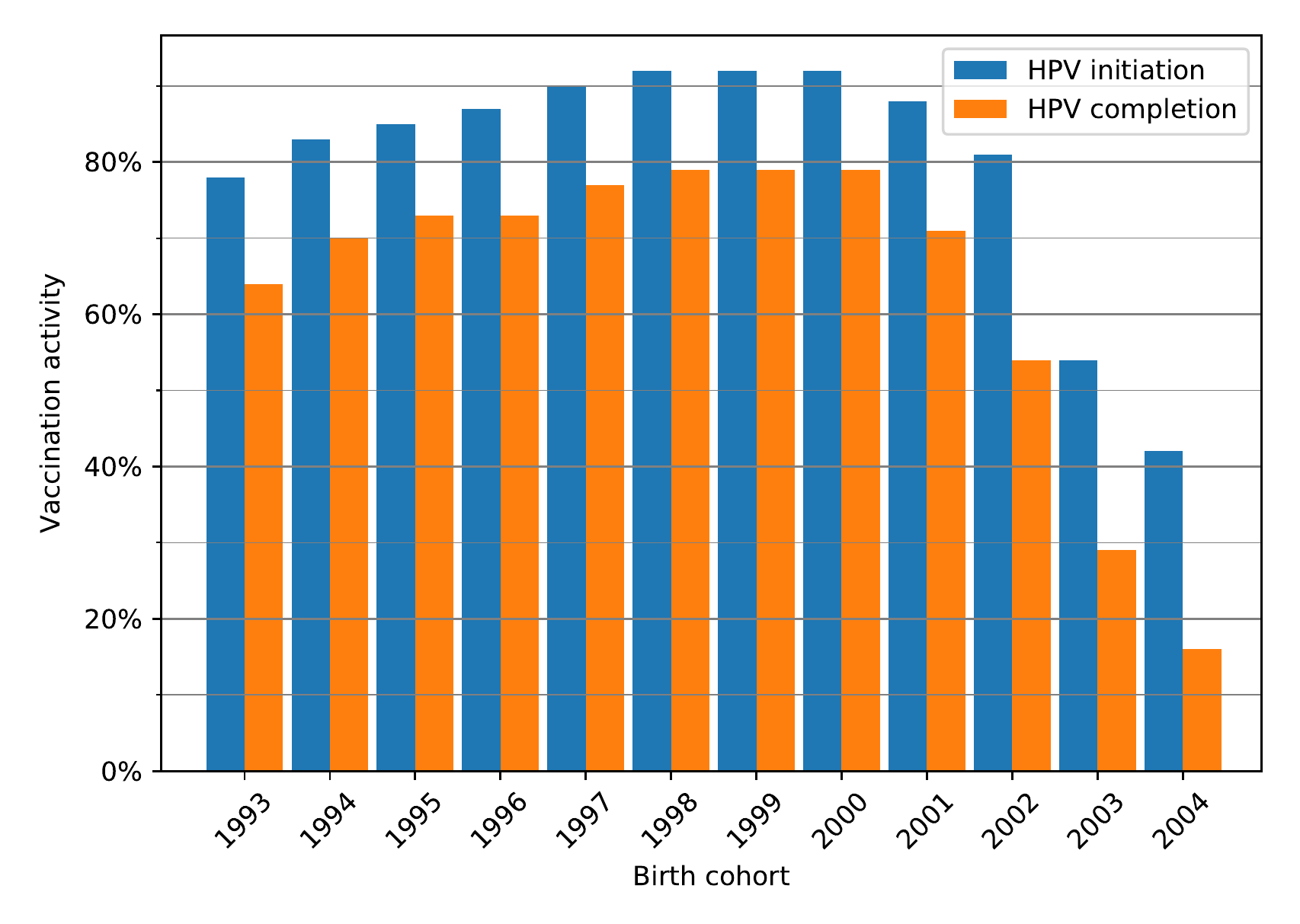} 
\caption{HPV-vaccination initiation and completion for girls in the childhood vaccination programme, Denmark birth cohorts 1993-2003. Three-dose vaccination schedule from 2009 until August 2014. Two-dose schedule from August 2014 until 14 October 2016. Data extracted June 2017.}
\label{fig:HPV_media_1}
\end{figure}

\subsection{Correlation between written media coverage and vaccination activity in Danish females}

Figure \ref{fig:HPV_media_2} shows HPV-vaccination activity and HPV-related media coverage (article percentage) from January 2009 to January 2016. As illustrated in the figure there was a high vaccination activity in the first part of 2009 after the programme was launched. After the launch, activity remained quite stable until mid-2013 where the vaccination coverage dropped to below 90\%. In 2013 there was a marked increase in media coverage (see table \ref{table:hpv_media_table_1} for a description of significant media events) which coincided with a drop in vaccination initiation.

\begin{figure}

\begin{subfigure}[b]{\textwidth}
\includegraphics[width=\textwidth]{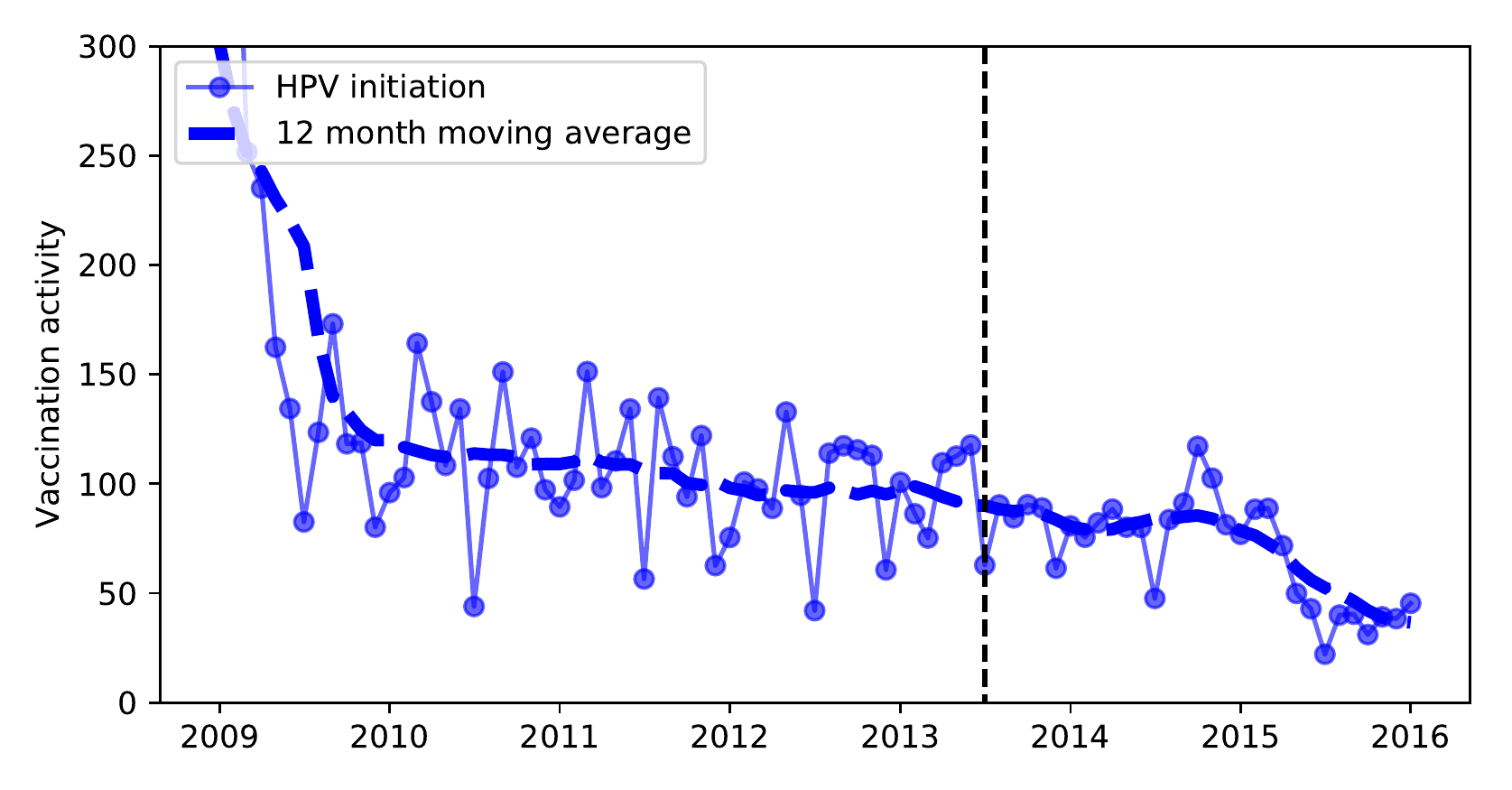} 
\caption{Monthly HPV-vaccination initiation (HPV1) activity and article percentage in Denmark 2009-2015. In the period 1 January 2009 to 1 January 2016 a total of 8,524 articles were retrieved, ranging from 1 to a maximum of 507 in November 2015. Vaccination activity $>$ 100, shows that more girls have been vaccinated than expected, this could be caused by a catch-up program or vaccinations that has previously been postponed.}
\label{fig:HPV_media_2a}
\end{subfigure}

\begin{subfigure}[b]{\textwidth}
	\includegraphics[width=\textwidth]{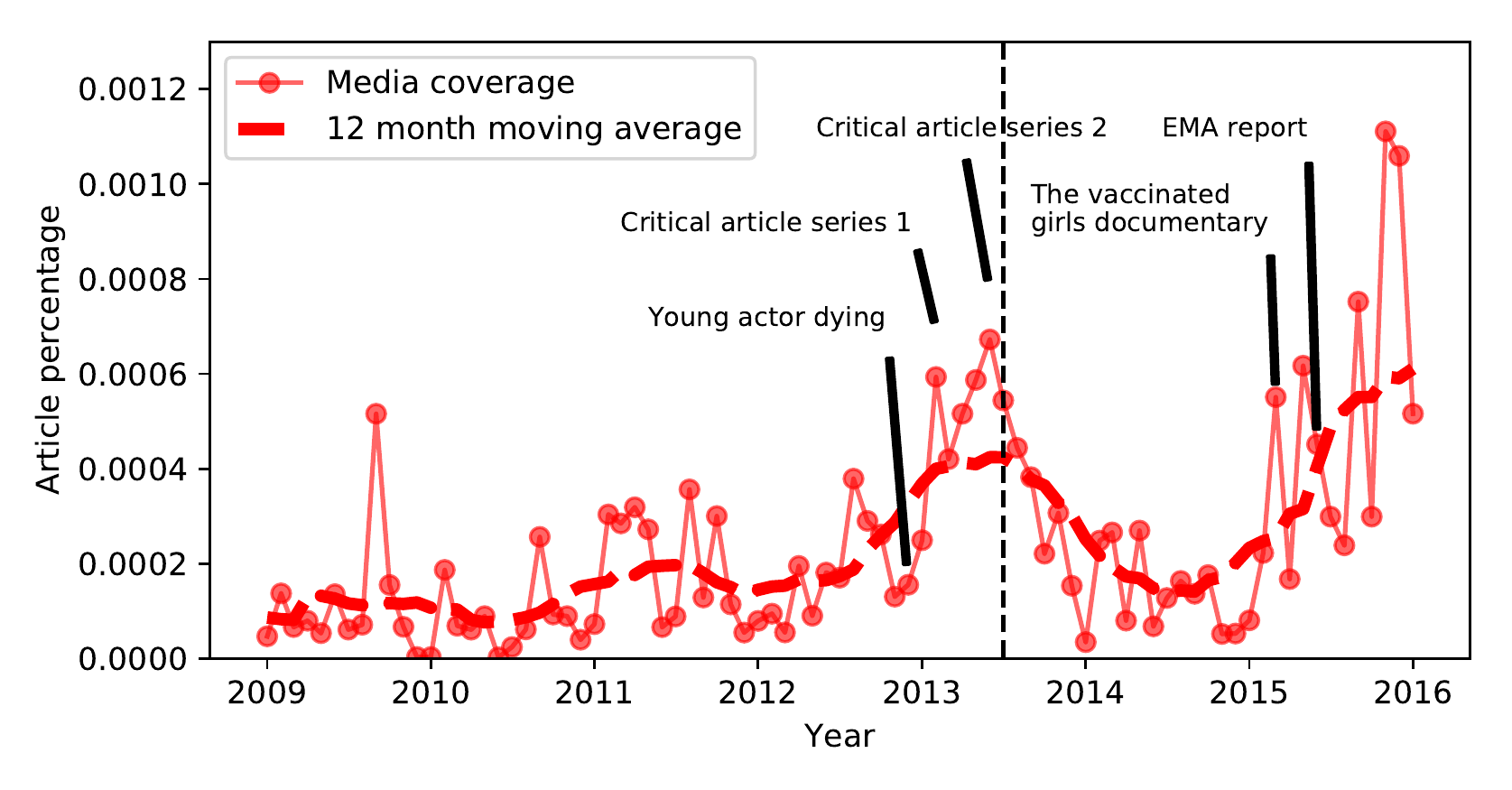} 
	\caption{Article percentage in Denmark 2009-2015 with events from table 2A highlighted. The vertical dotted line shows the tipping point in the correlation between vaccine activity and media coverage. The 120 annotated articles (overall raw agreement 85.4\%) showed a similar pattern. The number of articles focusing on benefits peaked in 2012 with 19/20 articles, after which it slowly dropped to 6/20 in 2015 with 11/20 articles focusing on a potential link to adverse reactions.}
	\label{fig:HPV_media_2b}
\end{subfigure}
\caption{Development in HPV-vaccinations and media coverage.}
\label{fig:HPV_media_2}
\end{figure}

We identified June 2013 as the tipping point in correlation between media coverage and vaccination activity. In the period from January 2009 until June 2013, there was a non-significant negative correlation of -0.10 (p-value=0.489) between vaccination uptake and media coverage, while for the period July 2013 until January 2016 there was a significant negative correlation of -0.54 (p-value=0.002). The fact that the tipping point occurs after the peak in media attention, i.e.\ June 2013, might indicate a delayed effect of media attention on vaccination activity.

For the regression analysis, we discarded the whole of 2009, since this initiation period deviates noticeably from the remaining time period. The regression analysis was performed for the periods January 2010 to June 2013 and July 2013 to January 2016. The first period showed an insignificant relation between article percentage and vaccination activity (coefficient 13,642, p=0.60). The second period showed a significantly negative relation (-44,165, p$<$0.001). Accordingly, for the first period an increase in article percentage of 0.005 corresponds to a change in vaccination activity of 6.8, meaning that if the media activity matching the HPV query increases with 0.005 percent point relative to the total media activity, then the monthly vaccination number will increase with what corresponds to 6.8 percent of the eligible 12 year-olds. In the second period, the same change in article percentage accounted for a reduction in vaccination number of -22.1. 

To evaluate the goodness of fit for the linear model, we calculated R-squared, which was 0.007 and 0.27 for the first and second period, respectively. This indicates that the model only explains parts of the variation in vaccination activity.

\begin{table}

\centering
\begin{tabularx}{\textwidth}{lXX}
\toprule
Period & Public initiatives and actions & Significant media events \\
\midrule
~2006 & First HPV-vaccine approved & \\\rule{0pt}{5ex}
2007 & Health technology assessment recommends the inclusion of HPV-vaccine in the childhood vaccination programme &  \\\rule{0pt}{5ex}
2008 & Catch-up programme (girls born 1993-1995) & \\\rule{0pt}{5ex}
2009 & 1 January start of routine vaccination & Campaign promoting vaccination run by the health authorities and the Danish Cancer society
 \\\rule{0pt}{5ex}
2012-13 & Catch-up programme targeting woman born 1985 to 1992 (27 August 2012 to 31 December 2013) 
\vspace{0.2cm} \newline The Danish safety signal raised to the European Medicines Agency (EMA) Pharmacovigilance Risk Assessment Committee (PRAC)  & The death of an actress due to cervical cancer
\vspace{0.2cm} \newline HPV critical articles linking vaccination and suspected adverse events and questioning the reliability of professional recommendations 
\vspace{0.2cm} \newline Critical article series 1 in a national newspaper and critical article series 2 in a national tabloid newspaper \\\rule{0pt}{5ex}
2015 & National Board of Health statement reaffirm the favourable risk/benefit balance of HPV vaccination 
\vspace{0.1cm} \newline EMA, PRAC report: no safety issue to alter recommendation of vaccination & ``The Vaccinated Girls'' TV-documentary describing perceived side effects from vaccination in 47 Danish girls 
\vspace{0.2cm} \newline The chairperson of the Parliament’s Health Committee voices mistrust in HPV-vaccine and EMA report 
\vspace{0.2cm} \newline 
Critical article series in MetroXpress newspaper running from 2015 until the beginning of 2016 \\
\bottomrule
\end{tabularx}
\caption{Public initiatives and significant media events regarding the HPV-vaccine in Denmark 2006-2013.}
\label{table:hpv_media_table_1}
\end{table}

The public initiatives and significant media events regarding the HPV-vaccine are presented in chronological order in table \ref{table:hpv_media_table_1}, while the vaccination activity is presented in figure \ref{fig:HPV_media_2a} and the article percentage with corresponding time points are indicated in figure \ref{fig:HPV_media_2b}.

Table \ref{table:hpv_media_table_1} describes the public initiatives and selected significant media events regarding the HPV-vaccine in Denmark 2006-2013. The peaks in media coverage (figure \ref{fig:HPV_media_2b}) correspond to increased attention after a public campaign for a catch- up programme announced in 2008 and 2010 for women born between 1993 and 1995. A steep rise in media coverage was seen after the death of a young actor from cervical cancer in December 2012. In 2013 two series of critical articles in nationwide newspapers questioned the reliability of professional recommendations and later suspected adverse events. Several patient organisations were formed. Among others, ``HPV-Update'' under the Danish Association of the Physical Disabled, comprising parents believing that their daughters suffered from adverse events after HPV-vaccination \cite{hpv_update}. A TV documentary from 2015, ``De vaccinerede piger'' (The Vaccinated Girls) displayed a group of girls reporting a diverse array of adverse reactions following HPV-vaccination \cite{devaccineredepiger}. This documentary was widely discussed in the media and on social media platforms. In the same period, the chairperson of the Health Committee in the Danish Parliament voiced great concern for the safety of the HPV-vaccine \cite{corfixen}. The issues of vaccination safety kept resurfacing in the media, an important driver was the freely available newspaper MetroXpress that had a daily feature page along with a webpage on HPV \cite{metroxepress_HPV}. The Danish safety signal was raised to the European Medicines Agency (EMA) Pharmacovigilance Risk Assessment Committee (PRAC) in September 2013. In July 2015, the European Commission requested PRAC to assess whether there was evidence of a causal association between HPV-vaccines and POTS and/or complex regional pain syndrome (CRPS). The suspicion of such an association had also been raised from Japan \cite{kinoshita2014peripheral}. PRAC concluded that no current evidence supported that HPV-vaccines causes CRPS or POTS \cite{ema_report_HPV}. In December 2015, the WHO Global Advisory Committee on Vaccine Safety reported that it had not found any safety issue that would alter its recommendation \cite{gbcvs}. The same year, the Danish National Board of Health reaffirmed the favourable risk/benefit balance of HPV-vaccination in a public statement \cite{danishHealth_hpv}. 

\section{Discussion}

\paragraph{Introduction of the HPV-vaccine in Denmark}
Our analysis of the correlation shows a change in the relation between the media coverage and vaccination uptake in June 2013. Subsequent regression analysis of the time-period prior to the time point shows no relation between the variables, while the period afterwards shows a significant negative relation. 

The introduction of the HPV-vaccine in Denmark in 2009 was the latest expansion of the national childhood vaccination schedule. The introduction was a response to the fact that despite a long history of organized cervical screening, and an observed decrease in the incidence of cervical cancer by 50-75\% over the last years \cite{iarc2005cervix}, we still experience 400 cases of cervical cancer and 100 deaths annually in a population of 5.7 million \cite{engholm2014nordcan}. After a successful launch of the vaccination programme, the vaccination uptake remained high for the next 6 years, after which we saw a decline. In this paper we describe the events leading up to the decline and focus on the potential correlation with media. 

Following the change in media content and the public debate about perceived adverse events, we found that there was a negative correlation between media activity and vaccination activity. Our study was ecological in its design, limiting the possibility of causal inference. However, the findings indicate that as media attention increased, the public confidence in the safety of the vaccine decreased, and accordingly, vaccination uptake was reduced. The situation was self-perpetuating with a continuous stream of news-stories describing the decline in vaccination uptake and suggesting fear of adverse events as leading explanation. 

The effect of media coverage on vaccination uptake has been studied in other contexts. Influenza vaccination rates has been shown to increase after significant media attention to a severe influenza season in the US \cite{ma2006influenza}. Similarly, articles suggesting a link between the MMR vaccine and autism has been linked to a drop in MMR vaccination in the US \cite{smith2008media} and the UK \cite{mason2000impact}. Similar to our hypothesis about the connection between increased adverse event reporting and media coverage, Eberth et al.\ reported that media coverage and internet search activity, in particular, may stimulate increased adverse event reporting \cite{Eberth2014}. The effect of singlehanded cases reported in the news has not only been seen in Denmark. In Ireland concerns are raised as fewer girls receive HPV-vaccine \cite{hpsc2016, cullen2016}. The Japanese government suspended their active recommendation of the HPV-vaccine in 2013 in response to unfounded fears about the safety profile. The decision is still unreversed despite declaration from the country's Vaccine Adverse Reactions Review Committee \cite{hanley2015hpv}. Apart from the withdrawal of the HPV vaccine in Japan, it is our impression that no other countries have experienced a HPV crisis as strong as in Denmark. This speaks against an influence from non-Danish media sources on the vaccination choices though this remains unknown.

Our data provide indirect insight into the decision-making process surrounding childhood vaccination. The decision to have one's child vaccinated is influenced by multiple factors, one of those being publications in the media \cite{sturm2005parental}. Media is part of shaping parents impressions of the safety of a vaccine, which is the largest concern parents have today regarding vaccines \cite{gamble2009factors}. A greater belief in the protection offered by childhood vaccines has been found to correlate with acceptance of HPV-vaccines \cite{de2008correlates}. Parent socio-economical \cite{brabin2006future} and social-environmental factors including cultural beliefs as well as social group norms may also play a role \cite{sturm2005parental}. Social group norms contain the social phenomena of ``Bandwagoning''; to vaccinate your children because this is norm \cite{hershey1994roles}. This phenomenon could explain the rapid decline in vaccination seen in our study. Both individual factors, as well as social group norms, can be influenced by massive media coverage and impact parental vaccination choices through different pathways. Furthermore, the concept of consumerized medicine entails that the general practitioner serves as an informant of possibilities, but it is the patients themselves that chooses how to act. This places the responsibility of any adverse events with the parents. In this situation, it is easy to understand that an omission act (waiting to vaccinate) which can be reversed, is preferred over a commission act (vaccinating) which can never be undone. This is also known as the omission bias \cite{ritov1999protected}.

In addition to the previously mentioned factors, physicians play a pivotal role in the parental vaccination decision pathway. Daley et al.\ has reported that physicians attitudes and intentions of recommending HPV-vaccination promote successful immunization delivery \cite{daley2006national}. HPV-vaccination being the latest addition to the Danish vaccination programme places it in a vulnerable position since physicians perceive the most recently introduced vaccines in the programme to be less important compared with the former \cite{flanagan2005dismissing}.

The majority of health professionals are in favour of vaccination. Available evidence continues to affirm a benefit/risk profile in favour of the HPV-vaccine and an association between the suspected adverse events have not been confirmed. This was not always reflected proportionally in the media coverage. In other words, the story told from Public Health authorities did not resonate with the population in the same sense as the stories in the press. At least not to such a degree that the vaccination uptake returned to the previous impressive levels.

Monitoring media activity and describing the effect of media coverage on vaccination may help to develop new approaches to reach and maintain the optimal vaccination uptake. This is also very important to keep in mind when planning the introduction of possible new vaccines into a childhood vaccination schedule. Listening to the public should be a fundamental element of any introduction of a new vaccine. Larsson et al.\ have constructed a communication model that envisions communication as integrating safety assessment and trust-building strategies \cite{larson2012globalization}. 

\section{Strength and limitations}

Using the DDV, we are able to closely monitor the development in vaccination uptake on population level using person identifiable data on a real time scale \cite{grove2011danish}. The reliable and timely updated vaccination activity provides a new opportunity for monitoring vaccination activity. The validity of DDV has previously been studied indicating a 3 percentage points underestimation \cite{wojcik2013validation}. The reporting to the DDV has since been made mandatory and automatized which should improve the completeness of the register \cite{epinytKrause}. 

While the media archive used for the media coverage analysis covers all the major Danish newspapers, it is still dynamically changing which might introduce a bias. The most vocal media in the HPV debate have been present in the archive throughout the study period and bias through changes in the database content should therefore be limited. The database does not include data on TV or radio broadcasts, which has influenced the parents and physicians directly. It is evident from the case of the documentary ``the vaccinated girls'' that a lot of spin-off stories were published in the written media. A large percentage of Danish parents may have sought their information online or in social networks since 92\% of Danes have access to the internet at home \cite{ds_itanvendelse}. Not having data on social media is a major limitation in this study and is an area for further research. Mollema et al.\ identified a strong correlation between the number of messages on social media and online news articles, indicating that public opinion is reflected on these platforms and that the written media determine themes discussed on social media \cite{mollema2015disease}. 

We analysed the content of a sample of the media stories and found the results to confirm our perception of increasingly negative media sentiment. Previous studies have shown that just the mention of a controversy in the media may impact public awareness regardless of specific content \cite{wilson2004reporting}. This paper is intended as a qualitative description of the events leading up to the decline in the HPV-vaccination uptake. We plan a qualitative study analysing the content of the news items using machine learning and natural language processing.

\section{Conclusion}

The HPV-vaccination uptake was very high in Denmark for the first years after the introduction in the childhood vaccination schedule. Following public concern and media attention, the uptake dropped to a dissatisfactory level of 54\% for the girls born in 2003. The drop in vaccination uptake of HPV correlated with written media attention from June 2013.

This study offer important information for the public health community as it continues to work for higher acceptance of present and emerging vaccines. Our findings suggest that media can have a potential negative impact on vaccination uptake. If providers and parents do become more cautious during periods of controversy, it will be very important to ensure that they have access to credible information. Vaccination decision-making is complex and providing recommendations might not be enough. Keeping the physician updated with information and strategies for discussing vaccination choices with parents may be the best investment to ensure the health of the child in a situation with increasing amounts of misleading information. This also includes public sentiment as reflected in media coverage. Monitoring media coverage is a way to gain insight into the population's concerns and inform future strategies. Media coverage has to be taking into account when planning a future public health intervention such as the introduction of a new vaccine. A strategy for handling a sudden change in public opinion reflected in written media could prove vital for reaching and maintaining the optimal vaccination uptake. WHO has developed an e-learning module for crisis communication including a case study on how a potential HPV vaccine crisis was averted in the UK \cite{who_hpv}. Allegations regarding vaccine-related adverse events needs to be dealt with rapidly and effectively to not undermine confidence in the vaccine. From a public trust standpoint, it is better to proactively take control of the story by communicating rapidly, accurately and provide transparency \cite{who_hpv}.

\vspace{0.5cm}
\noindent\textbf{\textit{List of abbreviations:}}

\begin{tabular}{ll}
CRPS & Chronic Regional Pain Syndrome \\

DDV & The Danish Vaccination Register \\

EMA &  European Medicines Agency  \\

HPV & Human Papillomavirus  \\

MMR & Measles, Mumps and Rubella \\

POTS & Postural Orthostatic Hypotensive Syndrome \\

PRAC & Pharmacovigilance Risk Assessment Committee \\

WHO &  World Health Organization \\

\end{tabular}

\vspace{0.5cm}

\noindent\textbf{\textit{Declarations:}}

\indent\textbf{Ethics and consent to participate:} The study is purely register-based and was notified to the Danish Data Protection Agency. Ethics approval and participant consent was not necessary according to Danish legislation.

\textbf{Consent for publication: }Not applicable. 

\textbf{Availability of data and materials:} The datasets generated and analysed during the current study are not publicly available due to Danish law as they contain information that could compromise research participant privacy but are available in aggregated form from the corresponding author on reasonable request.

\textbf{Competing interest:} The authors declare that they have no conflict of interest.

\textbf{Funding:} Camilla Hiul Suppli was founded through unrestricted grants by the following private foundations: Axel Muusfeldts fond, Christian Larsen og Dommer Ellen Larsens legat, Else og Mogens Wedell-Wedellsborgs fond, Familien Hede-Nielsens fond, AP Møller Fonden - Fonden til Lægevidenskabens fremme, Helsefonden, Illum Fondet, Ole Kirks fond and Rosalie Petersens fond.

Niels Dalum Hansen was funded by IBM Denmark and grant number 4135-00047B from the Innovation Fund Denmark. 

\textbf{Authors' contribution:} the idea and design of the study was executed by CHS, NDH and KM. The analysis was performed by NDH. The first draft of the manuscript was produced by CHS. All authors have contributed to the interpretation of the results and the wording of the manuscript. All authors of the manuscript have read and agreed to the content and are accountable for all aspects of the accuracy and integrity of the manuscript in accordance with ICMJE criteria. 

\textbf{Acknowledgements:} Not applicable.

\textbf{Authors' information:} CHS, KRM, TGV and PVB is employed at the Statens serum Institut responsible for the Danish national surveillance of infectious disease epidemiology and vaccination coverage of the Danish childhood vaccination programme.

\backmatter

\bibliographystyle{plain}
\bibliography{bibliography}

\end{document}